\documentclass[journal]{IEEEtran}
\usepackage{amsmath,graphicx, paralist,setspace}
\usepackage{amsthm}
\usepackage{xcolor}
\usepackage{amsmath, upgreek, amsfonts}
\usepackage{caption}
\usepackage[hidelinks]{hyperref}
\usepackage{multirow}
\usepackage{floatrow}
\usepackage[labelformat=simple]{subcaption}

\DeclareCaptionLabelSeparator{periodspace}{.\quad}
\usepackage{tabularx,booktabs}
\usepackage{float}
\usepackage{authblk}
\usepackage{tikz}
\usetikzlibrary{spy}
\usepackage{placeins}
\usepackage{stfloats}

\usepackage{booktabs}
\usepackage{multirow}
\usepackage{makecell}
\usepackage{xcolor}
\usepackage{colortbl}

\usepackage{threeparttable}

\def\idx#1{[#1]}

\newcommand{\ntau}{n_\tau}  

\newcommand{\X}{X}  
\newcommand{\C}{C}  
\newcommand{\N}{N}  

\newcommand{\dec}{\mathtt{Dec}}  
\newcommand{\CEnc}{\mathtt{CEnc}}  
\newcommand{\CEncmean}{\mathtt{CEnc_{mean}}} 
\newcommand{\NEnc}{\mathtt{NEnc}}  

\title{Coherent Spectral Feature Extraction Using Symmetric Autoencoders}
\author{Archisman Bhattacharjee, Pawan Bharadwaj
\thanks{
Archisman Bhattacharjee and Pawan Bharadwaj are with Centre for Earth Sciences, Indian Institute of Science, Bangalore, India.\\}}

\begin{document}
\maketitle
\begin{abstract}
Hyperspectral data acquired through remote sensing are invaluable for environmental and resource studies. While rich in spectral information, various complexities such as environmental conditions, material properties, and sensor characteristics can cause significant variability even among pixels belonging to the same material class. This variability poses nuisance for accurate land-cover classification and analysis. Focusing on the spectral domain, we utilize an autoencoder architecture called the symmetric autoencoder (SymAE), which leverages permutation invariant representation and stochastic regularization in tandem to disentangle class-invariant `coherent' features from variability-causing `nuisance' features on a pixel-by-pixel basis. This disentanglement is achieved through a purely data-driven process, without the need for hand-crafted modeling, noise distribution priors, or reference `clean signals'. Additionally, SymAE can generate virtual spectra through manipulations in latent space. Using AVIRIS instrument data, we demonstrate these virtual spectra, offering insights on the disentanglement. Extensive experiments across six benchmark hyperspectral datasets show that coherent features extracted by SymAE can be used to achieve state-of-the-art pixel-based classification. Furthermore, we leverage these coherent features to enhance the performance of some leading spectral-spatial HSI classification methods. Our approach especially shows improvement in scenarios where training and test sets are disjoint, a common challenge in real-world applications where existing methods often struggle to maintain relatively high performance. 
\end{abstract}
\captionsetup[subfigure]{skip=1pt}
\captionsetup[figure]{font=footnotesize,}
\captionsetup[table]{font=footnotesize}
\section{Introduction\label{sec:intro}}

Hyperspectral imaging has emerged as a powerful tool in remote sensing, offering detailed spectral information across hundreds of narrow contiguous bands. This rich spectral data enables precise material identification and characterization, crucial for applications ranging from land cover classification to environmental monitoring. In precision agriculture, for instance, it facilitates vegetation analysis for crop health assessment~\cite{cetin2005precision,LIANG2015123}, while in geological surveys, it enables mineral identification, enhancing the efficiency of exploration activities~\cite{bedini2017use,debba2005optimal}.

\begin{figure}[!hb]
\captionsetup{font=footnotesize}
    \vspace{0pt}
    \centering
        \begin{subfigure}{0.45\linewidth}
            \includegraphics[width=\linewidth]{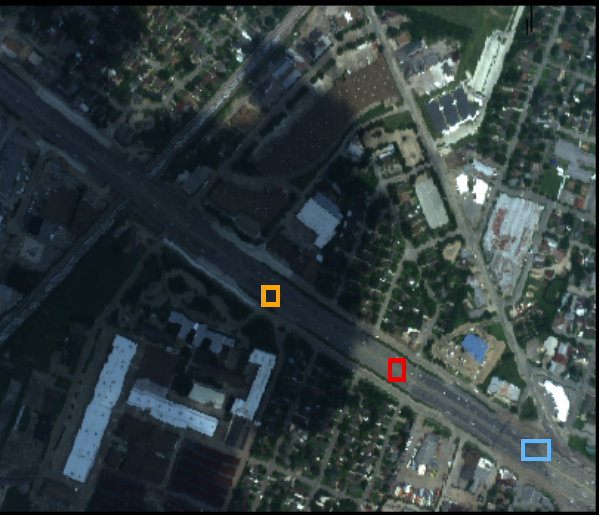}
            \caption{}

        \end{subfigure}
        \centering
        \begin{subfigure}{0.5\linewidth}
            \includegraphics[width=\linewidth]{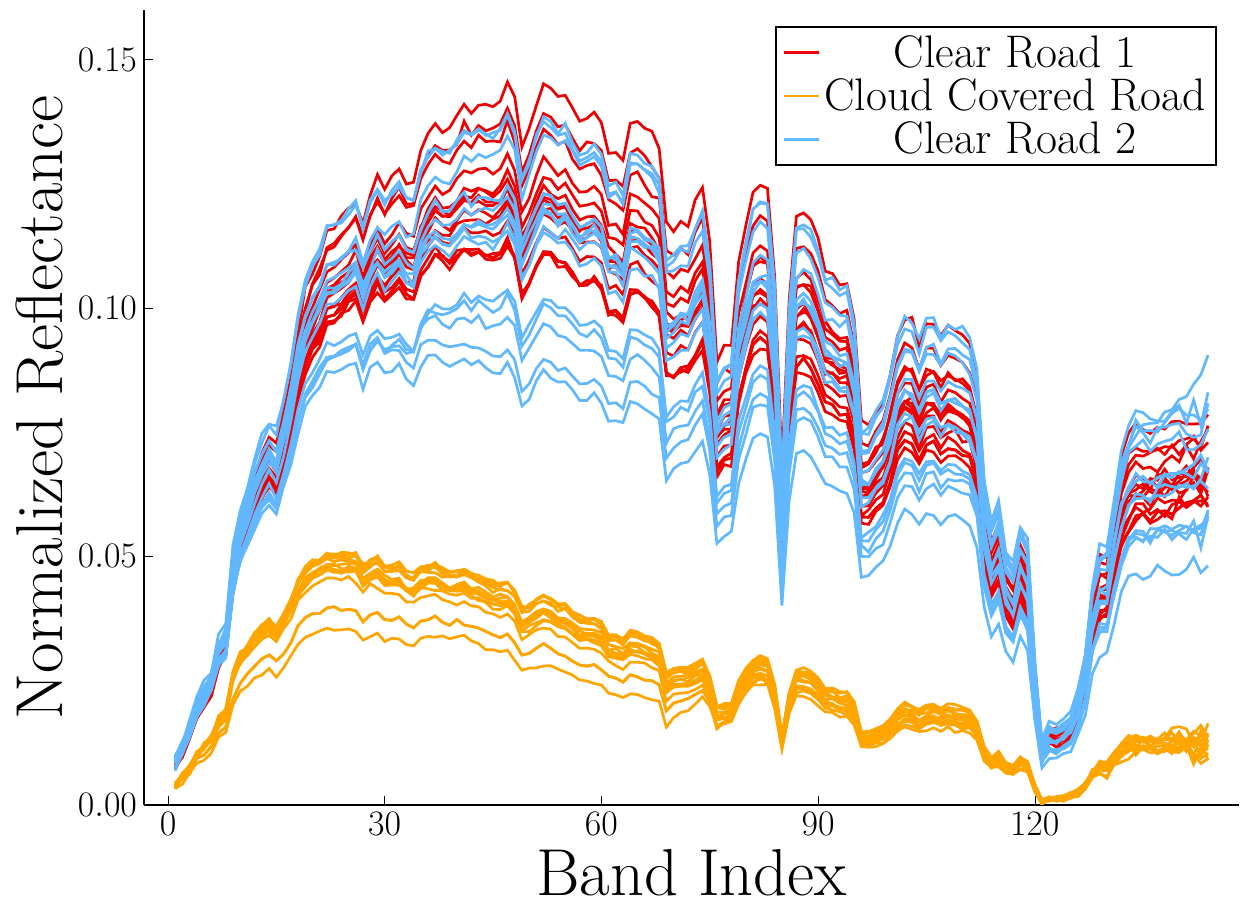}
        \caption{}
            
        \end{subfigure}

    \caption{ Demonstration of spectral variability within the road class. (a) False-color image of an urban area with samples from three road segments highlighted: clear road 1 (red), cloud-covered road (orange), and clear road 2 (sky-blue). (b) Normalized reflectance spectra of the highlighted road segments across different spectral bands, illustrating the intra-class variability due to different conditions. This variability complicates precise identification of surface features.
\label{fig:hfcloud}}
\end{figure}

Despite its capabilities, hyperspectral imaging faces challenges due to the complexity of its data. The high dimensionality of hyperspectral data, combined with various sources of spectral variability, introduces uncertainty in land cover classification and other inferential tasks. This variability stems from intrinsic material properties (e.g., intra-class variations in morphology or surface characteristics, even among nominally identical materials) and extrinsic factors (e.g., spatial heterogeneity within pixels, spectral mixing)~\cite{theiler2019spectral}. Environmental influences, such as atmospheric conditions and illumination variations, also contribute to these challenges (see Figure~\ref{fig:hfcloud}). Variations in sensor-target-illumination geometry can alter the observed spectra. The variability of noise across different scenes, whether due to instrument characteristics or environmental conditions, adds to the complexity~\cite{bioucas2013hyperspectral,shaw2003spectral}. These sources of variability introduce inconsistencies in the observed spectra, potentially leading to misclassification or erroneous analysis of land cover types. Furthermore, these factors can interact in non-linear ways~\cite{han2008investigation}, complicating the development of robust approaches to address them.  These interrelated aspects contribute to the multifaceted nature of hyperspectral data, necessitating sophisticated methods for analysis.

To address these challenges and improve land cover identification performance, we propose using an autoencoder architecture \cite{Kramer1991,Goodfellow2016}, called the Symmetric Autoencoder (SymAE) \cite{bharadwaj2022redatuming}. Our method focuses on extracting class-invariant spectral features, which we call \emph{coherent features}, disentangled from features representing variability within classes, which we term \emph{nuisance features}, in its latent space.
Our approach is motivated by the following premise:
\begin{itemize}
    \item For a given spectral class, there exists a subset of spectral characteristics that remain coherent despite various sources of spectral variability, including intrinsic, extrinsic and environmental factors. Isolating these coherent features could enhance spectral classification, as they are potentially more robust to spectral variability.\label{point:premise}
\end{itemize}
Traditional autoencoding ideas alone cannot perform this disentangled representation learning. To achieve disentanglement between coherent and nuisance features, we implement the permutation invariance symmetry in our encoder network (detailed in Section~\ref{sec:symae}). This symmetry is applied to groups of spectra that share common characteristics, such as those belonging to a specific spectral class, for instance, a particular mineral type, vegetation species, or urban surface material.
SymAE enables the generation of virtual spectra through manipulations in latent space. For example, we can extract coherent features from one pixel and nuisance features from another and then decode them to generate a virtual spectrum. 
In Section \ref{sec:KSCexp}, we show that these virtual spectra can be used as a sanity check for our coherent feature disentanglement process. 
Once our encoder is trained to extract coherent features, we use it for classification tasks. We conducted purely spectral classification experiments and also explored integrating these features into leading spectral-spatial HSI classification methods. Section \ref{sec:classification} details these experiments and evaluates the impact of coherent features on classification accuracy in different scenarios.

\subsection*{A brief history of feature extraction approaches}

Addressing the challenges due to spectral variability and improving land cover identification performance has been a focus of research for decades, leading to the development of various feature extraction techniques~\cite{9082155}. Early approaches relied on shallow feature extraction methods such as Principal Component Analysis (PCA), Independent Component Analysis (ICA), and Linear Discriminant Analysis (LDA) to reduce data dimensionality, extract informative features, and enhance class separability~\cite{rodarmel2002principal,wang2006independent,bandos2009classification}. These techniques were often coupled with statistical learning classifiers, such as Support Vector Machines (SVM), for land cover identification~\cite{melgani2004classification}. As the field progressed, more sophisticated methods emerged, including manifold learning and kernel-based approaches, which aimed to better capture the non-linear nature of hyperspectral data~\cite{lunga2013manifold,fauvel2009kernel}. The incorporation of spatial information through techniques like Extended Morphological Profiles (EMPs) and Composite Kernel Learning marked a significant advancement, recognizing the importance of spatial context in land cover classification~\cite{benediktsson2005classification,camps2006composite}.

Since the early 2010s, deep learning methods have gained traction in hyperspectral image analysis, marking a transition from shallow, handcrafted feature extraction to deep, data-driven approaches~\cite{li2019deep}. This shift has enabled the development of models capable of capturing complex patterns present in hyperspectral data. Unlike previous shallow feature extractors, these deep learning methods often extract features and perform classification in an end-to-end manner. Various architectures have been explored, including Convolutional Neural Networks (CNNs), Stacked Autoencoders (SAEs), Recurrent Neural Networks (RNNs), and Graph Convolutional Networks (GCNs), each addressing different aspects of hyperspectral feature learning~\cite{hu2015deep,chen2014deep,mou2017deep,hang2019cascaded,hong2020graph}. Transformer-based networks have more recently demonstrated high performance, leveraging their sequential understanding and attention mechanisms to model global spectral-temporal dependencies~\cite{dosovitskiy2020image,hong2021spectralformer,vaswani2017attention}.

The state-of-the-art in HSI classification has shifted toward spectral-spatial patch-based deep learning methods, which leverage both spectral signatures and spatial contexts for improved accuracy. These approaches process local neighborhoods to automatically learn robust and discriminative features and include models such as residual networks, attention-based architectures, and hybrid designs integrating transformers with convolutional methods~\cite{zhong2017spectral,zhong2021spectral,9508777,roy2020attention,wang2021fully,9684381,9142417,xue2022local,gao2023main}. Innovative research has explored graph-transformer combinations~\cite{yang2023gtfn,jiang2024graphgst}, optical flow features~\cite{liu2023hyperspectral}, and spatial relationship modeling~\cite{li2022exploring,li2023adaptive} to further enhance classification accuracy. Additionally, very recent developments have introduced state space models (SSMs)~\cite{gu2023mamba} for HSI classification, offering promising alternatives to transformer-based approaches by achieving efficient long-range modeling with linear computational complexity~\cite{li2024mambahsi,fu2024ssumamba,yang2024graphmamba,yao2024spectralmamba}.

\subsection*{Current Developments and Contributions}

The quest for more robust and discriminative features remains an active area of research in hyperspectral imaging. This continued exploration builds on the premise that more robust features can lead to better generalization in subsequent analysis. In hyperspectral imaging, spectral features provide the primary discriminative power for material identification and classification, with spatial context serving as valuable auxiliary information. However, the complex variability inherent in remote sensing data makes explicit modeling of each variation source challenging due to their numerous interactions.

Motivated by these characteristics, we employ SymAE to disentangle coherent features from unmodeled nuisance variability in a data-driven manner. Instead of modeling individual sources of variation explicitly, our approach learns to identify the persistent spectral characteristics that remain stable across variations. Our exploration is guided by our premise that certain inherent spectral features remain coherent within material classes, even as various factors introduce variability in the measured spectra.

Our experiments show that these coherent features can be leveraged to achieve state-of-the-art purely spectral classification accuracy, demonstrating improved resilience to spectral variability. Leading spectral-spatial methods have demonstrated high classification accuracy in standard scenarios. However, our experiments reveal that their performance degrades considerably when training and test areas are geographically disjoint. This is particularly concerning as such disjoint scenarios are common in real-world remote sensing applications. Notably, our method can complement these spectral-spatial approaches by leveraging coherent features, resulting in performance improvements in these challenging scenarios and suggesting better generalization to unseen regions.

The main contributions of this work can be summarized as:
\begin{itemize}
    \item An approach for extracting coherent (class-invariant) spectral features from hyperspectral data using SymAE, achieving disentanglement without explicit modeling of complex variability sources
    \item State-of-the-art performance in purely spectral classification, demonstrating the discriminative power of coherent features
    \item A demonstration of how coherent features can complement some leading spectral-spatial classification methods, showing particular benefits for scenarios with disjoint training and test regions
    \item Empirical validation of the utility of coherent spectral features across six benchmark hyperspectral datasets
\end{itemize}

The remainder of this paper is organized as follows: Section~\ref{sec:grouping} discusses the spectral grouping process; Section~\ref{sec:symae} details the SymAE architecture; Section~\ref{sec:KSCexp} presents insights into the disentanglement process through experiments; Section~\ref{sec:classification} describes HSI classification experiments; and Section~\ref{sec:discussion} discusses extending our approach to unsupervised grouping, identifies limitations and areas for improvement, and outlines future directions before concluding.

\section{Spectral Grouping and Datasets used\label{sec:grouping}}
In this section, we discuss the spectral grouping process.
The SymAE architecture is designed to disentangle features of data that remain consistent within groups from features that fluctuate within these groups. Consequently, the representations learned by SymAE are fundamentally dependent on how the input data is grouped.
To illustrate this concept in the context of hyperspectral imaging, consider an image of an agricultural area. If we group pixels based on crop types (e.g., corn, soybeans, wheat), SymAE would learn to separate features that are common within each group from those that vary. In this scenario, the `coherent' features learned might correspond to the consistent spectral features of each crop type, while the `nuisance' features could capture variations in appearance caused by factors like soil moisture, plant health, or illumination conditions. Importantly, SymAE performs this separation based solely on the provided grouping, without any prior knowledge of specific crop characteristics or variability factors.

The choice of grouping strategy is therefore crucial, as it directly influences the nature of the features disentangled by SymAE and, consequently, the performance of subsequent analysis tasks. In this study, we explore two types of grouping, leveraging a priori information derived from: 1) ground truth labeling and 2) spatial proximity.

    \begin{enumerate}
        \item Ground Truth Labeling: Groups are formed based on predefined class labels, such as specific land cover types or material classes identified in the hyperspectral image. Each group contains pixels assigned to the same class. SymAE is then trained to separate spectral features that are coherent within each group from those that vary, potentially corresponding to class-specific spectral signatures and instance-specific variations respectively. A subset of ground-truth labeled pixels is used to train SymAE, while the remaining labeled pixels are reserved for validation and further analysis of the learned features.
        \item Spatial Proximity: In cases where ground truth labels are limited, we group spatially proximate pixels (groups of 9 pixels in $3\times3$ patches). Here, SymAE is trained to extract spatially coherent features. This is an initial approach to explore SymAE's potential application to fully unsupervised cases where labeled data isn't available but natural groupings may exist. Section~\ref{subsec:3x3} delves into the preliminary findings and future prospects of unsupervised applications of SymAE.
    \end{enumerate}

Our study utilizes six popular hyperspectral datasets, each with distinct characteristics and challenges:
\begin{itemize}
    \item Kennedy Space Center (KSC): Acquired by AVIRIS~\cite{vane1993airborne} instrument over Florida, USA. The dataset covers 512×614 pixels at 18 m spatial resolution, with 176 spectral bands ranging from 400 to 2500 nm. It includes 13 upland and wetland classes, totaling 5,211 labeled pixels. KSC represents a complex coastal ecosystem with similar vegetation types. KSC serves as our primary dataset for demonstrating virtual spectra generation and analysis.
    \item Indian Pines (IP): Collected using AVIRIS over an agricultural landscape in Indiana, USA. It spans 145×145 pixels at 20 m resolution, with 200 spectral channels in the 400-2500 nm range. The dataset contains 16 vegetation classes and 10,249 labeled samples, notably featuring significant class imbalance.
    \item Pavia University (PU): Captured by ROSIS~\cite{kunkel1988rosis} sensor over Pavia, Italy, representing an urban environment. It covers 610×340 pixels at 1.3 m/pixel resolution, with 103 spectral bands between 430-860 nm. UP includes 9 urban land cover types, comprising 42,776 labeled pixels.
    \item Pavia Center (PC): Captured by the ROSIS sensor over Pavia, Italy, this dataset represents an urban environment with nine distinct land cover types. The image used in our study measures 1096 × 715 pixels, with a spatial resolution of 1.3 meters per pixel and includes 102 spectral bands. From the original 148,152 labeled pixels, we selected a subset of 19,800 to create disjoint and balanced train-test sets for the experiments.
    \item Houston 2013 (UH): Acquired by CASI-1500 sensor over the University of Houston campus and surrounding area. The dataset spans 349×1905 pixels at 2.5 m resolution, with 144 spectral bands ranging from 364 to 1046 nm. It features 15 urban classes with 15,029 labeled pixels, and uniquely offers spatially disjoint training and testing subsets.
    \item Longkou (LK): Acquired using a Headwall Nano-Hyperspec sensor mounted on a UAV platform over LongKou, China~\cite{zhong2020whu}. The dataset covers 550×400 pixels at 0.463 m spatial resolution, with 270 spectral bands spanning 400-1000 nm. It includes 9 land cover classes (including 6 crop species), comprising 204,542 labeled pixels, representing an agricultural landscape with high spatial detail.
\end{itemize}

\section{Symmetric Autoencoder\label{sec:symae}}

SymAE is a data-driven deep learning architecture designed to disentangle coherent features from nuisance variations in grouped datasets where multiple instances share common underlying information but differ due to variable factors.

An early rendition of SymAE was introduced by Bharadwaj et al. in 2020~\cite{bharadwaj2020symae} for passive time-lapse seismic monitoring. This initial version used two separate encoders, each implementing different known physical symmetries in seismic data to disentangle path effects (subsurface properties) from source effects (seismic signatures). In 2022, Bharadwaj et al.~\cite{bharadwaj2022redatuming} generalized SymAE beyond specific physical systems. The key innovation was to separate coherent information from unmodeled nuisance variations using a more flexible architecture with two encoders. They introduced a mechanism where permutation invariance and stochastic regularization work together to achieve feature disentanglement.

In this paper, we leverage the generalized SymAE framework for hyperspectral imaging, where the goal is to separate consistent spectral characteristics of different materials or land cover types from various sources of spectral variability. The domain-agnostic nature of the framework enables its application to hyperspectral data without architectural modifications.

\subsection{Problem Formulation}

In the context of HSI, we formulate each pixel spectrum as a function of two components:

\begin{enumerate}
    \item Coherent features ($C$): Latent features representing spectral characteristics that are consistent within the group the pixel belongs to.
    \item Nuisance features ($N$): Pixel-specific latent representation of features representing variability factors that cause individual pixels to deviate from the group's shared characteristics.
\end{enumerate}

Mathematically, we can express a pixel spectrum $P$ as:
\[
P = f(C, N)
\]
where $f$ is an unknown function combining coherent and nuisance features. Notably, $C$ is shared within a group while $N$ can vary for each pixel.
The goal of SymAE is to learn:
\begin{enumerate}
    \item Encoding functions that collectively approximate the inverse of $f$, separating $P$ into $C$ and $N$.
    \item A decoding function that approximates $f$, reconstructing $P$ from $C$ and $N$.
\end{enumerate}

\subsection{Data Structure}

The basic data structure that SymAE operates on is a collection of pixels belonging to the same group, which we call a datapoint $X$. A datapoint $X$ is formed by stacking $n_\tau$ pixel spectra from the same group:
$$
X = \begin{bmatrix} P_1 \\ P_2 \\ \vdots \\ P_{n_\tau} \end{bmatrix} = \begin{bmatrix} X[1] \\ X[2] \\ \vdots \\ X[n_\tau] \end{bmatrix}
$$
where $P_i$ represents the $i$-th pixel spectrum in the datapoint. Each $P_i$ is a vector of length $n_\text{band}$, equal to the number of spectral bands. In later parts of this section, we express the $i$-th pixel of datapoint $X$ as $X[i]$, which corresponds to $P_i$.
This structure allows SymAE to learn coherent features shared within each group while capturing pixel-specific nuisances. During training, datapoints are repeatedly generated for each group by sampling $n_\tau$ pixels with replacement from that group, ensuring diverse representations of group characteristics across different iterations while maintaining within-datapoint coherence. 

\begin{table}[h]
    \centering
    \scriptsize 
    \caption{Summary of Key Mathematical Notations \label{tab:notations}}
    \vspace{-10pt}
    \setlength{\tabcolsep}{3pt} 
    \renewcommand{\arraystretch}{1.1}
    \begin{tabular}{|c|p{3.8cm}|c|}
    \hline
    \textbf{Notation} & \textbf{Description} & \textbf{Dimension/Quantity} \\
    \hline
    \multicolumn{3}{|c|}{\textit{Data Representation}} \\
    \hline
    $P$ & Individual pixel spectrum & $\mathbb{R}^{n_\text{band}}$ \\
    $X$ & Datapoint: collection of $n_\tau$ pixel spectra from the same group & $\mathbb{R}^{n_\tau \times n_\text{band}}$ \\
    $X[i]$ & $i$-th pixel spectrum in datapoint $X$ & $\mathbb{R}^{n_\text{band}}$ \\
    $n_\tau$ & Number of pixel spectra in a datapoint & Scalar, $n_\tau \in \mathbb{N}$ \\
    $n_\text{band}$ & Number of spectral bands & Scalar, $n_\text{band} \in \mathbb{N}$ \\
    \hline
    \multicolumn{3}{|c|}{\textit{Feature Representation}} \\
    \hline
    $C$ & Coherent features: consistent spectral characteristics within a group & $\mathbb{R}^{d_c}$ \\
    $N$ & Nuisance features: pixel-specific variability & $\mathbb{R}^{d_n}$ \\
    \hline
    \multicolumn{3}{|c|}{\textit{SymAE Architecture}} \\
    \hline
    $\CEnc$ & Coherent Encoder: Designed to extract coherent features from pixels & $\CEnc: \mathbb{R}^{n_\text{band}} \rightarrow \mathbb{R}^{d_c}$ \\
    $\CEncmean$ & Applies $\CEnc$ to datapoints to create permutation-invariant features & $\CEncmean: \mathbb{R}^{n_\tau \times n_\text{band}} \rightarrow \mathbb{R}^{d_c}$ \\
    $\NEnc$ & Nuisance Encoder: Designed to extract pixel-specific nuisance features & $\NEnc: \mathbb{R}^{n_\text{band}} \rightarrow \mathbb{R}^{d_n}$ \\
    $\dec$ & Decoder: Reconstructs pixel spectra from coherent and nuisance features & $\dec: \mathbb{R}^{d_c} \times \mathbb{R}^{d_n} \rightarrow \mathbb{R}^{n_\text{band}}$ \\
    $d_c$ & Dimension of the coherent code & Scalar, $d_c \in \mathbb{N}$ \\
    $d_n$ & Dimension of the nuisance code & Scalar, $d_n \in \mathbb{N}$ \\
    \hline
    \multicolumn{3}{|c|}{\textit{Training and Evaluation}} \\
    \hline
    $\hat{X}$ & Reconstructed datapoint & $\mathbb{R}^{n_\tau \times n_\text{band}}$ \\
    $L$ & Expected reconstruction error & Scalar, $L \in \mathbb{R}_{\geq 0}$ \\
    \hline
    \end{tabular}
\end{table}

\subsection{SymAE Architecture \label{subsec:architecture}}
\begin{figure}[tp]
\captionsetup{font=footnotesize}
\centering
\begin{minipage}[b]{1\linewidth}
  \centering
  \centerline{\includegraphics[width=1\linewidth]{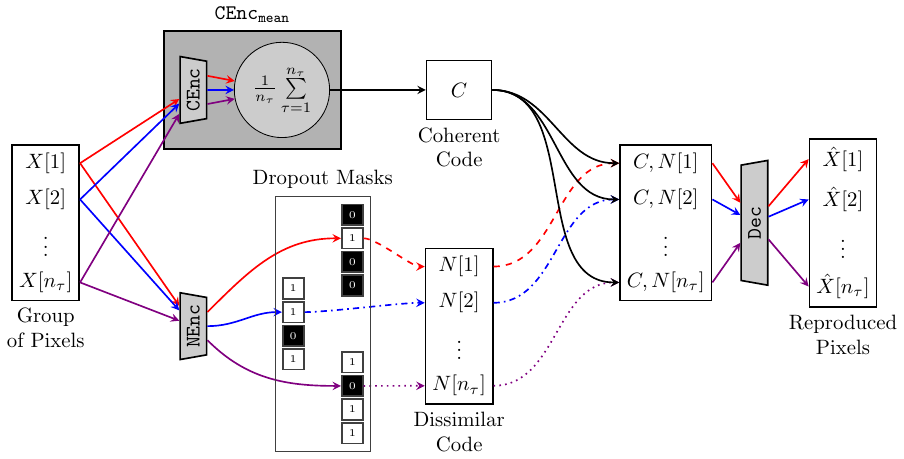}}
\end{minipage}
\caption{
The architecture of symmetric autoencoder (SymAE) disentangles coherent features from features representing variability in its latent space.
The coherent features are assumed to be consistent across the pixels in a group.
They propagate through the network via solid black arrows, extracted by a symmetric function (CEnc$_\text{mean}$) that is invariant to pixel ordering.
Colored arrows indicate the propagation of pixel-specific nuisance effects, processed by the nuisance encoder (NEnc).
Dropout masks are applied to the nuisance features to introduce stochastic regularization.
The decoder (Dec) combines coherent and nuisance features to reconstruct pixel spectra.
}
\label{fig:network}
\end{figure}

SymAE comprises three main functions that can be parameterized by any universal approximator~\cite{hornik1991approximation}. In our implementation, these are parameterized by dense feed-forward networks:
\begin{itemize}
    \item \textbf{Coherent Encoder ($\CEnc$)}: Extracts coherent features. Maps each input spectrum $X[i] \in \mathbb{R}^{n_\text{band}}$ to an intermediate feature space $\mathbb{R}^{d_c}$, where $d_c$ is the dimension of the coherent code.
    \item \textbf{Nuisance Encoder ($\NEnc$)}: Captures pixel-specific nuisance variations. Maps each input spectrum $X[i] \in \mathbb{R}^{n_\text{band}}$ to a nuisance feature space $\mathbb{R}^{d_n}$, where $d_n$ is the dimension of the nuisance code.
    \item \textbf{Decoder ($\dec$)}: Reconstructs the input spectra using both coherent and nuisance features. Maps a combined vector $[C, N[i]] \in \mathbb{R}^{d_c + d_n}$ to reconstruct an input spectrum $\hat{X}[i] \in \mathbb{R}^{n_\text{band}}$.
\end{itemize}
The processing of an input datapoint $X$ through SymAE involves two encoding paths, as illustrated in Figure \ref{fig:network}. The encoded features are combined in the latent space and subsequently decoded through $\dec$ to reconstruct $X$. 

The first path, via $\CEncmean$, extracts coherent features shared across pixels in $X$. To ensure this path encodes only shared features, we impose an invariance constraint. Specifically, for any permutation $\varPi$ of the pixel ordering:
\begin{equation}
\label{eqn:encr_sym}
\C = \CEncmean(\X) = \CEncmean(\X\idx{\varPi(1{:}\ntau)})
\end{equation}
This constraint guarantees that the extracted features $\C$ are independent of pixel ordering, thus focusing on characteristics shared across the entire datapoint $X$. The permutation invariance directly influences feature selection: since $\C$ must be used to reconstruct every pixel in the group, and must remain the same regardless of pixel ordering, encoding pixel-specific variations in $\C$ would be detrimental for reconstructing other pixels that don't share those variations. This constraint directs $\C$ to capture only features that are consistently useful across the group.
The coherent encoding path achieves permutation invariance following the approach of Zaheer et al.~\cite{Zaheer2017}. Such architectures employ pooling operations like $mean$ or $max$ across instances, ensuring permutation invariance while providing universal approximation guarantees for symmetric functions.
In our implementation, the spectrum of each pixel is transformed using $\CEnc$, and the mean is taken along the pixel dimension:
\begin{equation}
\label{eqn:enc_symmetric}
\C = \CEncmean(X) = \left(\frac{1}{n_\tau} \sum_{\tau=1}^{n_\tau} \CEnc(\X\idx{\tau})\right)
\end{equation}
The key aspect of this equation is that the aggregation of pixel-wise transformations $\CEnc(\X\idx{\tau})$ via the mean operation is symmetric with respect to their ordering, thereby ensuring the desired symmetry (Equation \ref{eqn:encr_sym}) is achieved.

The second encoding path in SymAE captures pixel-specific deviations for each pixel of the datapoint $X$. In this path, the datapoint passes through the nuisance encoder $\NEnc$, which, unlike $\CEnc$'s path, is unconstrained. This lack of constraints on $\NEnc$ presents a significant concern: the decoder $\dec$ might tend to ignore the coherent component $\C$ in favor of only using information from $\N$ for reconstruction.

As the purpose of $\NEnc$ is exclusively to encode pixel-specific nuisance information while disregarding coherent features, SymAE incorporates stochastic regularization to mitigate this issue. This regularization is implemented through dropout masks during training, utilizing Bernoulli dropout \cite{srivastava2014dropout} with a probability of $p=0.5$:
\begin{equation}
\N[\tau] = \mathtt{Dropout}(\NEnc(\X[\tau]))
\end{equation}
The dropout mechanism introduces random obfuscation to elements of $\N$, causing the decoder $\dec$ to perceive the codes as dissimilar and hindering the reconstruction of coherent information from $\N$. While traditional dropout primarily prevents co-adaptation of features, here it creates an intentionally unreliable path through $\NEnc$, directing coherent information through the $\CEnc$ path. This process creates a dichotomy in the information flow:
\begin{enumerate}
    \item A continuous stream of information from $\CEnc$
    \item Outputs from $\NEnc$ with randomly obfuscated features
\end{enumerate}
This dichotomy compels $\dec$ to extract maximal meaningful information from $\CEnc$, which inherently contains coherent data. Consequently, the architecture evolves such that:

-$\CEnc$ becomes adept at encoding stable, coherent features shared across pixels in datapoints

-$\dec$ becomes adept at capturing remaining variability from $\NEnc$, as these pixel-specific features cannot be encoded through $\CEnc$ without disrupting reconstruction\newline This design of SymAE allows for the simultaneous learning of coherent and nuisance features, with their separation emerging from the architectural constraints without requiring explicit prior knowledge of their characteristics or distributions.

\subsection{Training Process}

SymAE employs end-to-end training, simultaneously optimizing $\NEnc$, $\CEnc$, and $\dec$. The training process revolves around reconstructing the input datapoints using the extracted coherent and nuisance features.
The decoder combines the coherent code $C$ with each pixel's nuisance code $N[\tau]$ to reconstruct the original pixel spectra:
\[
\hat{X}[\tau] = \dec([\C, \N[\tau]]) \quad \text{for } \tau = 1 \text{ to } n_\tau,
\]
where $\hat{X}[\tau]$ is the reconstructed spectrum for the $\tau$-th pixel in the datapoint.
The model is trained by minimizing the reconstruction error:
\begin{equation}
\label{eqn:loss}
L = \frac{1}{n_X} \sum_{i=1}^{n_X} \|X_i - \hat{X}_i\|^2,
\end{equation}
where \( X_i \) is an input datapoint, \( \hat{X}_i \) is its full reconstruction, and \( n_X \) denotes the total number of datapoints.

It's important to note that while dropout is applied to the nuisance codes during training, at inference time, the full $N[\tau]$ code is used without dropout. Effective training of SymAE requires datapoints with diverse pixel-specific variations. This diversity is crucial for learning a robust representation in the latent space, where coherent and nuisance features are effectively separated.

\subsection{Feature Extraction and Virtual Spectra Generation}

Post-training, SymAE's encoding functions can be applied to individual pixel spectra for feature extraction. For a given pixel spectrum $P$, we can obtain:

\begin{enumerate}
    \item Coherent features: $C = \CEnc(P)$
    \item Nuisance features: $N = \NEnc(P)$
\end{enumerate}
Note that during feature extraction, the full $\NEnc$ output is used without dropout.
These extracted features can be used to generate virtual spectra through manipulations in the latent space. A basic formulation for this process is:
\begin{equation}
\label{eqn:virtual}
P_\text{virtual} = \dec([C_i, N_j])
\end{equation}
where $C_i$ and $N_j$ are coherent and nuisance features that may come from different pixels. This capability allows for the creation of hypothetical spectral signatures by combining features from various source pixels.

\subsection{Computational Complexity}

In this implementation of SymAE with dense feed-forward networks, let $n$ denote the input dimension to any layer, $d$ the hidden feature size, and $n_\tau$ the number of pixels in a datapoint.
Each dense layer has computational complexity $O(n \times d)$, with initial layers processing spectral inputs of dimension $n_\text{band}$. When operating on $n_\tau$ samples in parallel, the per-layer complexity becomes $O(n \times n_\tau \times d)$. The coherent encoding path additionally requires a mean operation with complexity $O(n_\tau)$ to achieve permutation invariance.
In terms of how the complexity scales with input dimension $n$ (assuming fixed hidden dimension $d$), the baseline architectures compared later in this paper have the following per-layer complexities:
\begin{itemize}
    \item Dense: $O(n)$ (our implementation in SymAE)
    \item CNN: $O(kn)$ where $k$ is the kernel width
    \item RNN: $O(n)$ for sequential processing
    \item Transformer: $O(n^2)$ due to self-attention operations
    \item miniGCN: $O(n)$ (plus $O(b^2n)$ for batchwise graph operations, where $b$ is the batch size)
    \item Mamba (SSM): $O(n)$ with efficient selective state space operations
\end{itemize}
The dense network implementation ensures computationally efficient operation with linear scaling. However, achieving desired levels of feature disentanglement may require the network to evolve gradually over extended training periods, a potential limitation we expand upon in Subsection~\ref{subsec:scope}.

\section{Disentanglement via SymAE: Analysis on KSC Dataset\label{sec:KSCexp}}
This section examines SymAE's feature extraction capabilities using the Kennedy Space Center (KSC) hyperspectral dataset. Through experiments on virtual spectra generation and latent space properties, we provide insights into the model's disentanglement process. This analysis serves as a sanity check for our approach, visualizing how SymAE separates class-specific information from instance-specific variations in spectral data.

\subsection{Experimental Setup \label{subsec:expsetup}}

The KSC dataset, acquired by NASA's AVIRIS~\cite{vane1993airborne} instrument on March 23, 1996, comprises 13 classes representing various upland and wetland land cover types, with a total of 5,211 labeled pixels. The hyperspectral image had been pre-processed to apparent surface reflectances using the ATREM program~\cite{adler1999atmospheric}, which applies radiative transfer modeling to retrieve surface reflectance by accounting for atmospheric absorption and scattering. From the original 224 spectral bands, the data underwent band reduction to remove water absorption and low signal-to-noise ratio bands, resulting in 176 effective spectral bands. Following these preprocessing steps, we normalized the reflectance values to the range $[0,1]$ for our analysis. A notable characteristic of this dataset is the presence of spectrally similar classes, particularly among vegetation types, which poses challenges for class discrimination using conventional spectral analysis methods. This spectral proximity makes the KSC dataset a relevant testbed for evaluating feature extraction methods.

\begin{table}
\tiny
\caption{Network Configurations of SymAE Components}
\label{tab:architecture}
\centering
\vspace{-10pt}
\begin{tabular}{lcl}
\hline
Component & Output Dimension & Details \\
\hline
\multicolumn{3}{l}{Coherent Encoder (\textbf{$\CEnc$})} \\
Dense + LeakyReLU($\alpha=0.5$) & 300 & Input: 176 (input spectrum) \\
Dense + LeakyReLU($\alpha=0.5$) & 300 & \\
Dense + LeakyReLU($\alpha=0.5$) & 300 & \\
Dense + LeakyReLU($\alpha=0.5$) & 150 & \\
Dense + LeakyReLU($\alpha=0.5$) & 64 & Output: Coherent code \\
\hline
\multicolumn{3}{l}{Nuisance Encoder (\textbf{$\NEnc$})} \\
Dense + LeakyReLU($\alpha=0.5$) & 300 & Input: 176 (input spectrum) \\
Dropout & 300 & $p=0.25$ \\
Dense + LeakyReLU($\alpha=0.5$) & 300 & \\
Dense + LeakyReLU($\alpha=0.5$) & 300 & \\
Dropout & 300 & $p=0.25$ \\
Dense + LeakyReLU($\alpha=0.5$) & 150 & \\
Dense + LeakyReLU($\alpha=0.5$) & 64 & Output: Nuisance code \\
\hline
\multicolumn{3}{l}{Decoder (\textbf{$\dec$})} \\
Dense + LeakyReLU($\alpha=0.5$) & 150 & Input: 128 (concatenated codes) \\
Dense + LeakyReLU($\alpha=0.5$) & 300 & \\
Dense + LeakyReLU($\alpha=0.5$) & 300 & \\
Dense + LeakyReLU($\alpha=0.5$) & 300 & \\
Dense + LeakyReLU($\alpha=0.5$) & 300 & \\
Dense + LeakyReLU($\alpha=0.5$) & 300 & \\
Dense & 176 & Output: reconstructed spectrum \\
\hline
\end{tabular}
\end{table}

For our experimental evaluation, we randomly split the labeled pixels into training and test sets at an approximate ratio of $1:9$. The SymAE components ($\CEnc$, $\NEnc$, $\dec$) are realized as feed-forward neural networks composed of dense layers with LeakyReLU activations. In our implementation, $\NEnc$ includes internal dropout layers ($p=0.25$) to moderate its learning capacity, separate from the architectural dropout ($p=0.5$) used for disentanglement (see Subsection~\ref{subsec:architecture}). The detailed configurations of these component networks are provided in Table~\ref{tab:architecture}. Both coherent and nuisance codes are set to dimension $d_c = d_n = 64$. The model was trained on datapoints consisting of $n_\tau = 8$ pixel spectra, grouped according to their respective ground truth categories. The training process ran for $3000$ epochs using the ADAM optimizer with a learning rate of $0.0001$, minimizing the mean squared error between input and reconstructed spectra. Each epoch comprised $2048$ batches with a batch size of $256$. The model was implemented using the Flux machine learning package~\cite{Flux.jl-2018} in Julia programming language and trained on a Linux workstation equipped with an AMD Ryzen Threadripper 3960X 24-core processor, 128 GB of RAM, and a 24 GB NVIDIA GeForce RTX 3090 GPU.

\subsection{Virtual Spectra and Redatuming\label{subsec:virtualspectra}}

As formulated in Equation \ref{eqn:virtual}, SymAE enables the generation of virtual spectra by combining coherent and nuisance codes from different pixels. In the context of HSI, we introduce \emph{redatuming}—drawing from seismic imaging terminology~\cite{mulder2005rigorous,bharadwaj2022redatuming}—as a specific application of virtual spectra generation that creates spectra with uniform nuisance conditions. For a set of pixels $\{P_i\}_{i=1}^n$ and a reference pixel $P_\text{ref}$, the redatuming process can be formulated as:

\begin{equation*}
\begin{array}{ll}
\textbf{Given:} & \{P_i\}_{i=1}^n, \; P_\text{ref} \\[5pt]
\textbf{Extract:} & C_i = \CEnc(P_i), \quad i = 1, \ldots, n, \\[5pt]
                  & N_\text{ref} = \NEnc(P_\text{ref}) \\[5pt]
\textbf{Redatum:} & P_{\text{redatumed},i} = \dec([C_i, N_\text{ref}]), \quad i = 1, \ldots, n
\end{array}
\end{equation*}

This process, by uniformizing the nuisance variations across samples to a common reference, is expected to reduce intra-class variability that could confound classifiers. Through this reduction in variability and preservation of class-specific coherent features, we anticipate more robust spectral classification.

\begin{figure}[b]
\captionsetup{font=footnotesize}
\centering
\begin{subfigure}{0.48\linewidth}
\includegraphics[width=\linewidth]{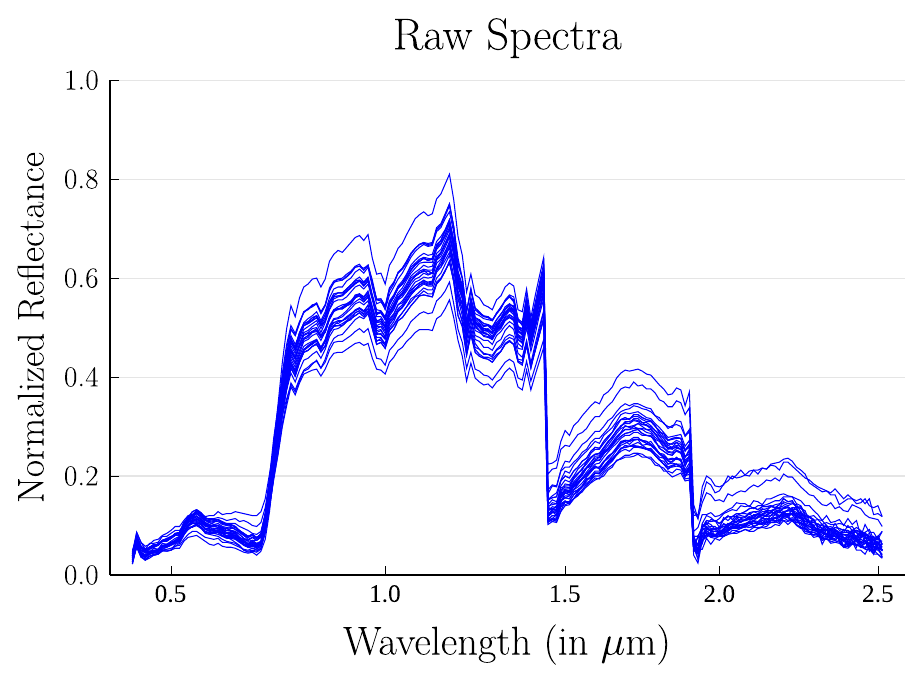}
\caption{}
\end{subfigure}
\begin{subfigure}{0.48\linewidth}
\includegraphics[width=\linewidth]{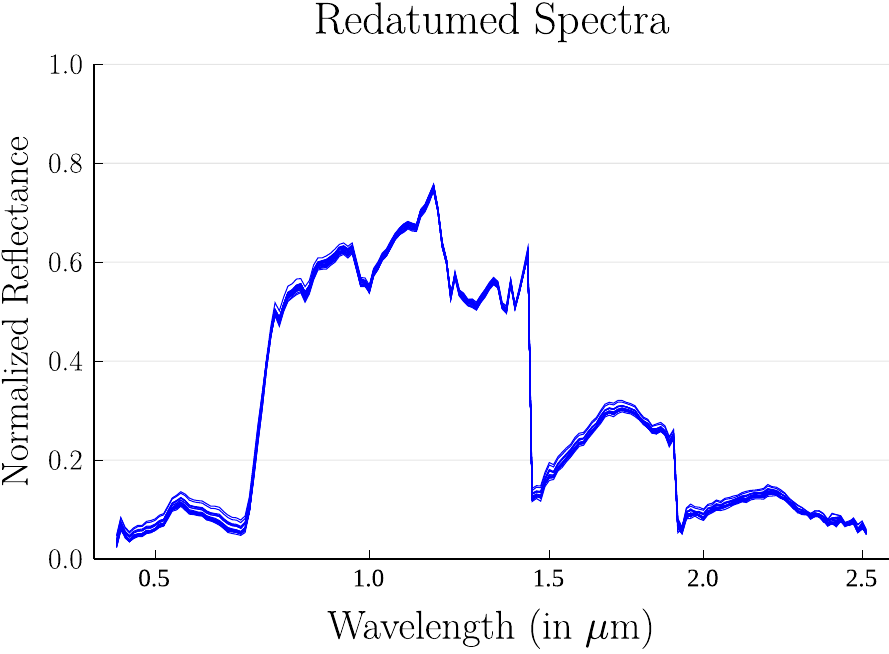}
\caption{}
\end{subfigure}
\caption{Demonstration of nuisance features in Oak Hammock test spectra and the effect of redatuming. (a) Original spectra where nuisance features manifest as natural intra-class variability across wavelengths, possibly arising from factors like canopy structure, vegetation density, moisture conditions, and illumination effects. (b) Redatummed spectra generated by SymAE using a single, consistent nuisance code obtained from a randomly selected Oak Hammock test set pixel as reference. Note how the redatuming process reduces these intra-class variations while preserving the characteristic spectral features of the Oak Hammock class.
\label{fig:redatumingoak}}
\end{figure}

\begin{figure*}
\captionsetup{font=footnotesize}
    \centering
    \begin{subfigure}{0.32\linewidth}
        \begin{tikzpicture}[spy using outlines={red,magnification=2.5,size=1.1cm, connect spies}]
\node {\includegraphics[width=\linewidth]{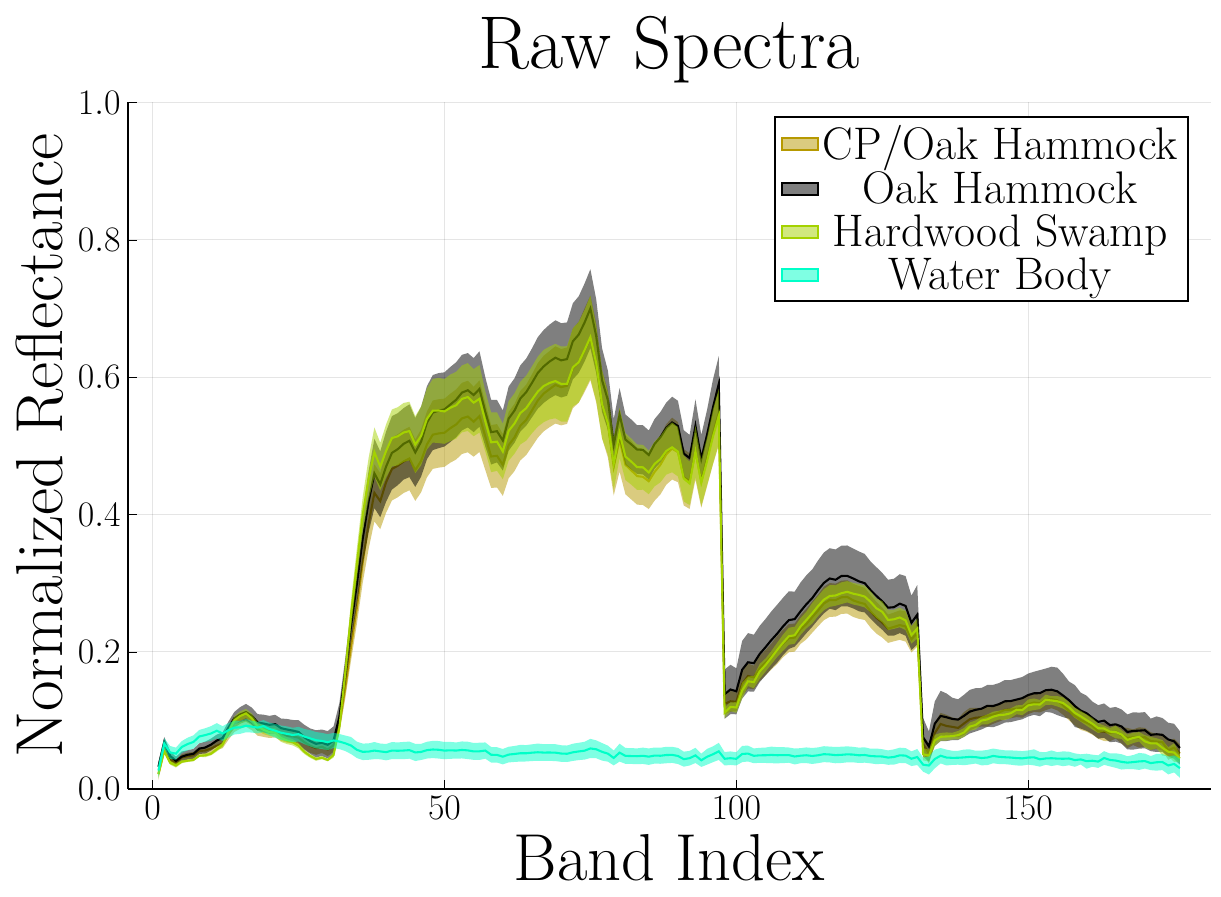}};
\spy on (-0.25,0.35) in node [left] at (-0.9,1);
\end{tikzpicture}
        \caption{}
    \end{subfigure}%
    \begin{subfigure}{0.32\linewidth}
        \begin{tikzpicture}[spy using outlines={red,magnification=2.5,size=1.1cm, connect spies}]
\node {\includegraphics[width=\linewidth]{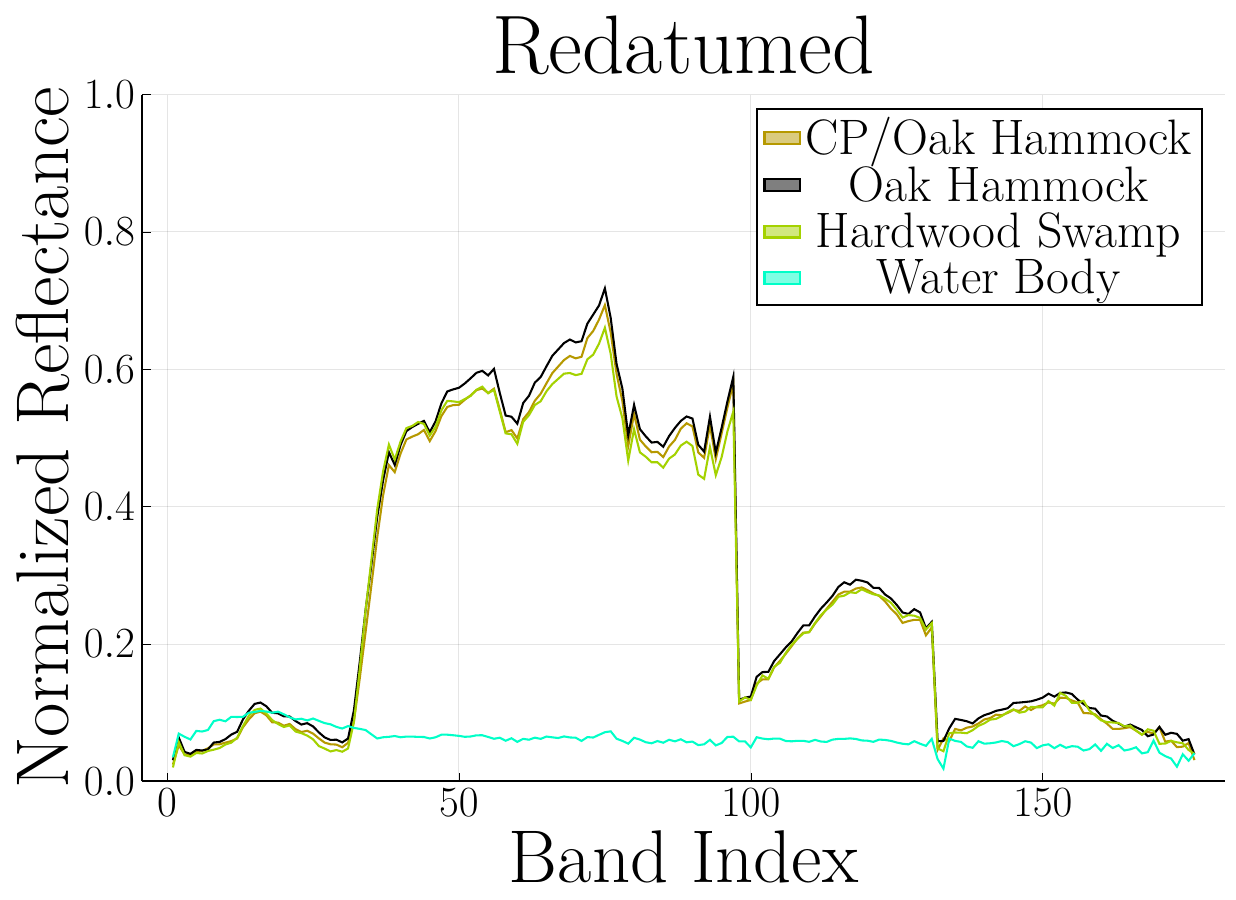}};
\spy on (-0.24,0.38) in node [left] at (-0.9,1);
\end{tikzpicture}
        \caption{}
    \end{subfigure}%
    \begin{subfigure}{0.32\linewidth}
        \includegraphics[width=\linewidth]{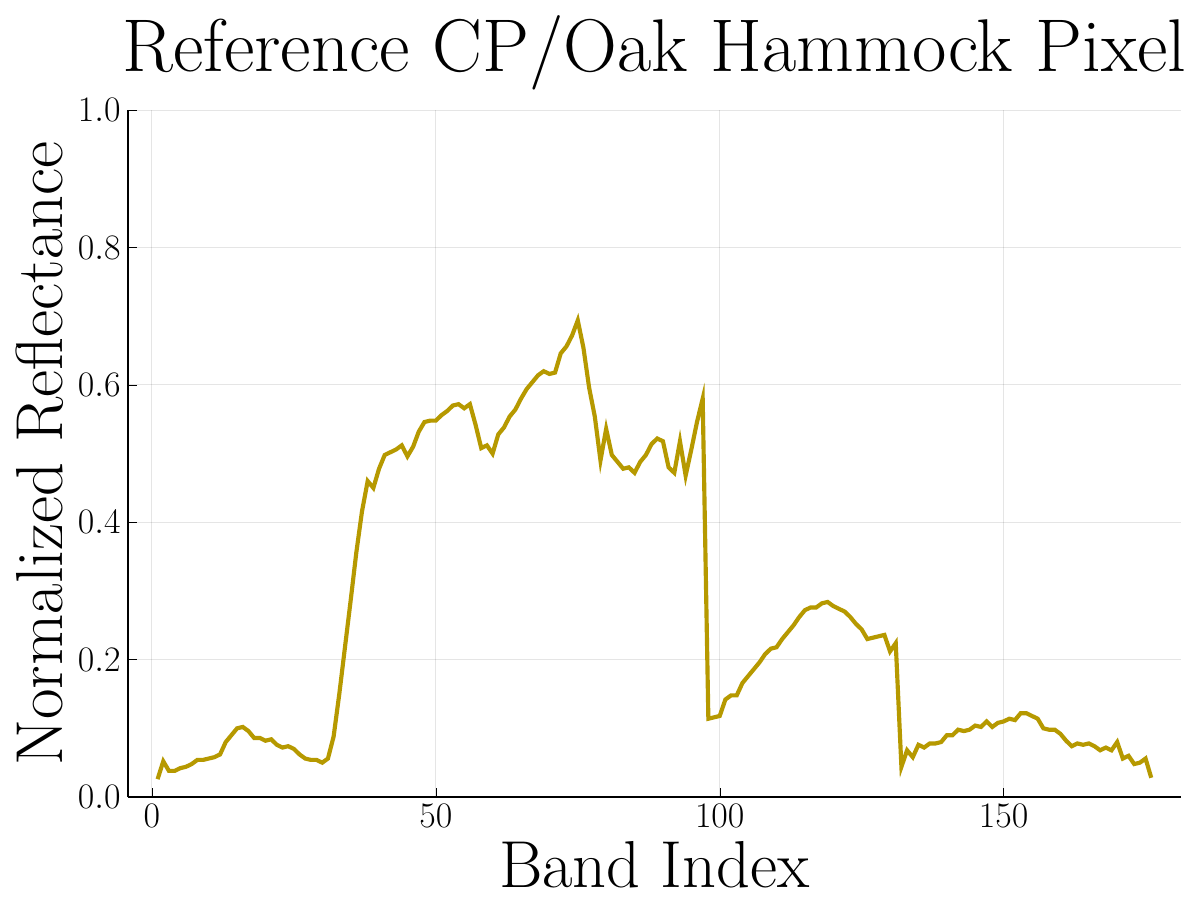}
        \caption{}
    \end{subfigure}
    \begin{subfigure}{0.32\linewidth}
        \begin{tikzpicture}[spy using outlines={black,magnification=2.5,size=1.1cm, connect spies}]
\node {\includegraphics[width=\linewidth]{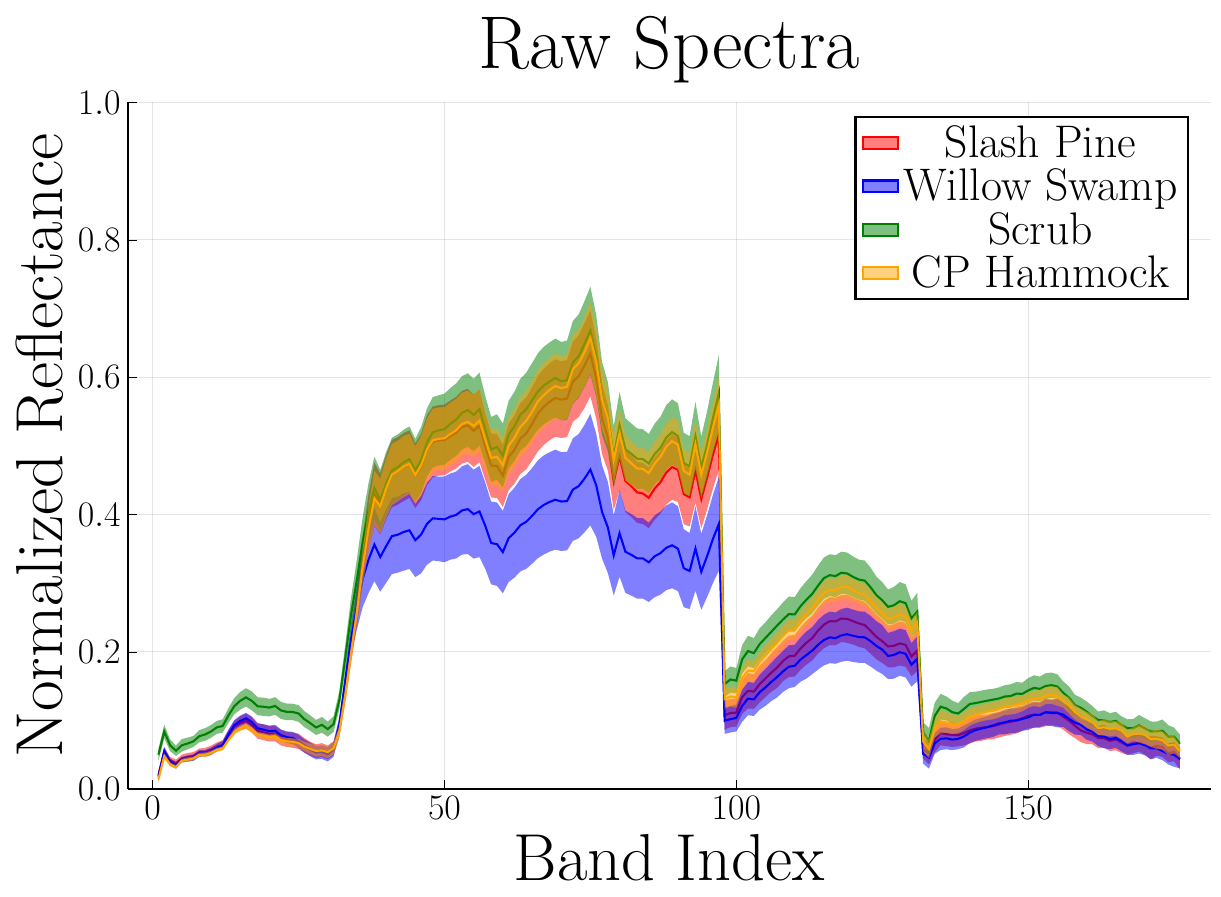}};
\spy on (1.98,-1.24) in node [left] at (2.75,-0.11);
\end{tikzpicture}
        \caption{}
    \end{subfigure}%
    \begin{subfigure}{0.32\linewidth}
        \begin{tikzpicture}[spy using outlines={black,magnification=3,size=1.1cm, connect spies}]
\node {\includegraphics[width=\linewidth]{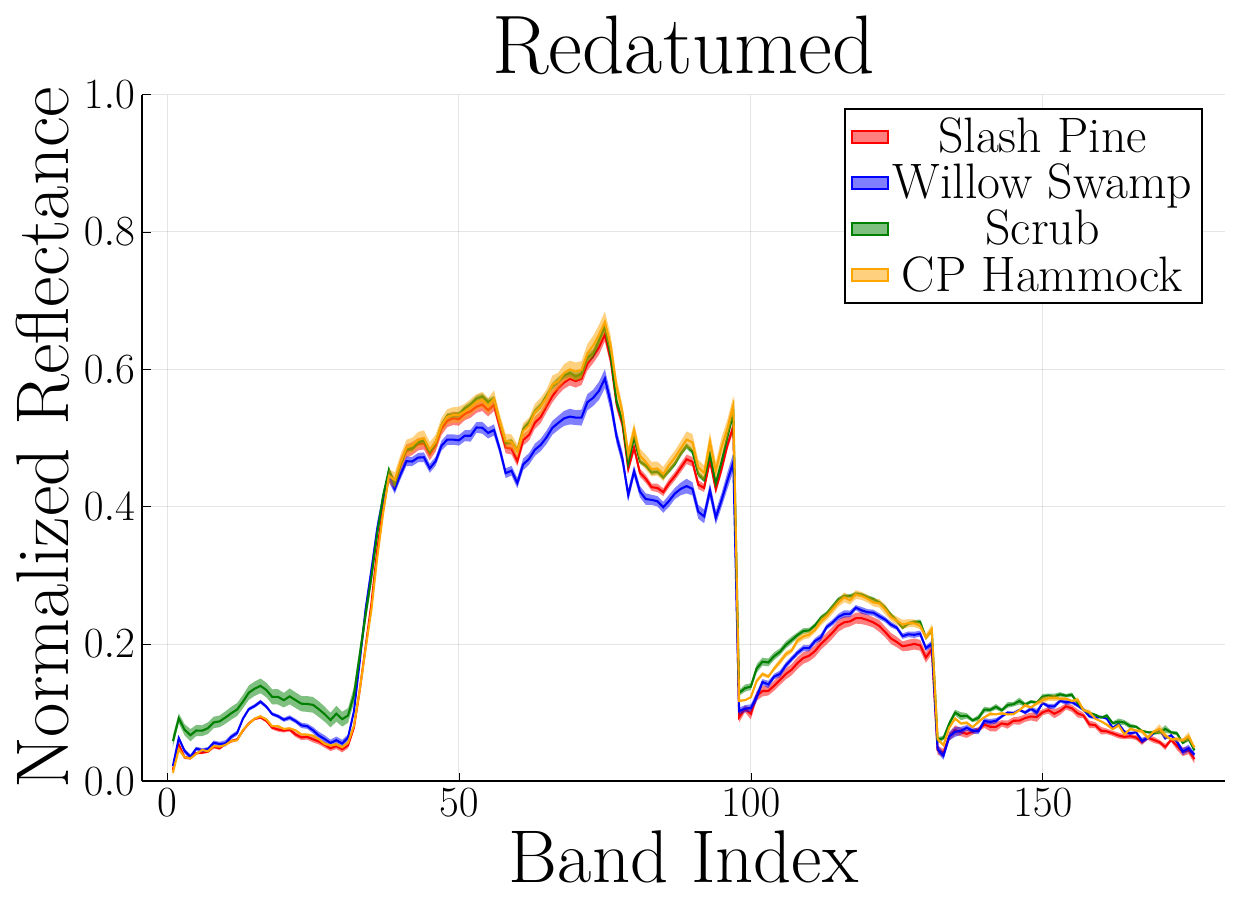}};
\spy on (1.98,-1.21) in node [left] at (2.75,-0.11);
\end{tikzpicture}
        \caption{}
    \end{subfigure}%
    \begin{subfigure}{0.32\linewidth}
        \includegraphics[width=\linewidth]{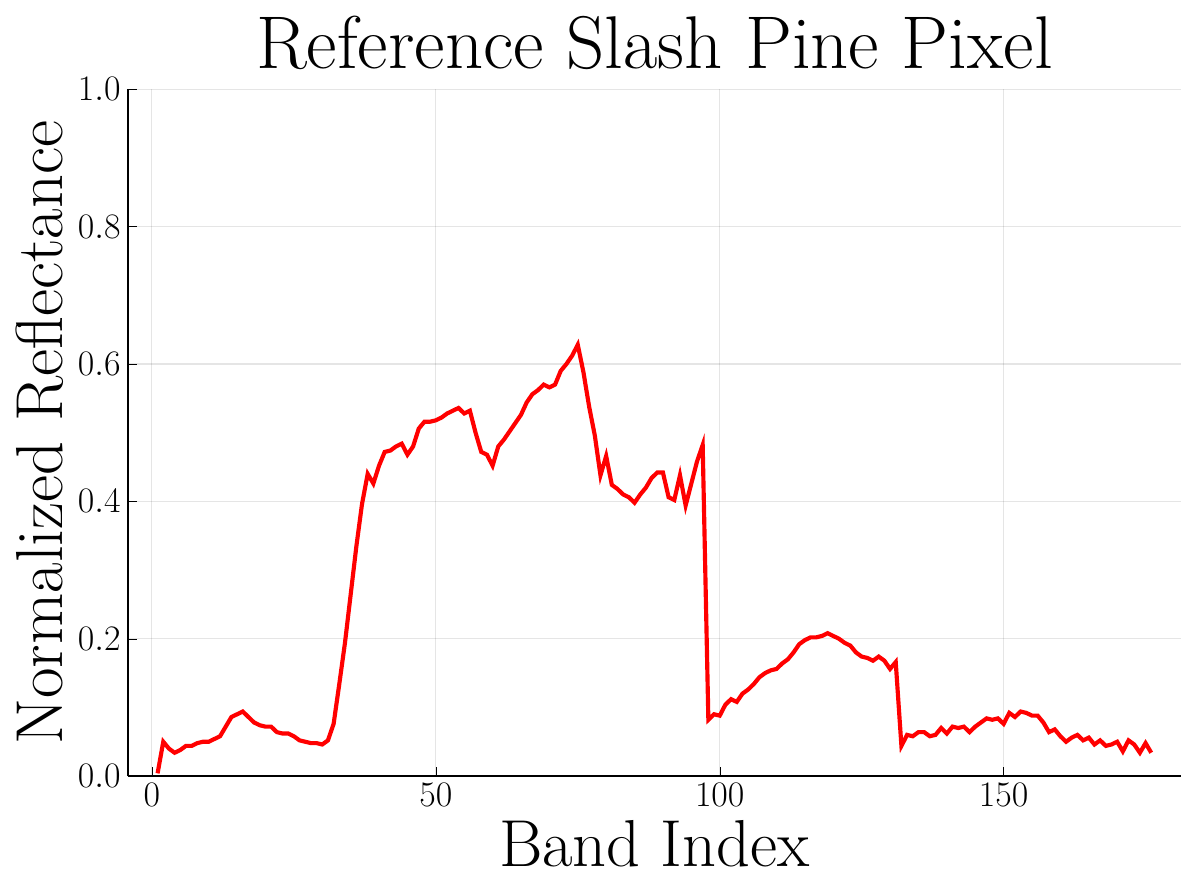}
        \caption{}
    \end{subfigure}
    \begin{subfigure}{0.32\linewidth}
            \begin{tikzpicture}[spy using outlines={blue,magnification=1.5,size=1.43cm, connect spies}]
\node {\includegraphics[width=\linewidth]{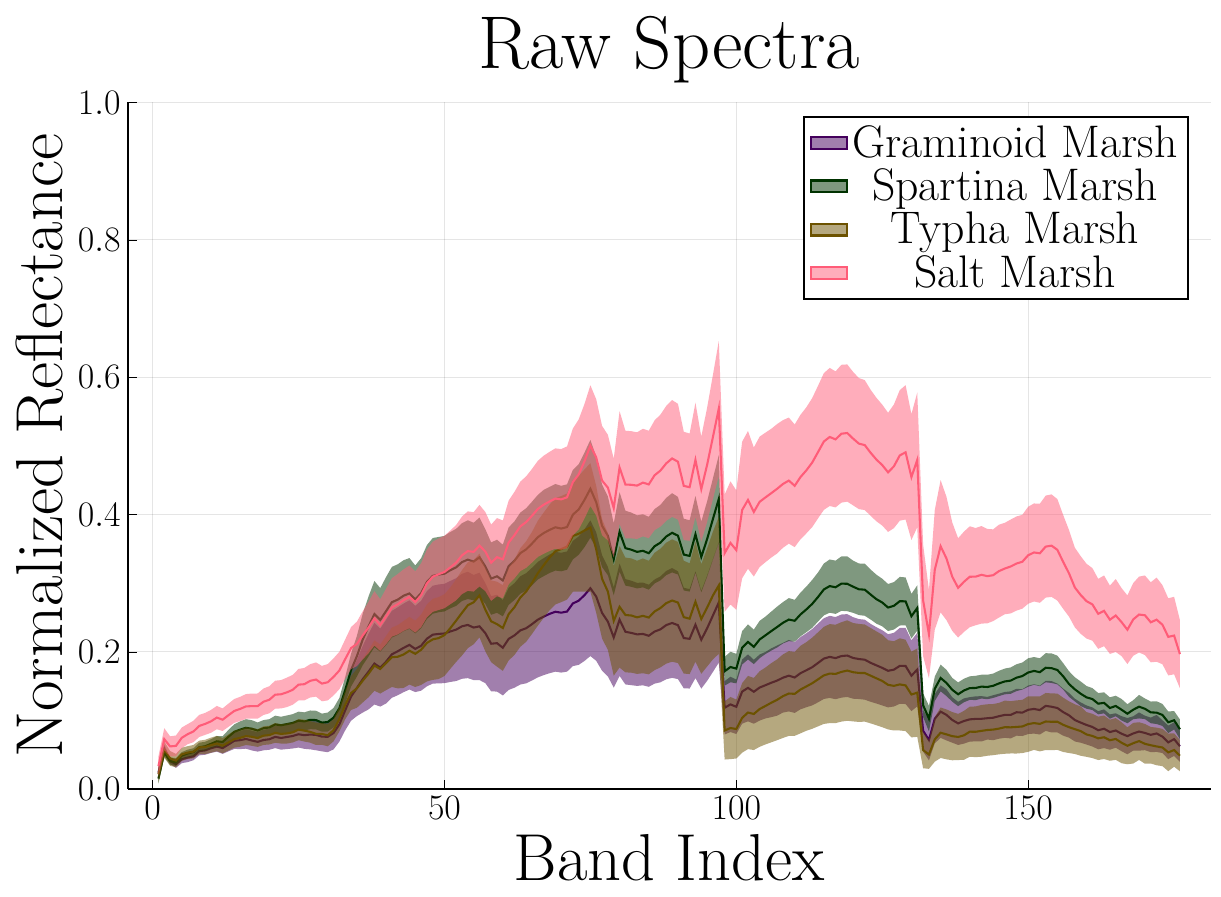}};
\spy on (-0.52,-0.72) in node [left] at (-0.66,0.77);
\end{tikzpicture}
        \caption{}
    \end{subfigure}%
    \begin{subfigure}{0.32\linewidth}
        \begin{tikzpicture}[spy using outlines={blue,magnification=3,size=1.43cm, connect spies}]
\node {\includegraphics[width=\linewidth]{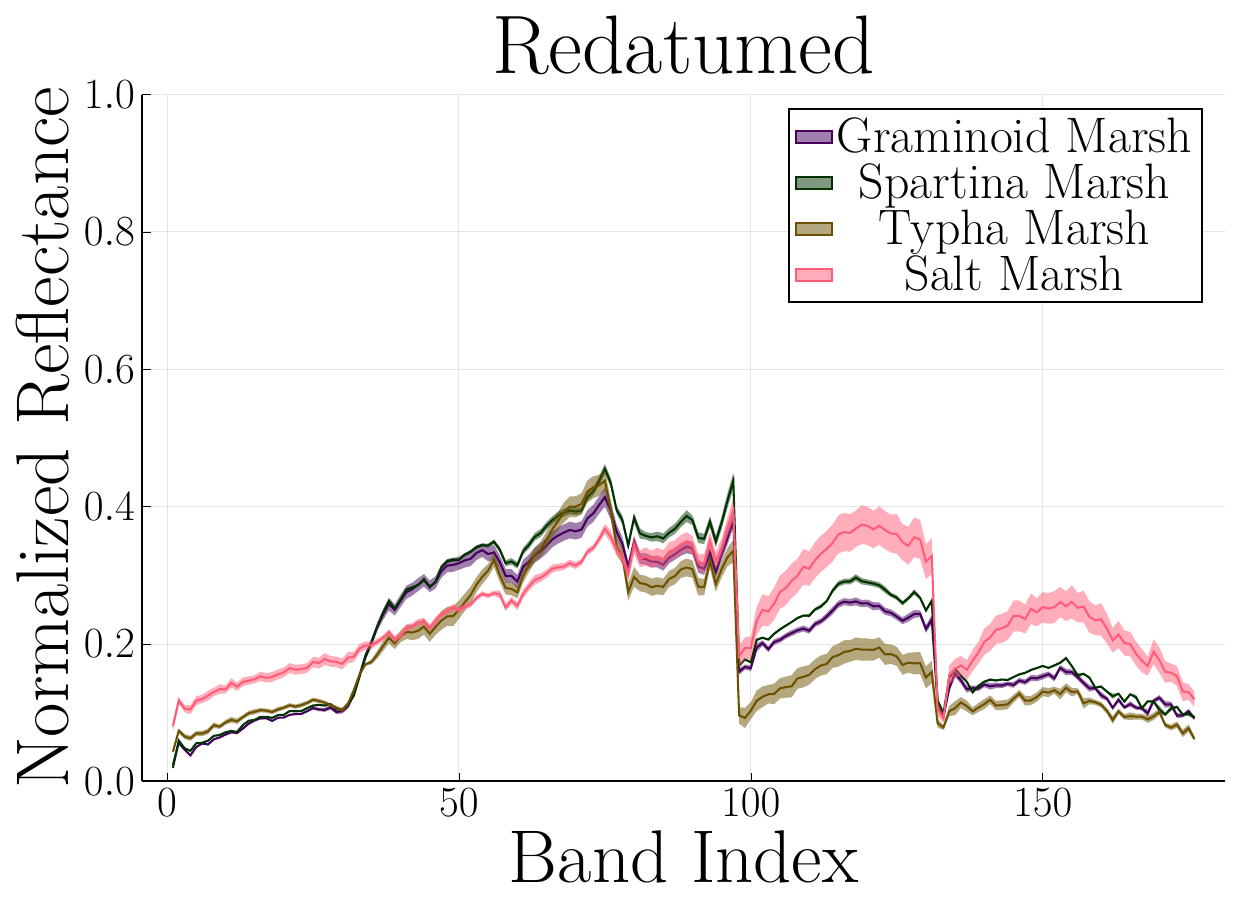}};
\spy on (-0.52,-0.57) in node [left] at (-0.66,0.77);
\end{tikzpicture}
        \caption{}
    \end{subfigure}%
    \begin{subfigure}{0.32\linewidth}
        \includegraphics[width=\linewidth]{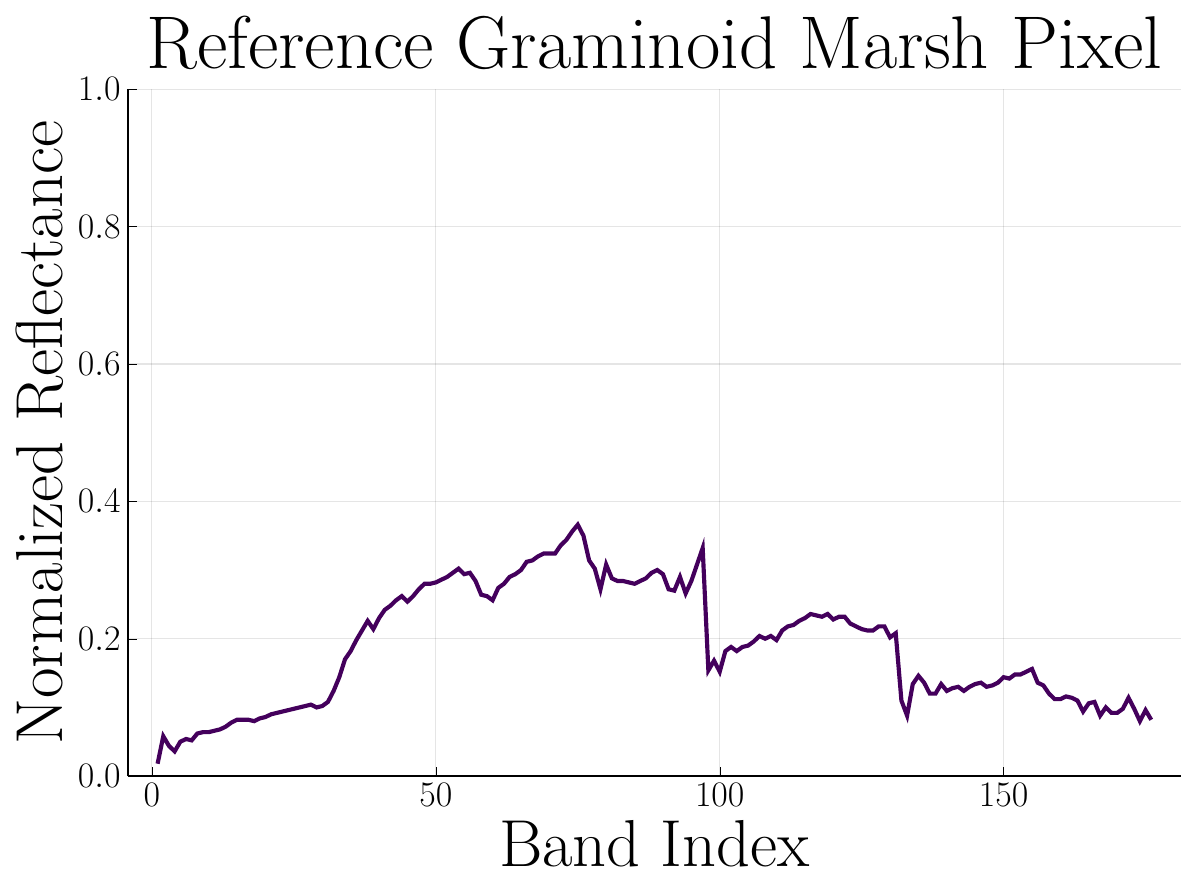}
        \caption{}
    \end{subfigure}
    \caption{Ribbon plots illustrate reduction in intra-class variance post-redatuming. 
    Each ribbon plot represents spectral distribution of distinct classes, with central line denoting mean and ribbon's width on either side indicating intra-class standard deviation. The three reference pixels used for redatuming were randomly sampled from test set. (a) Displays train set spectra from four distinct classes, while (b) shows their respective redatumed counterparts, wherein pixels from the same classes almost coincide, and (c) shows the CP/Oak Hammock reference pixel.
    (d)-(f) Show spectra from test set upland vegetation classes, following the pattern observed in (a)-(c), with (f) showing the Slash Pine reference pixel. 
    While not as pronounced as in (b), the redatumed test set pixels exhibit a discernible reduction in intra-class variance.
    (g)-(i) are same as (d)-(f) but for wetland classes, using a Graminoid Marsh reference pixel.
    \vspace{-10pt}
    }
    \label{fig:redatuming1}
\end{figure*}

Figures \ref{fig:redatumingoak} and \ref{fig:redatuming1} illustrate the resulting reduction in intra-class variance among vegetation spectra. The redatuming effect extends to test pixels excluded during training. To quantify intra-class variability, we define \emph{average variance} as the mean of per-band variances within a ground truth class. Table \ref{tab:class_variances} presents average variance values before and after redatuming with a random pixel. Post-redatuming, the residual average variance in test pixels falls below $5\%$ for most classes, demonstrating significant reduction in intra-class variability.

\begin{table*}[b]
    \centering
    \captionsetup{font=footnotesize}
     \caption{Class-wise Average Variance Reduction in KSC Test Set After Redatuming Using a Random Reference Pixel \label{tab:class_variances}}
     \vspace{-10pt}
     \scriptsize
    \setlength{\tabcolsep}{4pt}
    \renewcommand{\arraystretch}{1}
    \begin{tabular}{|>{\centering\arraybackslash}p{0.4cm}|>{\centering\arraybackslash}p{2.7cm}|>{\centering\arraybackslash}p{1.2cm}|>{\centering\arraybackslash}p{1.2cm}|>{\centering\arraybackslash}p{3.2cm}|>{\centering\arraybackslash}p{3.35cm}|>{\centering\arraybackslash}p{3.25cm}|}
    \hline
    No. & Class & Training Samples & Test Samples & Average Variance In Raw Spectra ($\times 10^{-6}$) & 
 Average Variance After Redatuming ($\times 10^{-6}$)&Residual Variance (\%) After Redatuming \\
        \hline
        1 & Scrub & 77 & 684 & 1173.9 & 24.6& 2.10 \%\\
        2 & Willow Swamp & 25 & 218 & 1938.3 & 17.0& 0.88 \% \\
        3 & CP Hammock & 26 & 230 & 591.9 & 38.0& 6.42 \% \\
        4 & CP/Oak Hammock & 26 & 226 & 1267.6 & 43.5& 3.43 \% \\
        5 & Slash Pine & 17 & 144 & 1315.4 & 22.2& 1.69 \% \\
        6 & Oak Hammock & 23 & 206 & 1346.5 & 61.5& 4.57 \% \\
        7 & Hardwood Swamp & 11 & 94 & 695.3 & 9.4&1.35 \% \\
        8 & Graminoid Marsh & 44 & 387 & 3466.7 & 51.0&1.47 \% \\
        9 & Spartina Marsh & 52 & 468 & 1530.9 & 132.3& 8.64 \% \\
        10 & Typha Marsh & 38 & 366 & 3086.0 & 141.1& 4.57 \% \\
        11 & Salt Marsh & 42 & 377 & 3986.9 & 415.4& 10.42\% \\
        12 & Mud Flats & 47 & 456 & 1529.5 & 325.1& 21.26\% \\
        13 & Water Body & 91 & 836 & 143.0 & 0.047& 0.03 \% \\
        \hline
    \end{tabular}
\end{table*}

\begin{figure}[h]
\captionsetup{font=footnotesize}
\scriptsize
    \begin{subfigure}{1\linewidth}
        \centering
        \includegraphics[width=\linewidth]{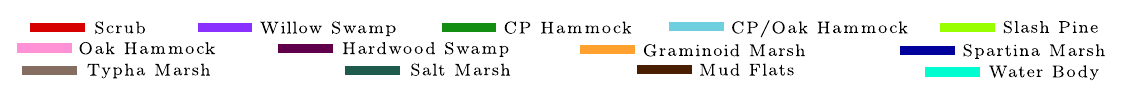}
    \end{subfigure}
    
    \centering
    \begin{subfigure}{0.32\linewidth}
        \centering
        \begin{tikzpicture}[spy using outlines={rectangle, lime, magnification=2, connect spies, line width=2pt}]
            \node {\includegraphics[width=\linewidth]{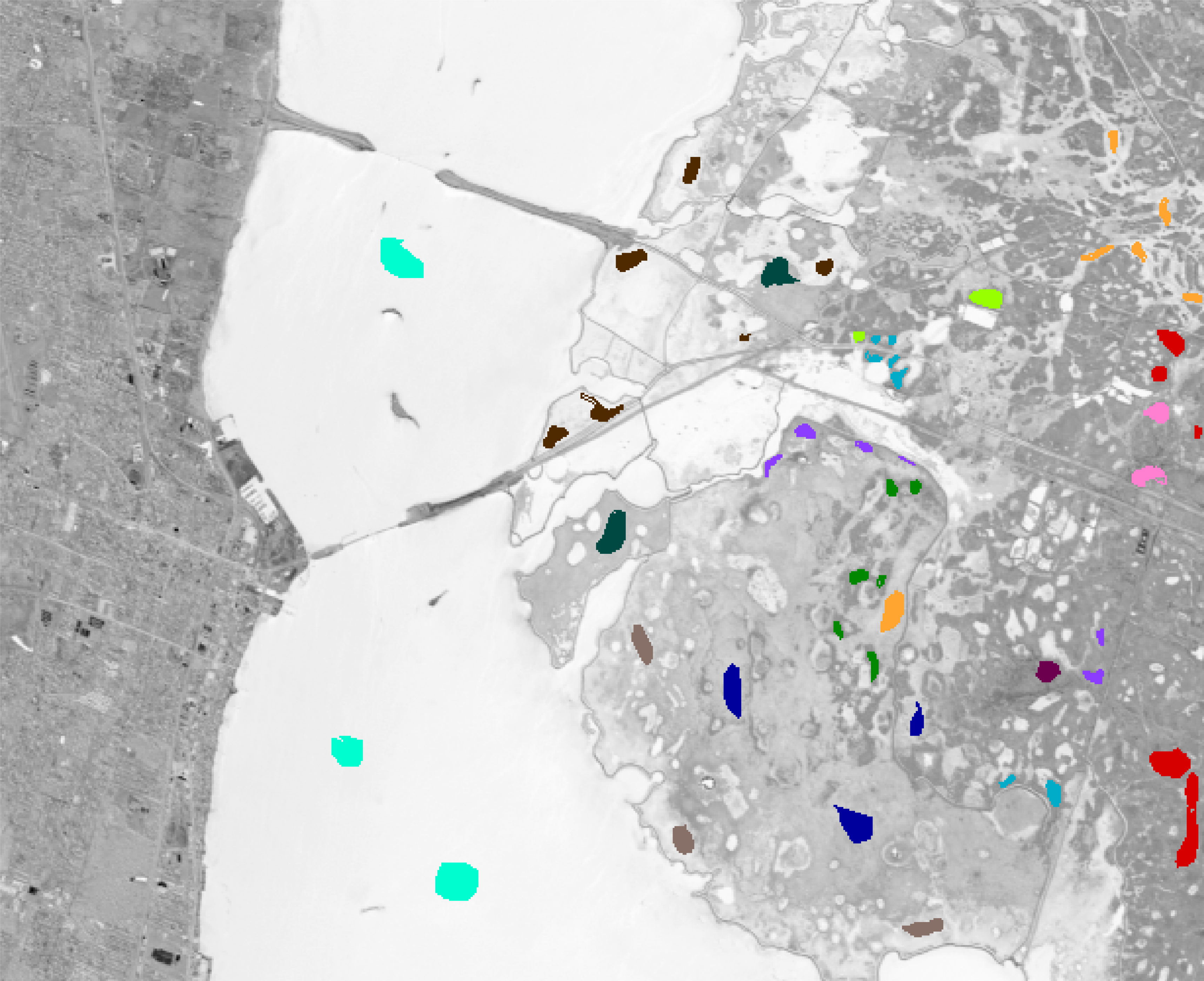}};
            \spy[rectangle, width=0.33cm, height=0.8cm, line width=2pt] on (1.31,0.19) in node [left] at (-1.05,0.32);
        \end{tikzpicture}
        \caption{}
    \end{subfigure}
    \begin{subfigure}{0.32\linewidth}
        \centering
        \begin{tikzpicture}[spy using outlines={rectangle, lime, magnification=2, connect spies, line width=2pt}]
            \node {\includegraphics[width=\linewidth]{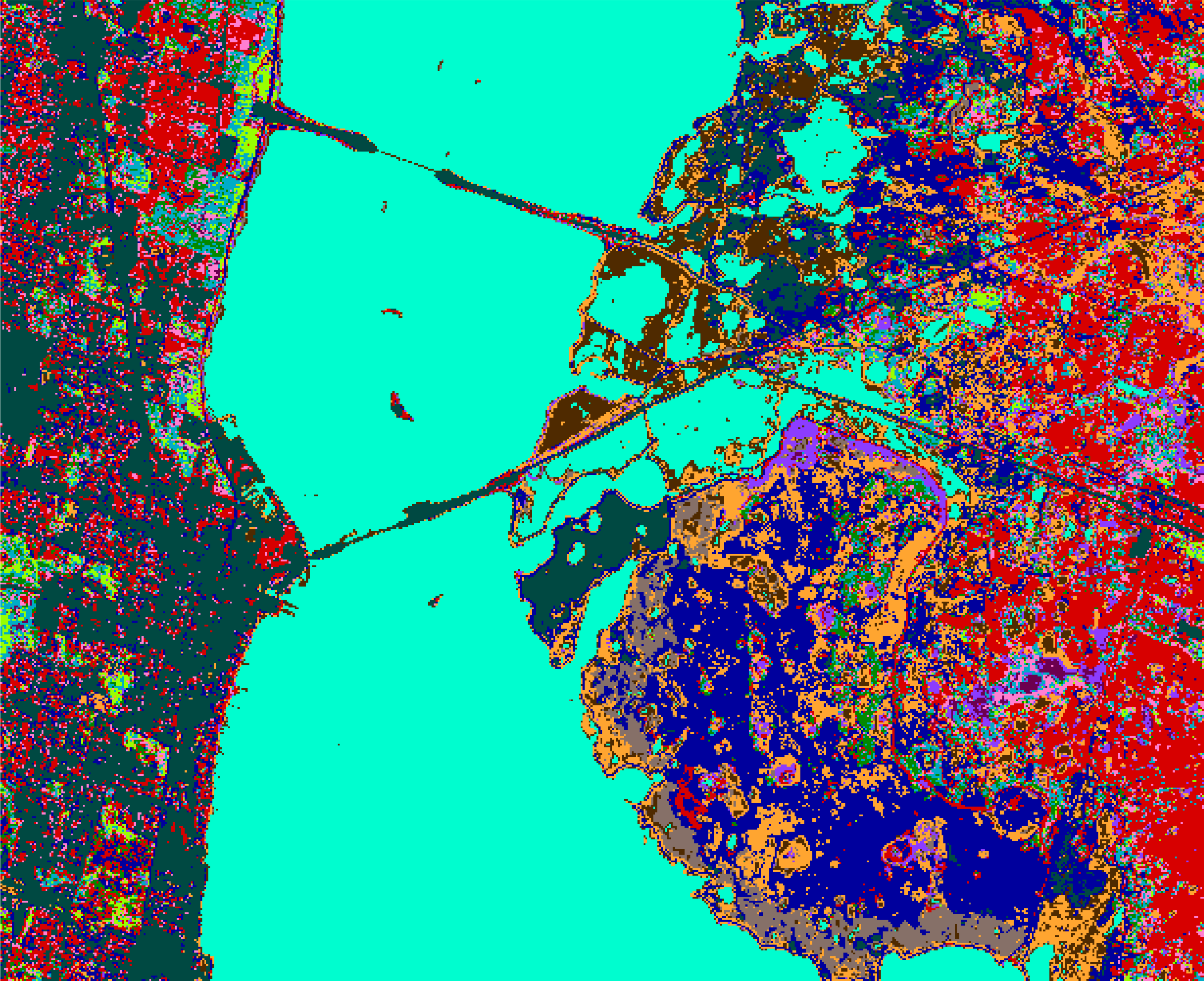}};
            \spy[rectangle, width=0.33cm, height=0.8cm, line width=1pt] on (1.31,0.19) in node [left] at (-1.05,0.32);
        \end{tikzpicture}
        \caption{}
    \end{subfigure}
    \begin{subfigure}{0.32\linewidth}
        \centering
        \begin{tikzpicture}[spy using outlines={rectangle, lime, magnification=2, connect spies, line width=2pt}]
            \node {\includegraphics[width=\linewidth]{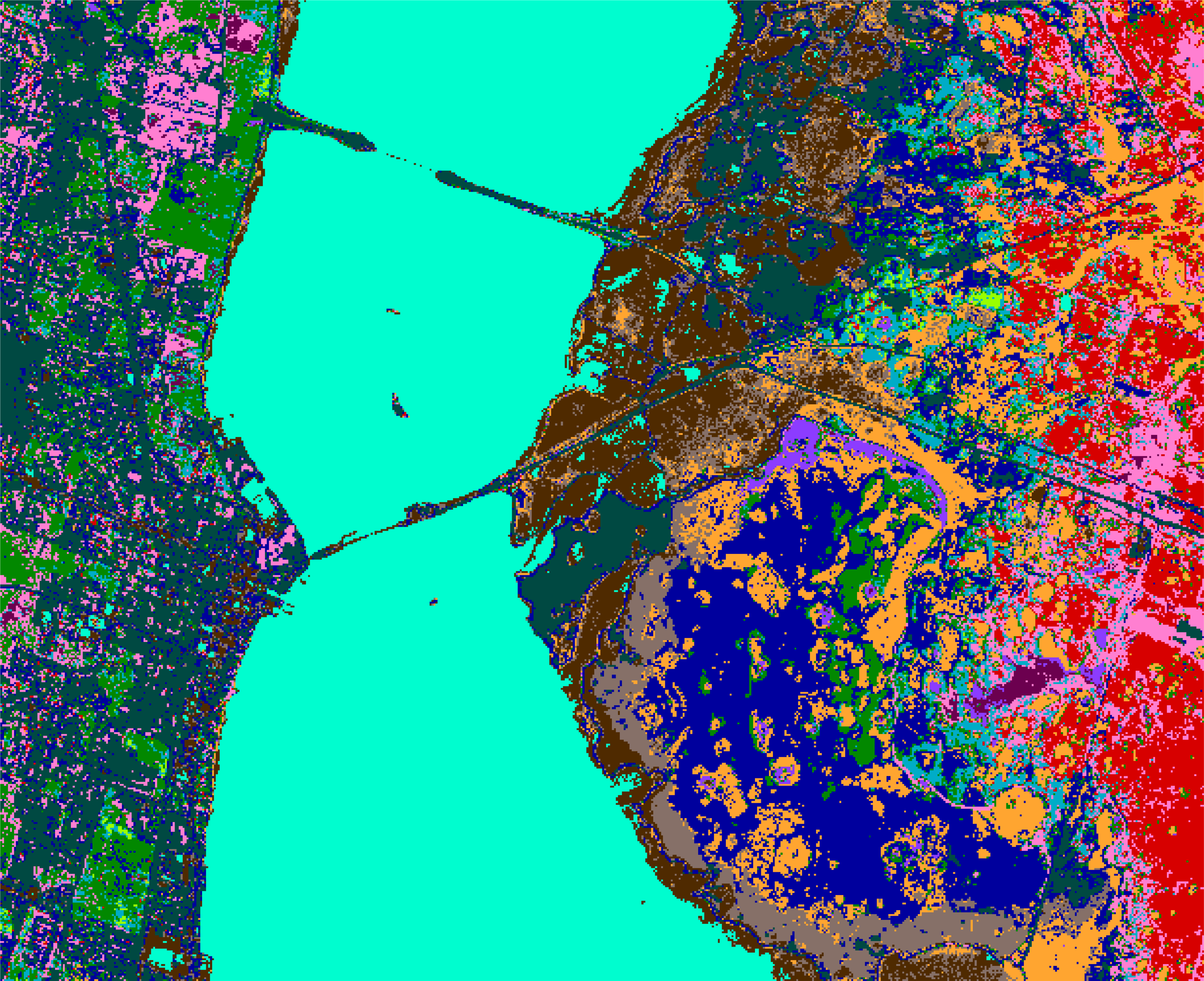}};
            \spy[rectangle, width=0.33cm, height=0.8cm, line width=2pt] on (1.31,0.19) in node [left] at (-1.05,0.32);
        \end{tikzpicture}
        \caption{}
    \end{subfigure}
    \caption{
    K-Nearest Neighbors (KNN) pixel classification results on KSC scene maps. (a) Ground truth map of the KSC scene, serving as the baseline. (b) Pixel-wise classification using KNN on the raw image, resulting in an overall accuracy of $81.6\%$ for the test set ground truth. (c) Pixel-wise classification conducted on a virtual image with uniformized nuisance, generated via the redatuming process, which elevated classification performance to an average overall accuracy of $92.8\%$. An example is the Oak Hammock class, which was heavily misclassified before redatuming but improved notably afterward, as shown in the insets.
    \label{fig:3knn}
    }
\end{figure}

To examine the effects of redatuming on classification performance, we applied three common classifiers—K-Nearest Neighbors (KNN, K=5), Random Forests, and linear Support Vector Machines (SVM)—to both the raw hyperspectral spectra and the virtual spectra generated through redatuming with $1000$ different reference pixels. Results of KNN classification are illustrated in Figure \ref{fig:3knn}. Table \ref{tab:classification_results} summarizes the overall accuracy (OA) for each classifier.

\begin{table}[h]
\renewcommand{\arraystretch}{1}
\scriptsize
\centering
\caption{Classification Accuracy Comparison\vspace{-10pt}}
\label{tab:classification_results}
\begin{tabular}{|c|c|c|}
\hline
Classifier & Raw Spectra OA (\%) & Redatumed Spectra OA (\%) \\
\hline
KNN (K=5) & 81.6 \% & 92.8 ± 0.9 \% \\
Random Forests & 86.2 \% & 93.0 ± 0.9 \% \\
Linear SVM & 74.0 \% & 85.8 ± 4.9 \% \\
\hline
\end{tabular}
\end{table}

The results show improvements in classification accuracy across all methods using virtual spectra generated through redatuming. While KNN and Random Forests demonstrate relatively consistent performance improvements, the linear SVM shows the largest gain in mean accuracy, albeit with higher variability. This sensitivity of the linear SVM, a simpler model, to the choice of reference pixel highlights the importance of reference pixel selection in the redatuming process.

All these observations align with the expected behavior of a disentanglement process, where class-specific information is separated from instance-specific variations. Section \ref{sec:classification} builds on these findings, presenting an approach that uses only the coherent features extracted by SymAE for classification tasks, thus avoiding the need for reference pixel selection.


\subsection{Distinction from Denoising and Unmixing}

SymAE's approach to hyperspectral data analysis fundamentally differs from traditional denoising and unmixing tasks. Denoising aims to estimate noise-free spectra~\cite{rasti2018noise}, while unmixing decomposes mixed pixel spectra into endmember abundances~\cite{keshava2002spectral}. In contrast, SymAE uses abstract representation learning to capture underlying coherent and variable patterns in grouped spectral data. As a result, direct comparisons between SymAE and these methods are not straightforward. Instead, SymAE provides a complementary perspective, potentially revealing patterns that traditional methods may overlook.

The key distinctions in the features extracted by SymAE are as follows:

\begin{itemize}
    \item SymAE learns abstract representations that may not directly correspond to physically interpretable spectra.
    \item Coherent features represent consistent patterns across groups of pixels but do not necessarily equate to `pure' or `clean' spectra.
    \item Nuisance features capture pixel-specific variations, which differ from traditional notions of `noise' or `impurities'.
    \item Reconstructing a virtual spectrum requires combining both coherent and nuisance features, unlike the clear-cut separation of signal and noise in denoising or the endmember-abundance model in unmixing.
\end{itemize}

 \begin{figure}
 \captionsetup{font=footnotesize}
 \scriptsize
    \centering
    \begin{subfigure}{0.49\linewidth}
        \centering
        \begin{tikzpicture}[spy using outlines={green,magnification=2,size=1.2cm, connect spies}]
\node {\includegraphics[width=\linewidth]{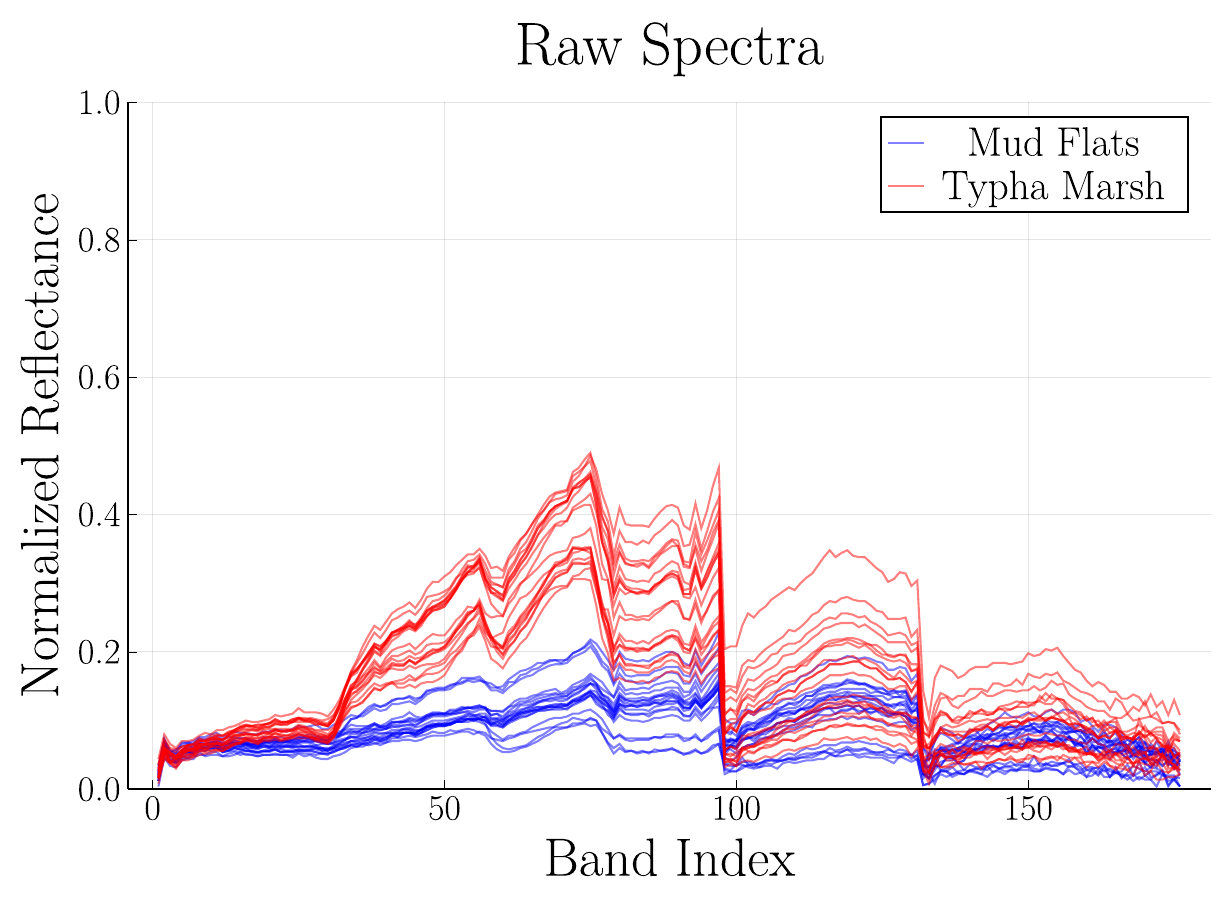}};
\spy on (1.71,-0.91) in node [left] at (-0.33,0.56);
\end{tikzpicture}
        \caption{\label{rawspectradae}}
    \end{subfigure}
    \begin{subfigure}{0.49\linewidth}
        \centering
        \begin{tikzpicture}[spy using outlines={green,magnification=2,size=1.2cm, connect spies}]
\node {\includegraphics[width=\linewidth]{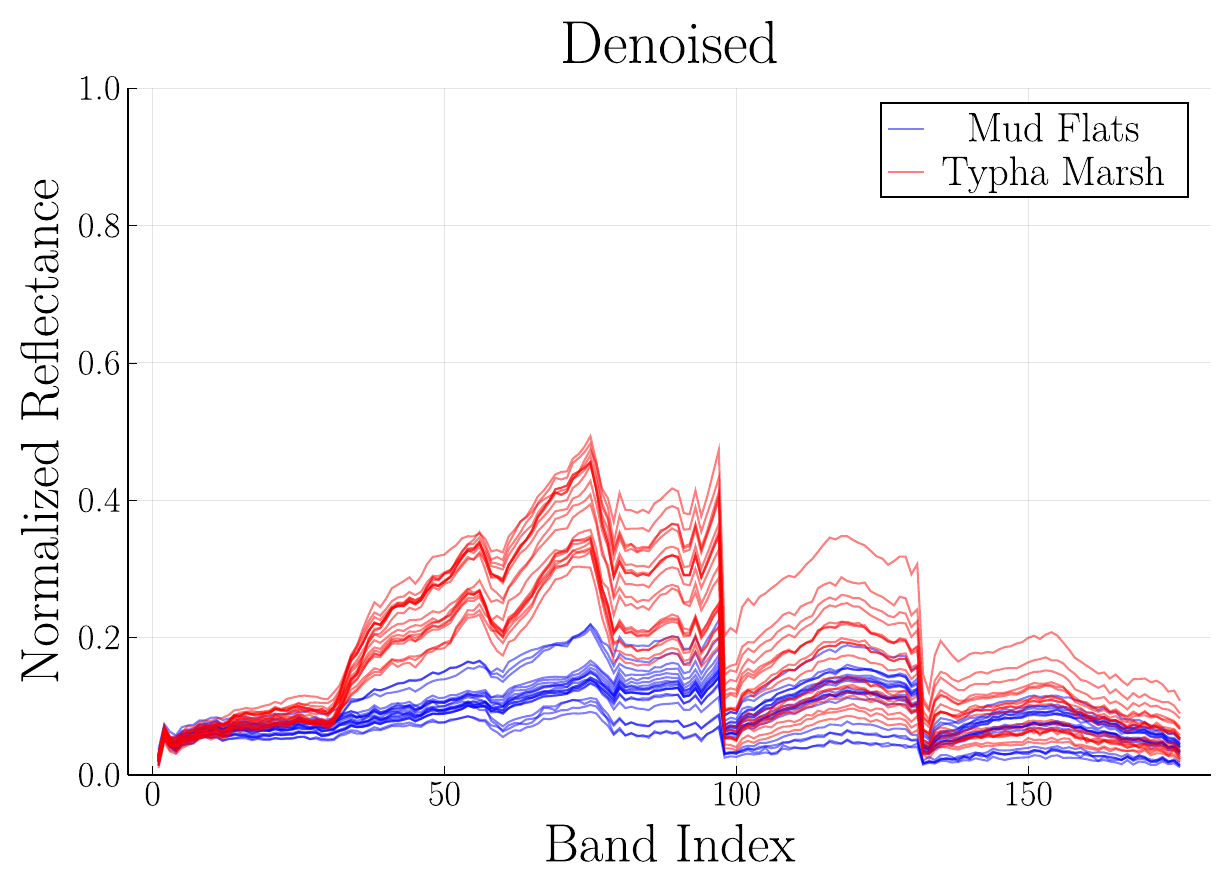}};
\spy on (1.715,-0.89) in node [left] at (-0.33,0.56);
\end{tikzpicture}
        \caption{\label{denoisedspectradae}}
    \end{subfigure}

    \begin{subfigure}{0.49\linewidth}
        \centering
        \begin{tikzpicture}[spy using outlines={green,magnification=2,size=1.1cm, connect spies}]
\node {\includegraphics[width=\linewidth]{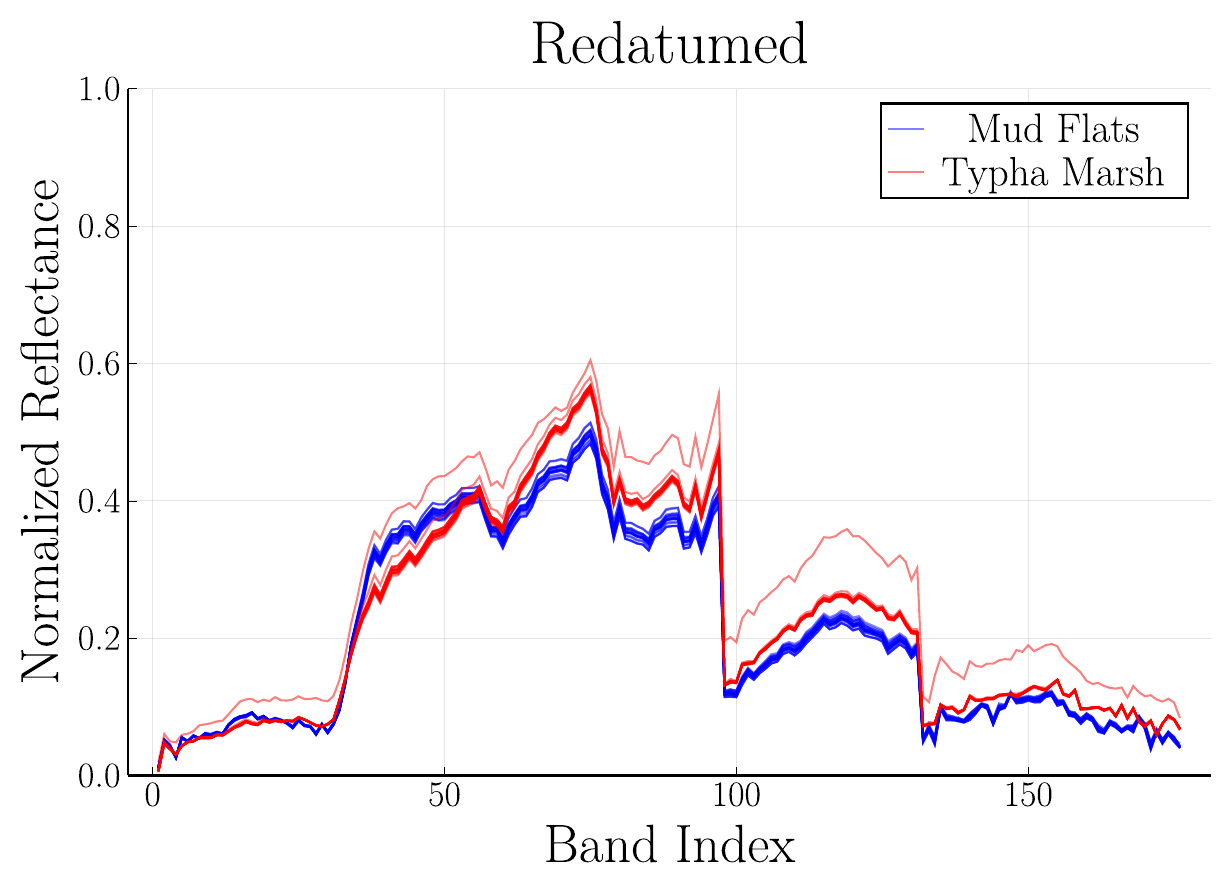}};
\spy on (1.71,-0.89) in node [left] at (-0.44,0.64);
\end{tikzpicture}
        \caption{\label{fig:redatumingdae}}
    \end{subfigure}
    \begin{subfigure}{0.49\linewidth}
        \centering
        \begin{tikzpicture}
        \node {\includegraphics[width=\linewidth]{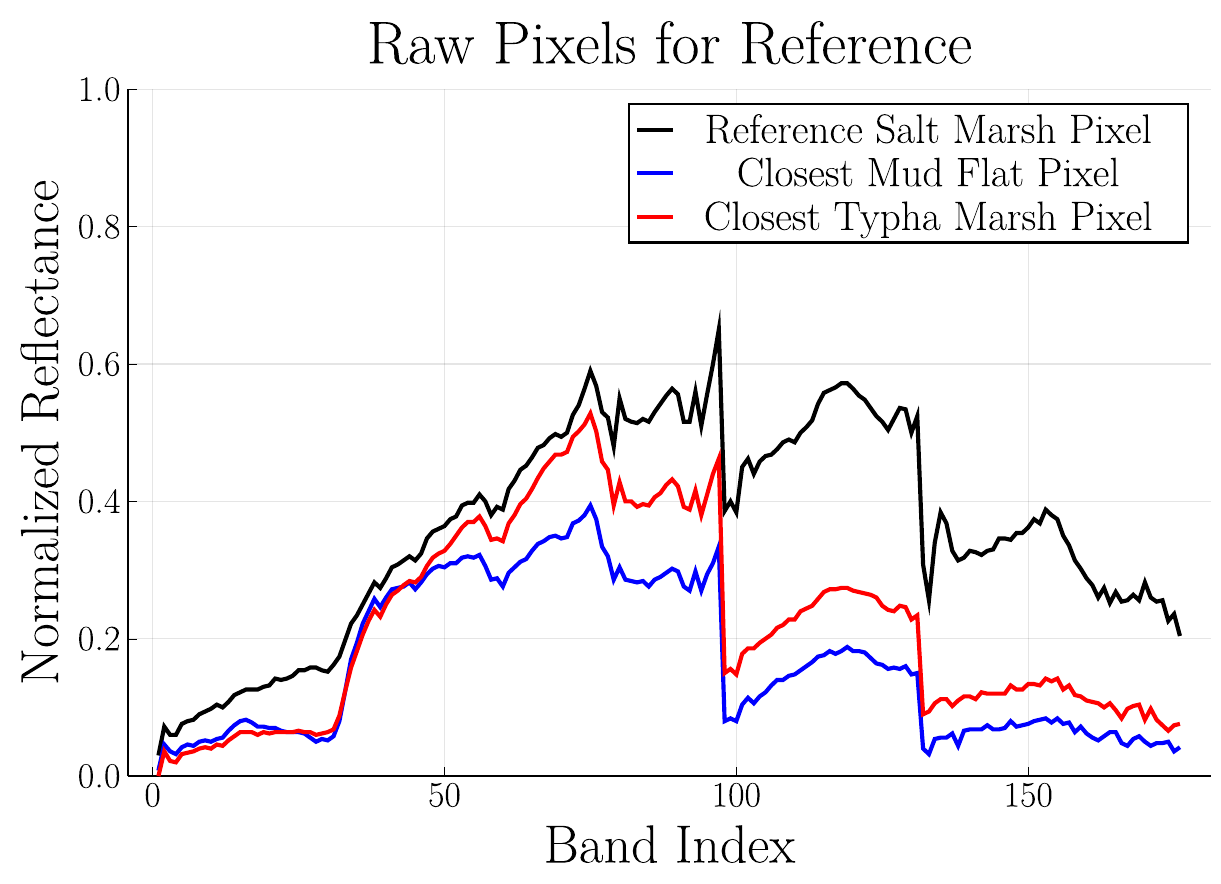}};
           \end{tikzpicture}
            \caption{\label{fig:refp}}
    \end{subfigure}
    
    \caption{ Comparative analysis of the application of DAE and SymAE on test data. (a) Raw spectra from two land-cover classes in the KSC scene. (b) DAE tends to smooth spectral data, but significant within-group variations remain. (c) Redatuming with SymAE more effectively reduces intra-class variance than denoising with DAE, though the redatumed spectra may differ considerably from the original raw spectra, in this case, exhibiting elevated energy levels. (d) Reference Salt Marsh pixel (with relatively high energy) used for redatuming, along with the real pixels from the respective ground truth classes that are closest (in the $\ell_1$ sense) to the redatumed spectra.}
    \label{fig:DAEvsSymAE}
\end{figure}

To illustrate these distinctions, we compared SymAE's performance with a deep learning-based denoising method, specifically Denoising Autoencoders (DAEs)~\cite{vincent2008extracting,xing2016stacked}. Our experiments show that DAEs smooth spectral data but do not significantly reduce intra-class variance (Figure \ref{denoisedspectradae}). In contrast, SymAE's redatuming more effectively mitigates intra-class variance (Figure \ref{fig:redatumingdae}). However, unlike conventional denoising techniques, the redatumed spectra may differ noticeably from the raw spectra, retaining influence from the reference pixel used in redatuming (Figure \ref{fig:refp}).
Additionally, we compared redatuming with sparse unmixing~\cite{iordache2011hyperspectral} for spectral decomposition and subsequent classification. Labeling pixels by maximum abundance from sparse unmixing achieved 83.14\% accuracy on the test set, while using SymAE's redatuming yielded significantly higher accuracies (see Table \ref{tab:classification_results}).

\begin{figure*}[!bp]
\vspace{0pt}
    \centering
    \begin{subfigure}{0.32\linewidth}
        \centering
        \includegraphics[width=\linewidth]{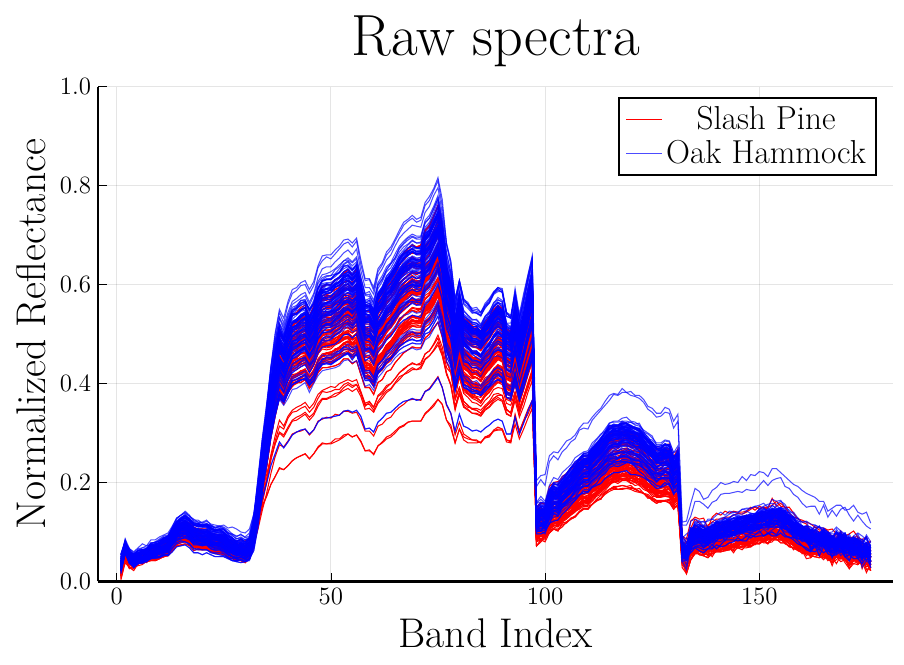}
        \caption{}
        \label{fig:slashoakraw}
    \end{subfigure}
    \begin{subfigure}{0.32\linewidth}
        \centering
        \includegraphics[width=\linewidth]{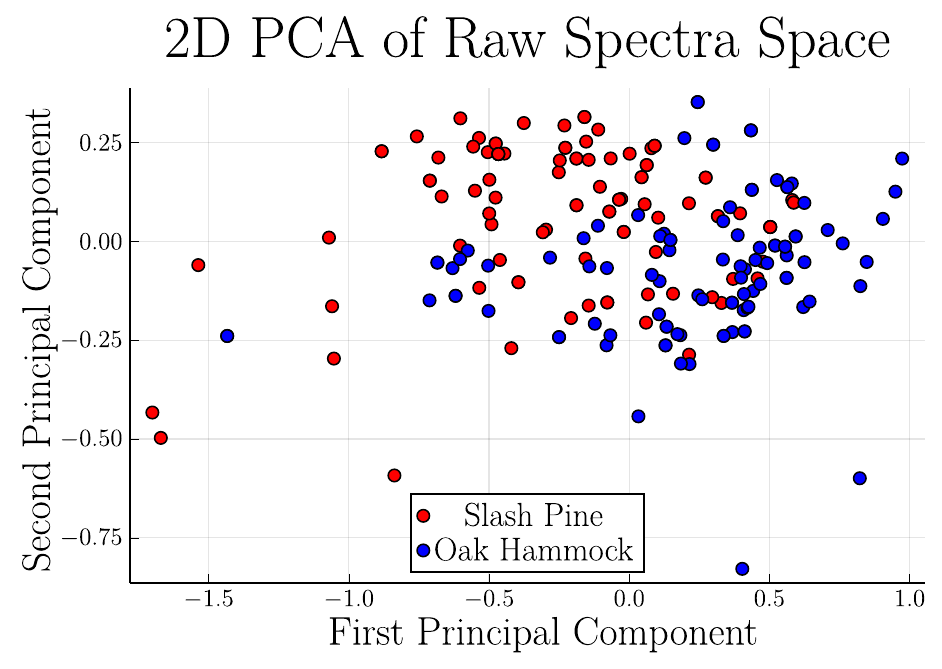}
        \caption{}
        \label{fig:slashoakraw2d}
    \end{subfigure}
    \begin{subfigure}{0.32\linewidth}
        \centering
        \includegraphics[width=\linewidth]{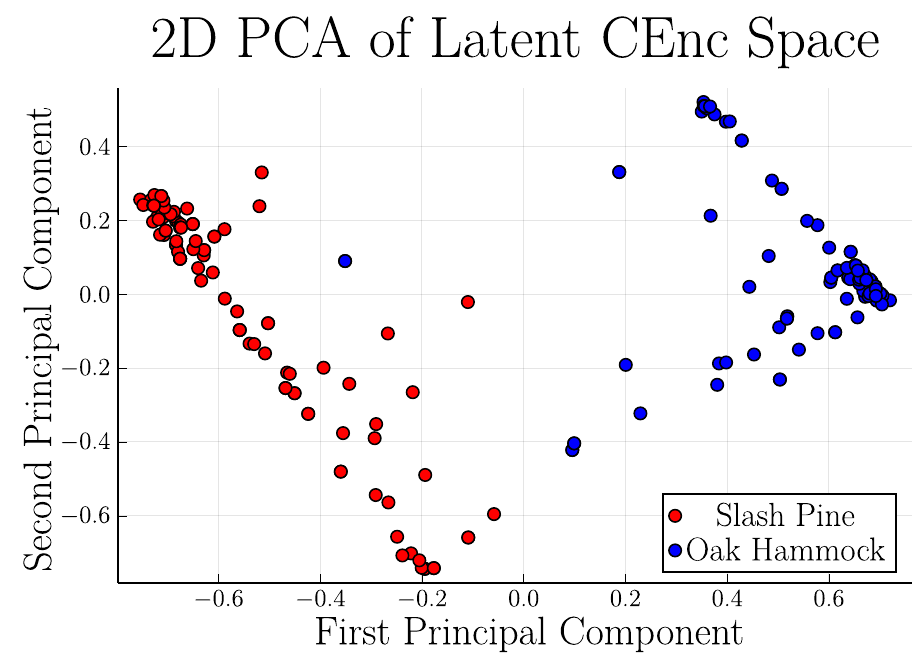}
        \caption{}
        \label{fig:slashoaklatent}
    \end{subfigure}
    \centering
    \begin{subfigure}{0.32\linewidth}
        \centering
        \includegraphics[width=\linewidth]{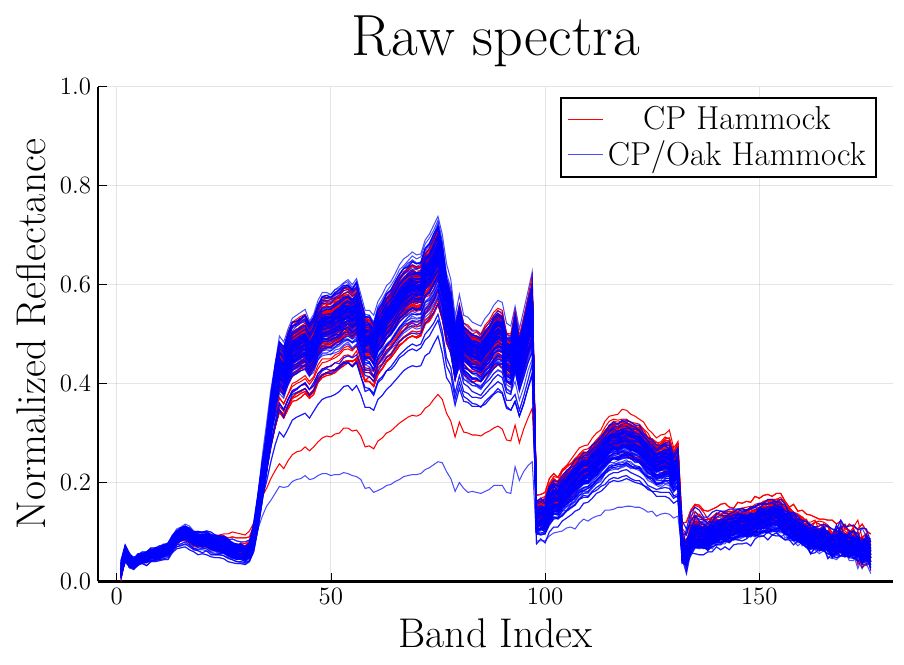}
        \caption{}
        \label{fig:cpoakraw}
    \end{subfigure}
    \begin{subfigure}{0.32\linewidth}
        \centering
        \includegraphics[width=\linewidth]{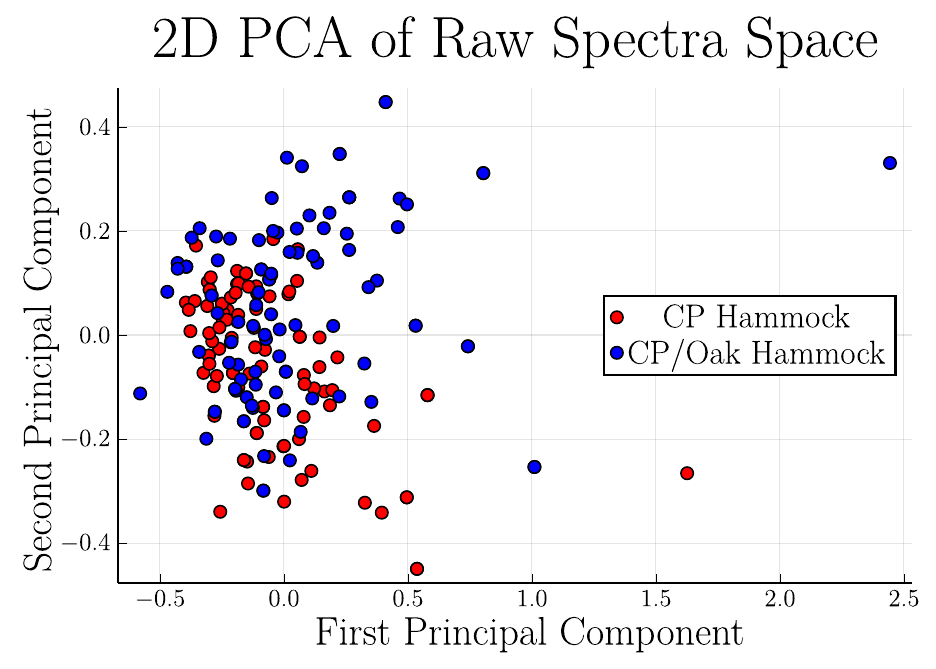}
        \caption{}
        \label{fig:cpoakraw2d}
    \end{subfigure}
    \begin{subfigure}{0.32\linewidth}
        \centering
        \includegraphics[width=\linewidth]{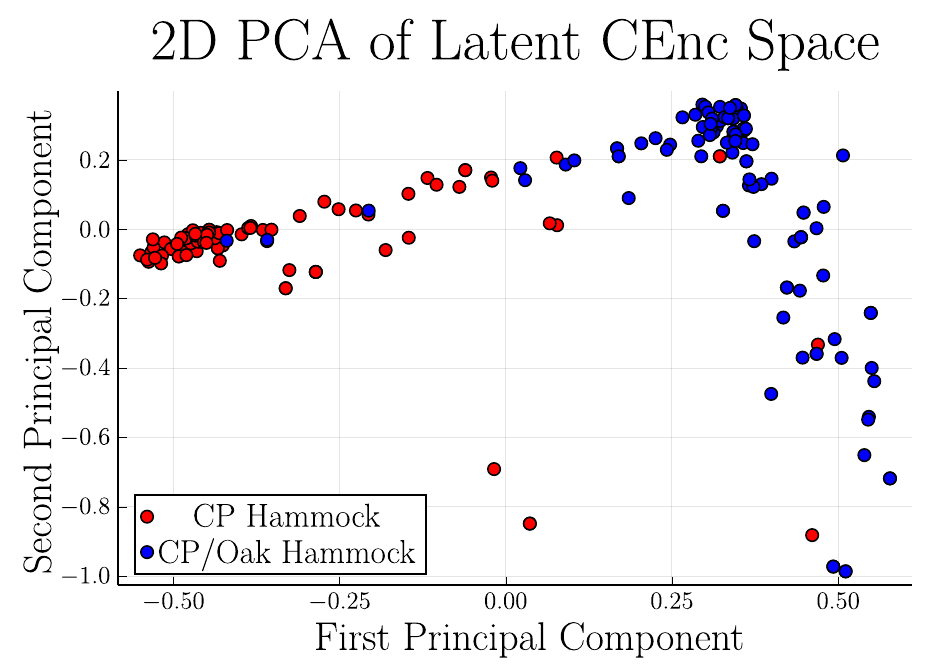}
        \caption{}
        \label{fig:cpoaklatent}
    \end{subfigure}
    \caption{SymAE extracts coherent features that enhance class separability, particularly for spectrally similar classes. (a,d) Raw spectra of spectrally close-by classes. (b,e) These classes are difficult to separate in 2D raw spectra space. (c,f) The classes with subtle differences in raw spectra are more easily discriminated in the latent coherent code space. The most significant improvement in the K-means clustering experiment is observed for classes with subtle differences, such as CP Hammock and CP/Oak Hammock depicted in (d), (e), and (f).
    \vspace{0pt}
    }
    \label{fig:gtclustering}
\end{figure*}

Redatuming transforms spectra in a more complex manner than simple denoising or unmixing. However, its abstract transformation can be challenging to interpret, especially since the choice of reference pixel heavily influences the redatumed virtual spectra. Interpretation may become difficult or nonsensical if the reference pixel belongs to a class with very different characteristics than the pixel being redatumed. For example, marsh classes may exhibit greater spectral variations due to surface water content than upland vegetation classes, while features related to crop ripeness may be irrelevant for water bodies. While Section \ref{subsec:scope} notes potential directions for addressing these interpretability challenges in future work, this paper focuses on the primary benefit of the approach: leveraging the coherent features for improved land cover or material identification.


\subsection{Clustering Analysis of Coherent Features\label{subsec:clustering}}

We designed the coherent features, $\CEnc(P)$, to capture consistent spectral characteristics while being robust to variability. To investigate whether these features improve discriminative power, we compare their performance against the original feature space. Our evaluation method involves applying K-means clustering (with Euclidean metric) to both the original and coherent feature spaces of the test set. We then assess how well the resulting clusters align with the known class labels, providing a measure of each space's ability to naturally separate classes.

Our analysis focuses on pairs of classes that are known to be challenging to distinguish in hyperspectral imagery. We initially examined two pairs of spectrally similar classes: Slash Pine vs. Oak Hammock, and CP Hammock vs. CP/Oak Hammock. For each pair, we sampled $100$ pixels per class and applied K-means clustering in both the raw spectral space and the coherent feature space.

For Slash Pine and Oak Hammock, K-means clustering in the raw spectral space achieved an accuracy of $75.5\%$, while in the coherent feature space, it reached $95.9\%$ - a substantial improvement of $20.4$ percentage points. Figure \ref{fig:slashoakraw} shows the raw spectra of these classes, while Figure \ref{fig:slashoakraw2d} illustrates their overlap in the first two principal components of the raw spectral space. In contrast, Figure \ref{fig:slashoaklatent} depicts the same pixels projected onto a 2-D linear subspace of the coherent feature space, where improved separation is visually apparent.

The improvement was even more pronounced for the challenging pair of CP Hammock and CP/Oak Hammock (Figures \ref{fig:cpoakraw}, \ref{fig:cpoakraw2d}). In this case, the accuracy in the raw spectral space was only $53.3\%$, but it increased to $89.9\%$ in the coherent feature space - an improvement of $36.6$ percentage points. This substantial enhancement for highly similar classes underscores the discriminative power of the coherent feature space.

\begin{figure}
    \centering
    \includegraphics[width=0.8\linewidth]{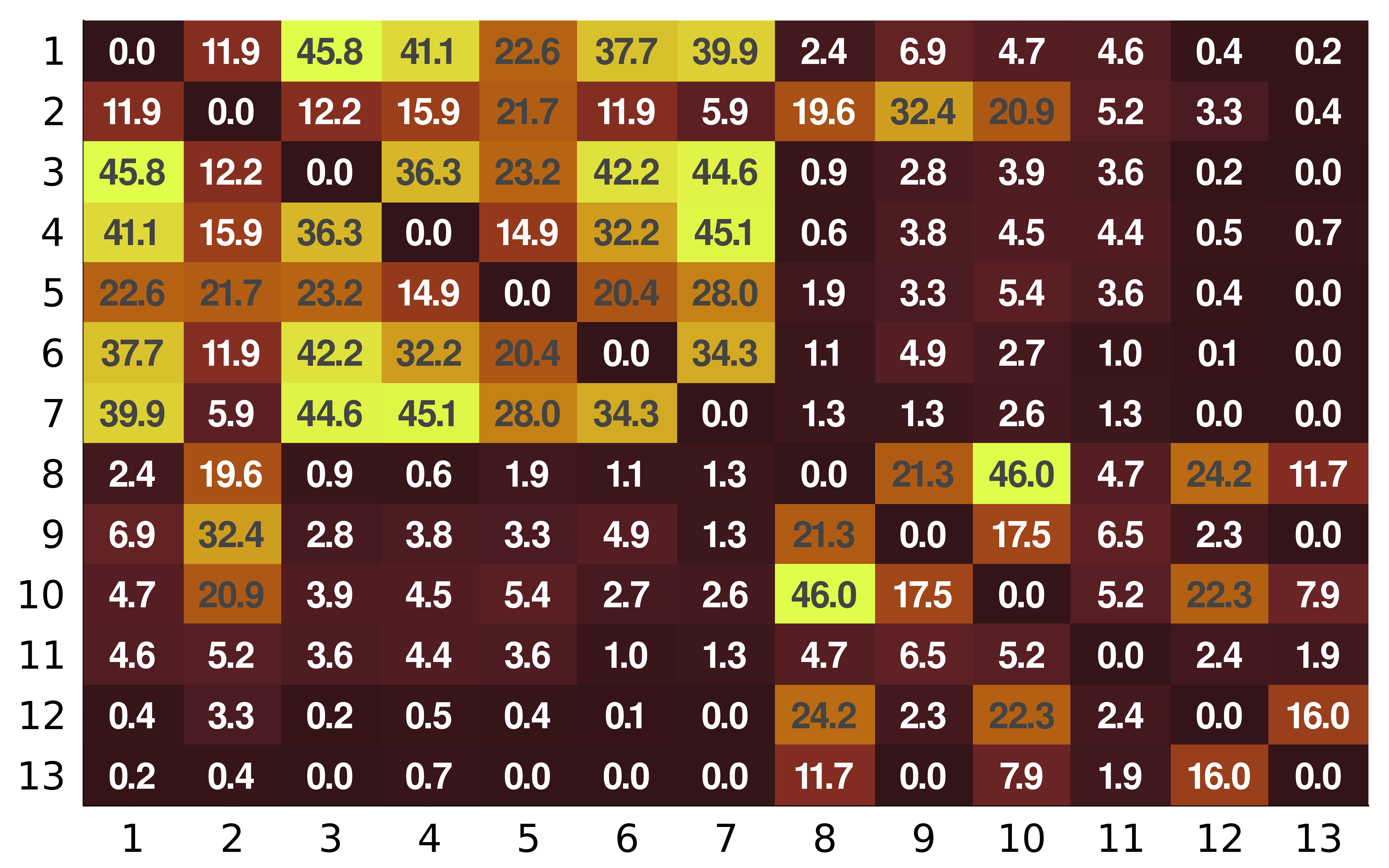}
    \caption{Heatmap illustrating improvement in clustering in KSC dataset. The matrix elements indicate the percentage accuracy difference between K-means clustering in the latent coherent code, $C$, and clustering in the raw spectral data while doing pairwise unsupervised clustering between land-cover classes. The numbers on axes indicate the class indices, following same ordering as in Table \ref{tab:class_variances}. This heatmap pertains to the ground truth-based training scenario and the clustering was done on test set. Pairs that show minimal improvement are those that already exhibit significant separation in raw spectra.\vspace{-0pt}}
    \label{fig:gtkmeans}
\end{figure}

We went on to conduct a comprehensive pairwise clustering experiment encompassing all ground truth classes within the scene. On average, we observed a $12.0$ percentage point improvement in clustering accuracy across all class pairs when using the coherent feature space. Notably, the most substantial improvements were evident among classes characterized by subtle spectral differences. Figure \ref{fig:gtkmeans} illustrates these pairwise improvements in a heatmap, where each element represents the percentage point accuracy difference between clustering in the coherent feature space and the raw spectral space. This visualization highlights the consistent enhancement in class separability achieved by SymAE's coherent features, particularly for spectrally similar classes.

These observations suggest improved discrimination power in the coherent feature space, particularly for spectrally similar classes. We exploit these coherent features for classification tasks in the following section.

\section{HSI Classification Experiments\label{sec:classification}}

Classification is a fundamental task in remote sensing, enabling the identification and mapping of targets, land cover types, and other relevant features from imagery. In this section, we utilize the coherent features extracted through SymAE for hyperspectral image classification. We first evaluate our approach in a pixel-based classification context, benchmarking it against other state-of-the-art methods. Then, we explore integrating these coherent features with some leading spectral-spatial classification techniques to provide insights into how they can complement these approaches.

\subsection{Purely Spectral Classification\label{subsec:spectralclassification}}

Spectral features are the primary source of discriminative information in hyperspectral imaging. To evaluate the coherent features extracted by SymAE, we focus on pixel-based classification, directly assessing the discriminative power of these spectral features.

We compare our method against several established techniques, including XGBoost~\cite{chen2016xgboost}, a leading tree-based model; the Support Vector Classifier (SVC), a widely used machine learning approach~\cite{melgani2004classification}; the Stacked Autoencoder (SAE)\cite{chen2014deep} followed by SVC classification, which leverages hierarchically learned features; the 1D Convolutional Neural Network (1D-CNN), which learns spectral patterns through one-dimensional filters\cite{hu2015deep}; the Cascaded Recurrent Neural Network (CasRNN), which exploits both redundant and complementary information in spectral bands~\cite{hang2019cascaded}; the miniGCN, which enables efficient graph-based learning~\cite{hong2020graph}; the Vision Transformer (ViT), which adapts transformer architectures for image data~\cite{dosovitskiy2020image}; SpectralFormer, a transformer-based model achieving state-of-the-art performance~\cite{hong2021spectralformer}. Additionally, state-space models~\cite{gu2023modeling} have emerged as efficient alternatives to transformers, offering competitive performance with lower computational complexity. For comparison, we adapt the Spectral Mamba (SpeMB) block from Li et al.'s MambaHSI~\cite{li2024mambahsi} framework for pixel-wise operation while keeping all other configurations unchanged\footnote{\label{MambaHSI}https://github.com/li-yapeng/MambaHSI}.

To evaluate the effectiveness of the extracted coherent features, we employ two classification strategies: CEnc + SVC and CEnc + Dense. Both strategies utilize the coherent encoder (CEnc) of trained SymAE model to extract features. In the CEnc + SVC strategy, these features are input to an SVC, while in the CEnc + Dense strategy, they are fed into a dense neural network classifier. The parameters and training configurations for the models compared in this study are as follows:

\begin{table*}[h]
\caption{Pixel-based Classification Results from Kennedy Space Center Dataset. \label{tab:KSC1}\vspace{-10pt}}
  \renewcommand{\arraystretch}{1}
  \setlength{\tabcolsep}{4pt}
  \scriptsize
    \centering
    \begin{tabular}{|c|c|>{\centering\arraybackslash}p{0.9cm}|>{\centering\arraybackslash}p{0.85cm}|>{\centering\arraybackslash}p{0.9cm}|>{\centering\arraybackslash}p{0.6cm}|>{\centering\arraybackslash}p{0.65cm}|>{\centering\arraybackslash}p{0.675cm}|>{\centering\arraybackslash}p{0.875cm}|>{\centering\arraybackslash}p{0.95cm}|>{\centering\arraybackslash}p{0.6cm}|>{\centering\arraybackslash}p{0.85cm}|>{\centering\arraybackslash}p{0.75cm}|>{\centering\arraybackslash}p{0.7cm}|>{\centering\arraybackslash}p{0.725cm}|}
        \hline
        No. & Class & Training Samples & Test Samples & XGBoost & SVC & SAE +SVC & 1D-CNN & CasRNN & miniGCN & ViT & Spectral -Former & Mamba & CEnc +SVC & CEnc +Dense\\
        \hline
        1 & Scrub & 77 & 684 & 95.32
& \textbf{97.22}& 96.64& 96.05 & 90.79& 89.04& 92.69 & 94.88 & 95.91& 96.49 & 95.61\\
        2 & Willow Swamp & 25 & 218 & 85.32
& 88.99& 89.91& 89.91 & 82.11& 84.4& 86.70 & 88.07 & 87.16& 95.87 & \textbf{96.79}\\   
        3 & CP Hammock & 26 & 230 & 77.39
& 87.39& 87.39& 87.83 & 86.52& 91.74& \textbf{96.09} & 87.39 & 90& 82.61 & 85.22\\   
        4 & CP/Oak Hammock & 26 & 226 & 55.31
& 74.78& 59.29& 69.03 & 60.62& 62.83& 49.56 & 66.37 & 68.14& \textbf{82.74} & 80.53\\   
        5 & Slash Pine & 17 & 144 & 52.08
& 72.92& 61.81& 62.50 & 67.36& 68.75& 61.81 & 69.44 & 65.97& 75.69 & \textbf{77.08}\\   
        6 & Oak Hammock & 23 & 206 & 43.2
& 67.48& 60.68& 65.53 & 54.37& 72.33& 42.23 & 53.40 & 68.45& 72.33 & \textbf{77.67}\\   
        7 & Hardwood Swamp & 11 & 94 & 64.89
& 79.79& 73.4& 70.21 & 81.91& \textbf{90.43}& 76.60 & 57.45 & 81.91& 87.23 & 88.23\\   
        8 & Graminoid Marsh & 44 & 387 & 84.75
& 95.09& 88.37& 90.44 & 87.6& 94.32& 85.27 & 91.21 & 89.92& 96.12 & \textbf{96.90}\\   
        9 & Spartina Marsh & 52 & 468 & 85.9
& 96.37& 94.02& 97.22 & 95.3& 86.32& 97.44 & 96.79 & 90.6& \textbf{97.65} &97.44\\   
        10 & Typha Marsh & 38 & 366 & 89.89
& 95.08& 78.42& 90.98 & 89.62& 95.9& 95.63 & 96.99 & 86.34& \textbf{97.54} & 97.27\\   
        11 & Salt Marsh & 42 & 377 & 96.55
& \textbf{98.14}& 98.67& 98.41 & 97.08& 98.14& 97.88 & 96.55 & 97.88& 97.88 & \textbf{98.14}\\   
        12 & Mud Flats & 47 & 456 & 89.04
& 97.37& 86.62& 96.27 & 81.14& 91.67& 87.06 & 94.08 & 95.39& 97.81 & \textbf{98.90}\\   
        13 & Water Body & 91 & 836 & 99.04
& \textbf{100.0}& \textbf{100.0}& \textbf{100.0} & 99.88& 99.88& \textbf{100.0} & \textbf{100.0} & \textbf{100.0}&\textbf{100.0} & \textbf{100.0}\\   
        \hline                                       
        & Overall Accuracy & & & 85.74
& 93.03& 88.38& 91.35 & 87.51& 89.98& 88.28 & 90.49 & 90.54& 94.27 & \textbf{94.65}\\   
        & Average Accuracy & & & 78.36
& 88.51& 82.71& 85.72 & 82.64& 86.6& 82.23 & 84.05 & 85.97& 90.77 & \textbf{91.53}\\   
        & $\kappa\times100$ & & & 84.1
& 92.23& 87.05& 90.35 & 86.09& 88.86& 86.95 & 89.40 & 89.46& 93.61 & \textbf{94.04}\\
        \hline
    \end{tabular}
    \begin{tablenotes}
      \tiny
      \item \hspace{0.4cm} Note: All accuracy values are reported in percentage (\%). Best results are shown in \textbf{bold}.
    \end{tablenotes}
\end{table*}

\begin{table*}[h]
\caption{Pixel-based Classification Results from Indian Pines Dataset. \label{tab:IP}
 \vspace{-10pt}}
  \renewcommand{\arraystretch}{1}
  \setlength{\tabcolsep}{4pt}
  \scriptsize
    \centering
    \begin{tabular}{|c|c|>{\centering\arraybackslash}p{0.9cm}|>{\centering\arraybackslash}p{0.85cm}|>{\centering\arraybackslash}p{0.9cm}|>{\centering\arraybackslash}p{0.6cm}|>{\centering\arraybackslash}p{0.65cm}|>{\centering\arraybackslash}p{0.675cm}|>{\centering\arraybackslash}p{0.875cm}|>{\centering\arraybackslash}p{0.95cm}|>{\centering\arraybackslash}p{0.6cm}|>{\centering\arraybackslash}p{0.85cm}|>{\centering\arraybackslash}p{0.75cm}|>{\centering\arraybackslash}p{0.7cm}|>{\centering\arraybackslash}p{0.725cm}|}
        \hline
        No. & Class & Training Samples & Test Samples & XGBoost & SVC & SAE +SVC & 1D-CNN & CasRNN & miniGCN & ViT & Spectral -Former & Mamba & CEnc +SVC & CEnc +Dense\\
        \hline
        1 & Alfalfa & 15 & 31 
& 77.42
& 74.19
& 48.39
& 77.42
& 87.10& 80.65
& 90.32
& \textbf{93.55}& 87.10& 90.32& 90.32\\
 2 & Corn-notill & 50 & 1378 
& 59.43
& 69.38
& 41.58
& 69.09
& 66.62
& \textbf{73.58}& 69.59
& 57.76& 73.15
& 71.34
&68.36\\
 3 & Corn-mintill & 50 & 780 
& 61.28
& 69.49
& 50.9
& 61.41
& 70.26
& 56.54
& 63.46
& 70.77& \textbf{71.54}& 69.49
&67.82\\
 4 & Corn & 50 & 187 
& 77.54
& 78.07
& 60.43
& 72.73
& 75.40& 67.91
& 80.75
& 82.35& 75.40& 80.75
&\textbf{84.49}\\
        5 & Grass-pasture & 50 & 433 
& 90.30& 90.99
& 82.22
& 84.30& 89.84
& 90.76
& 87.76
& 91.69& 89.61
& \textbf{92.61}& 90.99\\   
        6 & Grass-trees & 50 & 680 
& 82.21
& 92.50& 83.24
& 92.35
& 86.62
& 93.09
& 92.21
& \textbf{94.12}& 85.88
& 92.79
& 91.47\\   
        7 & Grass-pasture-mowed & 15 & 13 
& 84.62
& 84.62
& 84.62
& \textbf{92.31}& 84.62
& 84.62
& \textbf{92.31}& 84.62& 84.62
& \textbf{92.31}& \textbf{92.31}\\   
        8 & Hay-windrowed & 50 & 428 
& 96.26
& 97.66& 88.55
& 96.03
& \textbf{97.90}& 96.50& 97.20& 96.73& 95.33
& 96.73
& 97.66\\   
        9 & Oats & 15 & 5 
& 60.00& 80.00& 60.00& \textbf{100.0}& \textbf{60.00}& 60.00& \textbf{100.0}& 60.00& \textbf{100.0}& \textbf{100.0}& \textbf{100.0}\\   
        10 & Soybean-no-till & 50 & 922 
& 70.07
& 72.67
& 60.41
& 70.39
& 80.15
& 75.49
& 78.52
& \textbf{82.00}& 79.83
& 77.22
& 78.52\\   
        11 & Soybean-min-till & 50 & 2405 
& 60.75
& 53.14
& 51.89
& 46.24
& 62.16
& 53.80& 58.30& \textbf{65.03}& 63.87
& 62.00& 64.57\\   
        12 & Soybean-clean & 50 & 543 
& 71.64
& 71.09
& 39.04
& 65.38
& 70.9
& 67.96
& 69.98
& 73.30& 68.69
& 76.06
& \textbf{78.08}\\   
        13 & Wheat & 50 & 155 
& 94.19
& \textbf{99.35}& 96.13
& \textbf{99.35}& 97.42
& 96.13
& \textbf{99.35}& 98.71& 98.06
& \textbf{99.35}& \textbf{99.35}\\   
        14 & Woods & 50 & 1215 
& 84.44
& 85.02
& 82.55
& 86.58
& 81.48
& 86.75
& 89.96
& 88.07& 79.51
& 89.14
& \textbf{90.95}\\   
        15 & Buildings-Grass-Trees-Drives & 50 & 336 
& 74.40& 71.43
& 47.62
& 72.02
& 54.17
& \textbf{78.27}& 67.56
& 53.57& 66.96
& 69.64
& 70.54\\   
        16 & Stone-Steel-Towers & 50 & 43 
& 93.02
& \textbf{100.0}& \textbf{100.0}& 97.67
& \textbf{100.0}& 88.37
& 97.67
& \textbf{100.0}& \textbf{100.0}& 97.67
& 97.67\\   
        \hline                                       
        & Overall Accuracy & & 
& 71.17
& 72.5
& 60.55
& 69.27
& 73.58
& 72.46
& 74.26
& 74.94& 74.95
& 76.37
& \textbf{76.90}\\   
        & Average Accuracy & & 
& 77.35
& 80.6
& 67.35
& 80.2
& 79.04
& 78.15
& 83.43
& 80.77& 82.47
& 84.84
& \textbf{85.20}\\   
        & $\kappa\times100$ & & & 67.38
& 68.98
& 55.52
& 65.39
& 70.12
& 68.85
& 70.89
& 71.62 & 71.61
& 73.22
& \textbf{73.80}\\
        \hline
    \end{tabular}
    \begin{tablenotes}
      \tiny
      \item \hspace{0.4cm} Note: All accuracy values are reported in percentage (\%). Best results are shown in \textbf{bold}.
    \end{tablenotes}
\end{table*}

\begin{enumerate}
    \item XGBoost: Implemented with the xgboost package\footnote{\url{https://xgboost.readthedocs.io/en/stable/index.html}} using a multi-class objective (\verb|multi:softprob|), a learning rate of 0.1, a maximum depth of 10, and \verb|mlogloss| for evaluation, trained for 100 rounds.
    
    \item SVC: Implemented with the scikit-learn package\footnote{\url{https://scikit-learn.org/stable/modules/generated/sklearn.svm.SVC.html}} using a radial basis function (RBF) kernel, with \verb|gamma| set to `scale', probability estimates enabled, and \verb|C| parameter values varied logarithmically from 0.0001 to 1000.

    \item SAE + SVC: An autoencoder with 5 encoding and 5 decoding layers, with dimensions reducing linearly from the spectral dimension to 64 through the encoder and expanding back through the decoder, using SiLU activations except for the final layers. The model was trained for 5000 epochs with MSE loss, Adam optimizer (learning rate 0.001), and StepLR scheduler (step size 1000, gamma 0.5). Features from the encoded space were classified using the aforementioned SVC configuration.
    
    \item 1D-CNN: The model comprises two 1D convolutional layers (128 and 16 filters) with batch normalization, ReLU activation, and a fully connected classification layer. It was trained for 1000 epochs using cross-entropy loss and the Adam optimizer (learning rate 0.001, weight decay 0.001) with a batch size of 32.

    \item CasRNN: The model employs a cascaded Gated Recurrant Unit (GRU)~\cite{chung2014empirical} architecture with two layers (hidden sizes: 128 and 256), followed by batch normalization and a fully connected classification layer, following the official Github implementation\footnote{https://github.com/RenlongHang/CasRNN}. It was trained for 300 epochs using cross-entropy loss and the Adam optimizer (learning rate 0.001) with a batch size of 64.

    \item miniGCN: Implemented following the official Github repository\footnote{\url{https://github.com/danfenghong/IEEE_TGRS_GCN}}, the model features two graph convolutional layers with batch normalization and ReLU activation, followed by a fully connected classification layer. It was trained for 200 epochs using cross-entropy loss, Adam optimizer (learning rate 0.001, weight decay 0.001), and a StepLR scheduler (step size 50, gamma = 0.5) with mini-batches of size 32. Laplacian matrices were dynamically constructed using KNN ($K=10$) and a heat kernel ($\sigma = 1.0$).

    \item ViT: An encoder-only architecture with 5 attention blocks. For the PU, Houston, and IP datasets, we used the configuration suggested by Hong et al.~\cite{hong2021spectralformer} in their GitHub implementation\footnote{\label{spectralformer_github}\url{https://github.com/danfenghong/IEEE_TGRS_SpectralFormer}}. For the KSC dataset, we trained for 3000 epochs with a learning rate of 0.0005 and no weight decay. For the PC and LK datasets, the model was trained for 800 epochs with a learning rate and weight decay of 0.005.
    
    \item SpectralFormer: For the KSC dataset, the model was trained for 1000 epochs using 3-band patches, with a learning rate and weight decay of 0.0005. For the PC and LK datasets, it was trained for 600 epochs using 3-band patches, with a learning rate and weight decay of 0.005. For the other datasets, we followed the configurations suggested by Hong et al. in their GitHub implementation\footref{spectralformer_github}.

    \item Mamba: The model comprises a spectral embedding layer projecting to 128 dimensions, three stacked Mamba selective state space blocks~\cite{gu2023mamba} (with state dimension $d_{state}=16$ and convolution size $d_{conv}=4$) with residual connections, and a classification head with dropout (p=0.2). Training utilized cross-entropy loss with Adam optimizer (learning rate 0.0001) and StepLR scheduler (step size 100, gamma=0.9). The model was trained for 1000 epochs using batch size 64 and standardized input features.
    
    \item CEnc + SVC: SymAE was trained with $d_c = d_n = 64$ for 3000 epochs on KSC and Pavia datasets, and $d_c = d_n = 128$ for 4000 epochs on IP, LK, and UH datasets. All other SymAE configurations followed the experimental setup detailed in Subsection~\ref{subsec:expsetup}. The SVC used the previously specified parameters.
    
    \item CEnc + Dense: Using the same SymAE models as CEnc + SVC, the dense classifier ($d_c \to 1024 \to 512 \to 64 \to \text{No. of classes}$) used Leaky ReLU and Bernoulli dropout (0.5). Training used cross-entropy loss, Adam optimizer ( learning rate 0.0001), batch size 32, and ran for 1000 epochs. The code for training SymAE and the classifier have been made available on github\footnote{\label{symae_github}\url{https://github.com/archieb1999/SymAE-for-HSI}}.
    
\end{enumerate}

\begin{table*}[t]
\caption{Pixel-based Classification Results from Pavia Center Dataset. \label{tab:PC1}
\vspace{-10pt}}
  \renewcommand{\arraystretch}{1}
  \setlength{\tabcolsep}{4pt}
  \scriptsize
    \centering
    \begin{tabular}{|c|c|>{\centering\arraybackslash}p{0.9cm}|>{\centering\arraybackslash}p{0.85cm}|>{\centering\arraybackslash}p{0.9cm}|>{\centering\arraybackslash}p{0.6cm}|>{\centering\arraybackslash}p{0.65cm}|>{\centering\arraybackslash}p{0.675cm}|>{\centering\arraybackslash}p{0.875cm}|>{\centering\arraybackslash}p{0.95cm}|>{\centering\arraybackslash}p{0.6cm}|>{\centering\arraybackslash}p{0.85cm}|>{\centering\arraybackslash}p{0.75cm}|>{\centering\arraybackslash}p{0.7cm}|>{\centering\arraybackslash}p{0.725cm}|}
        \hline
        No. & Class & Training Samples & Test Samples & XGBoost & SVC & SAE +SVC & 1D-CNN & CasRNN & miniGCN & ViT & Spectral -Former & Mamba & CEnc +SVC & CEnc +Dense\\
        \hline
        1 & Water& 200& 2000
& 96.65& 98.80& 98.10& 99.50& 98.60& 98.80& 99.10& \textbf{99.60}& 98.40& 99.35& 99.55\\
        2 & Trees& 200& 2000
& 91.25& 93.75& 89.15
& \textbf{96.65}& 93.45
& 92.05
& 85.45& 88.50& 93.80& 93.30& 92.85\\   
        3 & Asphalt& 200& 2000
& 53.90& 76.30& 62.85
& 74.20& 78.05
& 88.45
& 85.25& \textbf{91.70}& 74.60& 85.15& 88.50\\   
        4 & Self-Blocking Bricks& 200& 2000
& 71.70& 77.75& 75.90& 85.80& 76.65
& 79.35
& 79.45& \textbf{86.5}& 76.20& 84.00& 83.60\\   
        5 & Bitumen& 200& 2000
& \textbf{89.95}& 89.10& 87.25
& 87.55& 89.40& 87.85
& 83.80& 83.70& 88.75
& 88.75& 89.00\\   
        6 & Tiles& 200& 2000
& 87.85& 95.55& 92.25
& 97.50& 92.05
& 97.00& 94.60& 93.15& 91.20& 97.80& \textbf{97.90}\\   
        7 & Shadows& 200& 2000
& 72.15& 81.25& 76.30& \textbf{81.35}& 80.15
& 78.45
& 79.75& 80.05& 81.00& 80.95& 81.30\\   
        8 & Meadows& 200& 2000
& 94.90& 97.60& 97.75
& 98.55& 97.90& 98.70& 98.50& 98.45& 97.70& 98.50& \textbf{99.10}\\   
        9 & Bare Soil& 200& 2000& 83.65& 99.80& 99.10& 99.95& 99.95
& \textbf{100.0}& \textbf{100.0}& \textbf{100.0}& 99.8
& \textbf{100.0}& \textbf{100.0}\\   
        \hline                                       
        & Overall Accuracy & & & 82.44& 89.99& 86.52
& 91.23& 89.58
& 91.18
& 89.54& 91.29& 89.05
& 91.98& \textbf{92.42}\\   
        & Average Accuracy & & & 82.44\%& 89.99& 86.52
& 91.23& 89.58
& 91.18
& 89.54& 91.29& 89.05
& 91.98& \textbf{92.42}\\   
        & $\kappa\times100$ & & & 80.25& 88.74& 84.83
& 90.13& 88.28
& 90.08
& 88.24& 90.21& 87.68
& 90.98& \textbf{91.48} \\
        \hline
    \end{tabular}
    \begin{tablenotes}
      \tiny
      \item \hspace{0.4cm} Note: All accuracy values are reported in percentage (\%). Best results are shown in \textbf{bold}.
    \end{tablenotes}
\end{table*}

\begin{table*}[t]
\caption{Pixel-based Classification Results from Pavia University Dataset. \label{tab:PU1}
\vspace{-10pt}}
  \renewcommand{\arraystretch}{1}
  \setlength{\tabcolsep}{4pt}
  \scriptsize
    \centering
    \begin{tabular}{|c|c|>{\centering\arraybackslash}p{0.9cm}|>{\centering\arraybackslash}p{0.85cm}|>{\centering\arraybackslash}p{0.9cm}|>{\centering\arraybackslash}p{0.6cm}|>{\centering\arraybackslash}p{0.65cm}|>{\centering\arraybackslash}p{0.675cm}|>{\centering\arraybackslash}p{0.875cm}|>{\centering\arraybackslash}p{0.95cm}|>{\centering\arraybackslash}p{0.6cm}|>{\centering\arraybackslash}p{0.85cm}|>{\centering\arraybackslash}p{0.75cm}|>{\centering\arraybackslash}p{0.7cm}|>{\centering\arraybackslash}p{0.725cm}|}
        \hline
        No. & Class & Training Samples & Test Samples & XGBoost & SVC & SAE +SVC & 1D-CNN & CasRNN & miniGCN & ViT & Spectral -Former & Mamba & CEnc +SVC & CEnc +Dense\\
        \hline
        1 & Asphalt & 548 & 6304 & 79.93& 84.52& 81.42
& 80.25& 77.38
& 85.07
& 77.19& 87.34& 87.17
& 87.12& \textbf{88.06}\\
        2 & Meadows & 540 & 18146 & 57.76 & 68.81 
& 76.28
& 65.31 
& 68.11
& 74.98
& 67.32 & 77.12 
& 70.76
& 79.58 & \textbf{79.61}
\\   
        3 & Gravel & 392 & 1815 & 51.90& 68.98& 67.33
& 69.15& 49.59
& 61.27
& 67.77& 54.93& 70.58
& 69.15& \textbf{71.96}\\   
        4 & Trees & 524 & 2912 & 98.63& 98.35& 97.63
& 81.63& \textbf{98.80}& 87.84
& 95.95& 97.73& 95.95
& 95.05& 94.78\\   
        5 & Painted metal sheets & 265 & 1113 & 99.46& 99.37& 98.92
& \textbf{99.64}& 99.01
& \textbf{99.64}& 99.37& 99.37& 99.37
& 99.55& \textbf{99.64}\\   
        6 & Bare soil & 532 & 4572 & 83.95& 94.14& 88.95
& 94.27& 73.53
& 88.54
& 89.85& 92.67& 90.73
& 94.75& \textbf{96.50}\\   
        7 & Bitumen & 375 & 981 & 84.20& 90.32& 86.85
& 87.05& 87.46
& 88.58
& 87.46& 86.14& \textbf{92.56}& 91.64& 91.03\\   
        8 & Self blocking bricks & 514 & 3364 & 91.05& 92.42& 92.36
& 86.56& 89.57
& \textbf{96.14}& 85.17& 93.73& 90.99
& 91.26& 90.49\\   
        9 & Shadows & 231 & 795 & 97.11& 99.37& 97.36
& 98.62& 90.06
& \textbf{100.0}& 97.36& 67.55& 95.60& 96.35& 98.74\\   
        \hline                                       
        & Overall Accuracy & & & 72.35& 80.31& 82.35
& 76.28& 75.16
& 81.73
& 77.04& 83.05& 80.98
& 85.32& \textbf{85.76}\\   
        & Average Accuracy & & & 82.67& 88.48& 87.46
& 84.72& 81.50& 86.90& 85.37& 84.06& 88.19
& 89.38& \textbf{90.09}\\   
        & $\kappa\times100$ & & & 65.64 & 75.07 & 77.3
& 70.10 & 68.35
& 76.37
& 70.99 & 78.11 & 75.78
& 80.96 & \textbf{81.53}\\
        \hline
    \end{tabular}
    \begin{tablenotes}
      \tiny
      \item \hspace{0.4cm} Note: All accuracy values are reported in percentage (\%). Best results are shown in \textbf{bold}.
    \end{tablenotes}
\end{table*}

\begin{table*}[h]
\caption{Pixel-based Classification Results from Houston2013 Dataset. \label{tab:H1}
\vspace{-10pt}}
  \renewcommand{\arraystretch}{1}
  \setlength{\tabcolsep}{4pt}
  \scriptsize
    \centering
    \begin{tabular}{|c|c|>{\centering\arraybackslash}p{0.9cm}|>{\centering\arraybackslash}p{0.85cm}|>{\centering\arraybackslash}p{0.9cm}|>{\centering\arraybackslash}p{0.6cm}|>{\centering\arraybackslash}p{0.65cm}|>{\centering\arraybackslash}p{0.675cm}|>{\centering\arraybackslash}p{0.875cm}|>{\centering\arraybackslash}p{0.95cm}|>{\centering\arraybackslash}p{0.6cm}|>{\centering\arraybackslash}p{0.85cm}|>{\centering\arraybackslash}p{0.75cm}|>{\centering\arraybackslash}p{0.7cm}|>{\centering\arraybackslash}p{0.725cm}|}
        \hline
        No. & Class & Training Samples & Test Samples & XGBoost & SVC & SAE +SVC & 1D-CNN & CasRNN & miniGCN & ViT & Spectral -Former & Mamba & CEnc +SVC & CEnc +Dense\\
        \hline
        1 & Healthy Grass& 198& 1053& \textbf{89.74}& 83.67& 82.72
& 89.08& 83.86
& 87.84
& 83.76& 86.13& 81.96
& 84.24& 84.05\\
 2 & Stressed Grass& 190& 1064& 97.09& 96.90& 94.08
& 97.37& 98.12
& 97.93
& 97.18& 96.52& 97.93
& 98.59&\textbf{98.68}\\
 3 & Synthetic Grass& 192& 505& 97.43& 99.80& \textbf{100.0}& 99.80& 99.41
& 98.81
& 99.80& 99.80& 99.80& 99.80&\textbf{100.0}\\
        4 & Tree& 188& 1056& 94.60& 98.30& 97.44
& 91.76& 96.97
& 96.4
& \textbf{98.77}& 97.63& 98.01
& 97.54& 96.88\\   
        5 & Soil& 186& 1056& 93.94& 97.73& 97.25
& 97.44& 97.35
& 98.77
& 98.01& \textbf{98.86}& 97.44
& \textbf{98.86}& \textbf{98.86}\\   
        6 & Water& 182& 143& 96.50& 95.10& 94.41
& 93.71& \textbf{98.60}& 89.51
& \textbf{98.60}& 96.50& 95.1
& 95.10& \textbf{98.60}\\   
        7 & Residential& 196& 1072& 81.53& 83.40& 74.72
& 72.67& 81.62
& \textbf{86.85}
& 77.52& 86.01& 85.17
& 82.84& 85.35\\   
        8 & Commercial& 191& 1053& 49.76& 51.76& 59.16
& 59.92& 60.97
& 64.58
& 56.98& 52.90& 50.90& \textbf{77.97}& 60.78\\   
        9 & Road& 193& 1059& 68.65& 76.68& 71.95
& 77.71& 65.06
& \textbf{87.44}& 67.52& 64.59& 69.41
& 80.64& 82.63\\   
        10 & Highway& 191& 1036& 61.10& 81.37& 61.68
& 77.03& 48.84
& 66.41
& 66.70& \textbf{93.34}& 79.34
& 78.76\% & 86.20\\   
        11 & Railway& 181& 1054& 77.42& \textbf{91.37}& 84.35
& 84.35& 73.43
& 90.89
& 68.41& 73.15& 71.63
& 87.95& 87.86\\   
        12 & Parking Lot1& 192& 1041& 55.33& 68.78& 69.84
& 72.53& 57.54
& 66.09
& 50.82& 58.50& 70.80& \textbf{83.57}& 83.38\\   
        13 & Parking Lot2& 184& 285& 68.42& 69.12& 68.42
& 70.53& 67.37
& 49.82
& 63.16& 72.98& 70.18
& 77.19& \textbf{78.25}\\   
        14& Tennis Court& 181& 247& 97.57& \textbf{100.0}& \textbf{100.0}& 98.38& 99.19
& 97.98
& 99.19& 98.79& 99.60& 98.79& 98.79\\   
        15& Running Track& 187& 473& 95.56& 97.89& 97.25
& \textbf{98.73}& 95.98
& 98.31
& 98.52& \textbf{98.73}& 91.33
& 97.46& 97.67\\   
        \hline                                       
        & Overall Accuracy & & & 79.01& 84.46& 81.26
& 83.59& 78.69
& 85.06
& 78.85& 82.64& 81.86
& \textbf{88.14}& 87.68\\   
        & Average Accuracy & & & 81.64& 86.12& 83.55
& 85.40& 81.62
& 85.18
& 81.66& 84.96& 83.91
& \textbf{89.29}& 89.20\\   
        & $\kappa\times100$ & & & 77.28 & 83.16 & 79.70& 82.20 & 76.95
& 83.77
& 77.10 & 81.20 & 80.36
& \textbf{87.14} & 86.62\\
        \hline
    \end{tabular}
    \begin{tablenotes}
      \tiny
      \item \hspace{0.4cm} Note: All accuracy values are reported in percentage (\%). Best results are shown in \textbf{bold}.
    \end{tablenotes}
\end{table*}

\begin{table*}[h]
\caption{Pixel-based Classification Results from LongKou Dataset. \label{tab:LK1}\vspace{-10pt}}
  \renewcommand{\arraystretch}{1}
  \setlength{\tabcolsep}{4pt}
  \scriptsize
    \centering
    \begin{tabular}{|c|c|>{\centering\arraybackslash}p{0.9cm}|>{\centering\arraybackslash}p{0.85cm}|>{\centering\arraybackslash}p{0.9cm}|>{\centering\arraybackslash}p{0.6cm}|>{\centering\arraybackslash}p{0.65cm}|>{\centering\arraybackslash}p{0.675cm}|>{\centering\arraybackslash}p{0.875cm}|>{\centering\arraybackslash}p{0.95cm}|>{\centering\arraybackslash}p{0.6cm}|>{\centering\arraybackslash}p{0.85cm}|>{\centering\arraybackslash}p{0.75cm}|>{\centering\arraybackslash}p{0.7cm}|>{\centering\arraybackslash}p{0.725cm}|}
        \hline
        No. & Class & Training Samples & Test Samples & XGBoost & SVC & SAE +SVC & 1D-CNN & CasRNN & miniGCN & ViT & Spectral -Former & Mamba & CEnc +SVC & CEnc +Dense\\
        \hline
        1 & Corn & 54 & 34457 & 85.79 & 91.28 & 84.82 & 93.01 & 84.28 & 90.48 & 70.83 & 81.70 & 93.41 & 95.95 & \textbf{96.25}\\
        2 & Cotton & 53 & 8321 & 40.06 & 76.54 & 61.00 & 16.72 & 40.21 & \textbf{91.37} & 34.82 & 39.86 & 68.32 & 88.82 & 90.39\\   
        3 & Sesame & 54 & 2977 & 51.46 & 60.43 & 52.44 & 64.46 & 51.66 & \textbf{78.54}& 55.76 & 57.71 & 53.48 & 71.11 & 71.78 \\   
        4 & Broad-Leaf Soybean & 53 & 63159 & 70.23 & 77.47 & 70.98 & 76.00 & 75.28 & 82.93 & 72.52 & 77.97 & 76.12 & 85.29 & \textbf{85.66}\\   
        5 & Narrow-Leaf Soybean & 54 & 4097 & 70.42 & 92.95 & 88.41 & 95.78 & 71.05 & 91.07 & 80.82 & 89.55 & 85.53 & 96.22 & \textbf{96.34}\\   
        6 & Rice & 50 & 11804 & 72.37 & 91.69 & 94.89 & 93.88 & 80.79 & \textbf{98.03} & 85.55 & 85.17 & 88.93 & 97.05 & 98.01\\   
        7 & Water & 53 & 67003 & 97.06 & 99.94 & 99.97 & \textbf{99.99}& \textbf{99.99}& 99.96 & 99.98 & \textbf{99.99}& 99.95 & 99.73 & 99.98 \\   
        8 & Roads and Houses & 53 & 7071 & 71.84 & 76.04 & 65.51 & 56.67 & 73.50 & 79.73& \textbf{87.72} & 60.09 & 72.78 & 79.73& 74.77 \\   
        9 & Mixed Weed & 51 & 5178 & 55.47 & 57.69 & 51.95 & 70.03 & 46.06 & 52.41 & 44.09 & 55.00 & 64.48 & \textbf{81.48}& 79.47 \\
        \hline                                       
        &Overall Accuracy & & & 79.97 & 87.47 & 83.22 & 84.77 & 82.57 & 90.23 & 80.20 & 83.43 & 86.73 & 92.38 & \textbf{92.53}\\   
        & Average Accuracy & & & 68.30 & 80.45 & 74.44 & 74.06 & 69.20 & 84.95 & 70.23 & 71.89 & 78.11 & \textbf{88.38}& 88.07 \\   
        & $\kappa\times100$ & & & 74.56 & 83.93 & 78.69 & 80.50 & 77.70 & 87.34 & 74.85 & 78.67 & 82.98 & 90.14 & \textbf{90.32}\\
        \hline
    \end{tabular}
    \begin{tablenotes}
      \tiny
      \item \hspace{0.4cm} Note: All accuracy values are reported in percentage (\%). Best results are shown in \textbf{bold}.
    \end{tablenotes}
\end{table*}

\begin{figure}[!bp]
\captionsetup{font=footnotesize}
  \centering
    \includegraphics[width=0.8\linewidth]{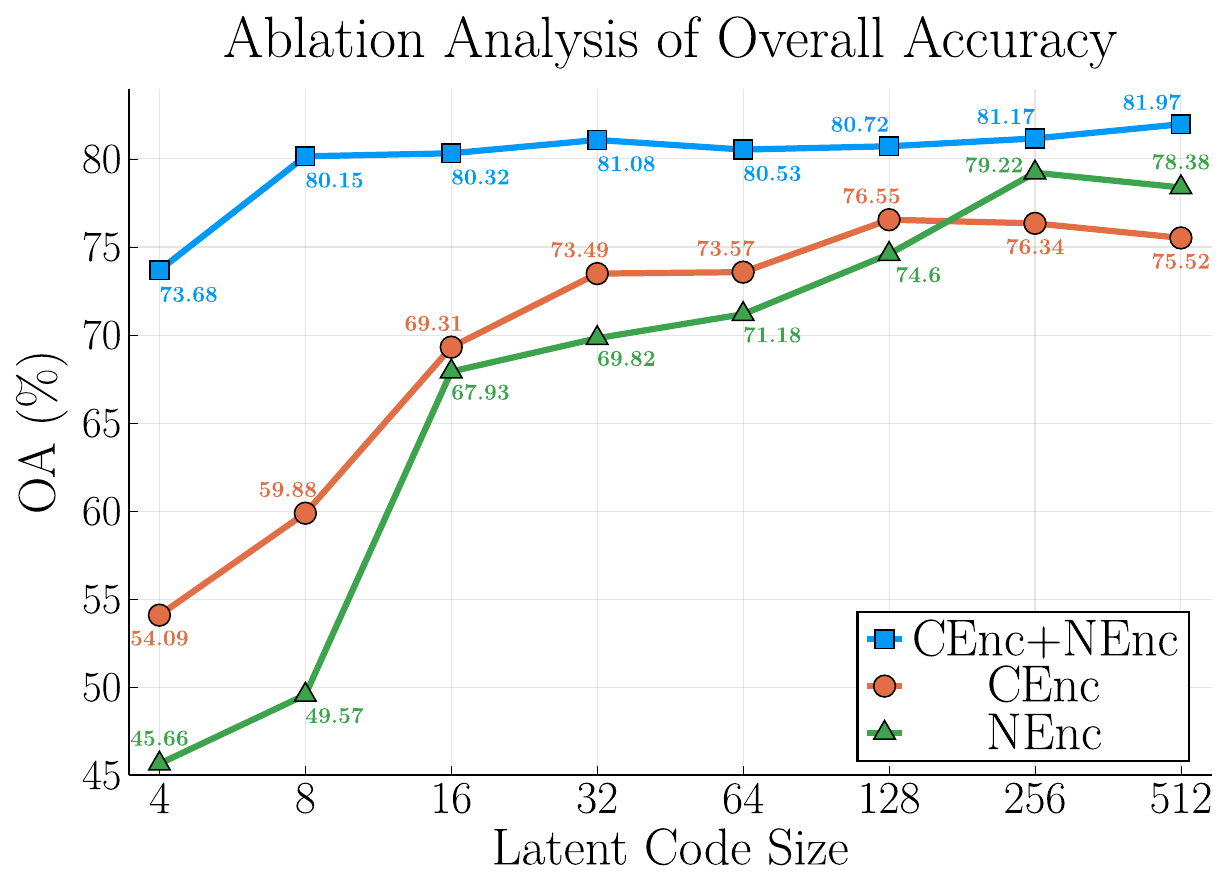}
  \caption{Ablation study results for classes 3-6 in the KSC scene, showing the impact of using both $\CEnc$ and $\NEnc$ encoders on classification performance. The results demonstrate reduced sensitivity to latent code size variations and improved accuracy when both encoders are employed concurrently.}
  \label{fig:Ablation}
\end{figure}

The results of the classification experiments on the KSC, PU, PC, IP, UH and LK datasets are presented in Tables~\ref{tab:KSC1},\ref{tab:PU1},\ref{tab:PC1},\ref{tab:IP}, \ref{tab:H1} and~\ref{tab:LK1} respectively. Across all datasets, the $\CEnc$-based methods consistently deliver superior performance, with the best $\CEnc$-based classifier achieving, on average, 2.13 and 2.36 percentage points higher Overall Accuracy (OA) and Average Accuracy (AA), respectively, and a 2.51 point increase in the scaled Kappa coefficient ($\kappa \times 100$), compared to the best method not using CEnc. These improvements support our premise (Premise~\ref{point:premise}, Section~\ref{sec:intro}) that isolating coherent features can enhance spectral classification. The results suggest SymAE's ability to extract features that are more robust to spectral variability compared to the other methods tested in this study.

\subsubsection*{Ablation Analysis}

We conducted an ablation study focusing on classes 3-6 from Table \ref{tab:KSC1} of the KSC scene, identified as the most challenging to discriminate. For this analysis, all networks underwent training for 4000 epochs, each containing 64 batches with a batch size of 256. Post-training, classification was conducted using latent codes in a similar dense neural network with dropout regularization as previously used. In configurations where both $\CEnc$ and $\NEnc$ were employed, the latent code size was equally divided between them. Note that when only $\NEnc$ is used, it effectively functions as a stacked autoencoder with stochastic regularization. The results, depicted in Figure~\ref{fig:Ablation}, indicate that utilizing both encoders concurrently results in reduced sensitivity to variations in latent code size and consistently yields higher accuracy.
This improvement, along with the observed superiority of SymAE + SVC over SAE + SVC in pixel-based classification experiments, can be attributed to the structured disentanglement enforced in SymAE, which enables more effective separation of spectral features.

\begin{figure}[!bp]
    \centering
    \begin{subfigure}{0.142\textwidth}
        \centering
        \includegraphics[width=\linewidth]{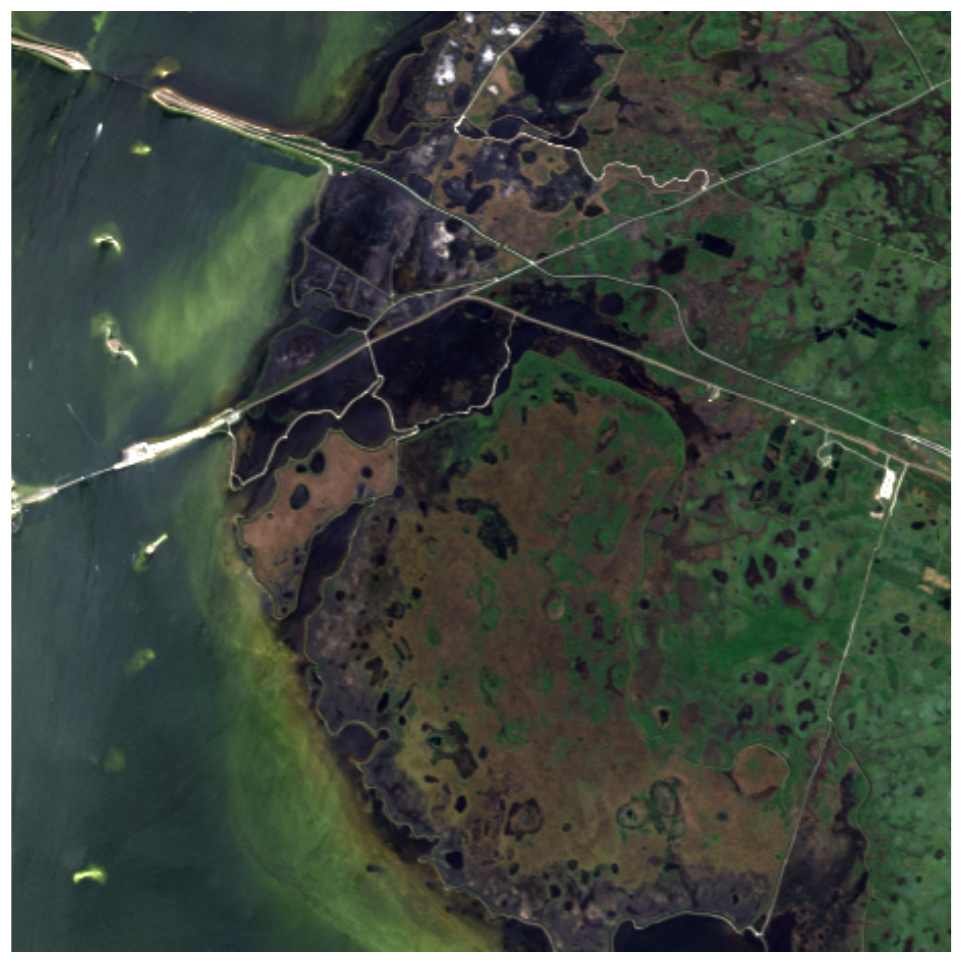}
        \caption{}
    \end{subfigure}%
    \begin{subfigure}{0.142\textwidth}
        \centering
        \includegraphics[width=\linewidth]{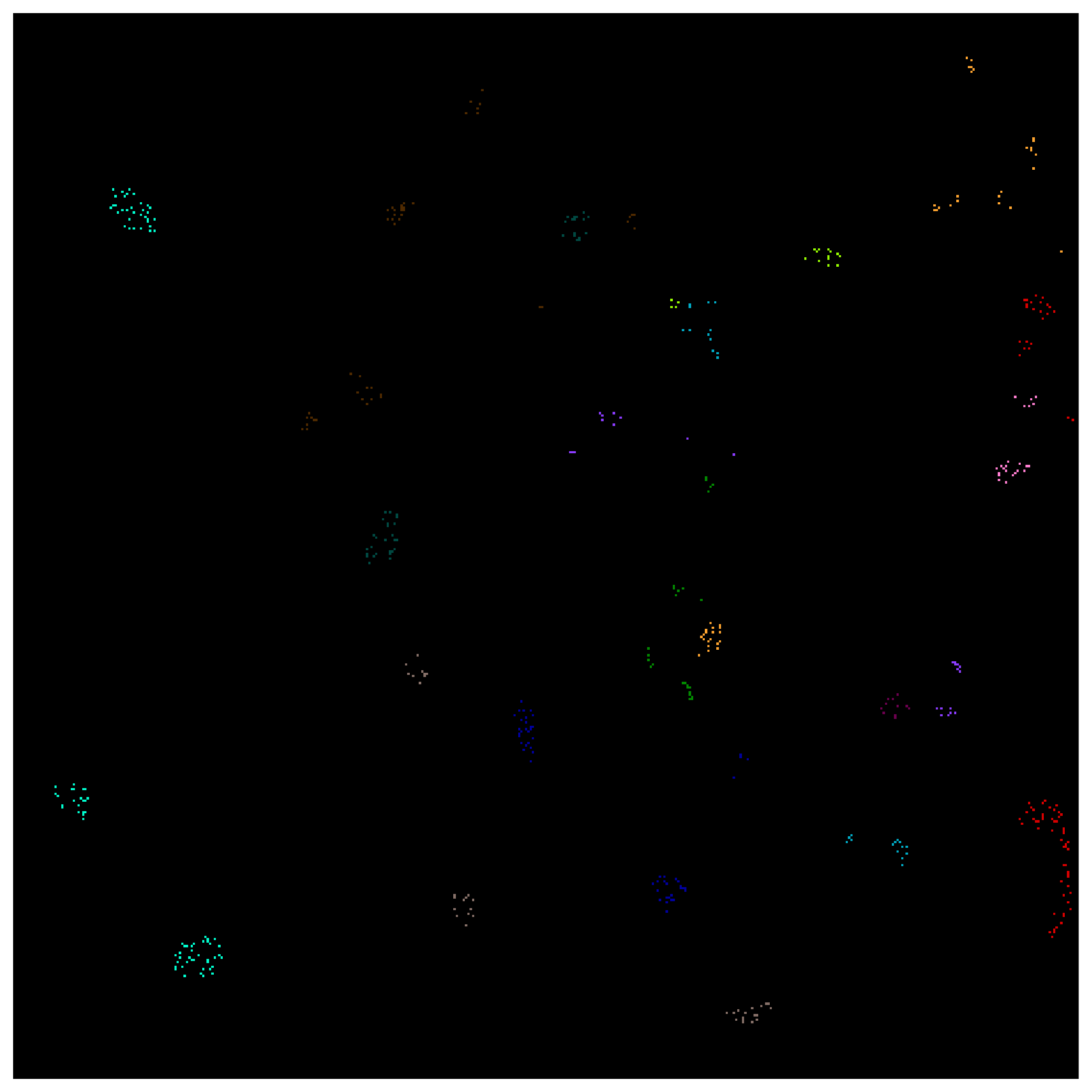}
        \caption{}
    \end{subfigure}%
    \begin{subfigure}{0.142\textwidth}
        \centering
        \includegraphics[width=\linewidth]{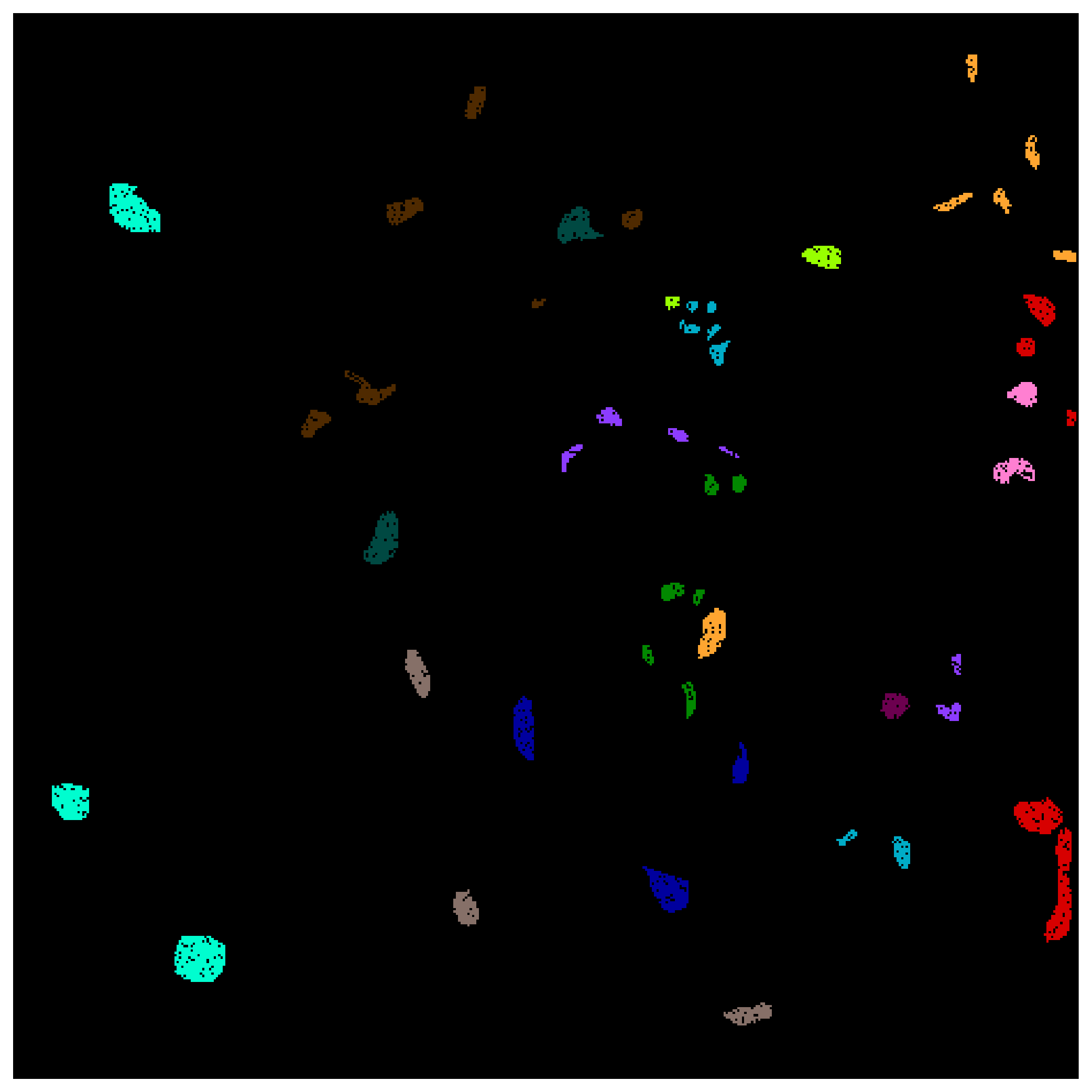}
        \caption{}
    \end{subfigure}%
    \begin{subfigure}{0.142\textwidth}
        \centering
        \includegraphics[width=\linewidth]{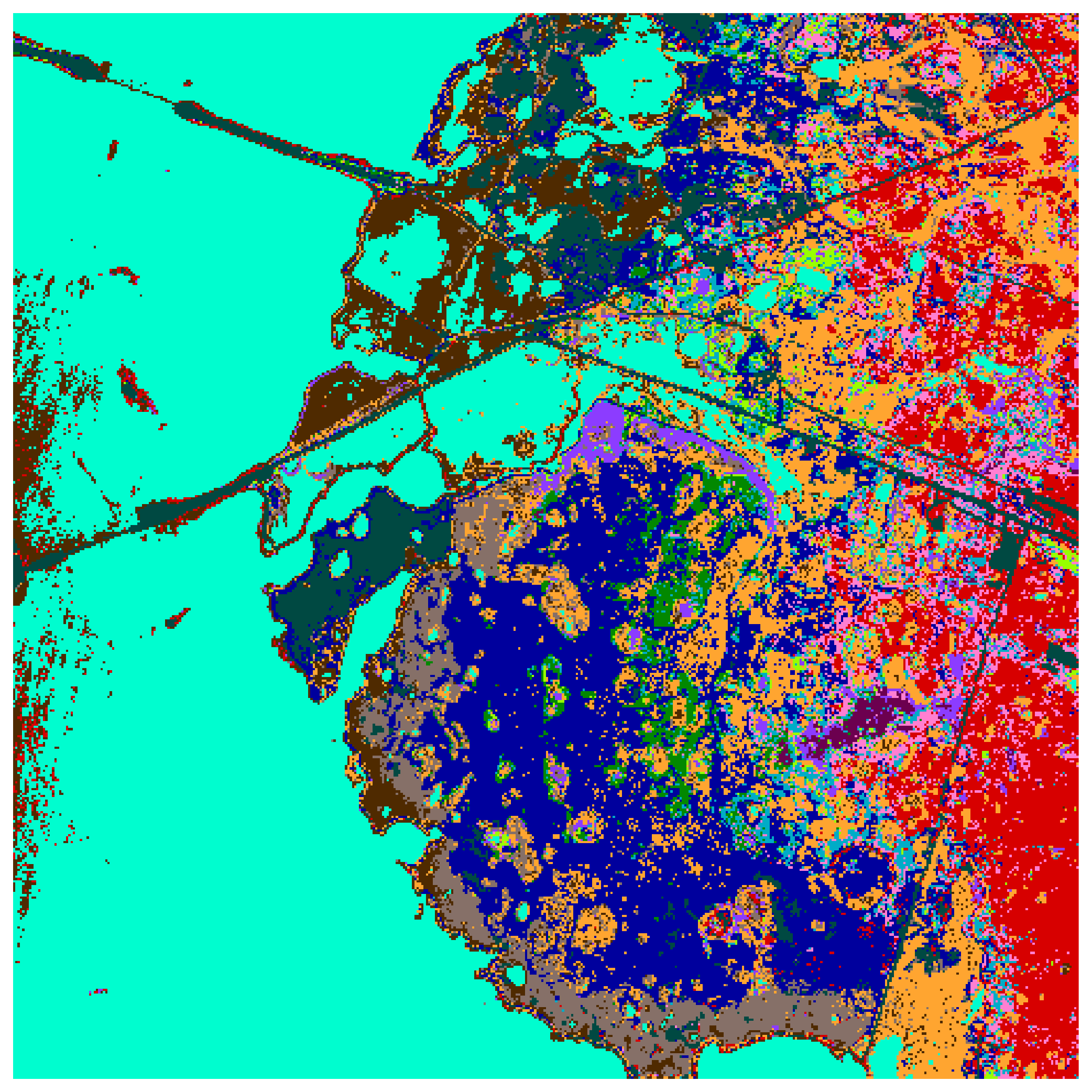}
        \caption{}
    \end{subfigure}%
    \begin{subfigure}{0.142\textwidth}
        \centering
        \includegraphics[width=\linewidth]{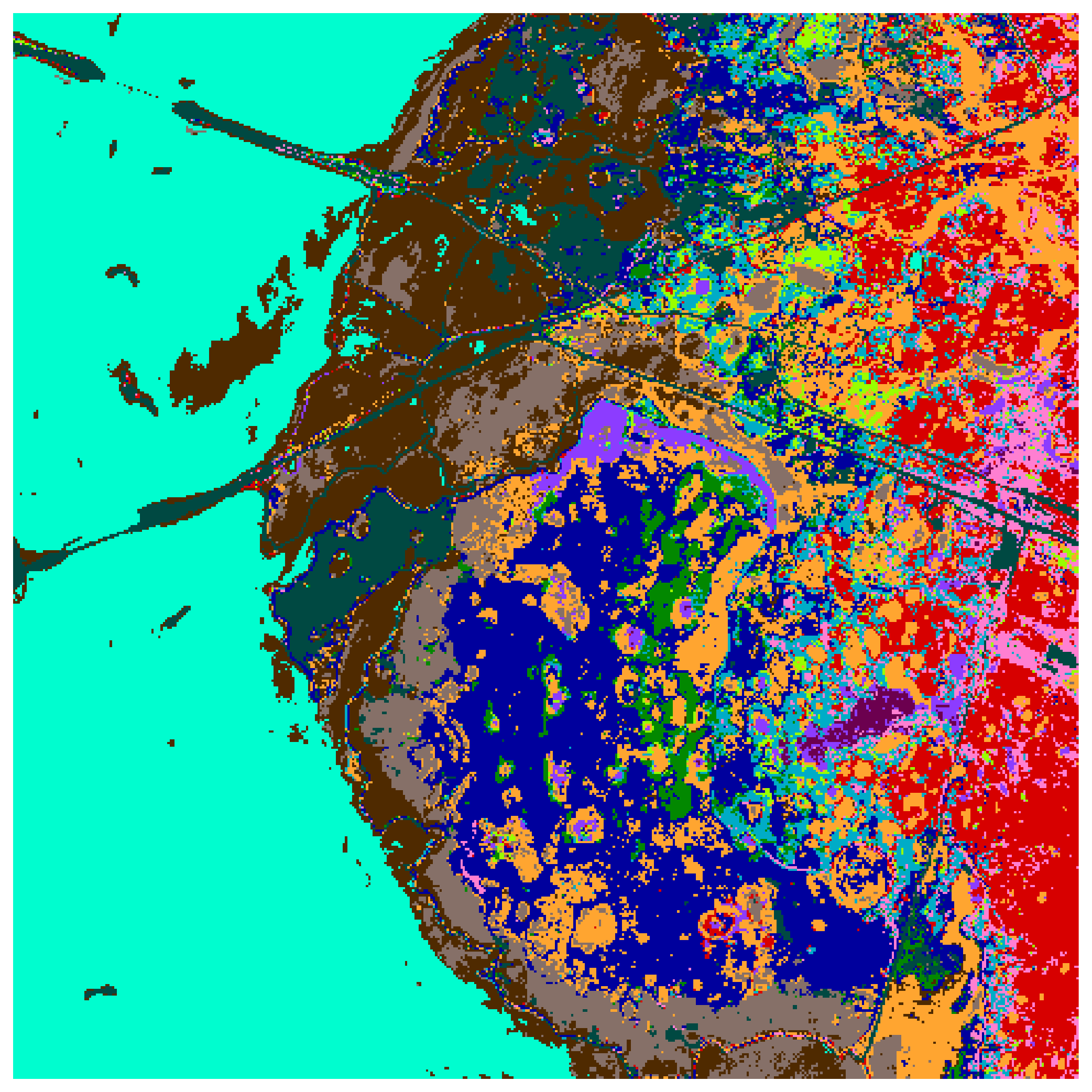}
        \caption{}
    \end{subfigure}%
    \begin{subfigure}{0.142\textwidth}
        \centering
        \includegraphics[width=\linewidth]{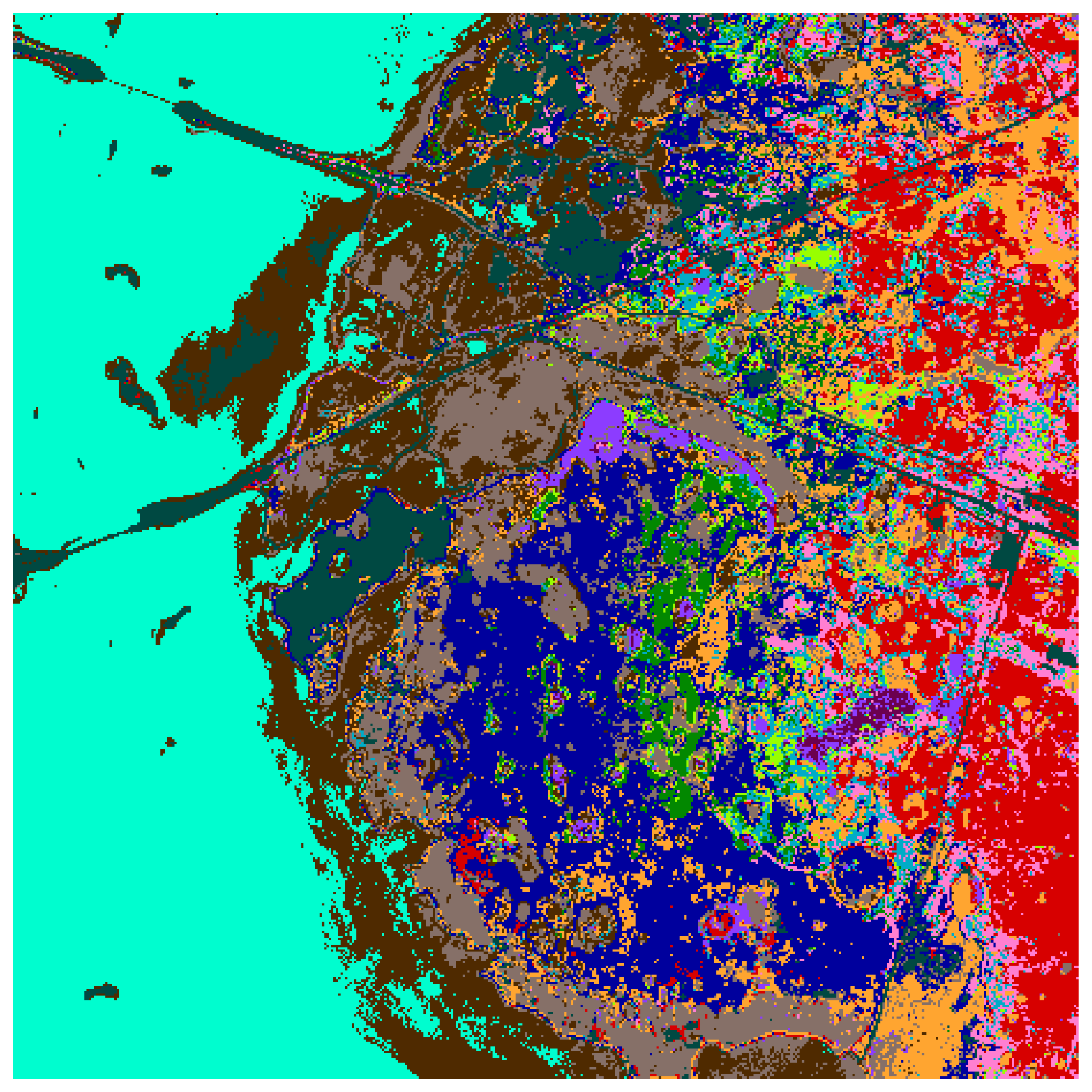}
        \caption{}
    \end{subfigure}%
    \begin{subfigure}{0.142\textwidth}
        \centering
        \includegraphics[width=\linewidth]{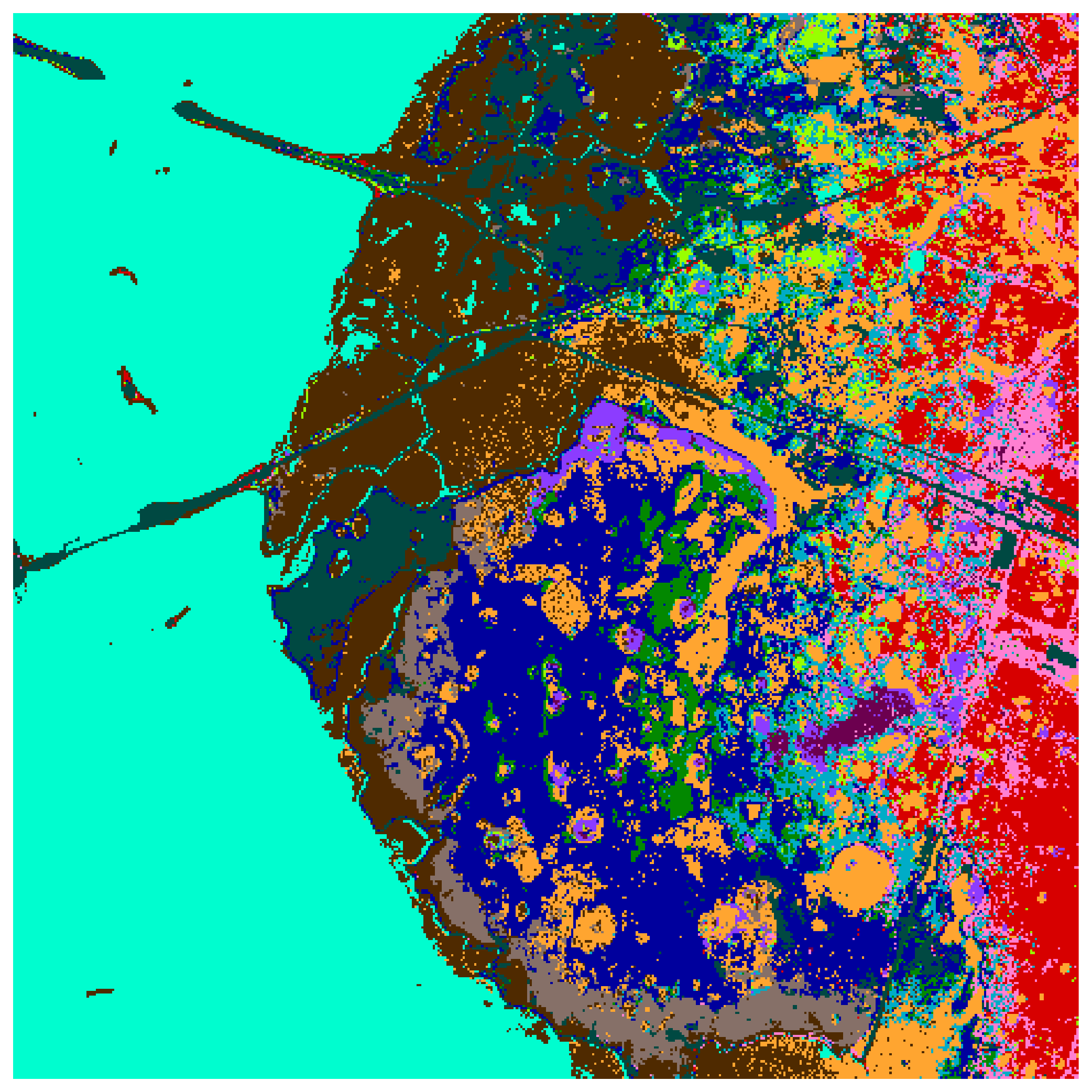}
        \caption{}
    \end{subfigure}%
    \vspace{0cm} 
    \begin{subfigure}{0.142\textwidth}
        \centering
        \includegraphics[width=\linewidth]{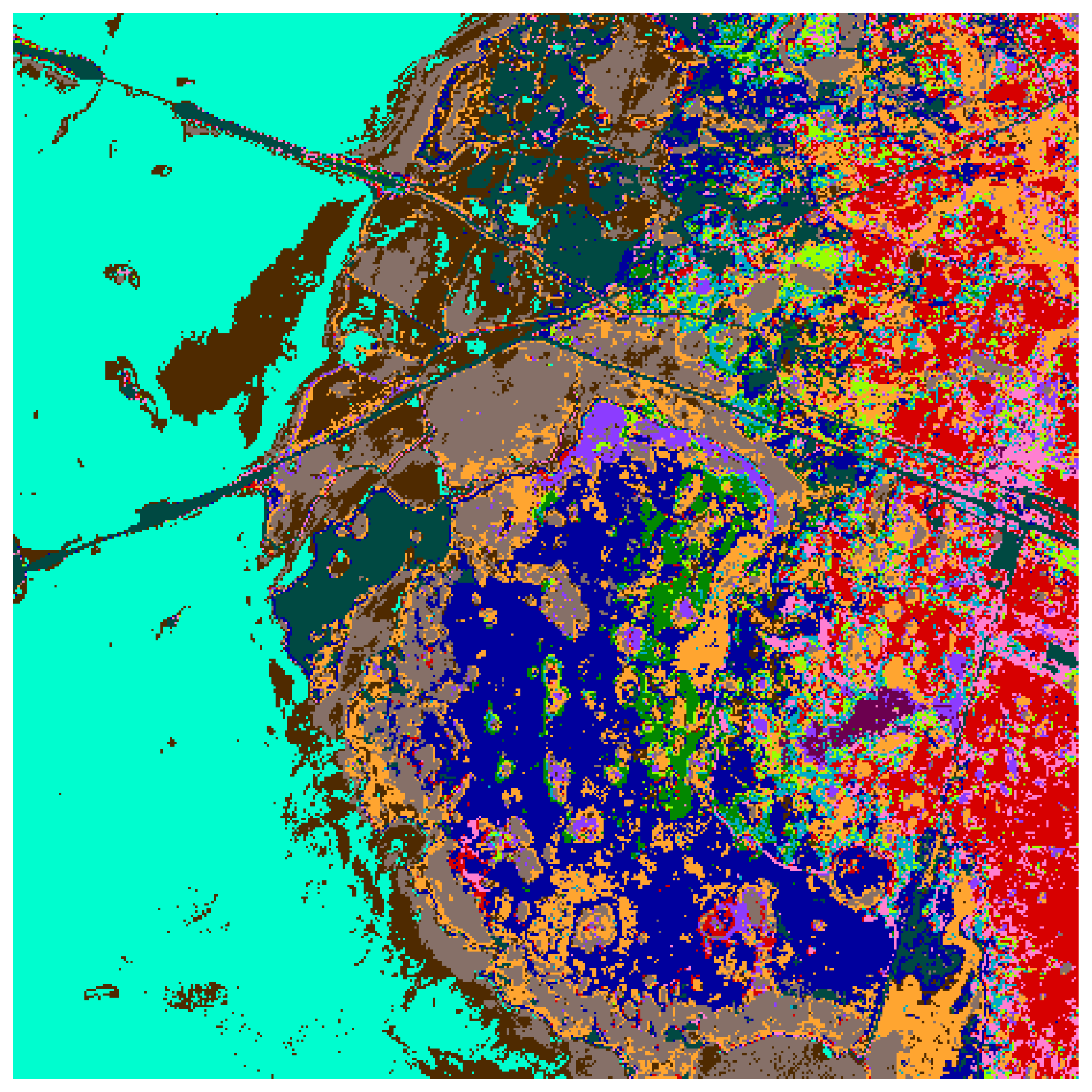}
        \caption{}
    \end{subfigure}%
    \begin{subfigure}{0.142\textwidth}
        \centering
        \includegraphics[width=\linewidth]{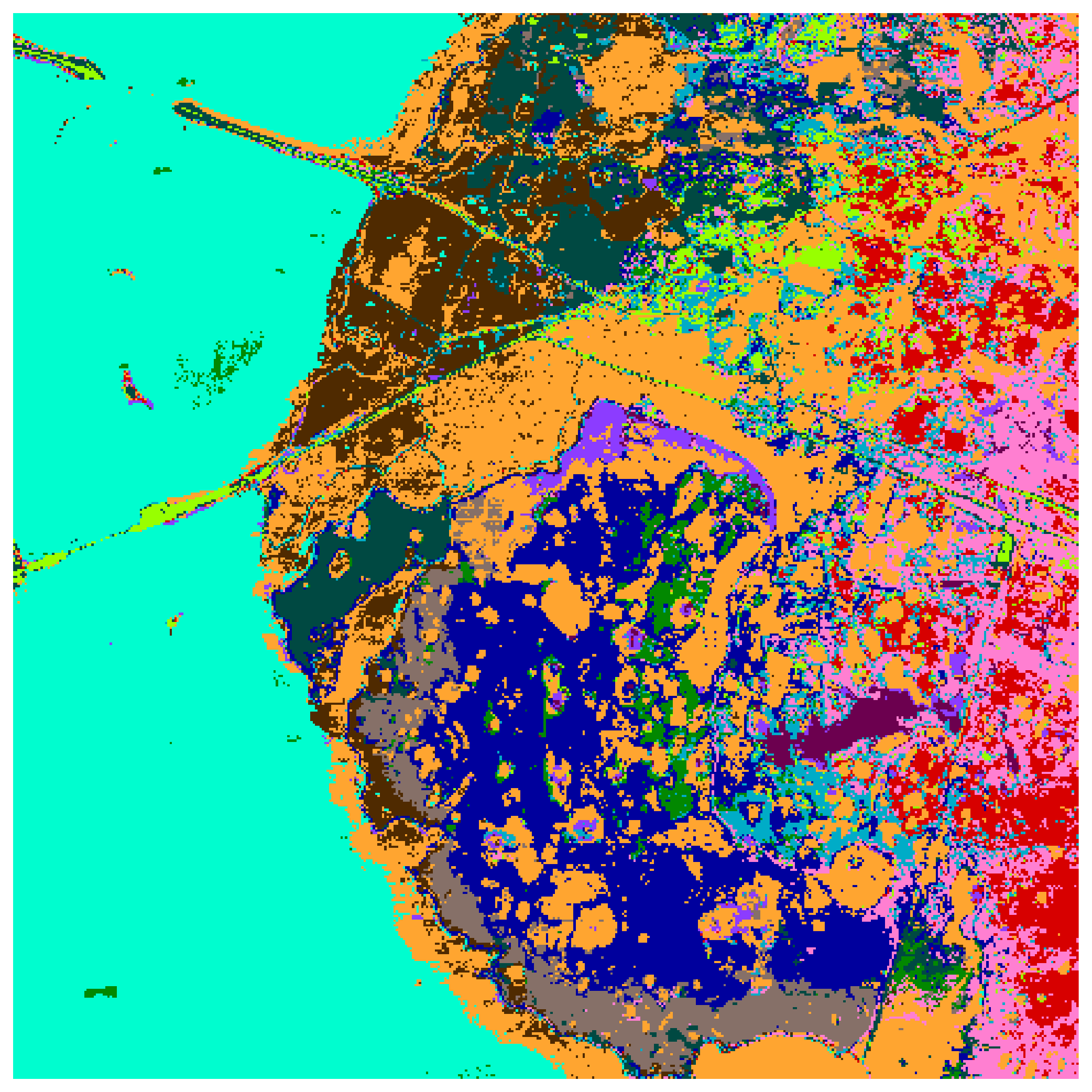}
        \caption{}
    \end{subfigure}%
    \begin{subfigure}{0.142\textwidth}
        \centering
        \includegraphics[width=\linewidth]{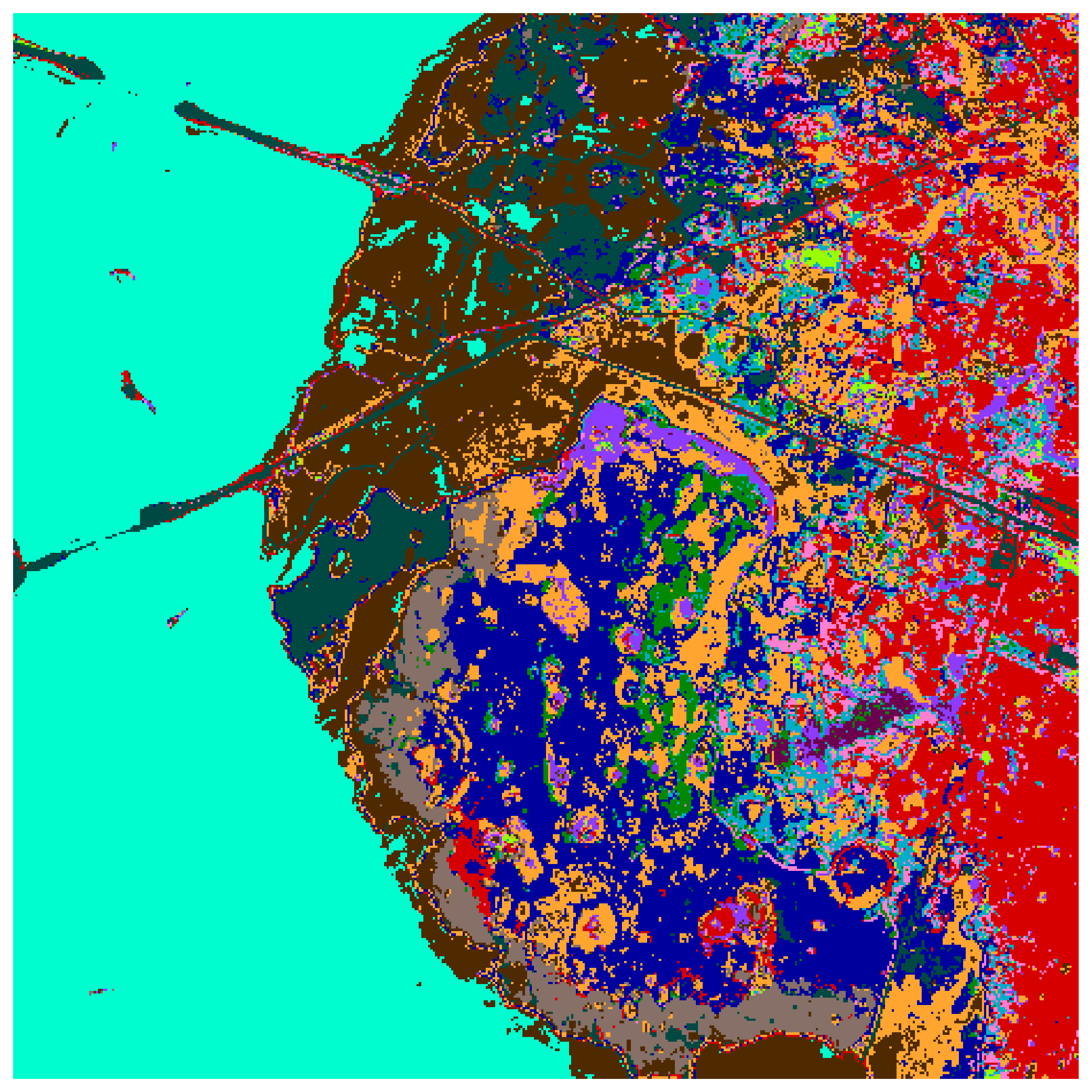}
        \caption{}
    \end{subfigure}%
    \begin{subfigure}{0.142\textwidth}
        \centering
        \includegraphics[width=\linewidth]{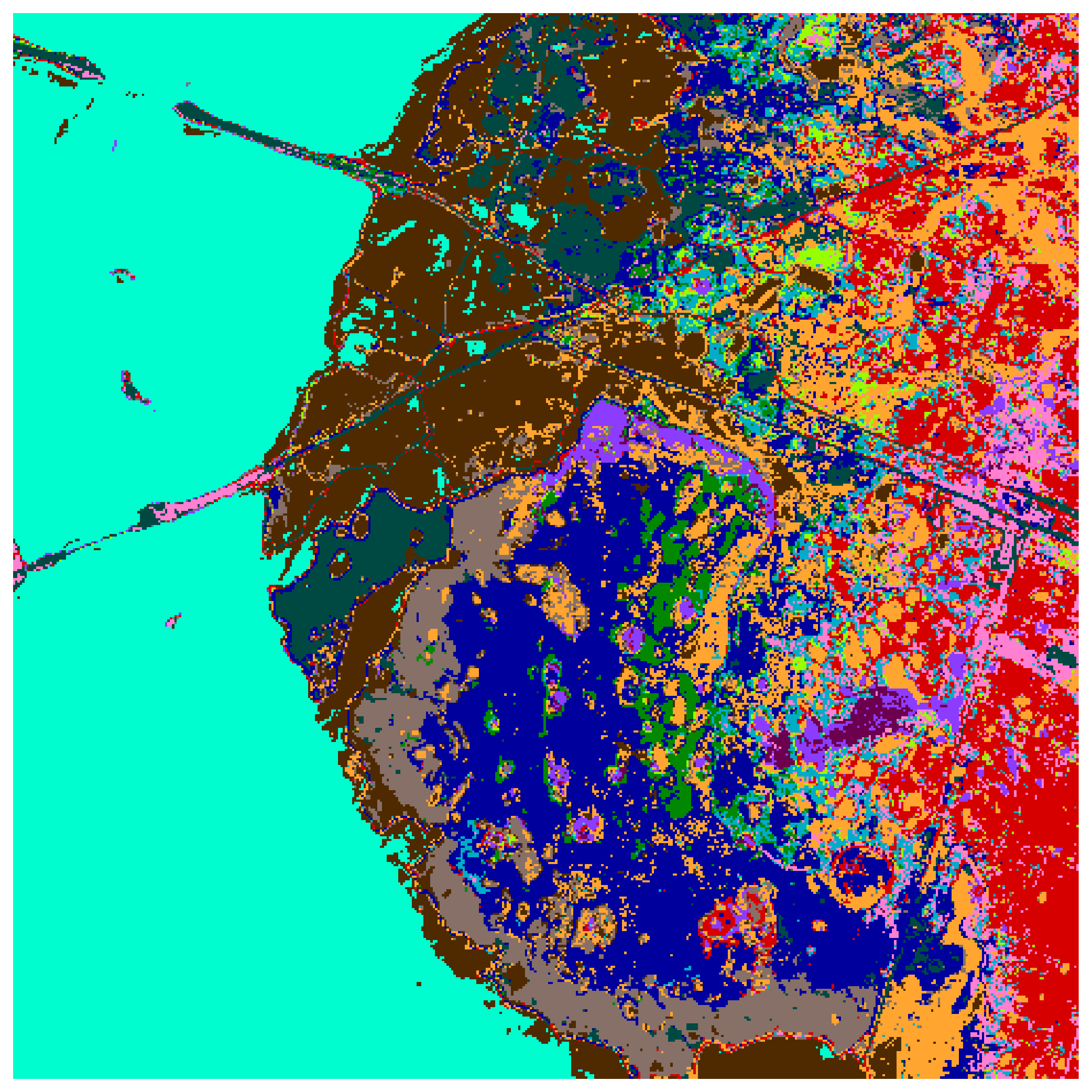}
        \caption{}
    \end{subfigure}%
    \begin{subfigure}{0.142\textwidth}
        \centering
        \includegraphics[width=\linewidth]{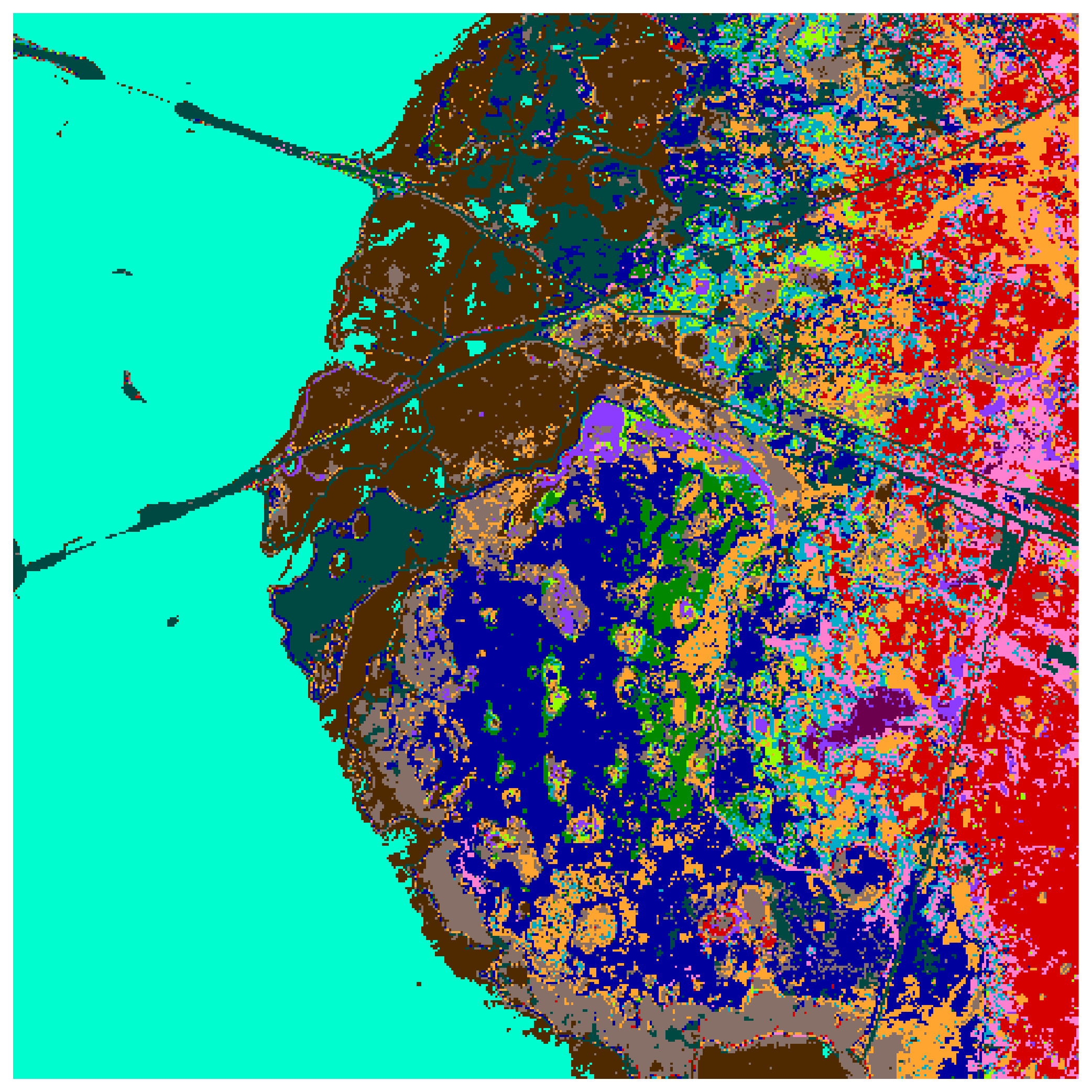}
        \caption{}
    \end{subfigure}%
    \begin{subfigure}{0.142\textwidth}
        \centering
        \includegraphics[width=\linewidth]{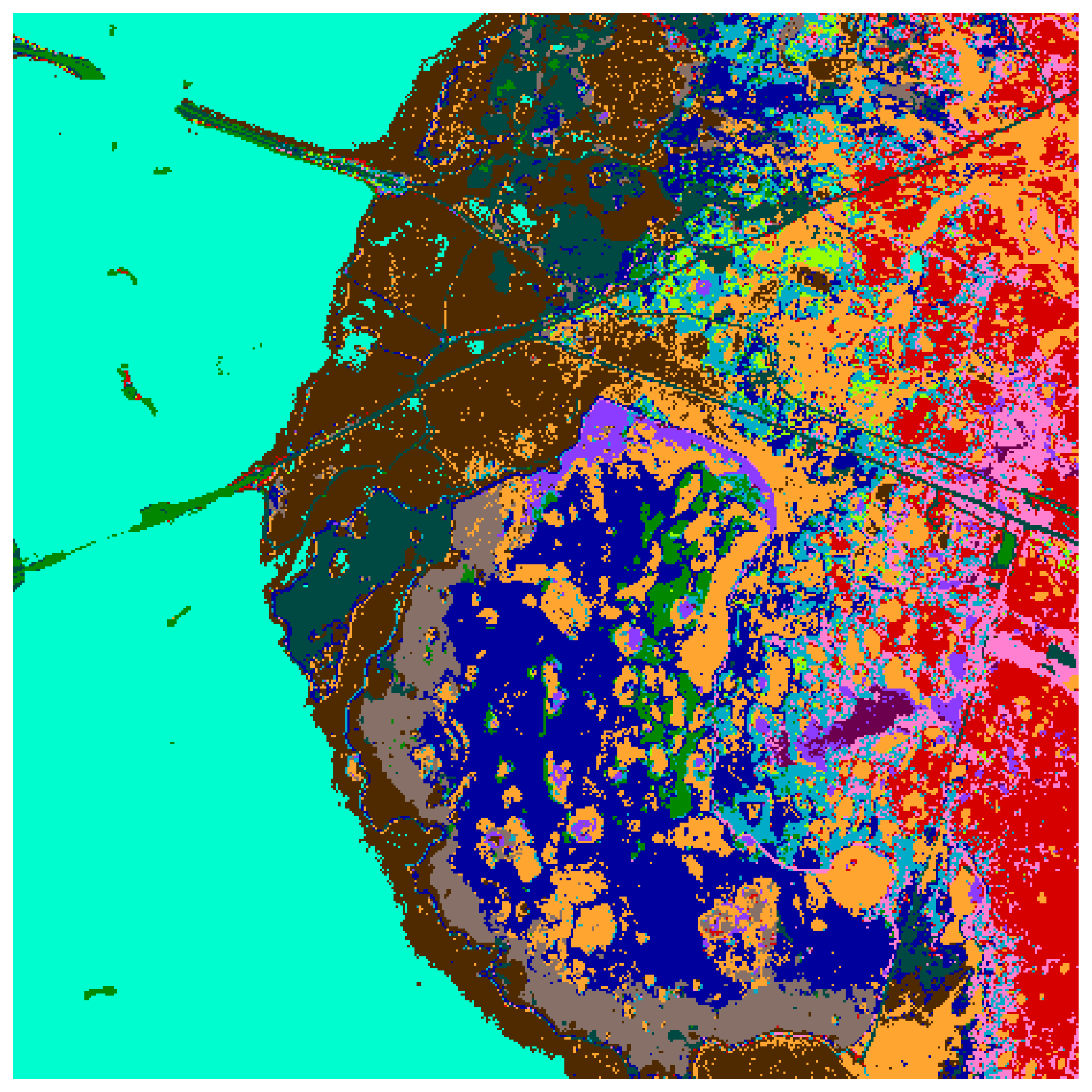}
        \caption{}
    \end{subfigure}%
    \begin{subfigure}{0.142\textwidth}
        \centering
        \includegraphics[width=\linewidth]{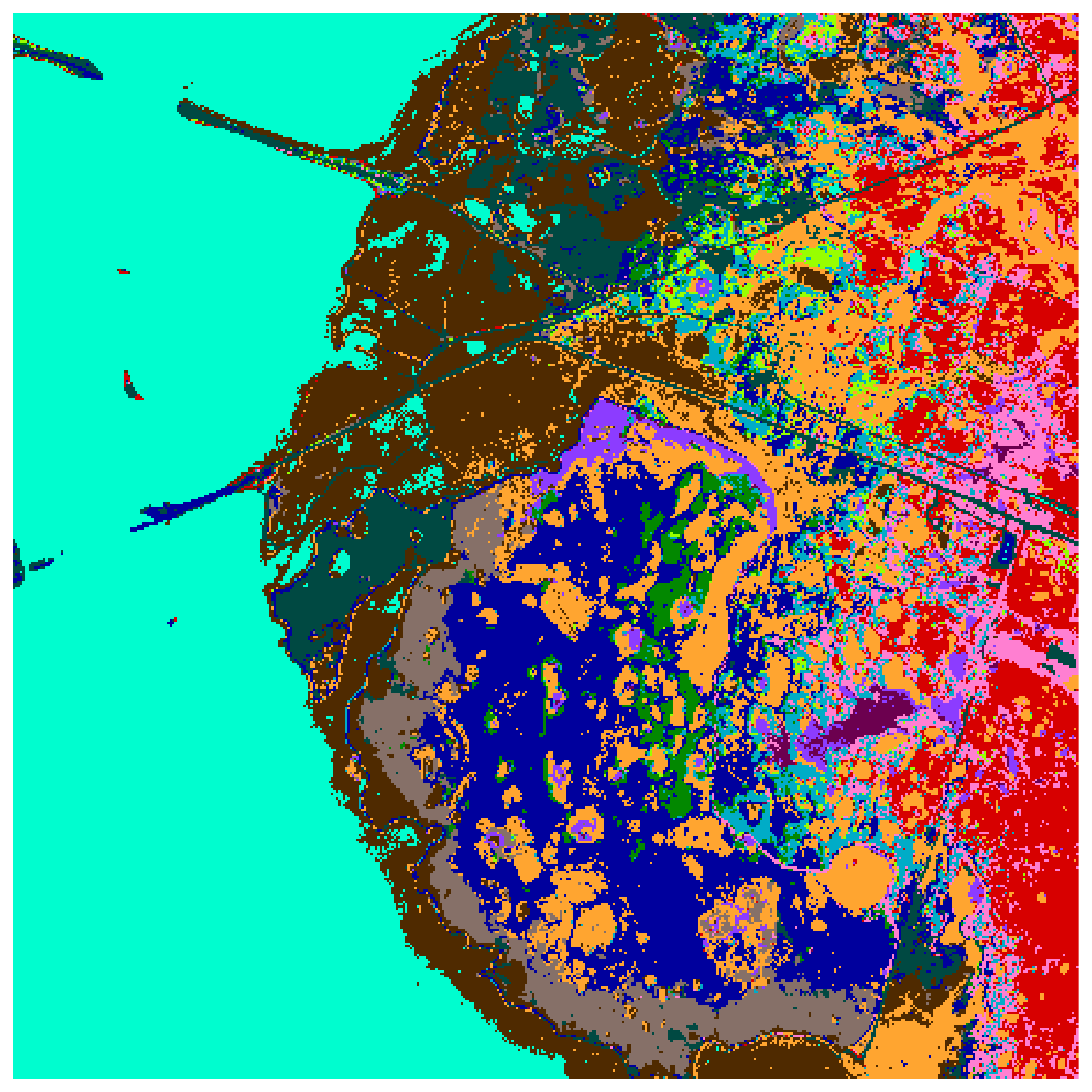}
        \caption{}
    \end{subfigure}%
    \vspace{0cm} 
    \begin{subfigure}{\textwidth}
        \centering
        \includegraphics[width=\linewidth]{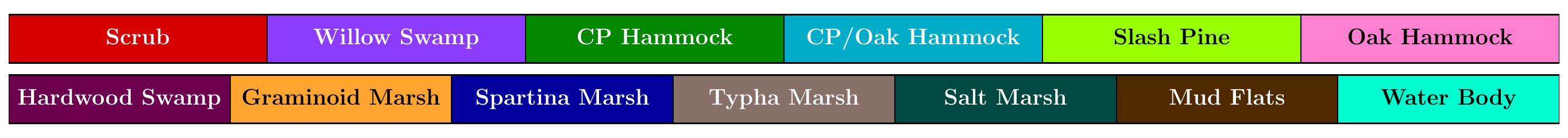}
    \end{subfigure}
    \caption{KSC scene visualization and pixel-based classification maps: (a) False color map, (b) training set, (c) test set, (d) XGBoost, (e) SVC, (f) SAE + SVC (g) 1D-CNN, (h) CasRNN (i) miniGCN, (j) ViT, (k) SpectralFormer, (l) Mamba, (m) CEnc + SVC, (n) CEnc + Dense}
    \label{fig:KSC_spectral}
\end{figure}

\begin{figure}[!htp]
    \centering
    \begin{subfigure}{0.142\textwidth}
        \centering
        \includegraphics[width=\linewidth]{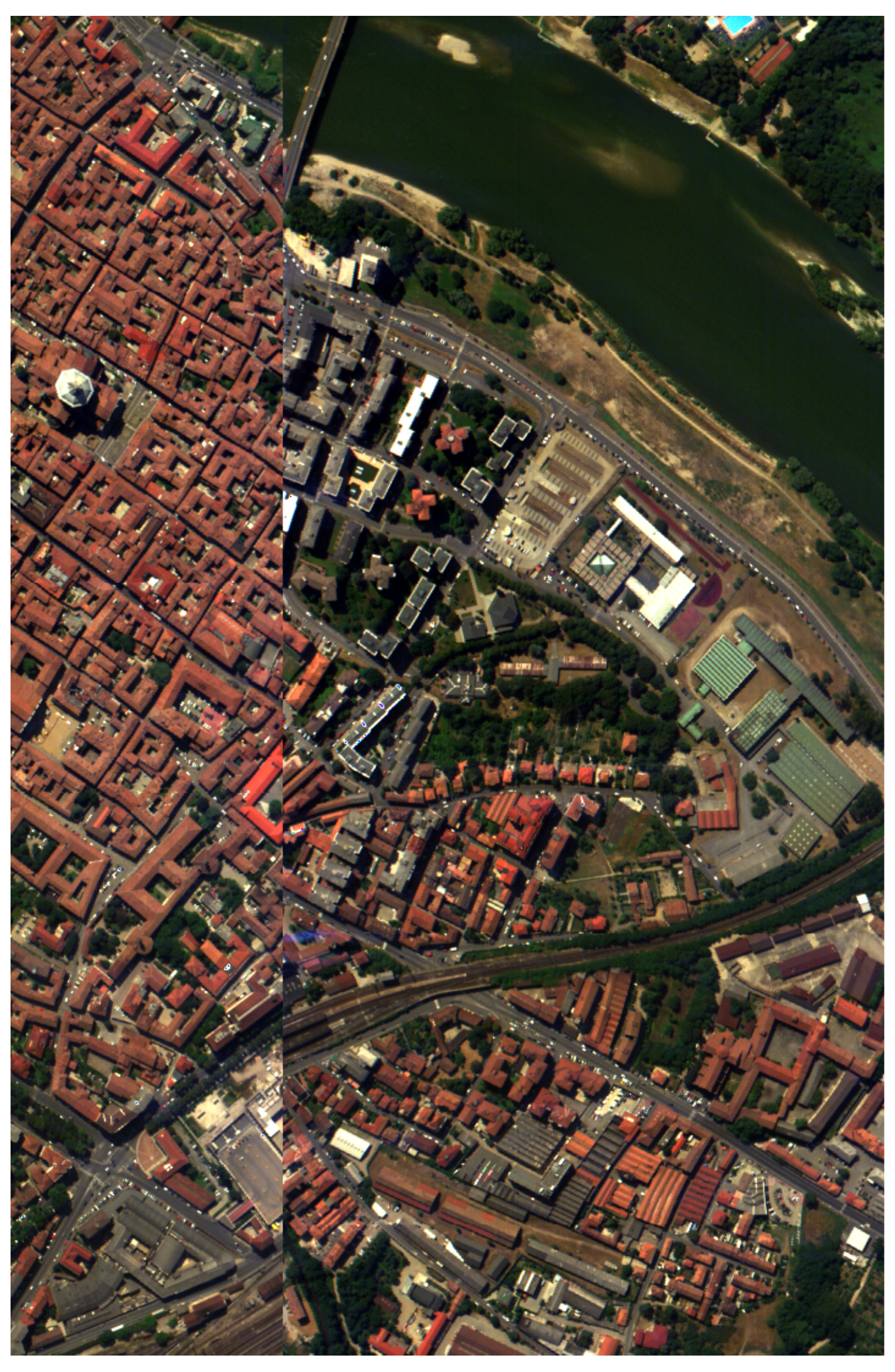}
        \caption{}
    \end{subfigure}%
    \begin{subfigure}{0.142\textwidth}
        \centering
        \includegraphics[width=\linewidth]{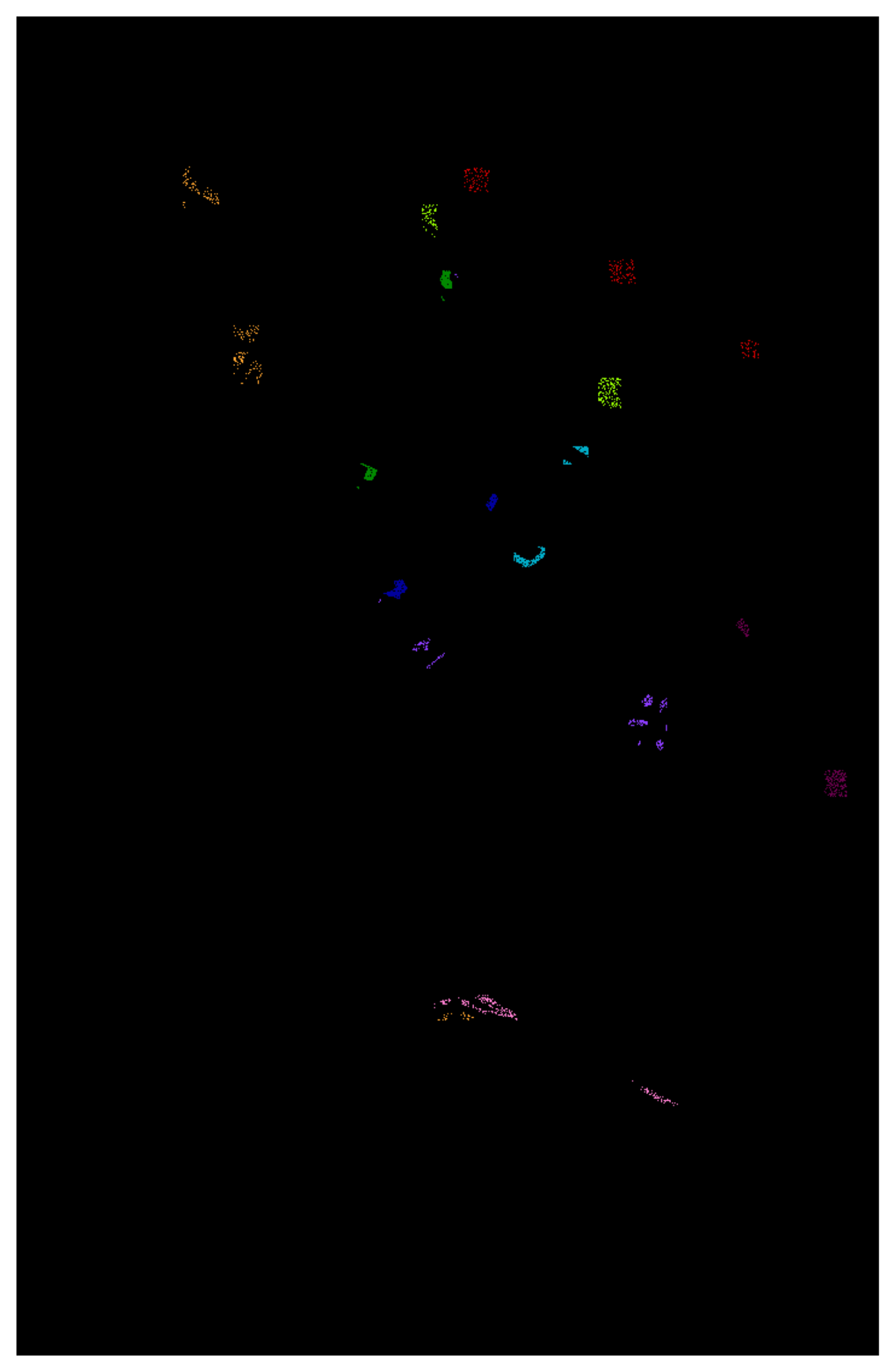}
        \caption{}
    \end{subfigure}%
    \begin{subfigure}{0.142\textwidth}
        \centering
        \includegraphics[width=\linewidth]{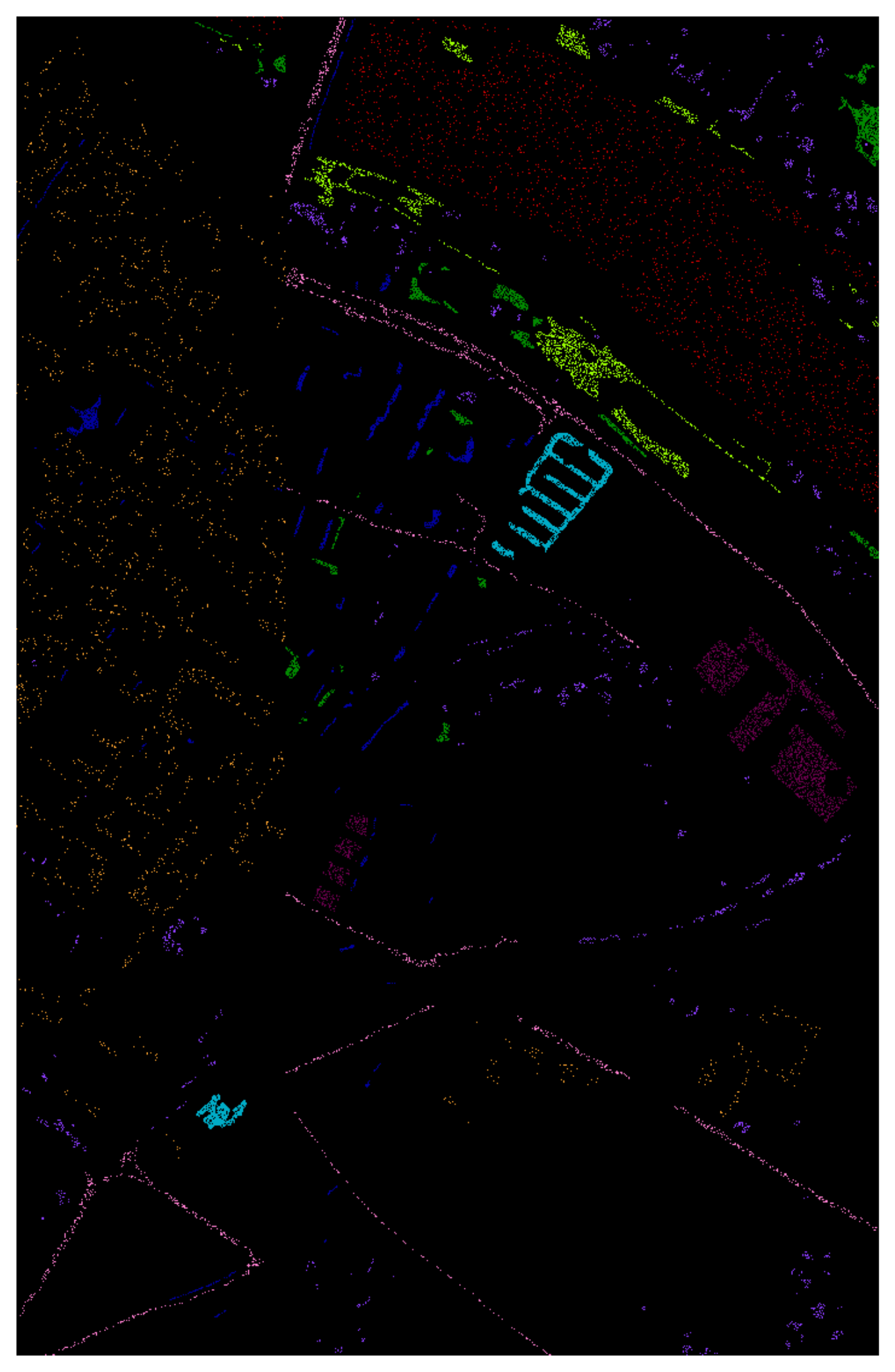}
        \caption{}
    \end{subfigure}%
    \begin{subfigure}{0.142\textwidth}
        \centering
        \includegraphics[width=\linewidth]{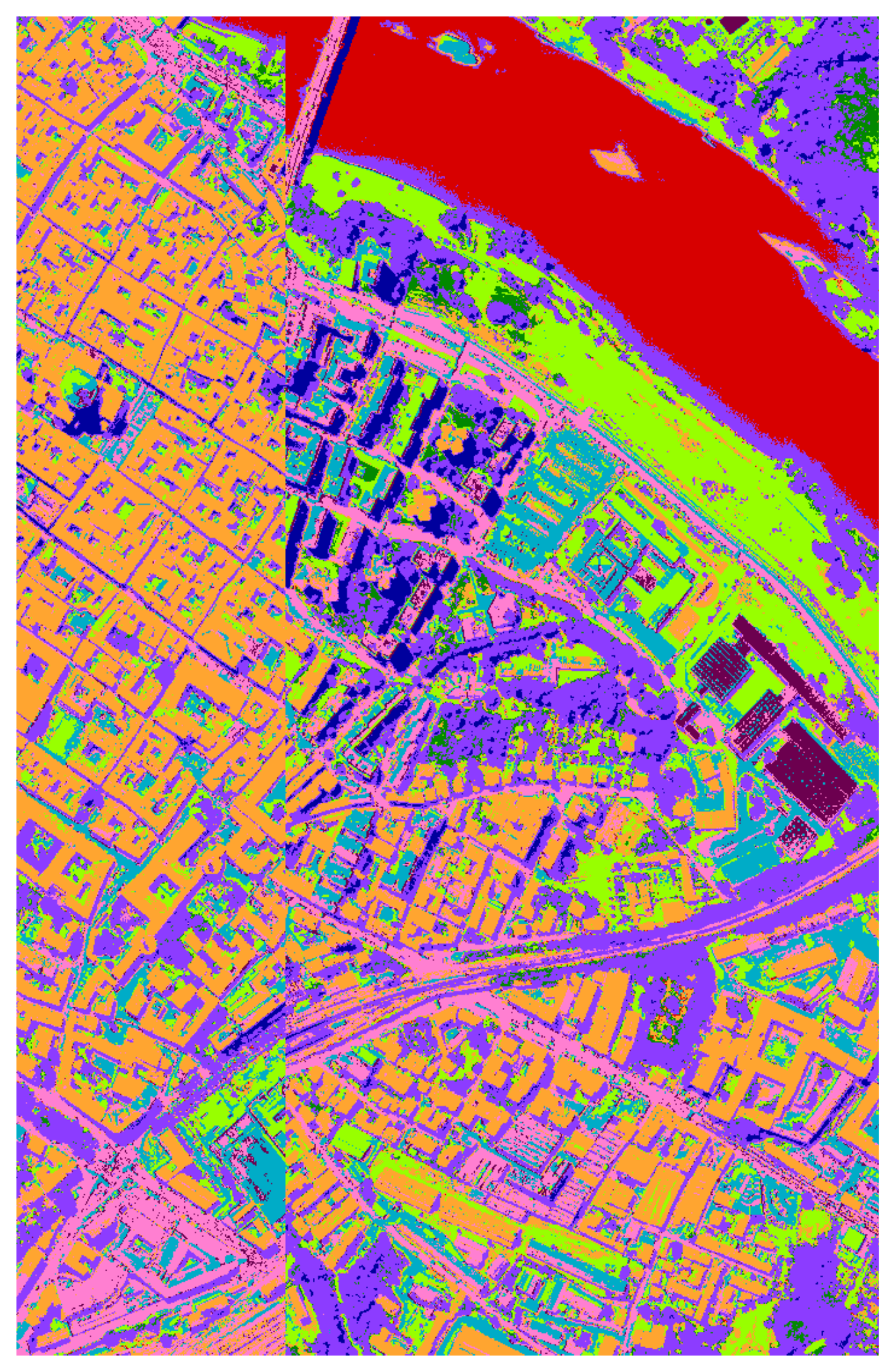}
        \caption{}
    \end{subfigure}%
    \begin{subfigure}{0.142\textwidth}
        \centering
        \includegraphics[width=\linewidth]{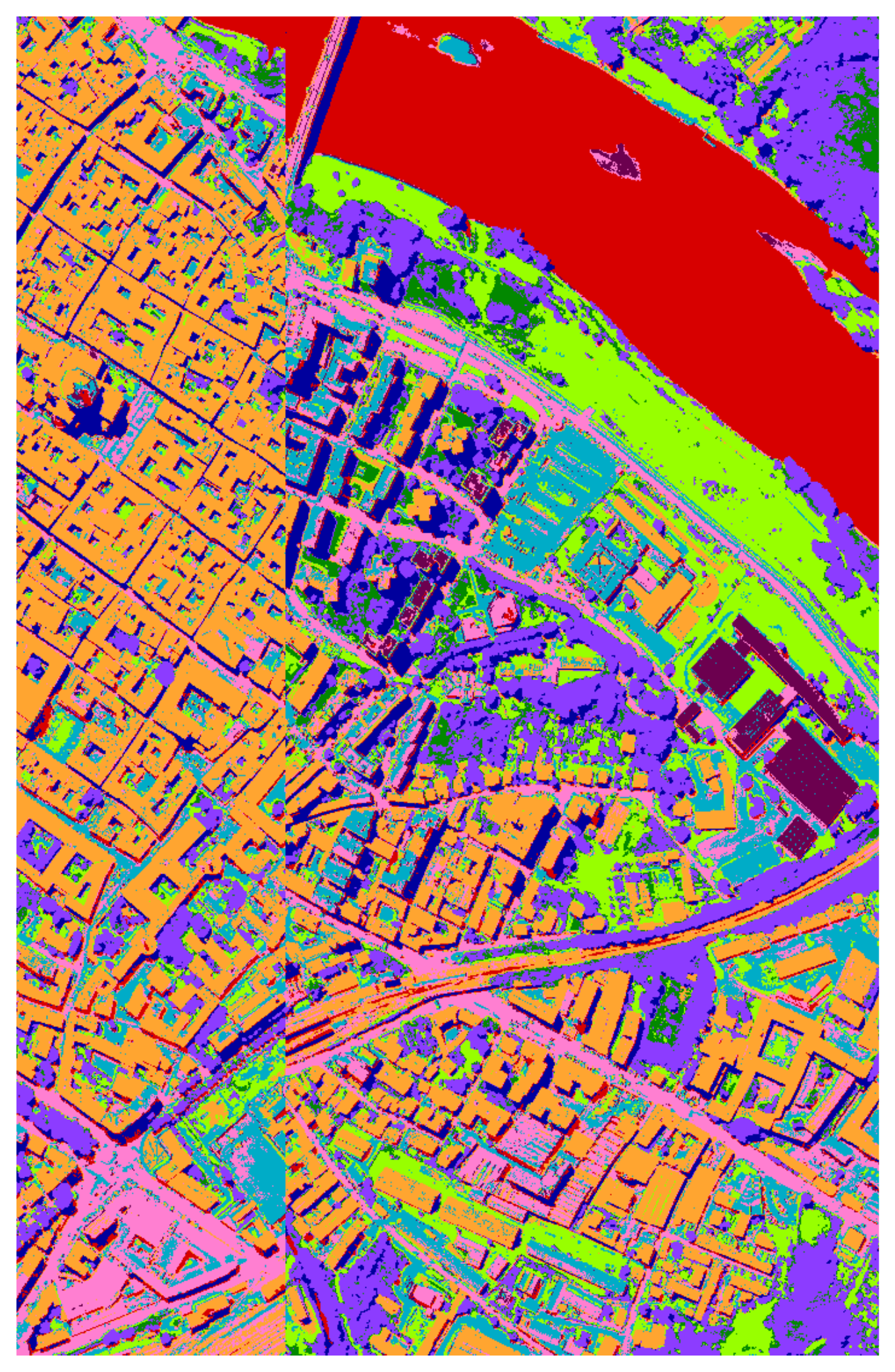}
        \caption{}
    \end{subfigure}%
    \begin{subfigure}{0.142\textwidth}
        \centering
        \includegraphics[width=\linewidth]{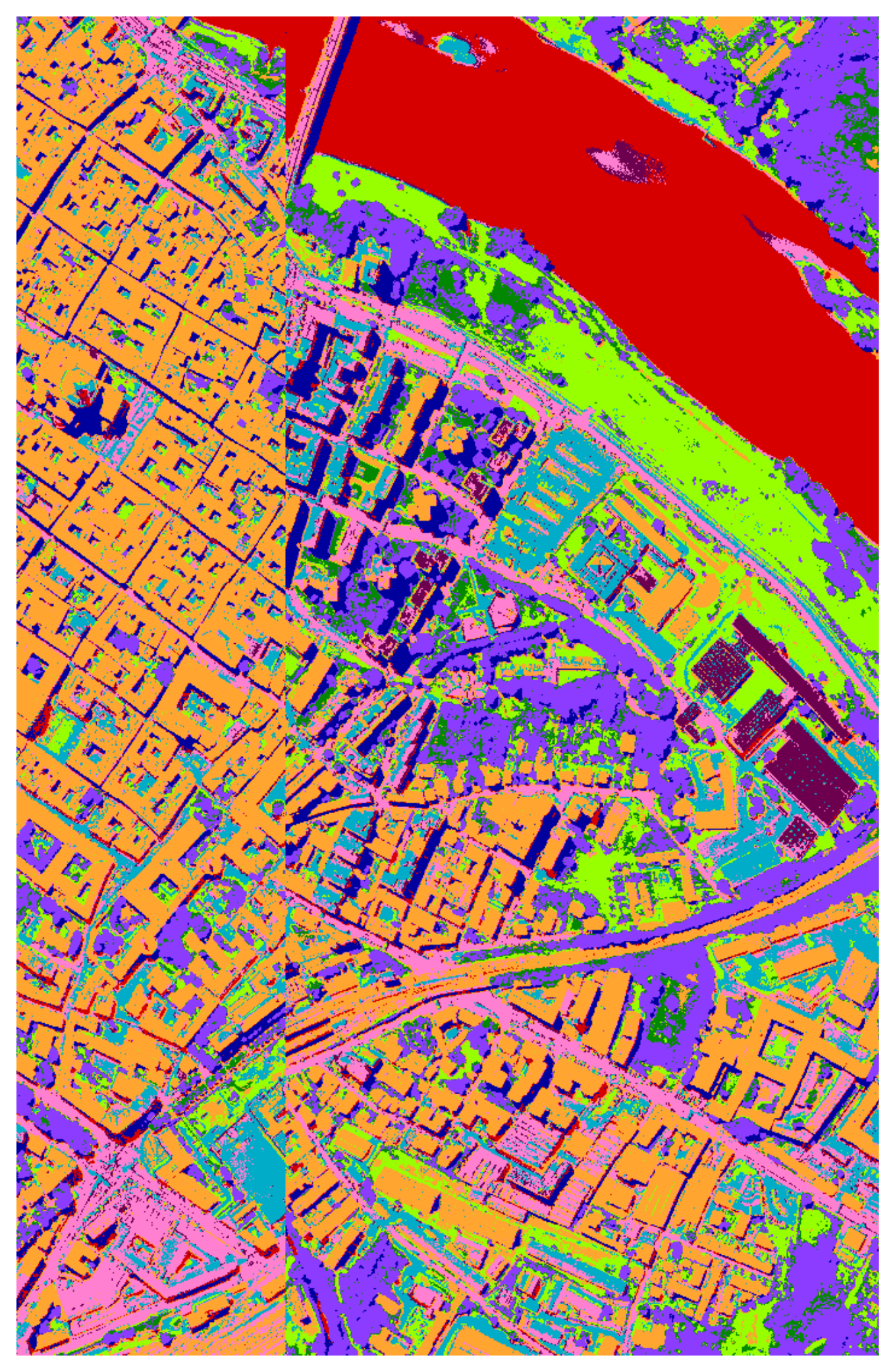}
        \caption{}
    \end{subfigure}%
    \begin{subfigure}{0.142\textwidth}
        \centering
        \includegraphics[width=\linewidth]{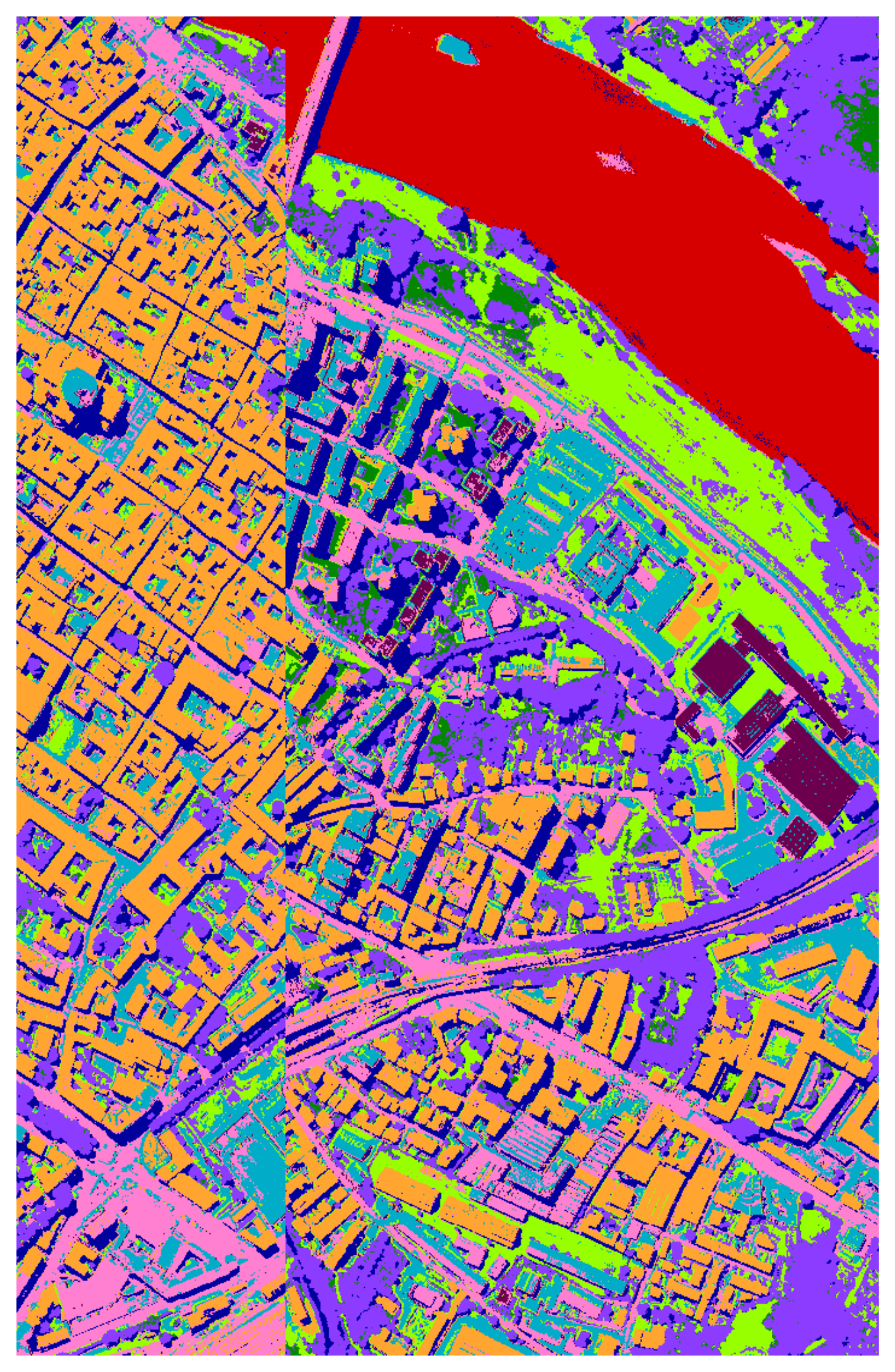}
        \caption{}
    \end{subfigure}%
    \vspace{0cm} 
    \begin{subfigure}{0.142\textwidth}
        \centering
        \includegraphics[width=\linewidth]{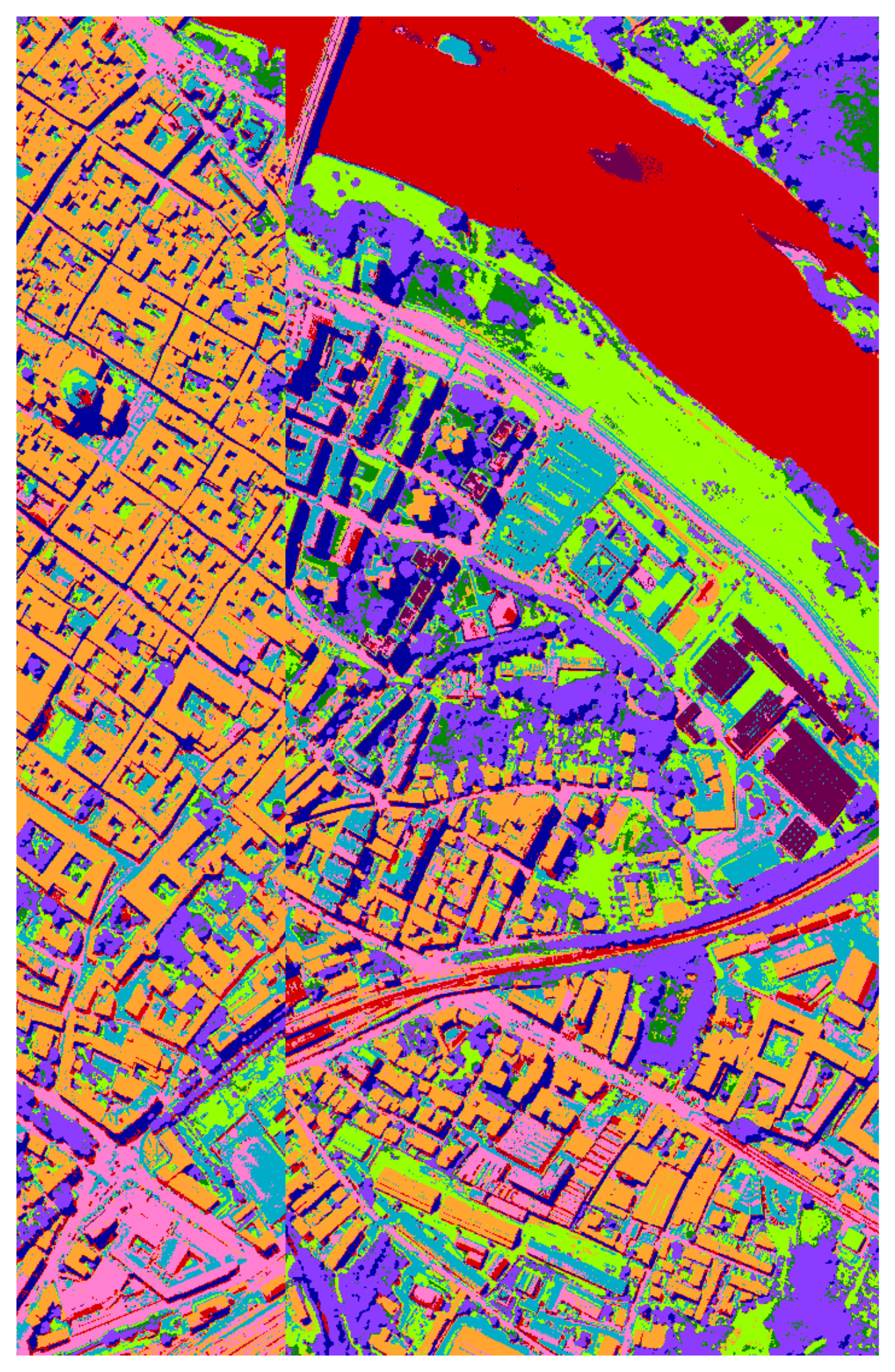}
        \caption{}
    \end{subfigure}%
    \begin{subfigure}{0.142\textwidth}
        \centering
        \includegraphics[width=\linewidth]{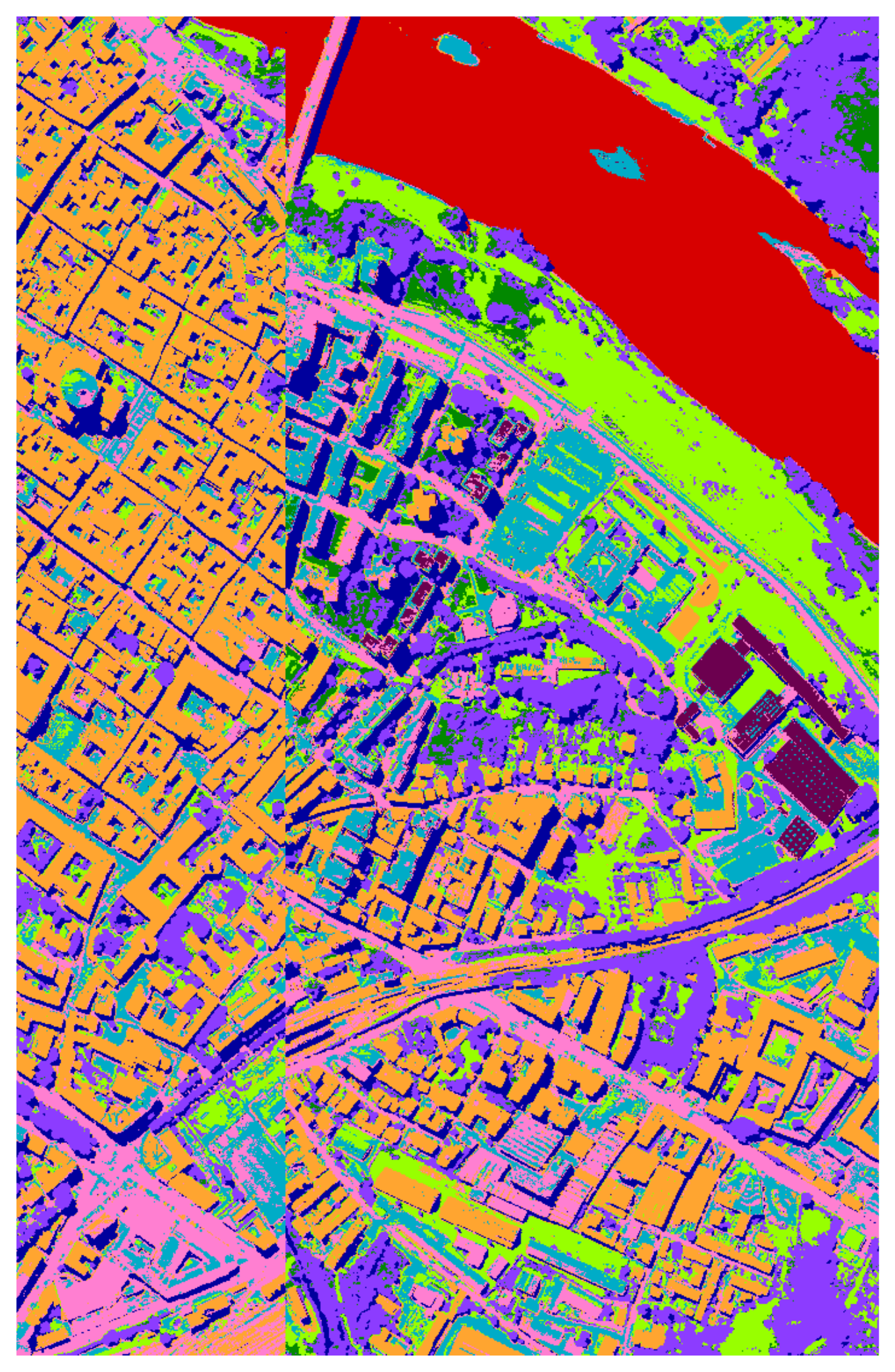}
        \caption{}
    \end{subfigure}%
    \begin{subfigure}{0.142\textwidth}
        \centering
        \includegraphics[width=\linewidth]{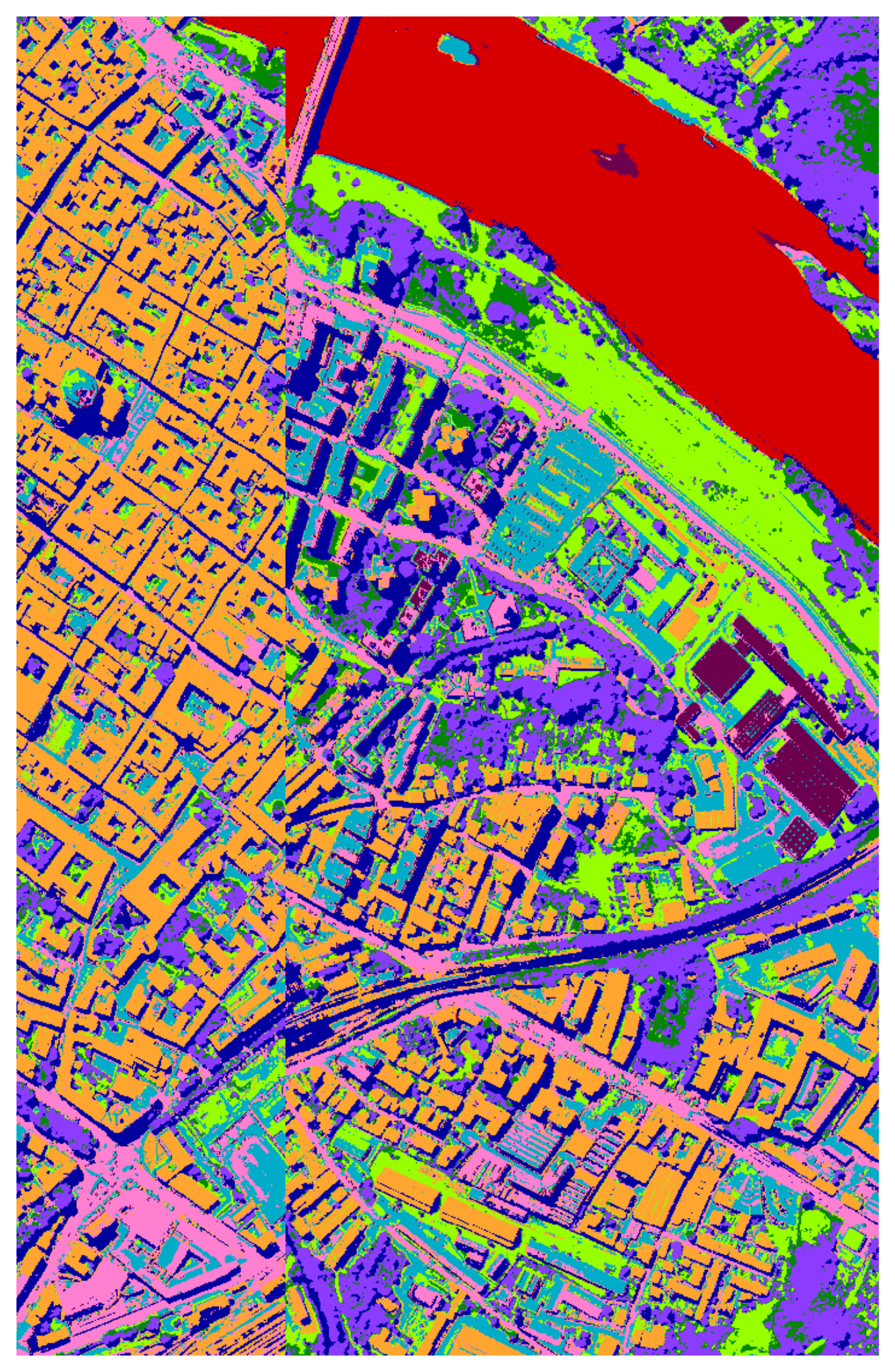}
        \caption{}
    \end{subfigure}%
    \begin{subfigure}{0.142\textwidth}
        \centering
        \includegraphics[width=\linewidth]{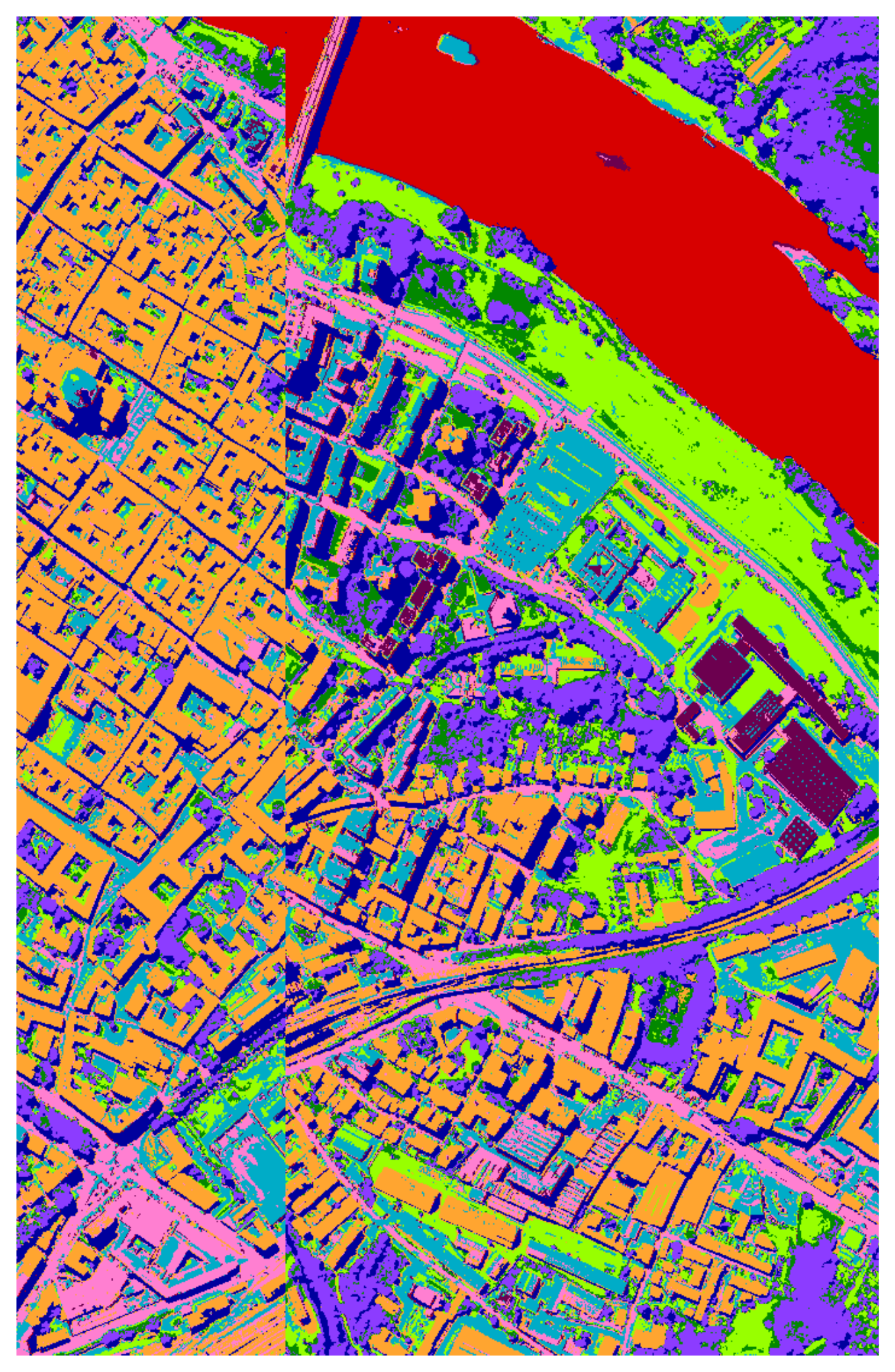}
        \caption{}
    \end{subfigure}%
    \begin{subfigure}{0.142\textwidth}
        \centering
        \includegraphics[width=\linewidth]{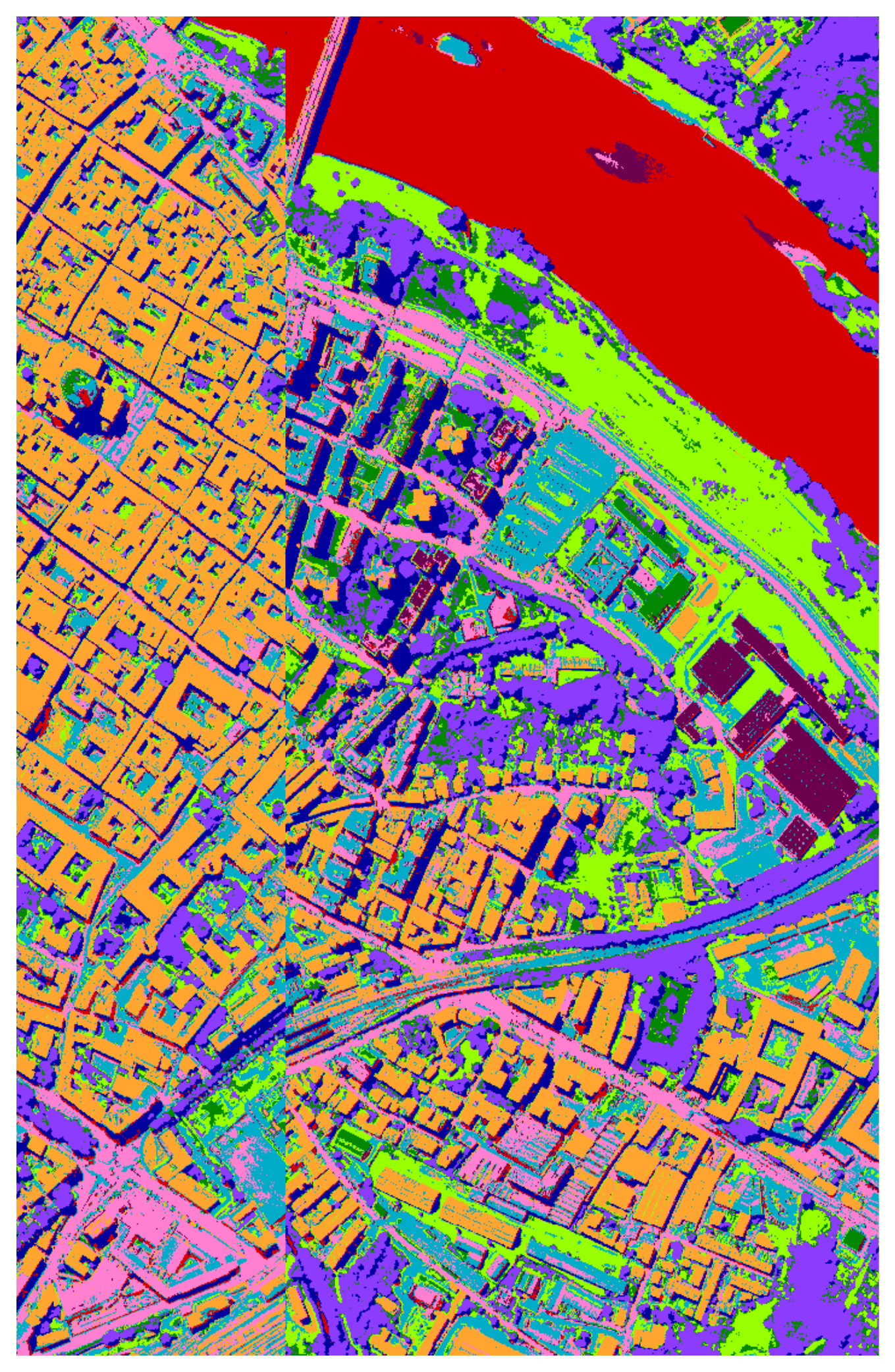}
        \caption{}
    \end{subfigure}%
    \begin{subfigure}{0.142\textwidth}
        \centering
        \includegraphics[width=\linewidth]{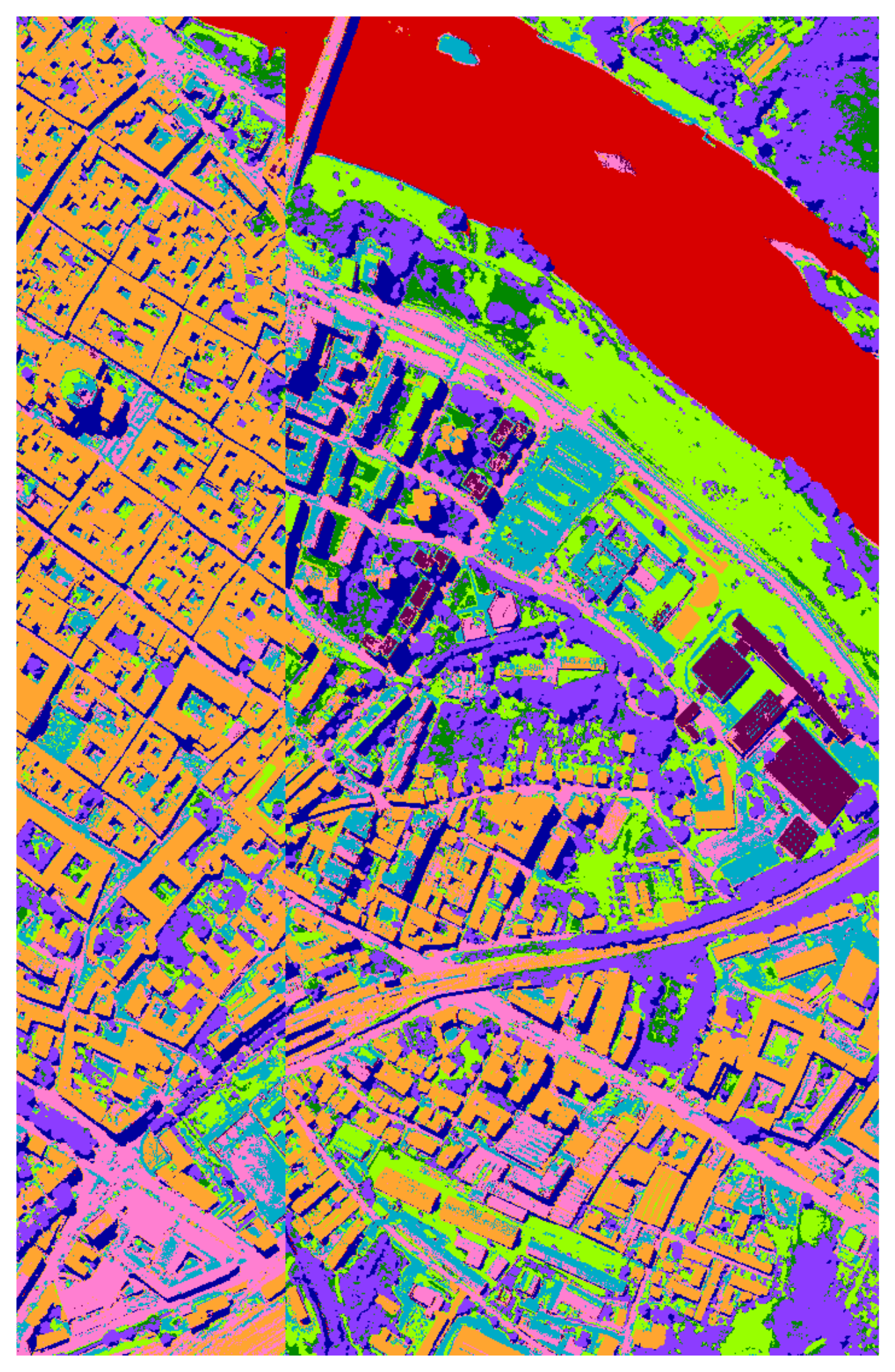}
        \caption{}
    \end{subfigure}%
    \begin{subfigure}{0.142\textwidth}
        \centering
        \includegraphics[width=\linewidth]{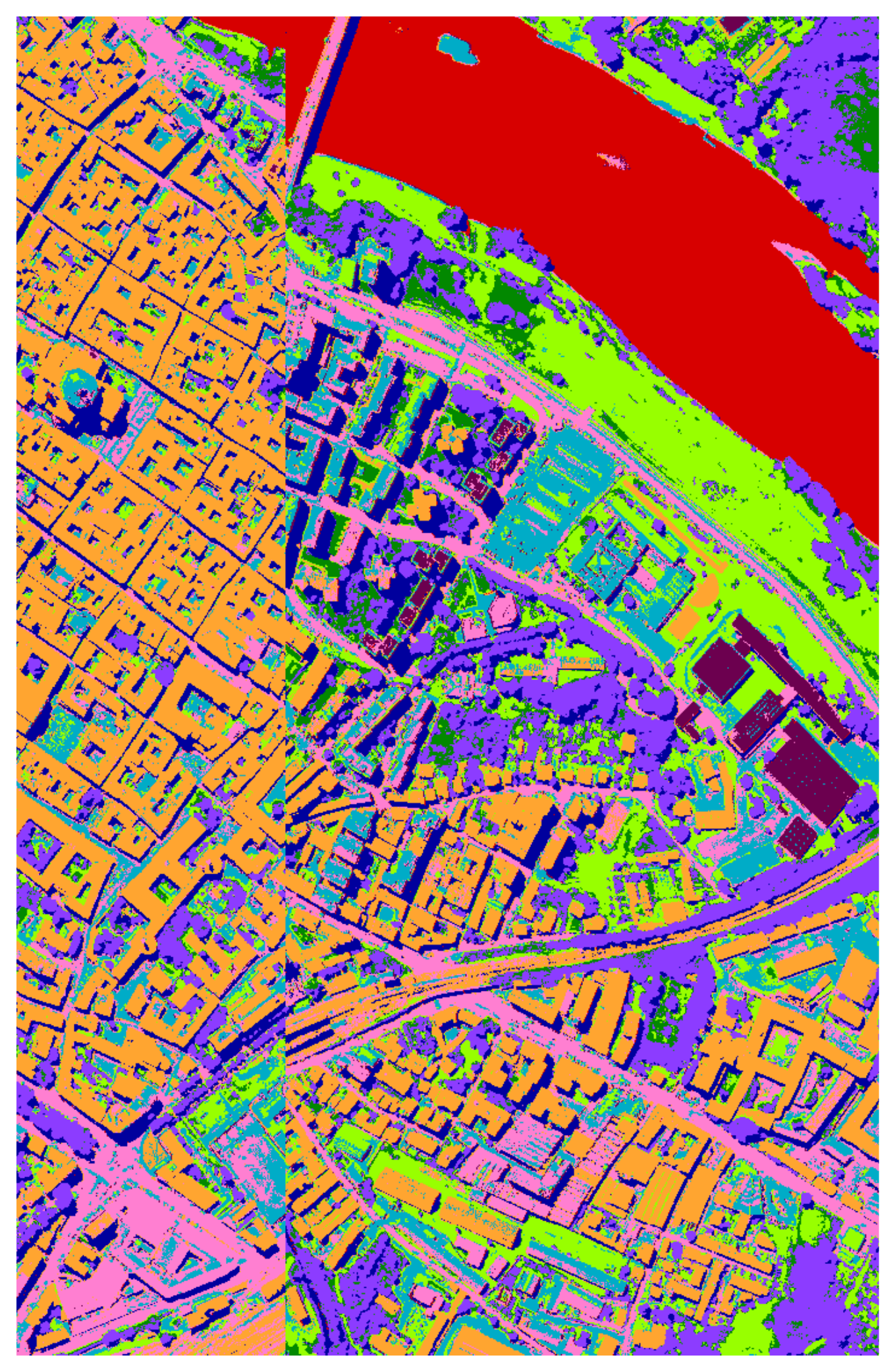}
        \caption{}
    \end{subfigure}%
    \vspace{0cm} 
    \begin{subfigure}{\textwidth}
        \centering
        \includegraphics[width=\linewidth]{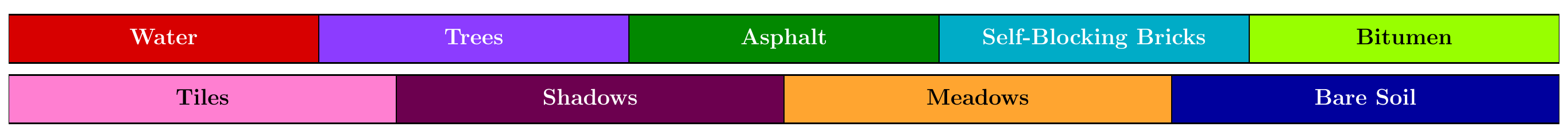}
    \end{subfigure}
    \caption{PC scene visualization and pixel-based classification maps: (a) False color map, (b) training set, (c) test set, (d) XGBoost, (e) SVC, (f) SAE + SVC (g) 1D-CNN, (h) CasRNN (i) miniGCN, (j) ViT, (k) SpectralFormer, (l) Mamba, (m) CEnc + SVC, (n) CEnc + Dense}
    \label{fig:PC_spectral}
\end{figure}

\begin{figure}[!htp]
    \centering
    \begin{subfigure}{0.142\textwidth}
        \centering
        \includegraphics[width=1\linewidth]{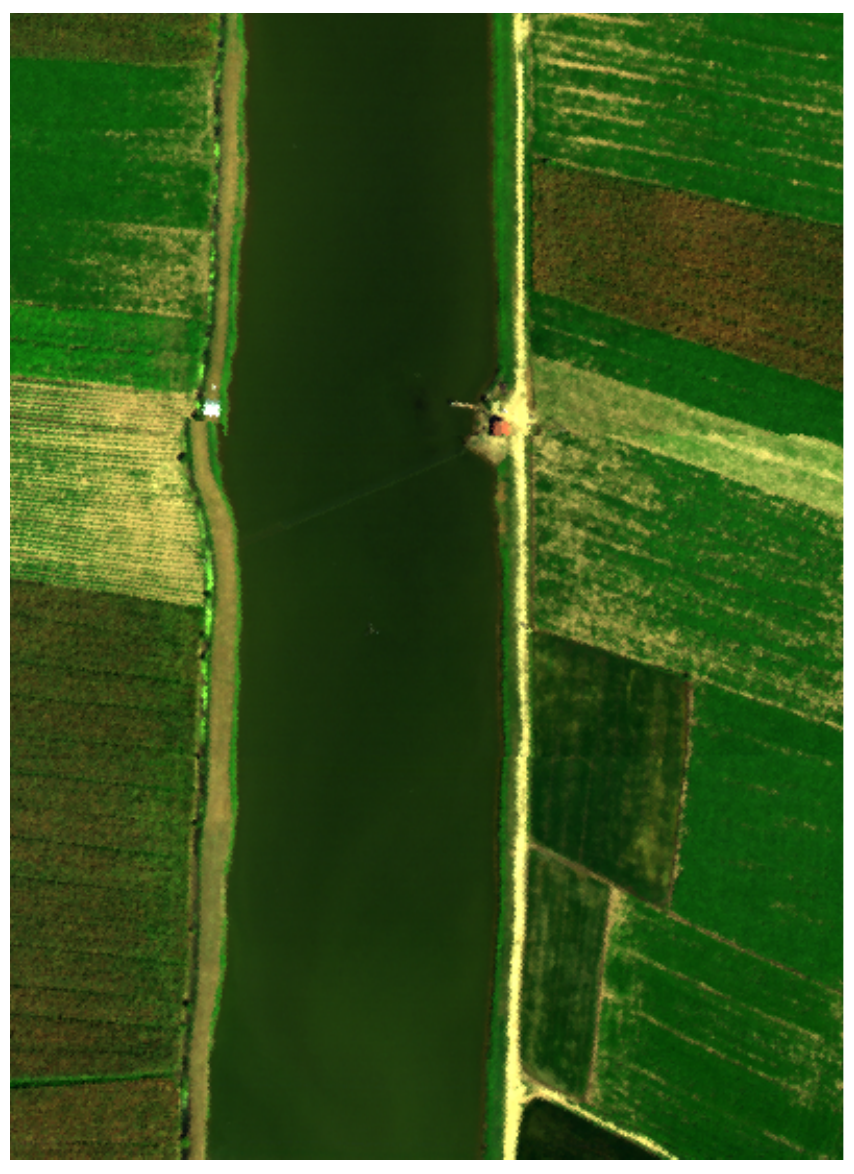}
        \caption{}
    \end{subfigure}%
    \begin{subfigure}{0.142\textwidth}
        \centering
        \includegraphics[width=\linewidth]{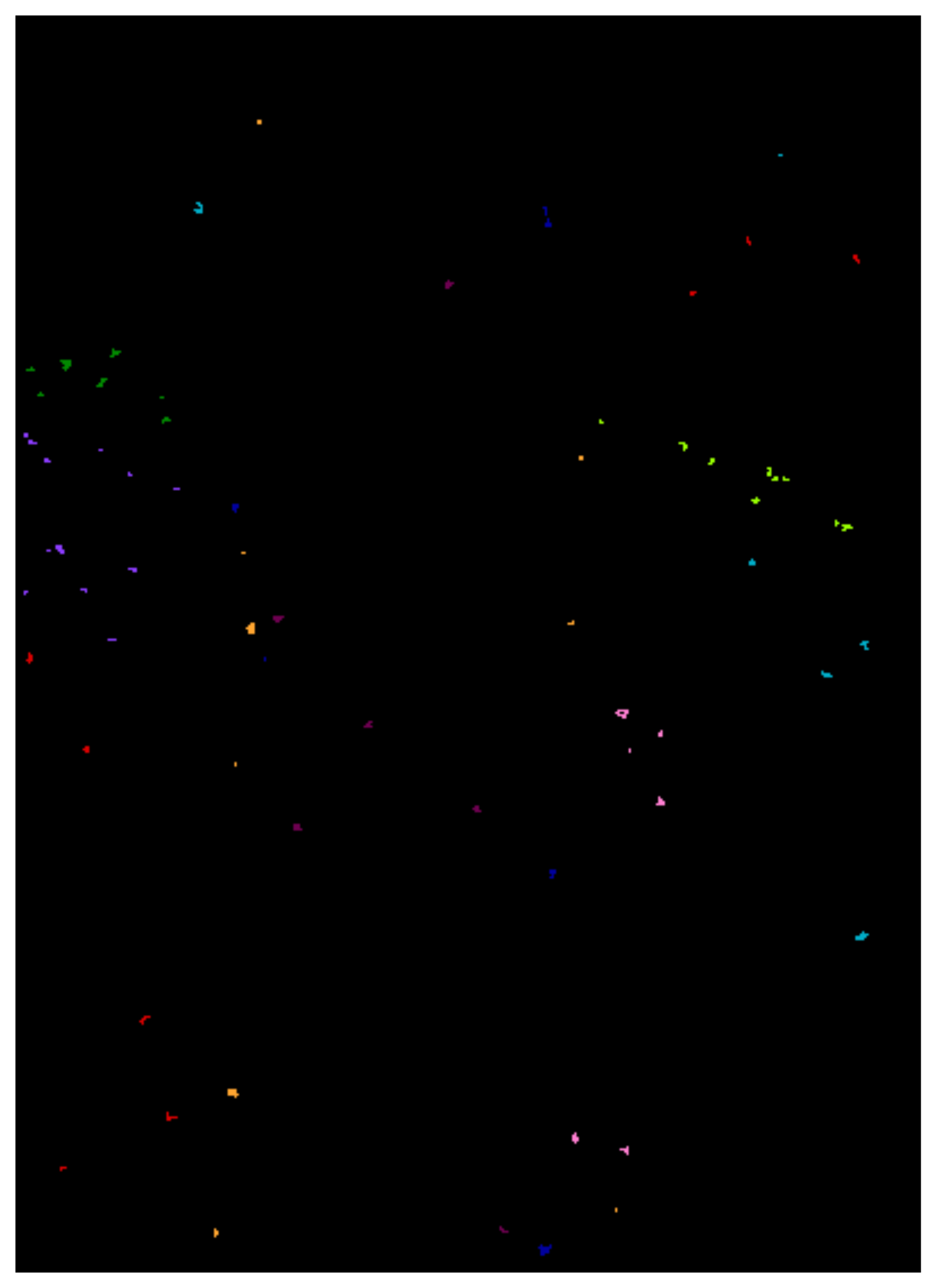}
        \caption{}
    \end{subfigure}%
    \begin{subfigure}{0.142\textwidth}
        \centering
        \includegraphics[width=\linewidth]{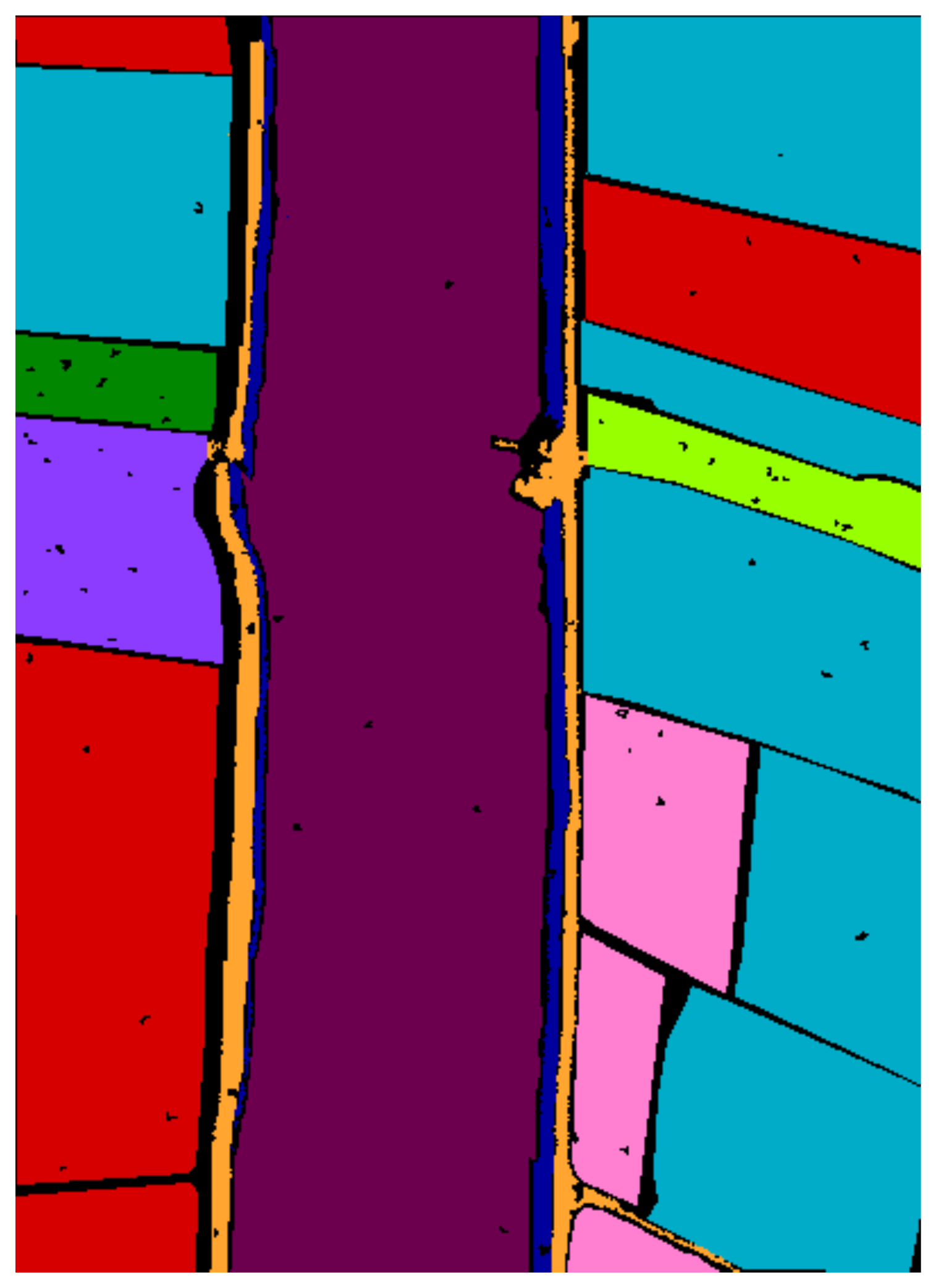}
        \caption{}
    \end{subfigure}%
    \begin{subfigure}{0.142\textwidth}
        \centering
        \includegraphics[width=\linewidth]{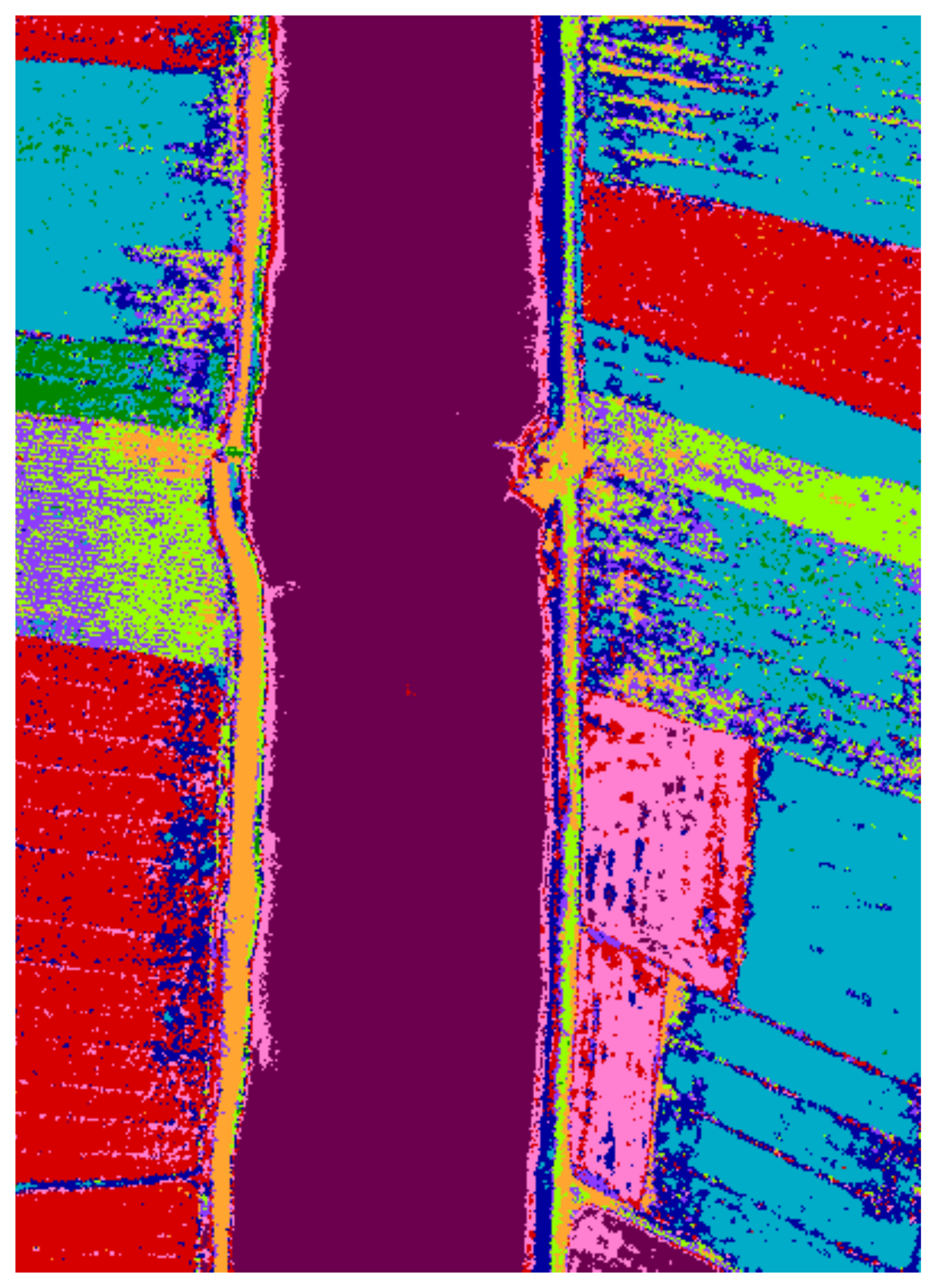}
        \caption{}
    \end{subfigure}%
    \begin{subfigure}{0.142\textwidth}
        \centering
        \includegraphics[width=\linewidth]{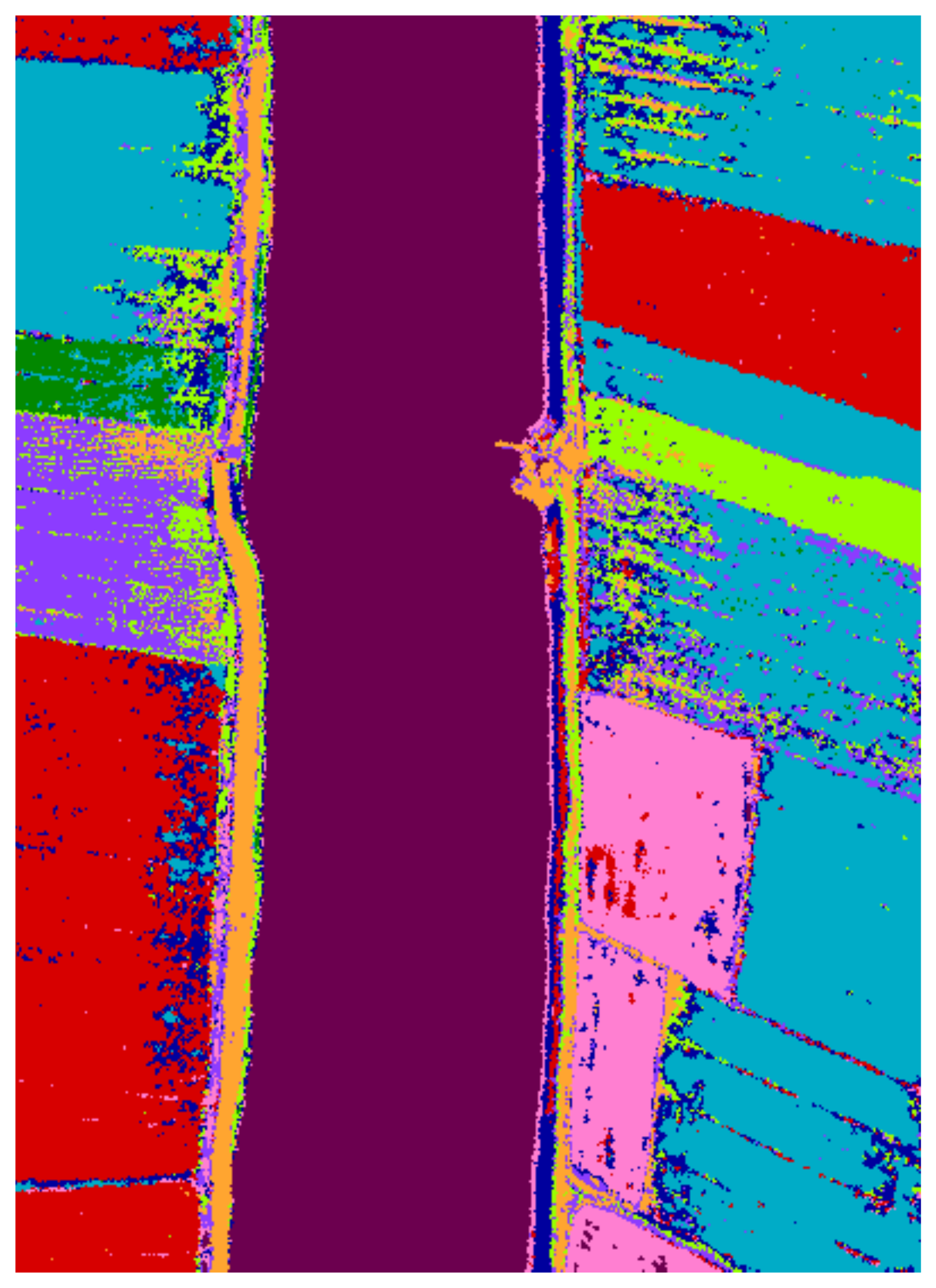}
        \caption{}
    \end{subfigure}%
    \begin{subfigure}{0.142\textwidth}
        \centering
        \includegraphics[width=\linewidth]{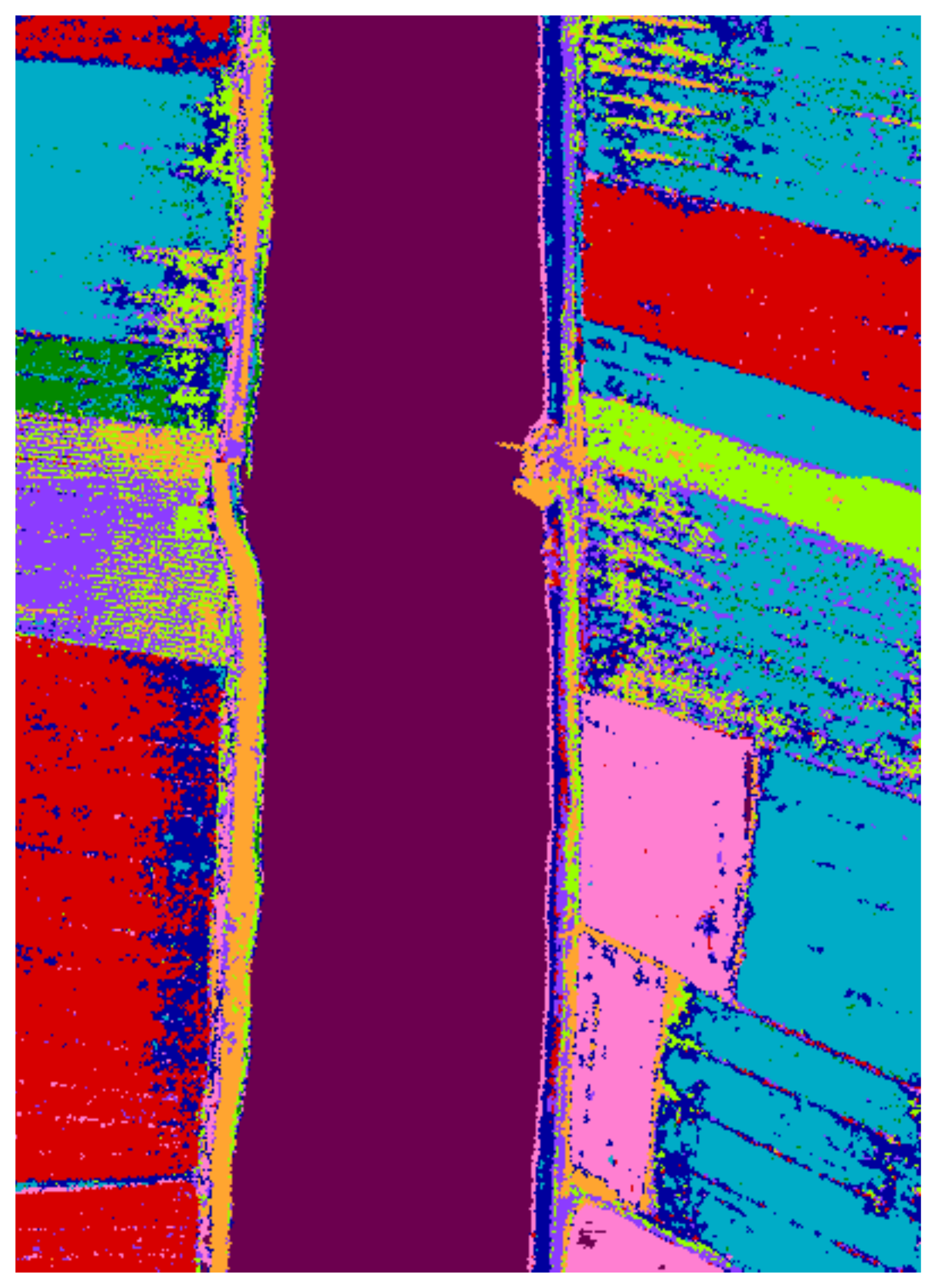}
        \caption{}
    \end{subfigure}%
    \begin{subfigure}{0.142\textwidth}
        \centering
        \includegraphics[width=\linewidth]{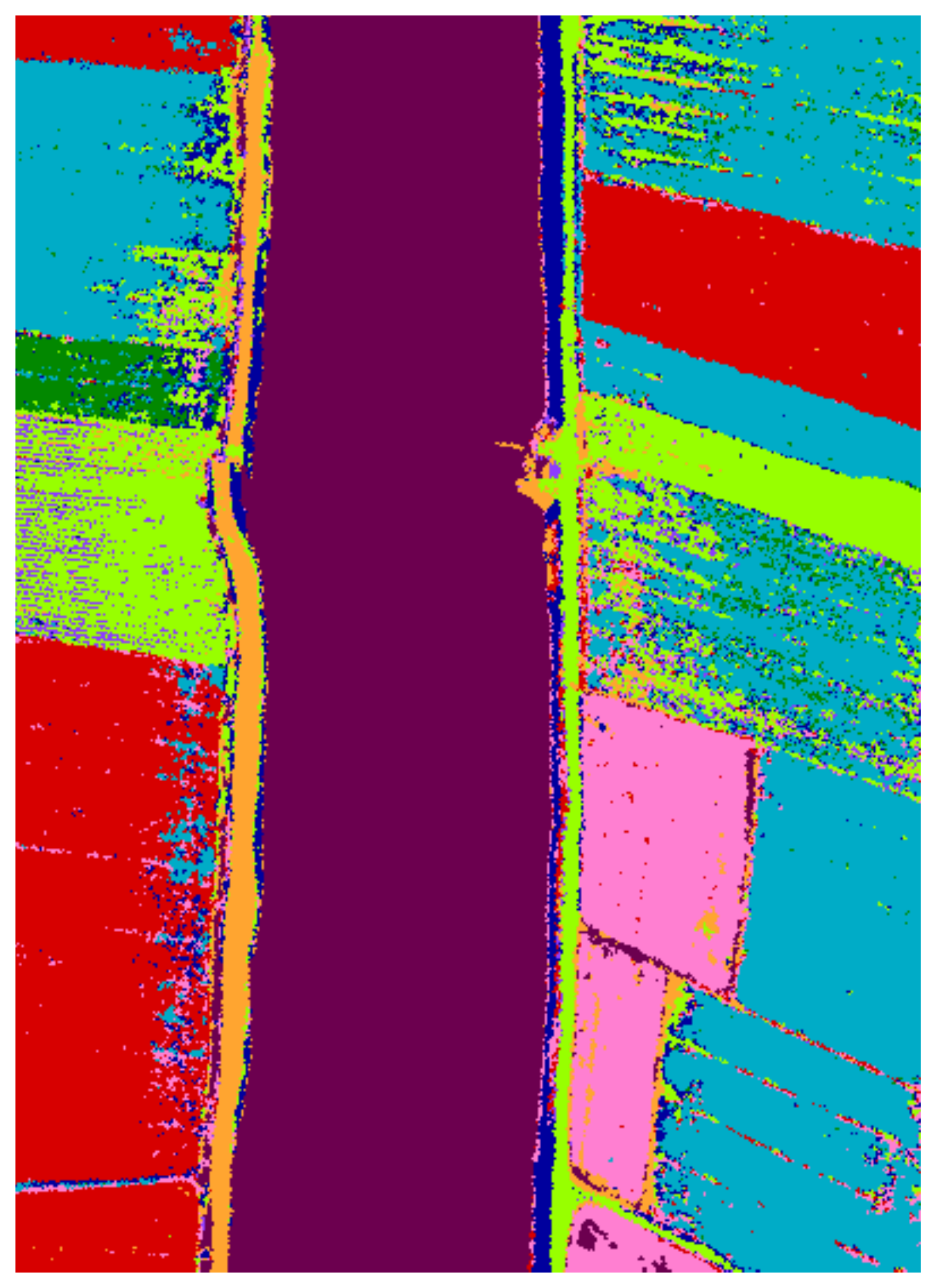}
        \caption{}
    \end{subfigure}%
    \vspace{0cm} 
    \begin{subfigure}{0.142\textwidth}
        \centering
        \includegraphics[width=\linewidth]{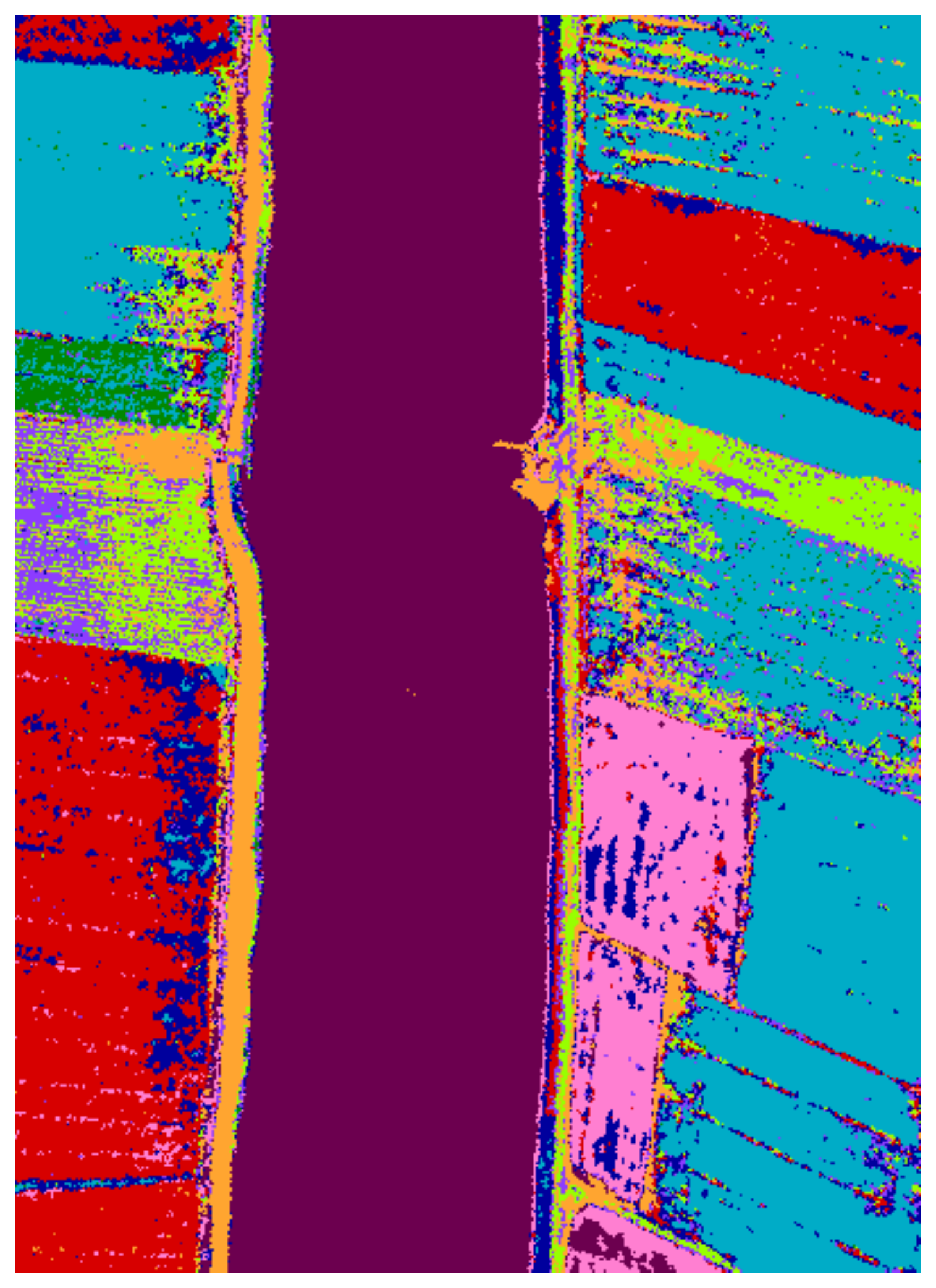}
        \caption{}
    \end{subfigure}%
    \begin{subfigure}{0.142\textwidth}
        \centering
        \includegraphics[width=\linewidth]{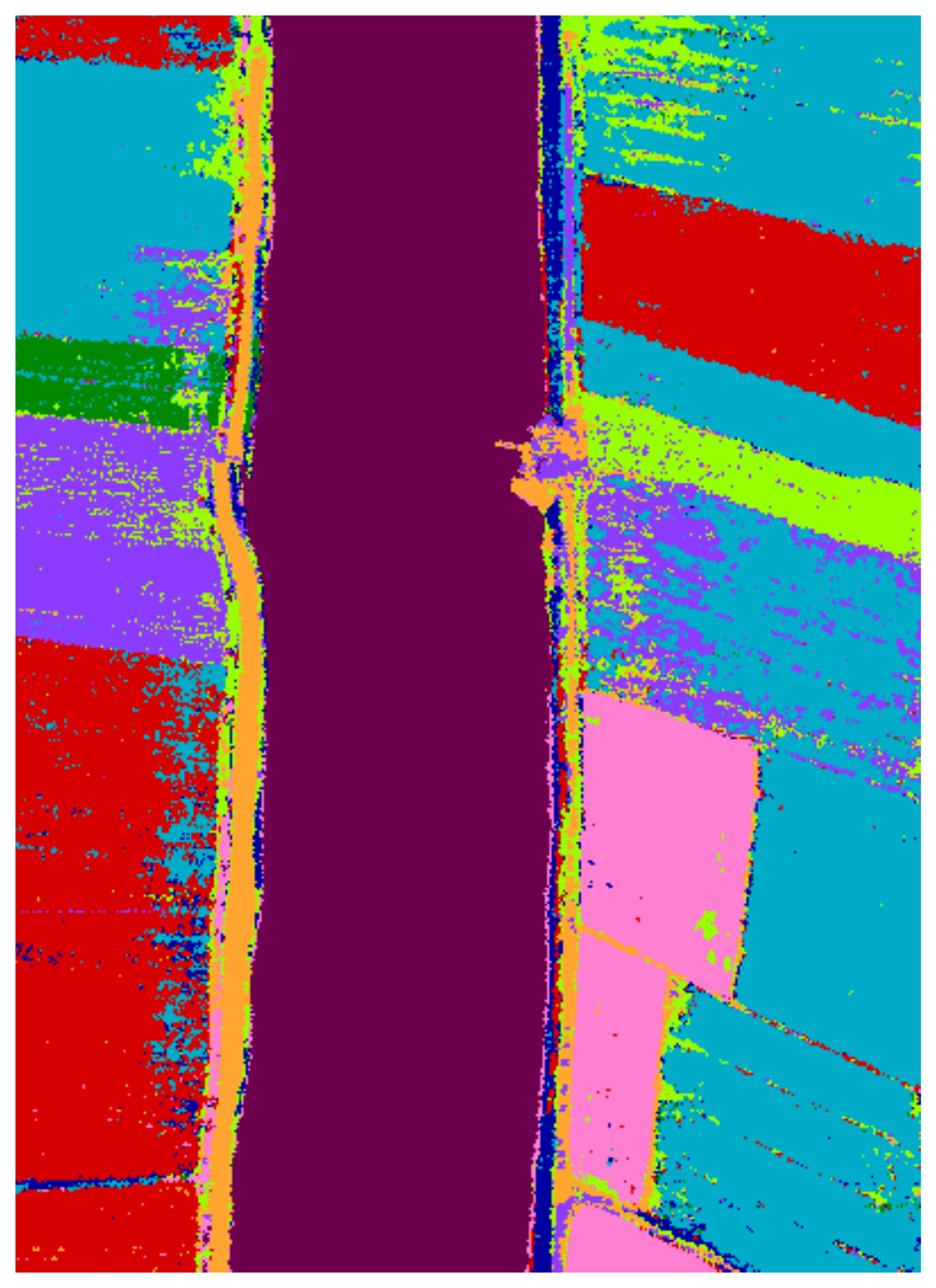}
        \caption{}
    \end{subfigure}%
    \begin{subfigure}{0.142\textwidth}
        \centering
        \includegraphics[width=\linewidth]{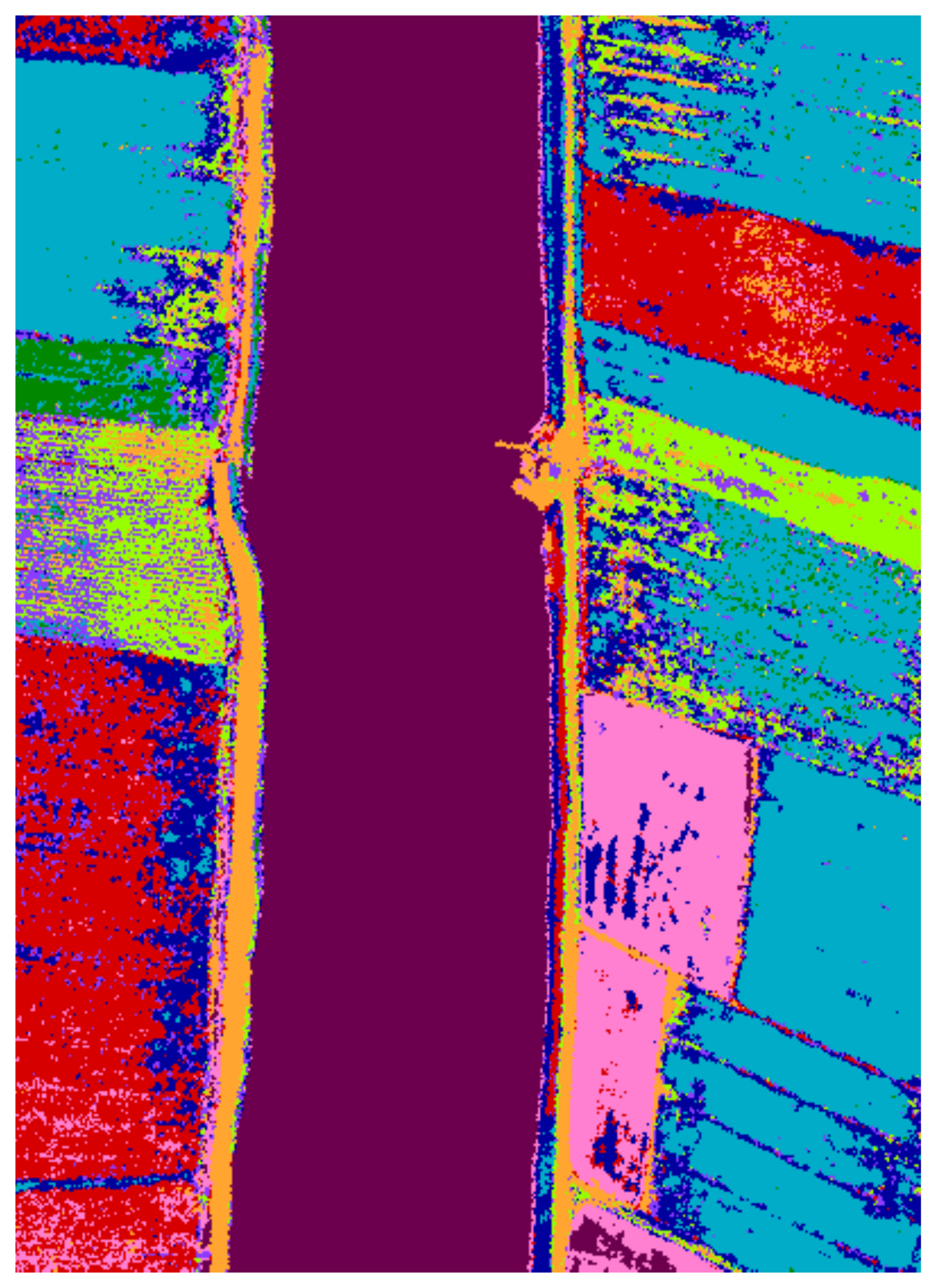}
        \caption{}
    \end{subfigure}%
    \begin{subfigure}{0.142\textwidth}
        \centering
        \includegraphics[width=\linewidth]{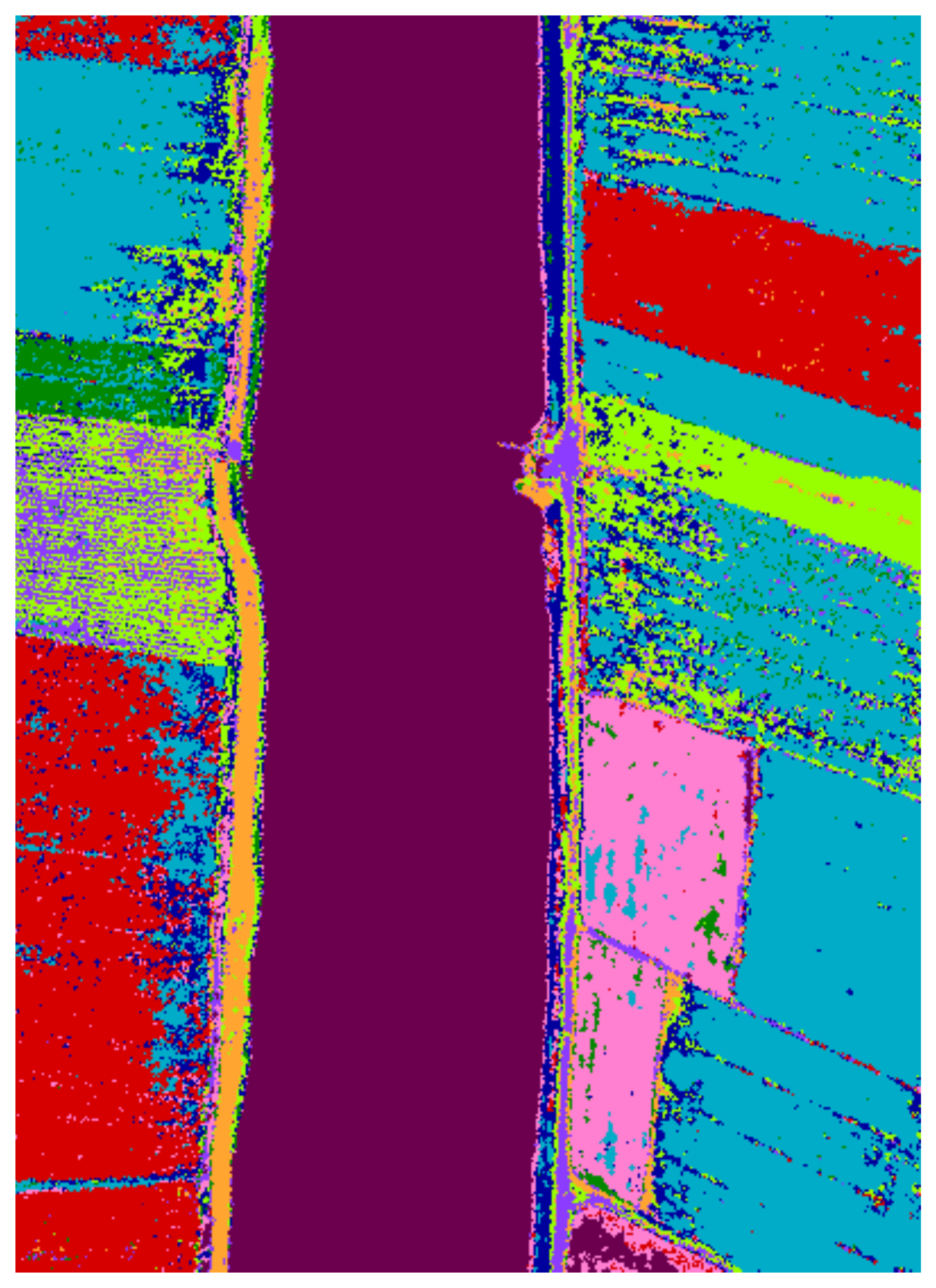}
        \caption{}
    \end{subfigure}%
    \begin{subfigure}{0.142\textwidth}
        \centering
        \includegraphics[width=\linewidth]{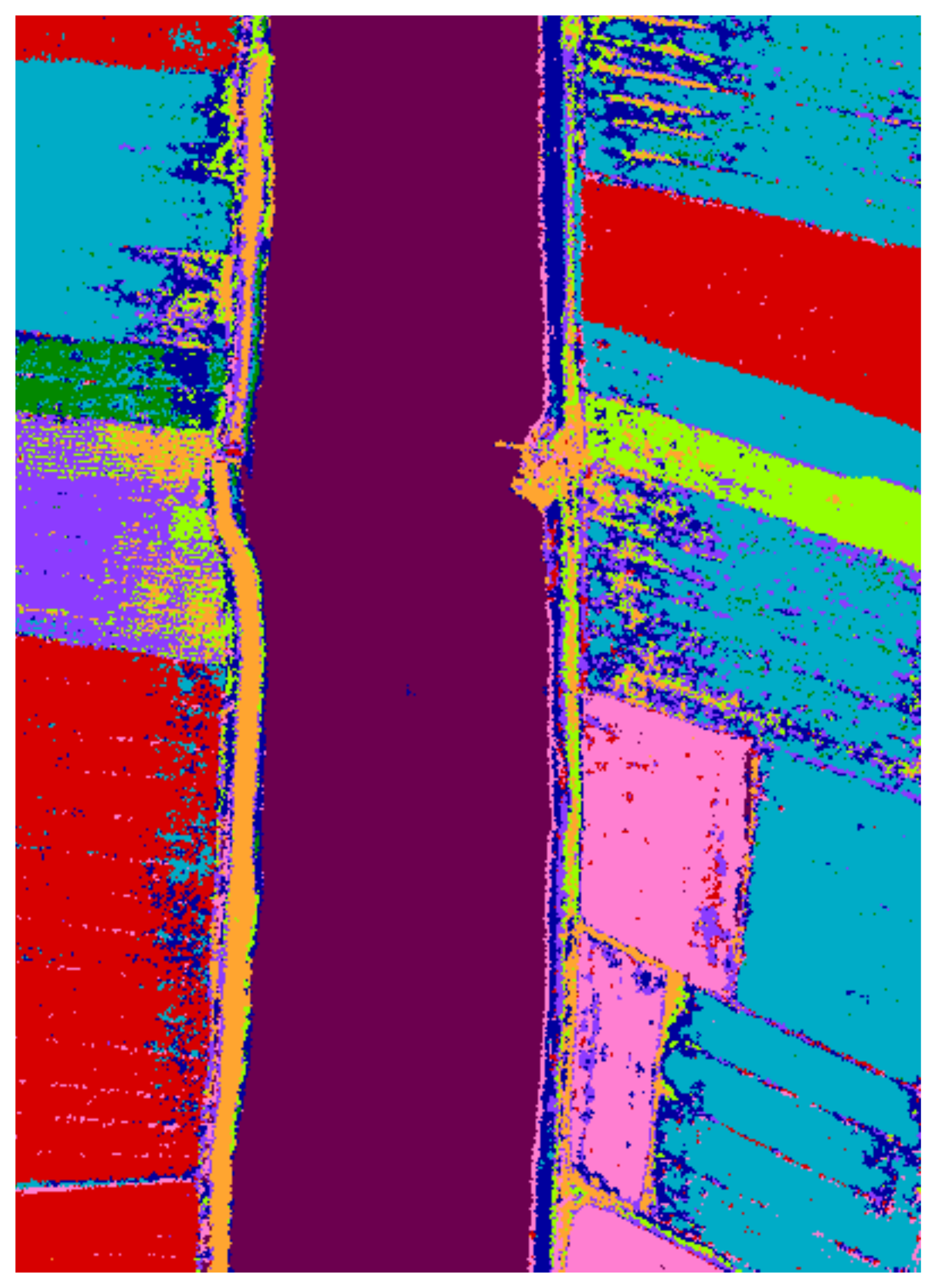}
        \caption{}
    \end{subfigure}%
    \begin{subfigure}{0.142\textwidth}
        \centering
        \includegraphics[width=\linewidth]{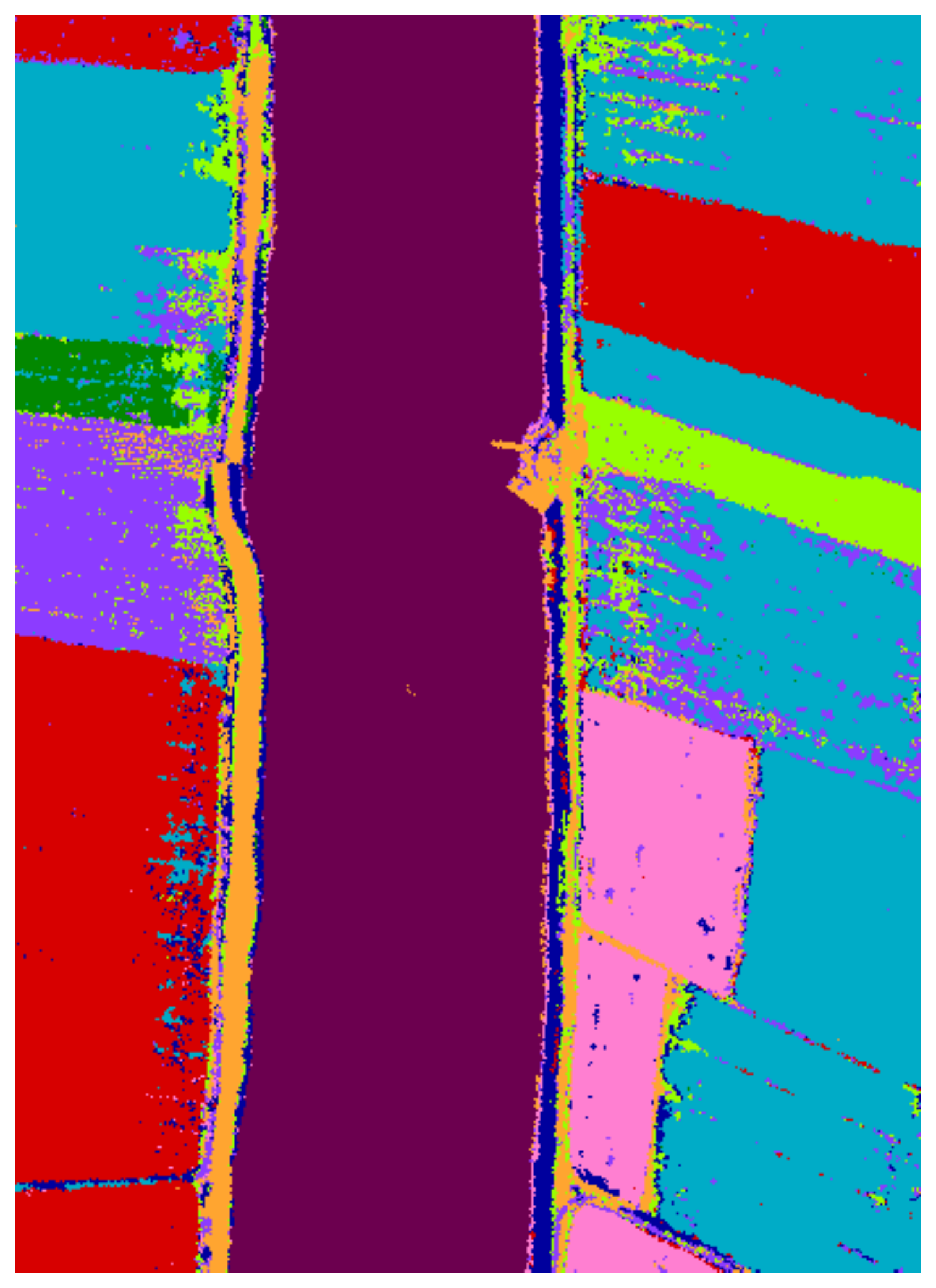}
        \caption{}
    \end{subfigure}%
    \begin{subfigure}{0.142\textwidth}
        \centering
        \includegraphics[width=\linewidth]{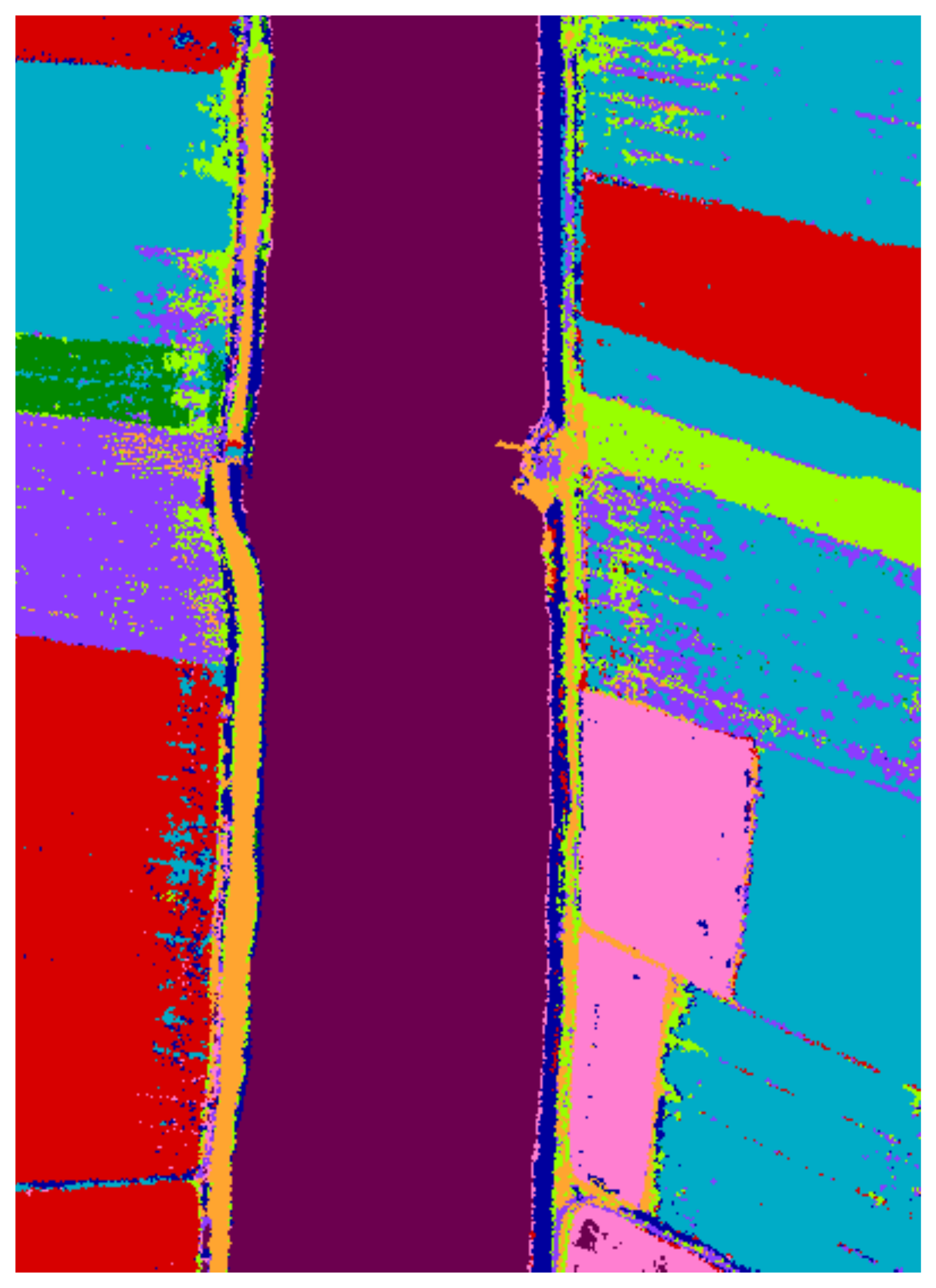}
        \caption{}
    \end{subfigure}%
    \vspace{0cm} 
    \begin{subfigure}{\textwidth}
        \centering
        \includegraphics[width=\linewidth]{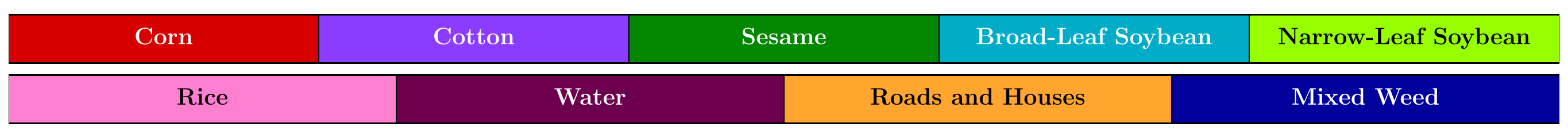}
    \end{subfigure}
    \caption{LK scene visualization and pixel-based classification maps: (a) False color map, (b) training set, (c) test set, (d) XGBoost, (e) SVC, (f) SAE + SVC (g) 1D-CNN, (h) CasRNN (i) miniGCN, (j) ViT, (k) SpectralFormer, (l) Mamba, (m) CEnc + SVC, (n) CEnc + Dense}
    \label{fig:LK_spectral}
\end{figure}

\subsection{Complementing Spectral-Spatial Methods\label{subsec:spectralspatial}}  

While our approach using coherent features demonstrated leading performance for a purely spectral method, it is important to acknowledge that state-of-the-art HSI classification methods are predominantly based on spectral-spatial patch-based deep learning techniques. These methods, such as SSRPnet~\cite{cheng2021hyperspectral}, morphFormer~\cite{roy2023spectral}, SMF-UL~\cite{sun2022perceiving}, and ESSAN~\cite{wang2023expansion}, leverage local contextual information by capitalizing on the spatial continuity often observed in land cover classes. Consequently, these methods generally exhibit robustness against salt-and-pepper noise, thereby enhancing their accuracy. By incorporating both spectral and spatial information, they have achieved notable performance on benchmark datasets, particularly in cases where classes exhibit high contiguity.

\subsubsection*{Limitations of Current Evaluation Practices\label{subsubsec:limit_ss}}

Despite the success of spectral-spatial methods, a critical examination of common evaluation practices reveals a potential overestimation of their real-world efficacy. Many studies employ random splits between training and test sets without considering spatial adjacency, potentially leading to test pixels being included within training patches. This overlap allows networks to learn the spatial context of test samples. For instance, a $7\times7$ patch centered on a training pixel may inadvertently include neighboring test pixels, providing the model with indirect access to test data during training. This issue can lead to an overestimation of the model's generalization capabilities, as it does not accurately reflect real-world applications where overlap between training and test areas is typically unlikely, such as in large-scale surveys or when models are applied to novel, unseen regions.

\begin{figure}[h]
    \centering
    \begin{minipage}{\linewidth}
        \centering
        \begin{subfigure}{0.138\linewidth}
            \centering
            \includegraphics[width=\linewidth]{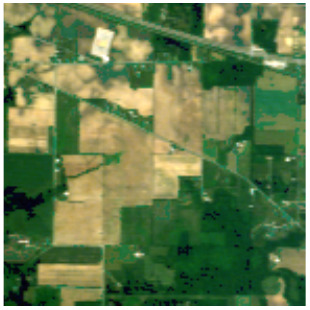}
            \caption{}
        \end{subfigure}%
        \hfill
        \begin{subfigure}{0.138\linewidth}
            \centering
            \includegraphics[width=\linewidth]{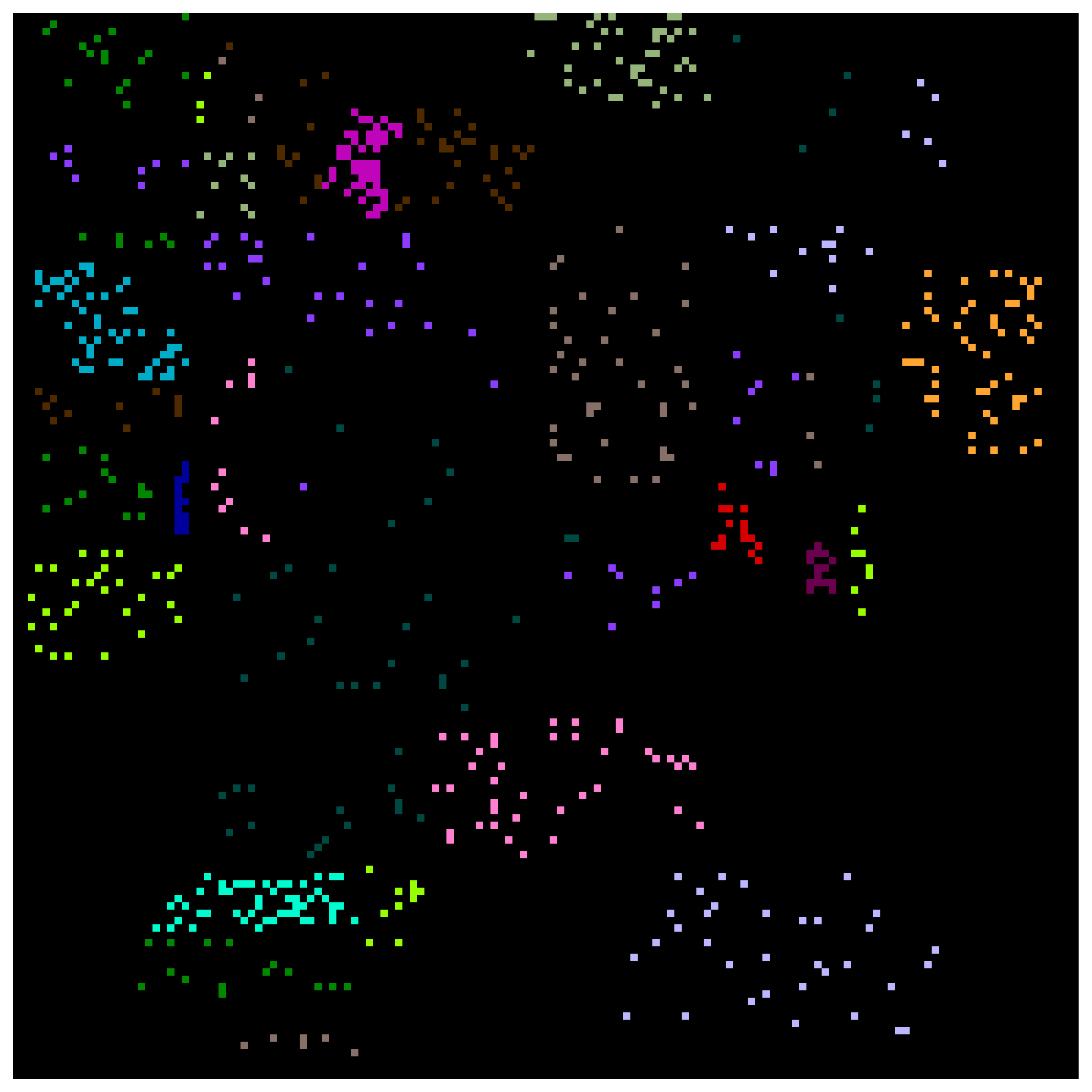}
            \caption{}
        \end{subfigure}%
        \hfill
        \begin{subfigure}{0.138\linewidth}
            \centering
            \includegraphics[width=\linewidth]{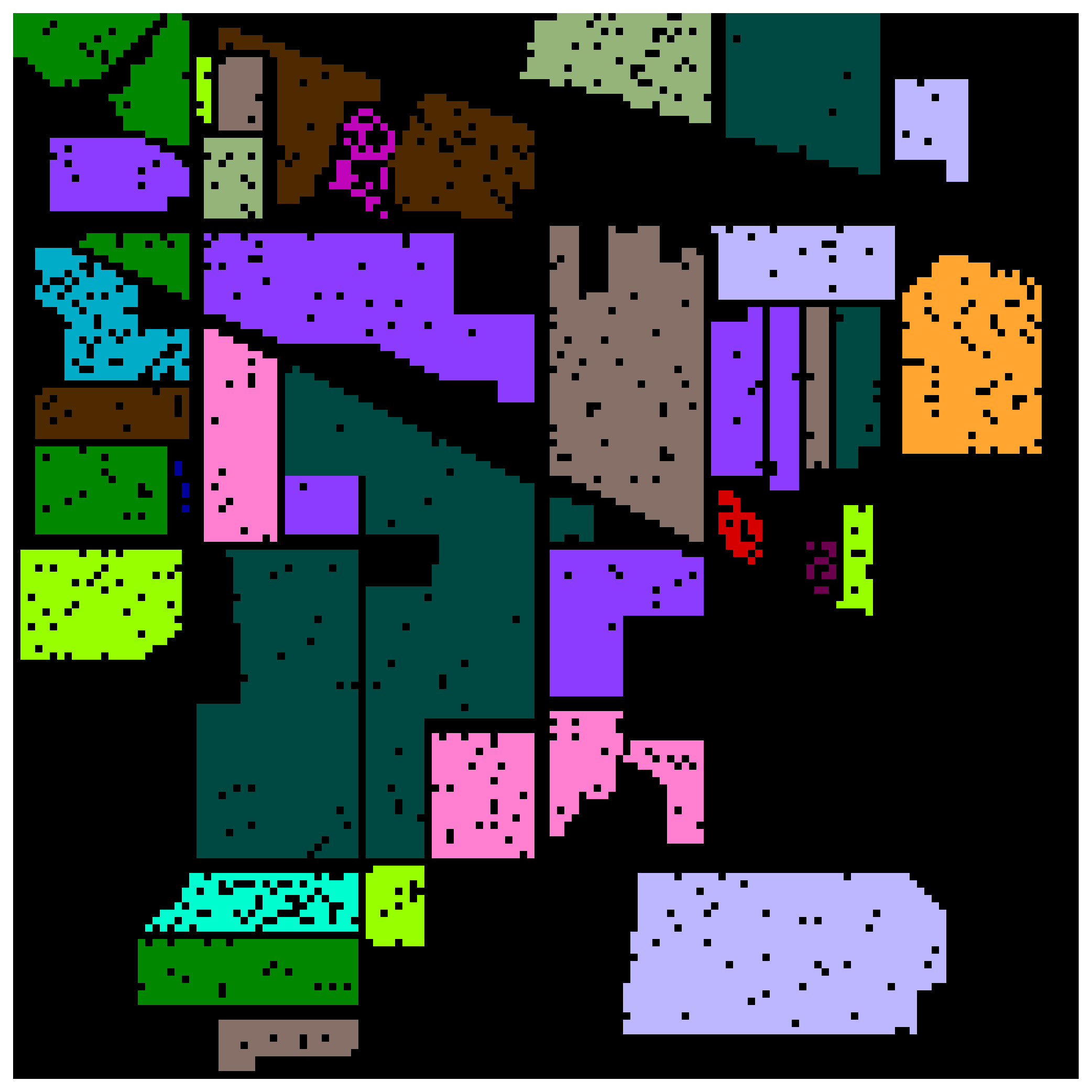}
            \caption{}
        \end{subfigure}%
        \hfill
        \begin{subfigure}{0.138\linewidth}
            \centering
            \includegraphics[width=\linewidth]{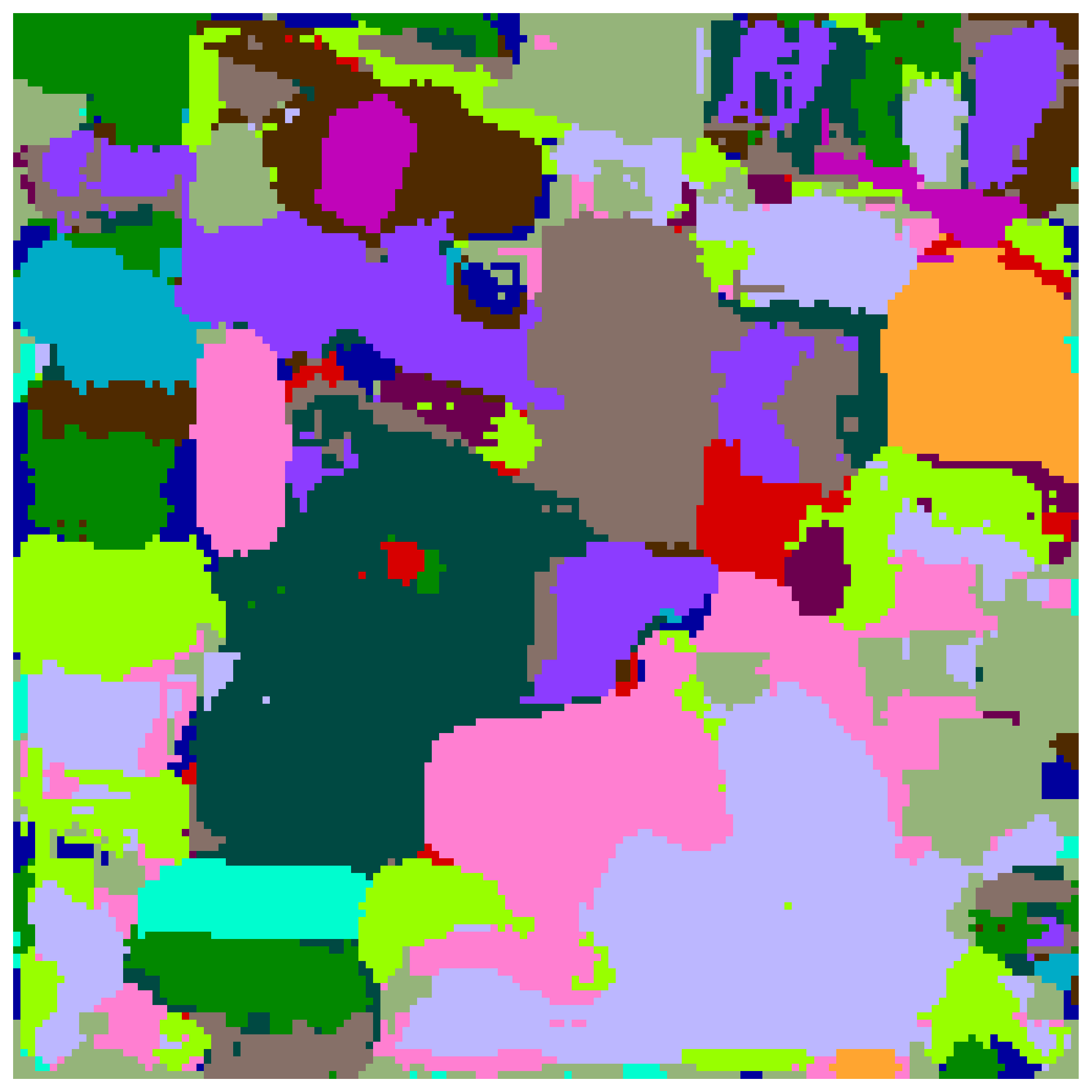}
            \caption{}
        \end{subfigure}%
        \hfill
        \begin{subfigure}{0.138\linewidth}
            \centering
            \includegraphics[width=\linewidth]{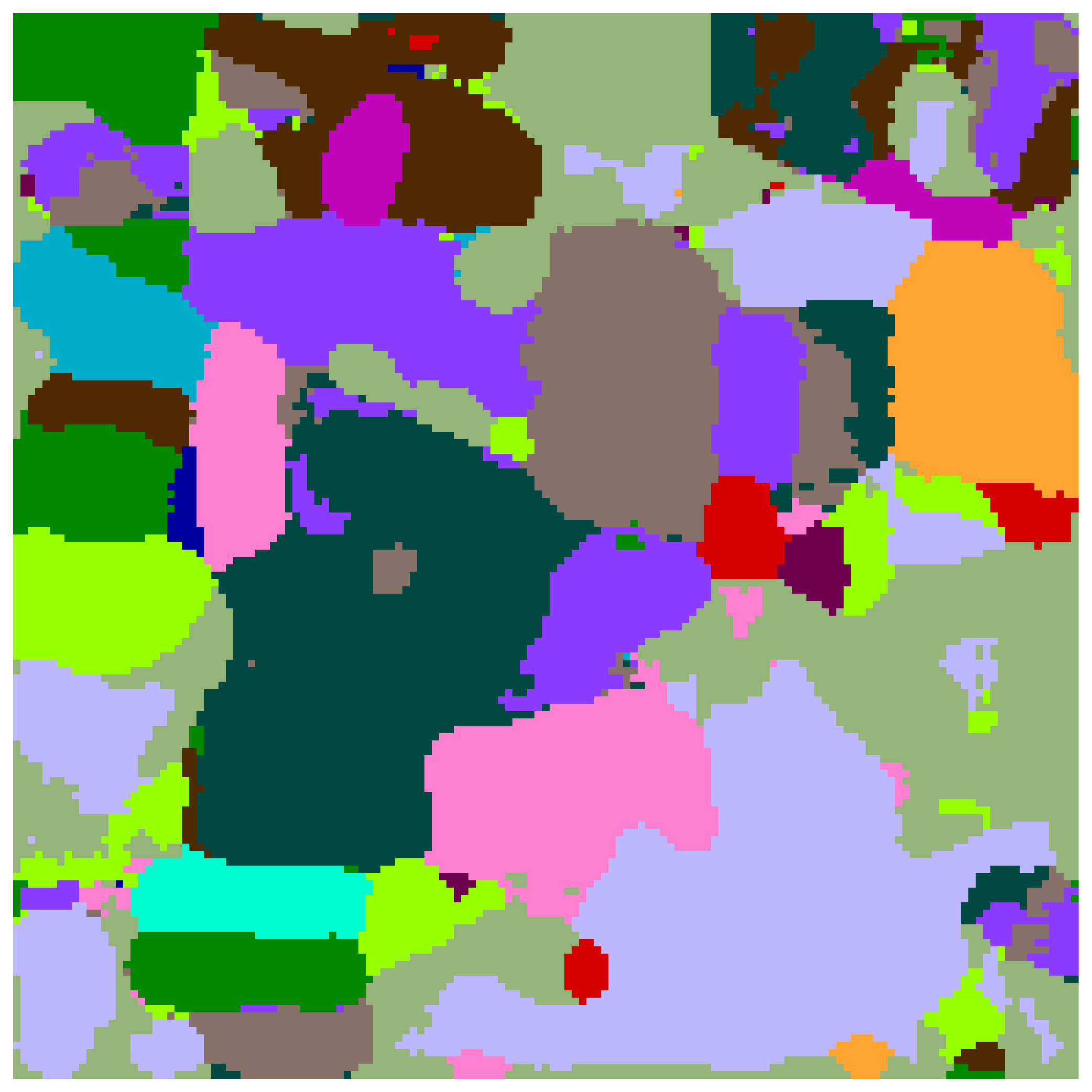}
            \caption{}
        \end{subfigure}%
        \hfill
        \begin{subfigure}{0.138\linewidth}
            \centering
            \includegraphics[width=\linewidth]{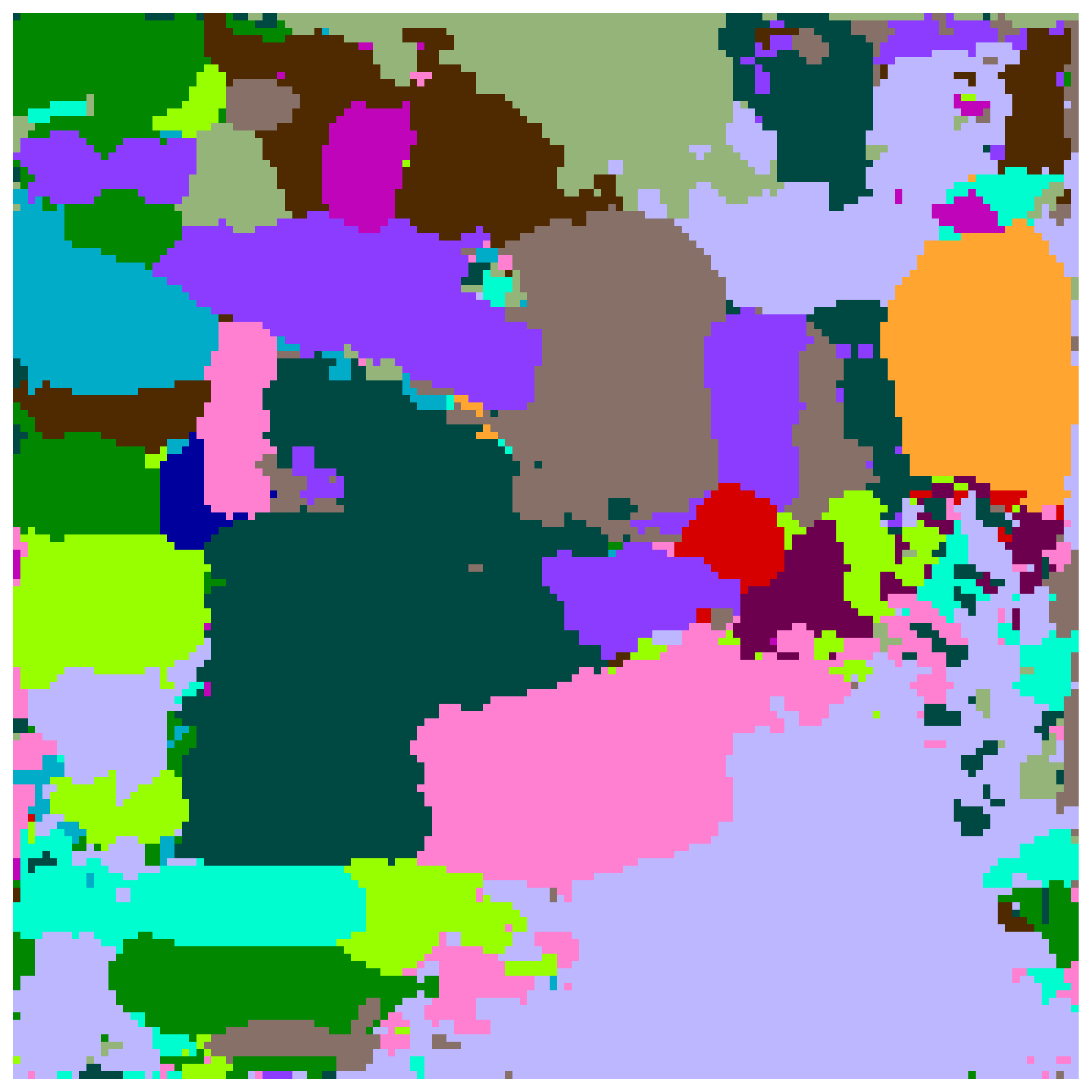}
            \caption{}
        \end{subfigure}%
        \hfill
        \begin{subfigure}{0.138\linewidth}
            \centering
            \includegraphics[width=\linewidth]{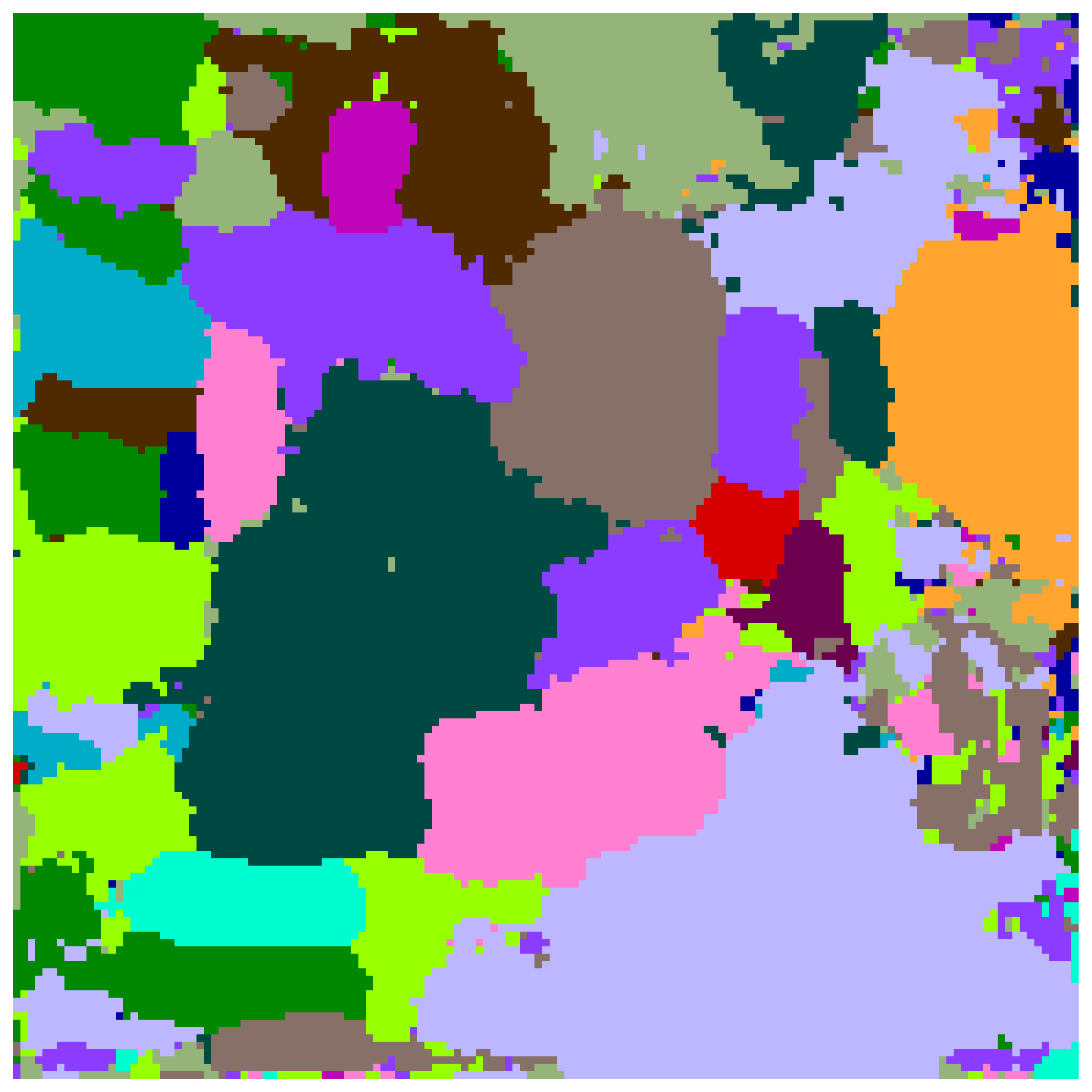}
            \caption{}
        \end{subfigure}
    \end{minipage}
    
    \vspace{0.1cm}
    
    \begin{minipage}{\linewidth}
        \centering
        \begin{subfigure}{0.138\linewidth}
            \centering
            \includegraphics[width=\linewidth]{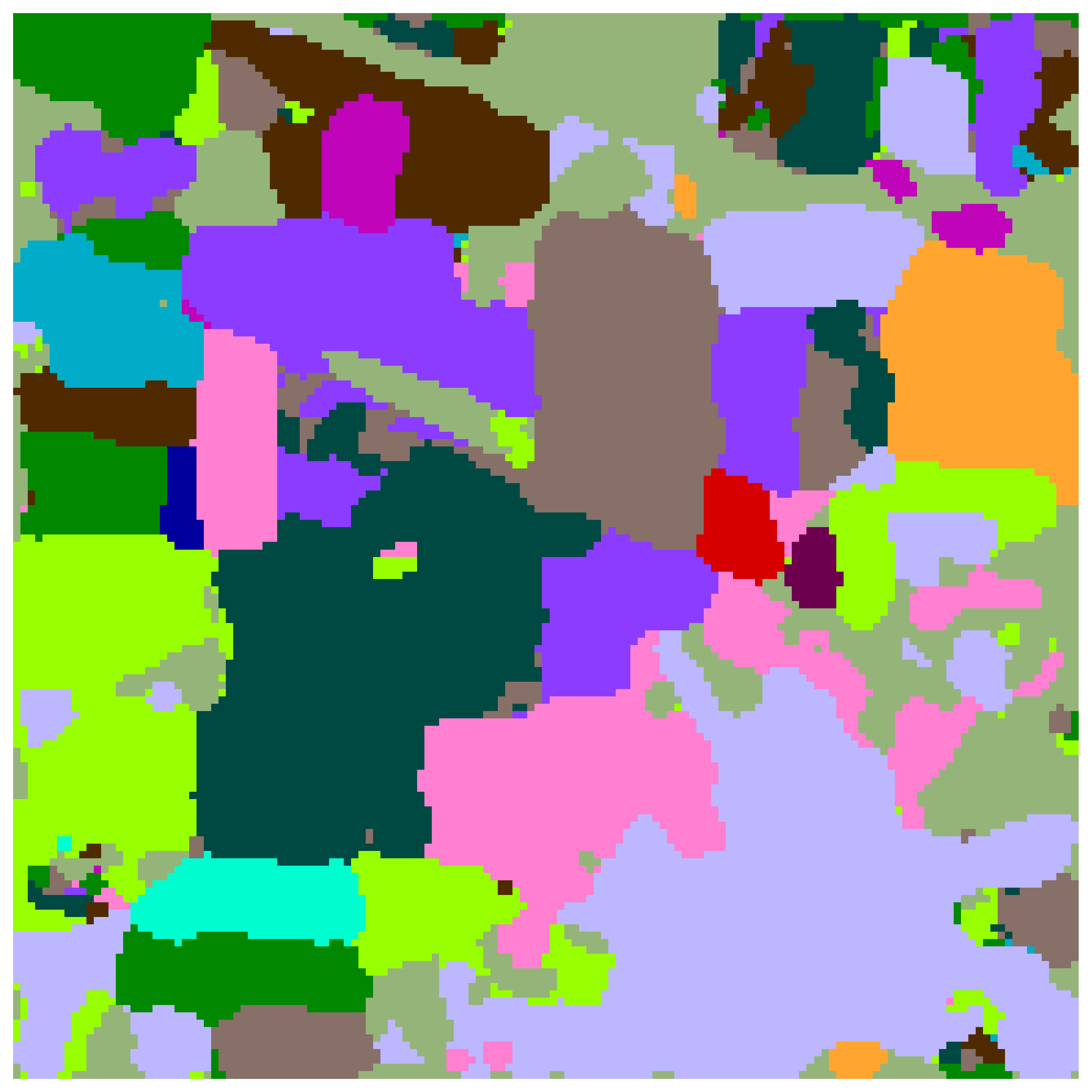}
            \caption{}
        \end{subfigure}%
        \hspace{0.0057\linewidth}%
        \begin{subfigure}{0.138\linewidth}
            \centering
            \includegraphics[width=\linewidth]{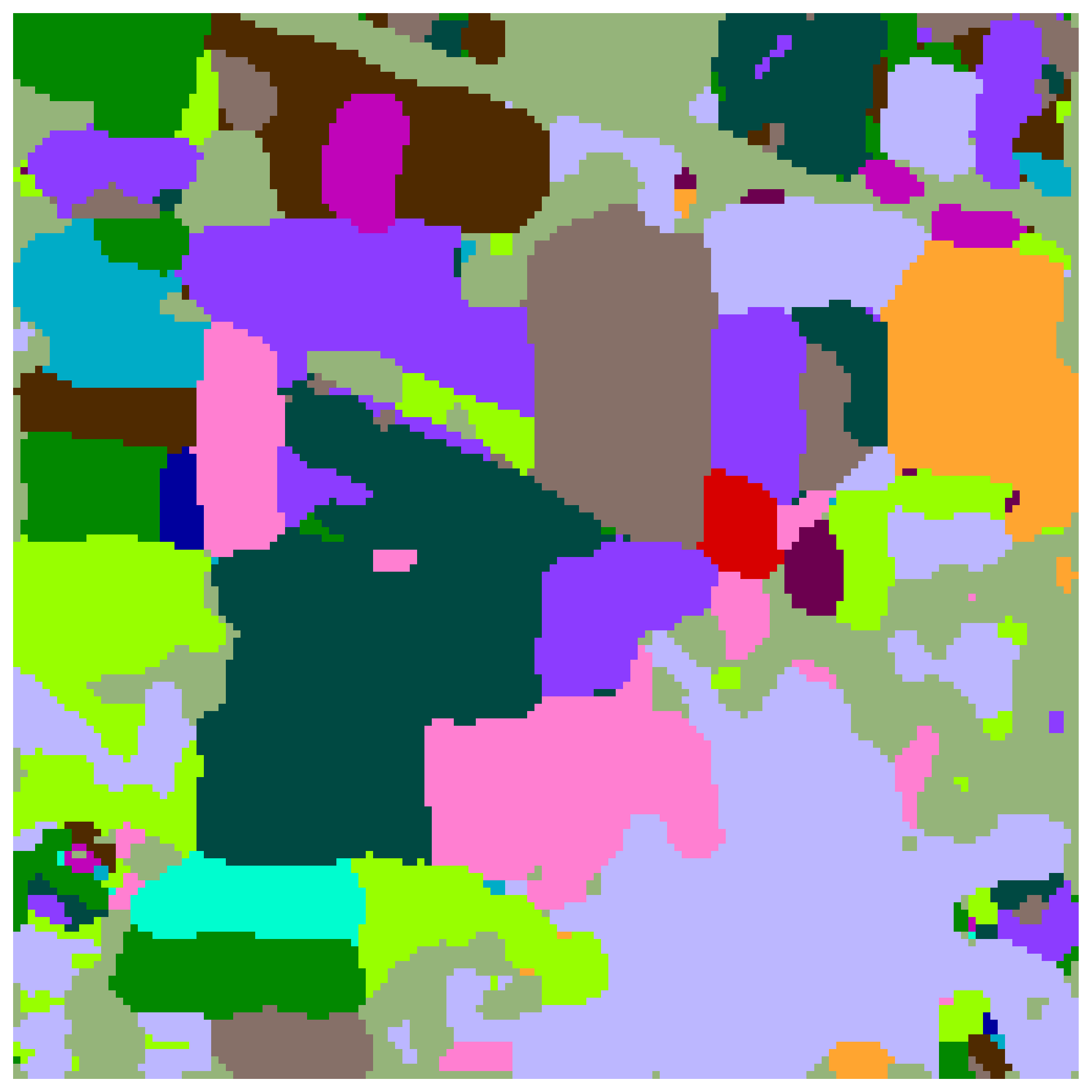}
            \caption{}
        \end{subfigure}%
        \hspace{0.0057\linewidth}%
        \begin{subfigure}{0.138\linewidth}
            \centering
            \includegraphics[width=\linewidth]{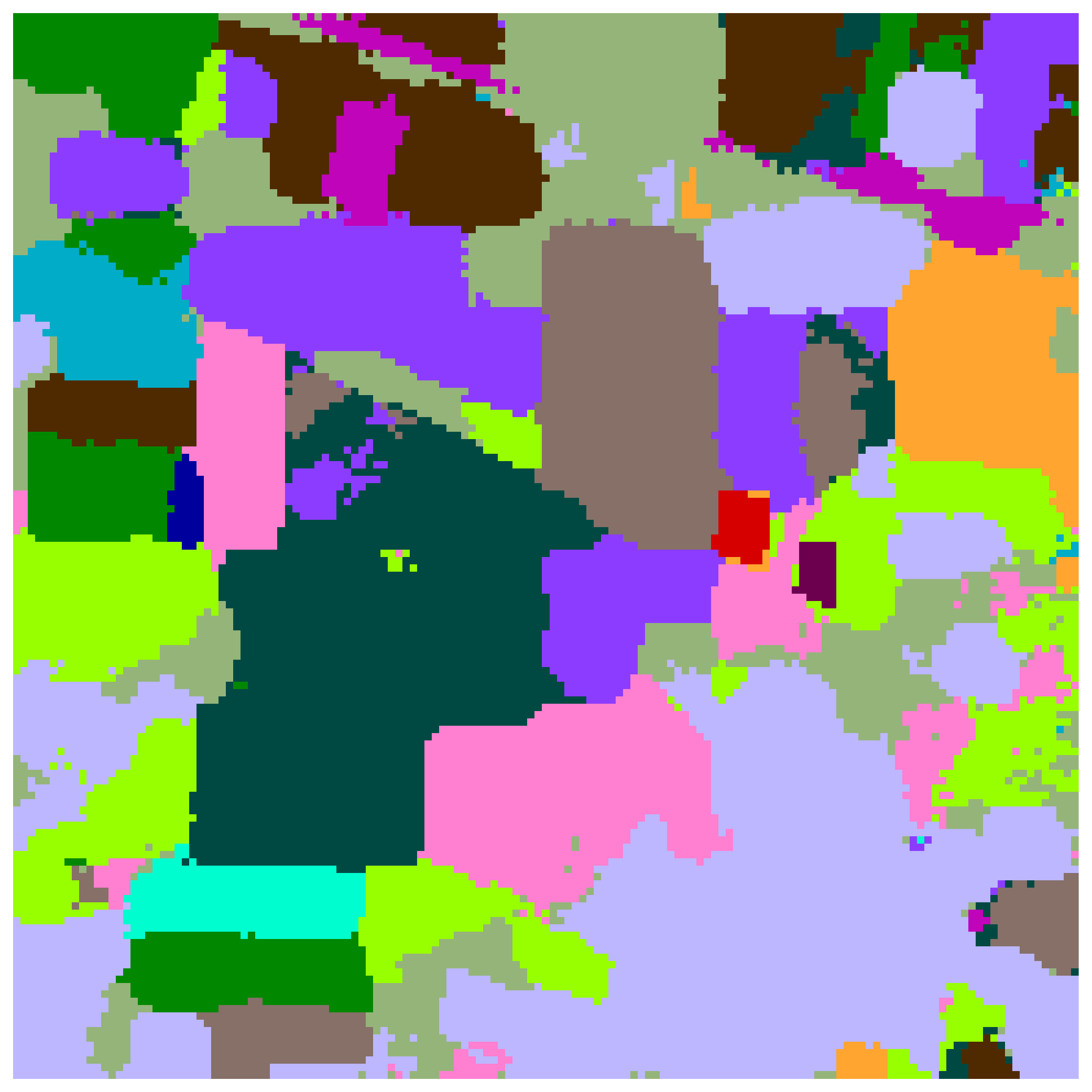}
            \caption{}
        \end{subfigure}%
        \hspace{0.0057\linewidth}%
        \begin{subfigure}{0.138\linewidth}
            \centering
            \includegraphics[width=\linewidth]{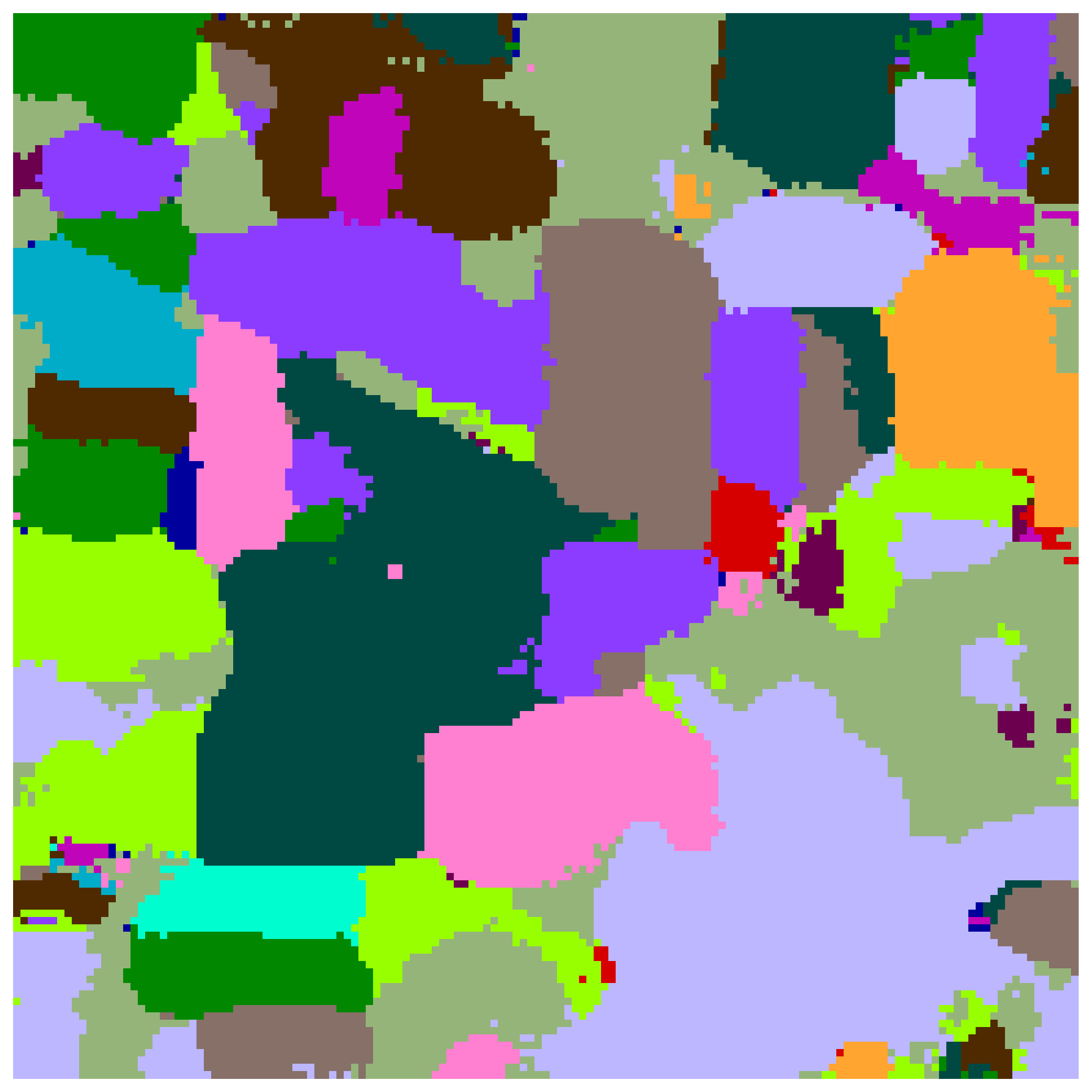}
            \caption{}
        \end{subfigure}%
        \hspace{0.0057\linewidth}%
        \begin{subfigure}{0.138\linewidth}
            \centering
            \includegraphics[width=\linewidth]{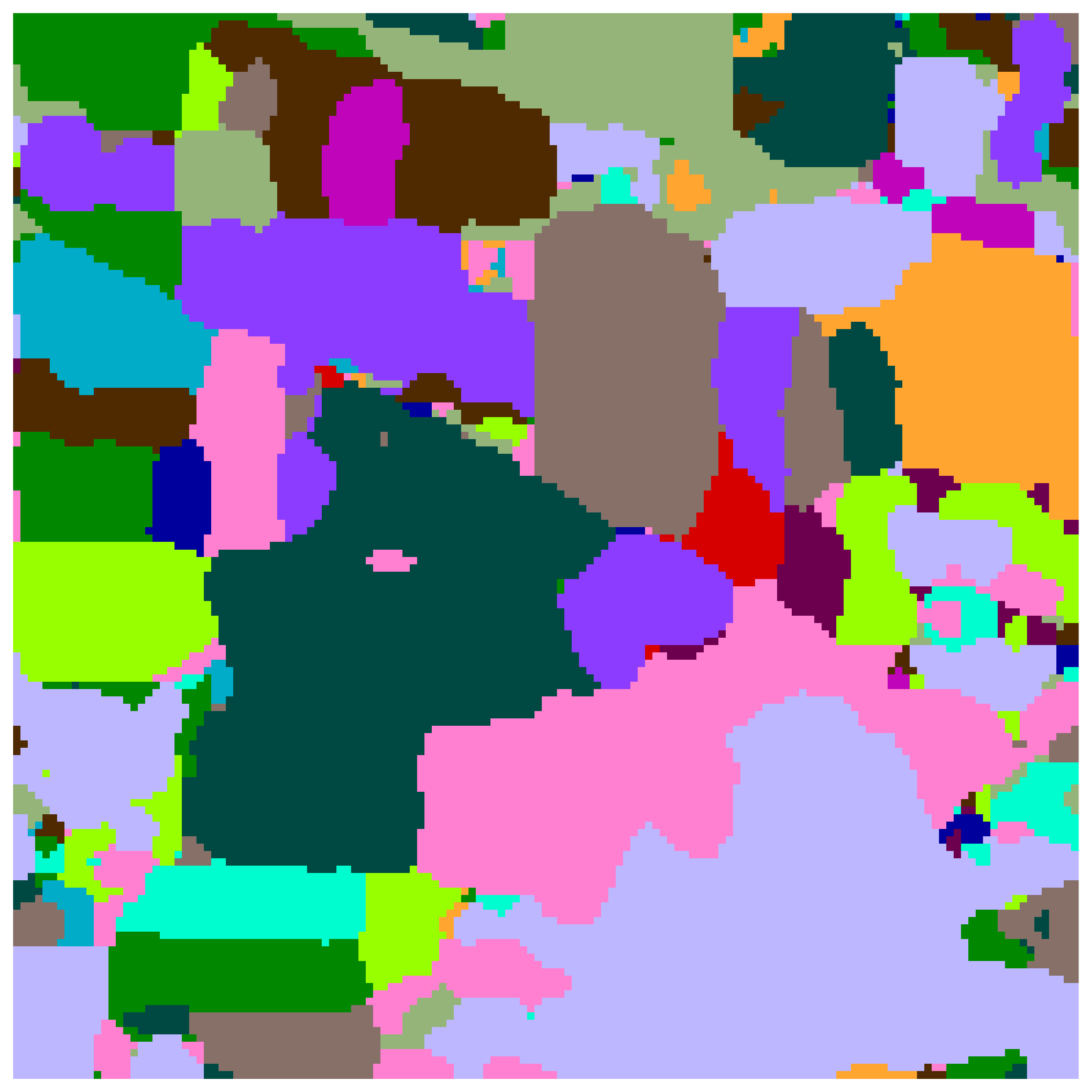}
            \caption{}
        \end{subfigure}%
        \hspace{0.0057\linewidth}%
        \begin{subfigure}{0.138\linewidth}
            \centering
            \includegraphics[width=\linewidth]{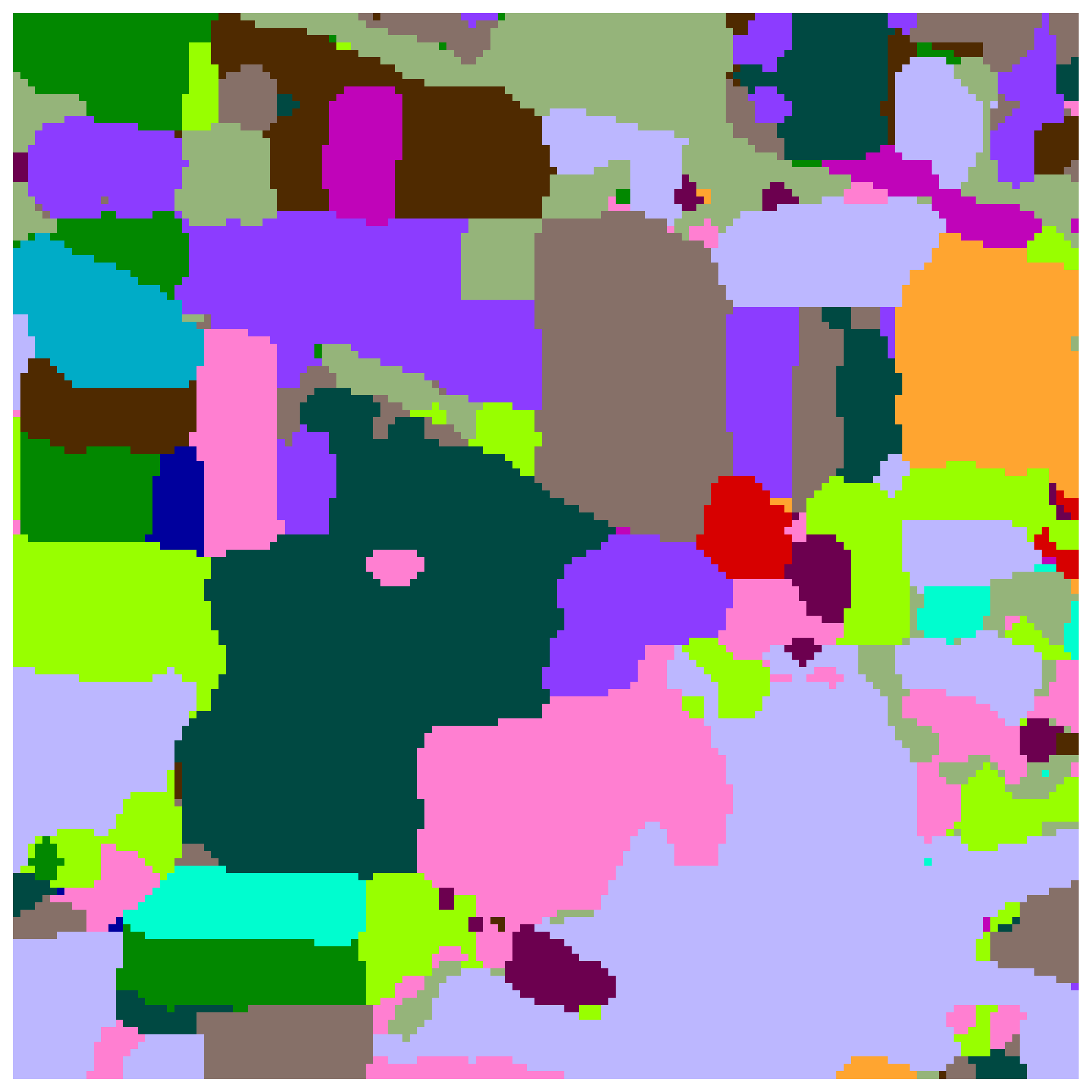}
            \caption{}
        \end{subfigure}
    \end{minipage}
    
    \vspace{0.1cm}
    
    \begin{subfigure}{1\linewidth}
        \centering
        \includegraphics[width=\linewidth, height=1cm]{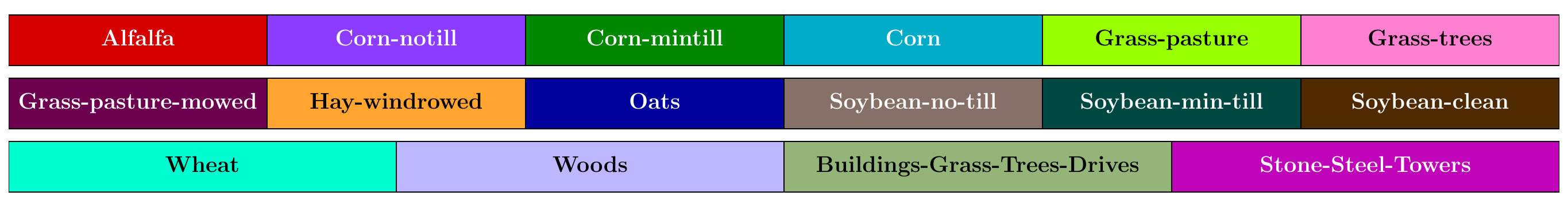}
    \end{subfigure}
    
    \caption{Classification maps and train-test sets for IP scene. (a) False color map, (b) Train set, (c) Test set, (d) SSTN, (e) SSTN + CEnc, (f) FCN, (g) FCN + CEnc, (h) A2S2K, (i) A2S2K + CEnc, (j) AMS-M2ESL, (k) AMS-M2ESL + CEnc, (l) MambaHSI, (m) MambaHSI + CEnc}
    \label{fig:classification_maps_IP}
\end{figure}

\begin{figure}[h]
    \centering
    \begin{minipage}{\linewidth}
        \centering
        \begin{subfigure}{0.138\linewidth}
            \centering
            \includegraphics[width=\linewidth]{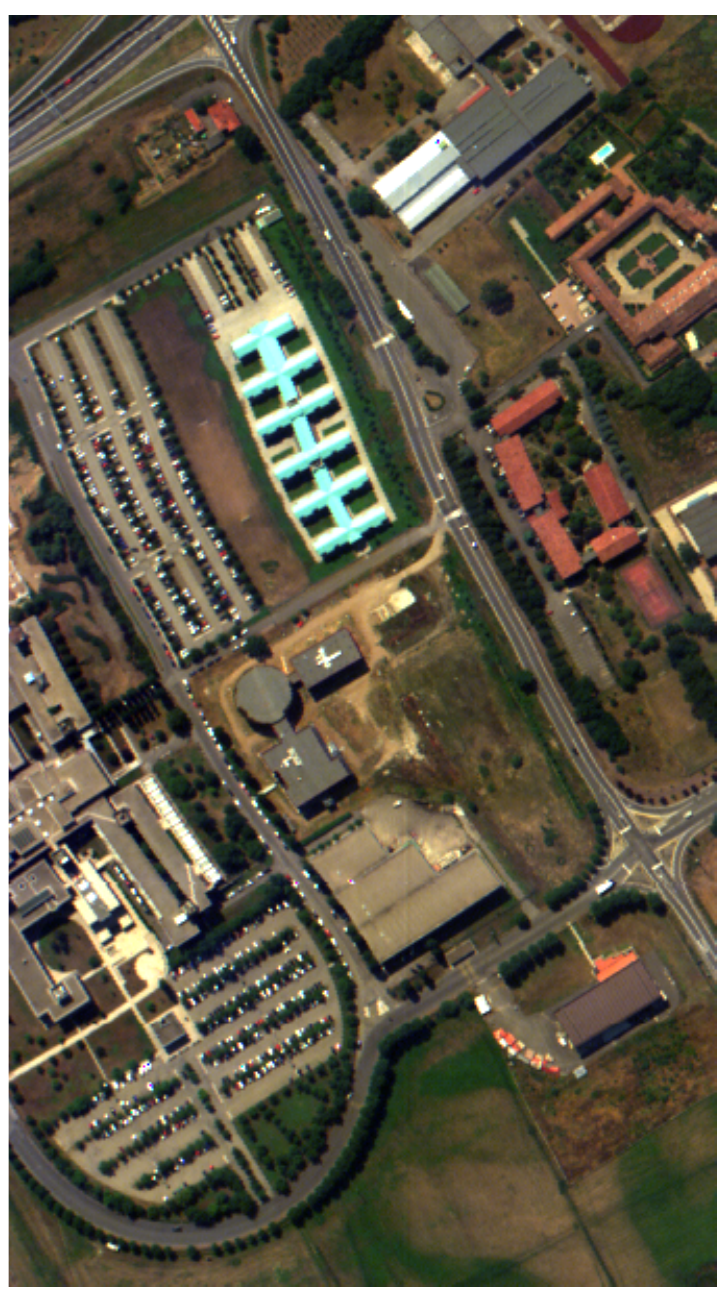}
            \caption{}
        \end{subfigure}%
        \hfill
        \begin{subfigure}{0.138\linewidth}
            \centering
            \includegraphics[width=\linewidth]{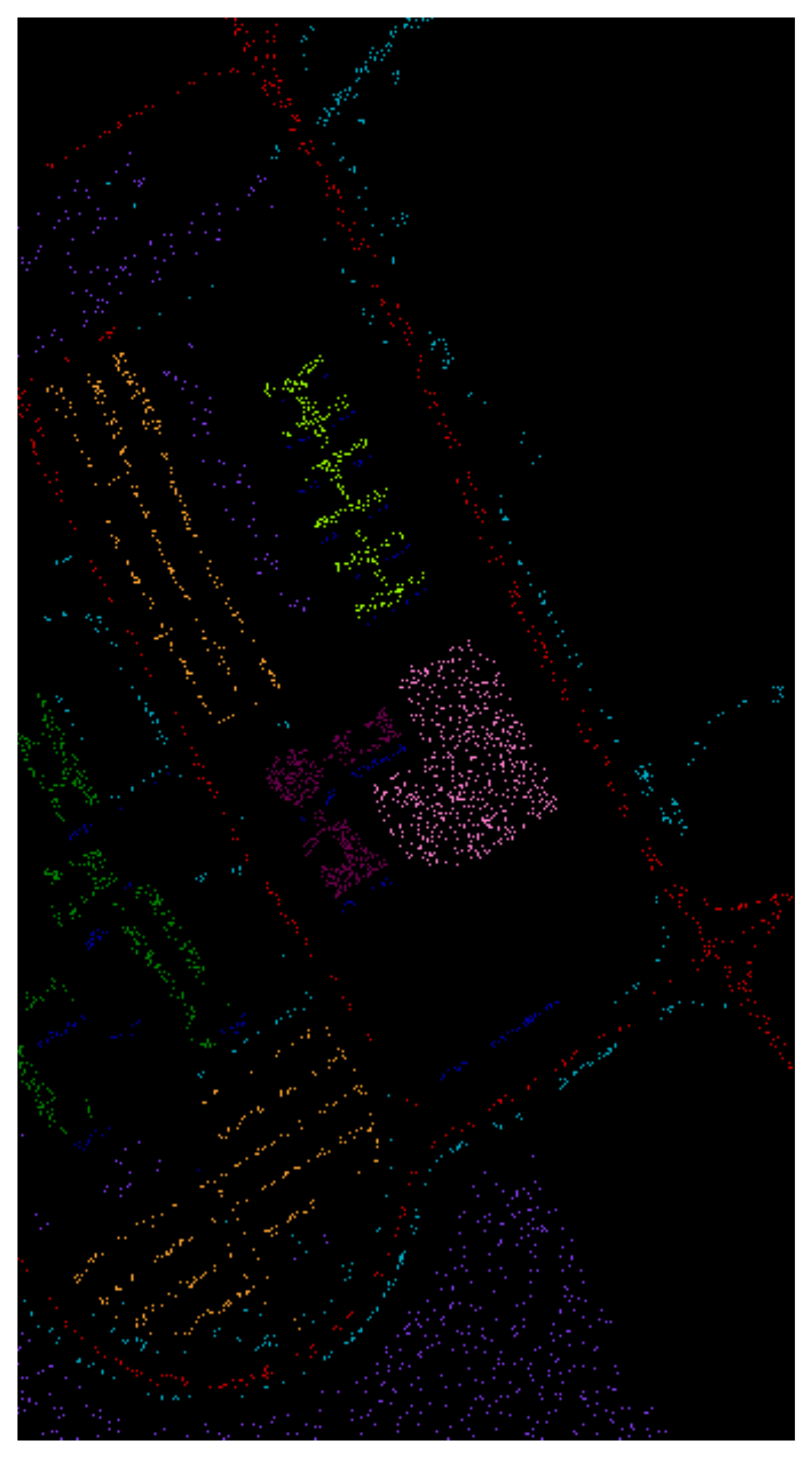}
            \caption{}
        \end{subfigure}%
        \hfill
        \begin{subfigure}{0.138\linewidth}
            \centering
            \includegraphics[width=\linewidth]{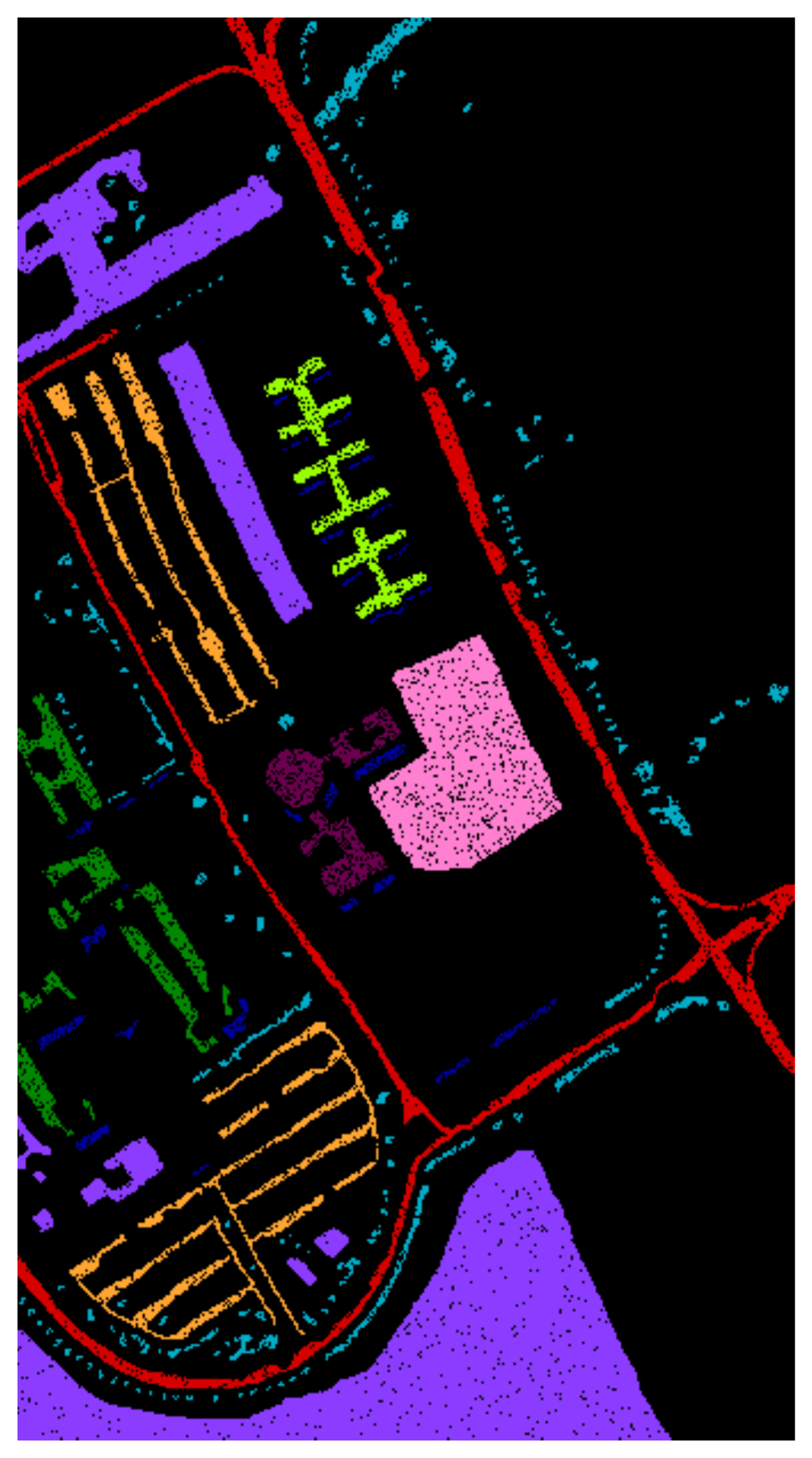}
            \caption{}
        \end{subfigure}%
        \hfill
        \begin{subfigure}{0.138\linewidth}
            \centering
            \includegraphics[width=\linewidth]{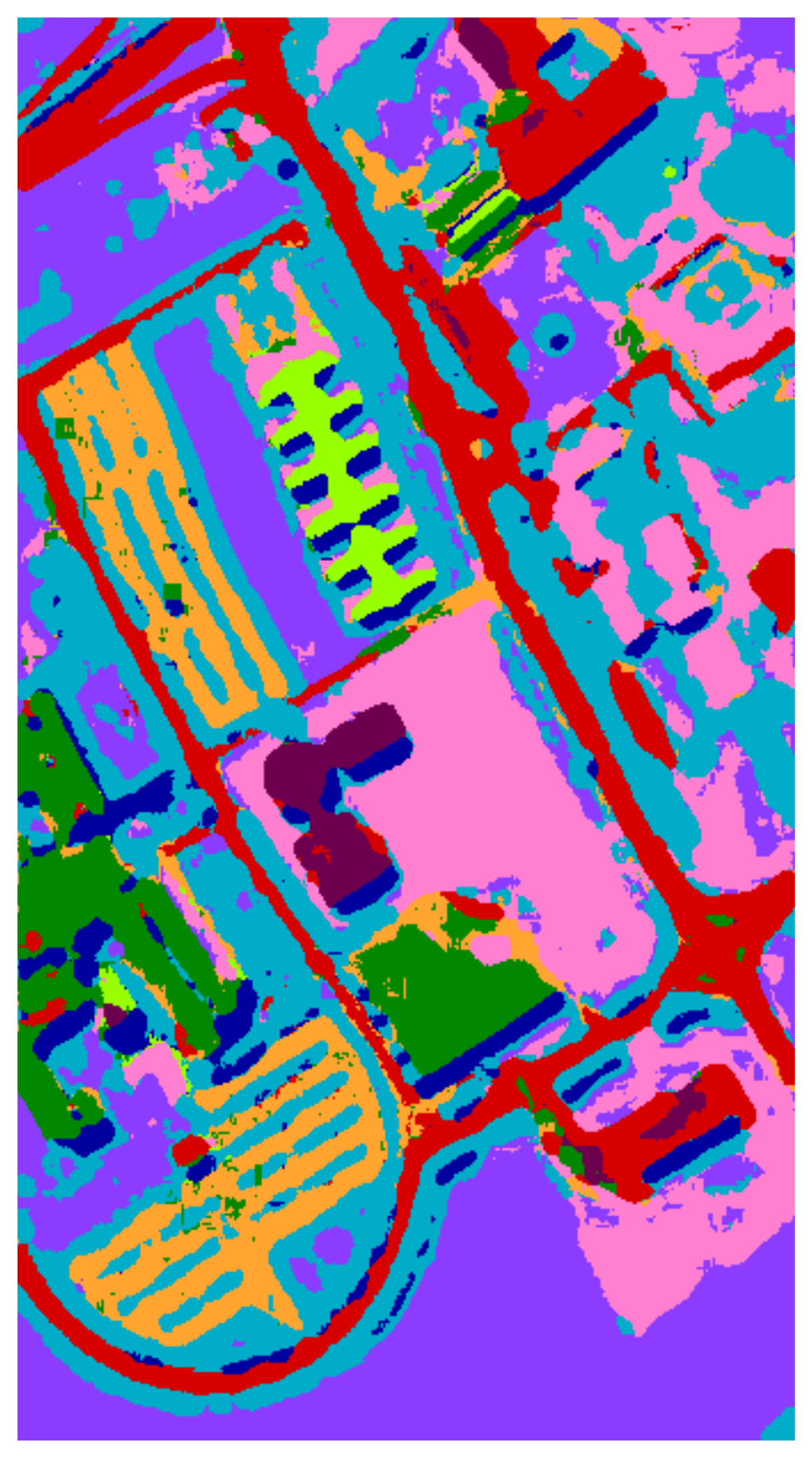}
            \caption{}
        \end{subfigure}%
        \hfill
        \begin{subfigure}{0.138\linewidth}
            \centering
            \includegraphics[width=\linewidth]{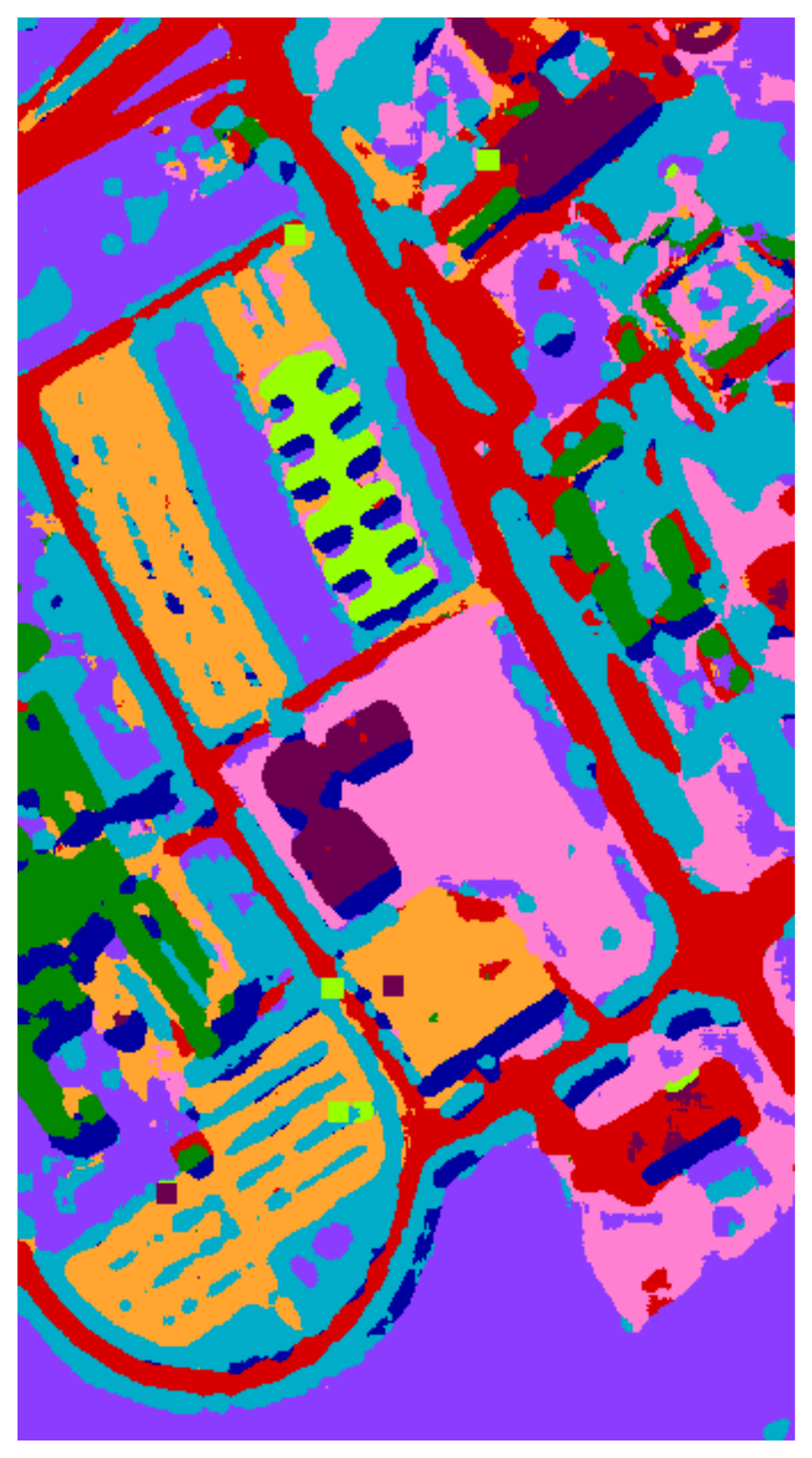}
            \caption{}
        \end{subfigure}%
        \hfill
        \begin{subfigure}{0.138\linewidth}
            \centering
            \includegraphics[width=\linewidth]{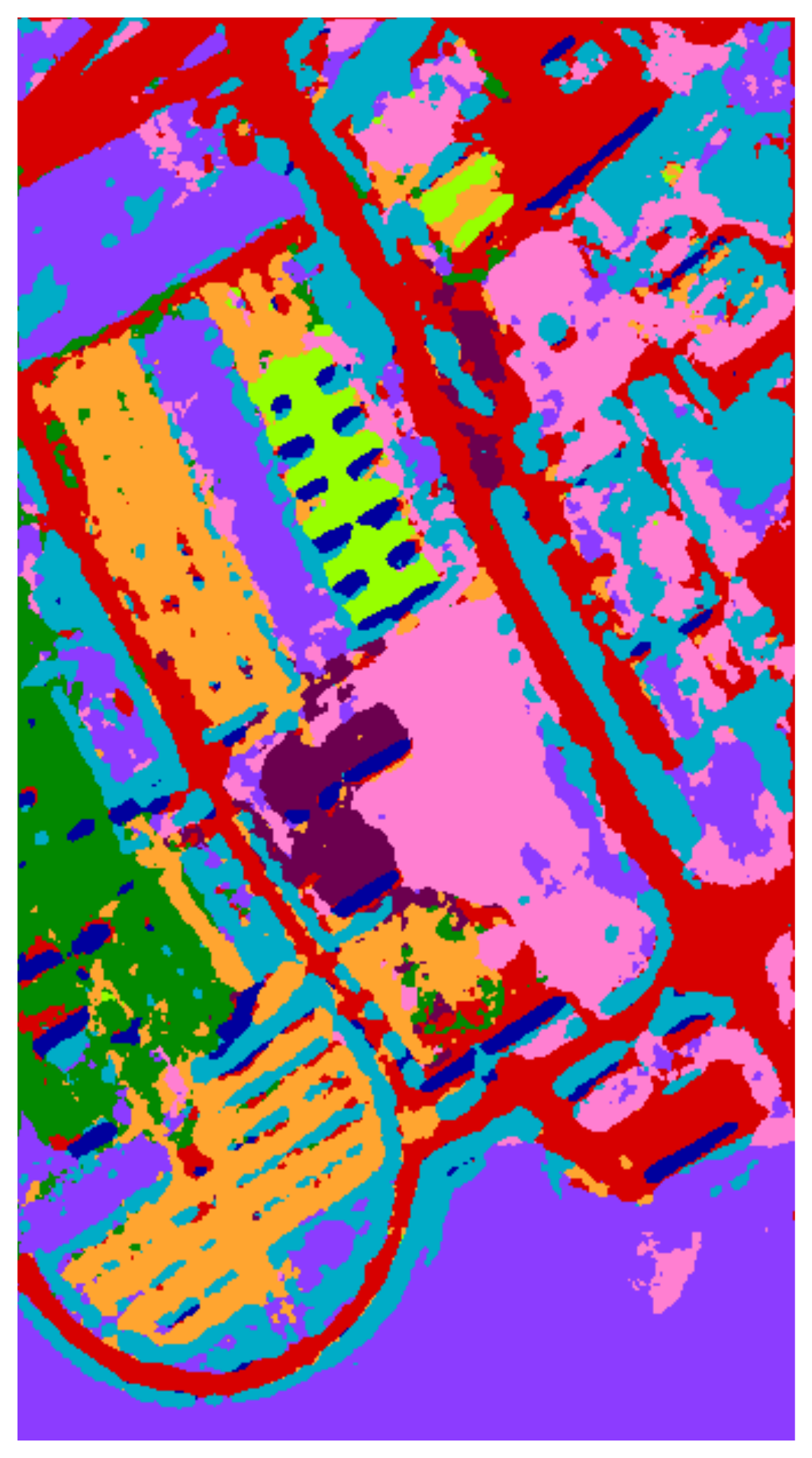}
            \caption{}
        \end{subfigure}%
        \hfill
        \begin{subfigure}{0.138\linewidth}
            \centering
            \includegraphics[width=\linewidth]{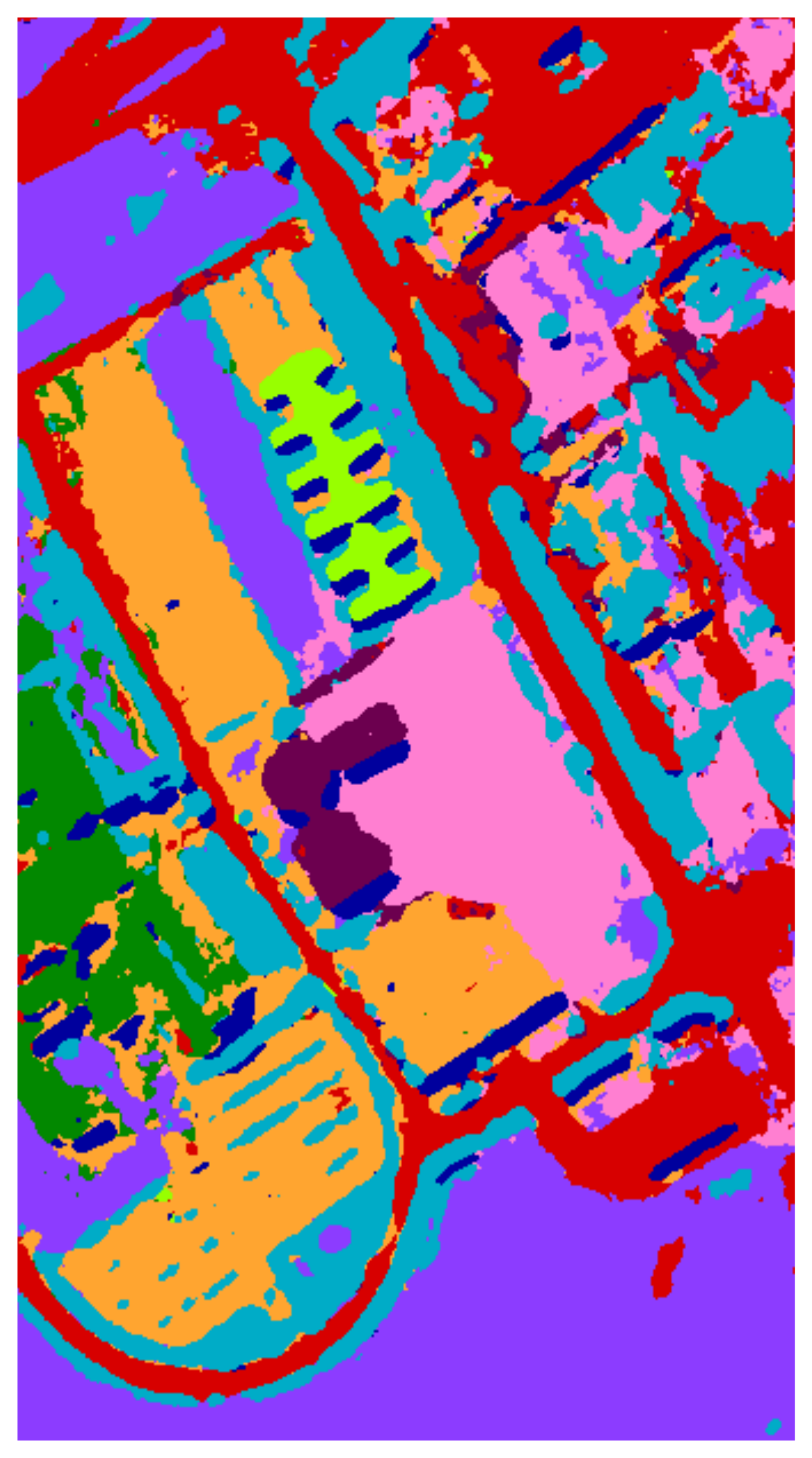}
            \caption{}
        \end{subfigure}
    \end{minipage}
    
    \vspace{0.1cm}
    
    \begin{minipage}{\linewidth}
        \centering
        \begin{subfigure}{0.138\linewidth}
            \centering
            \includegraphics[width=\linewidth]{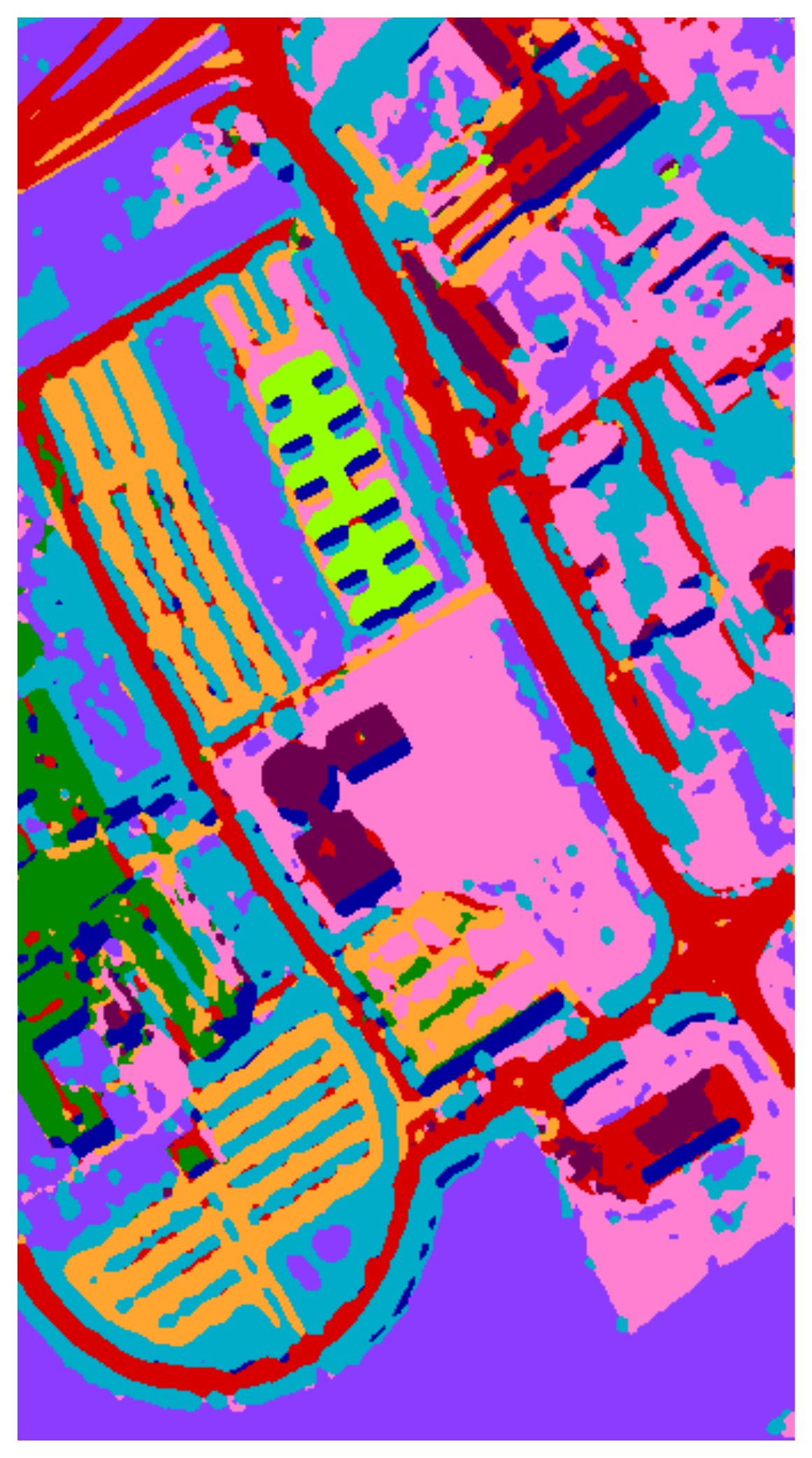}
            \caption{}
        \end{subfigure}%
        \hspace{0.0057\linewidth}%
        \begin{subfigure}{0.138\linewidth}
            \centering
            \includegraphics[width=\linewidth]{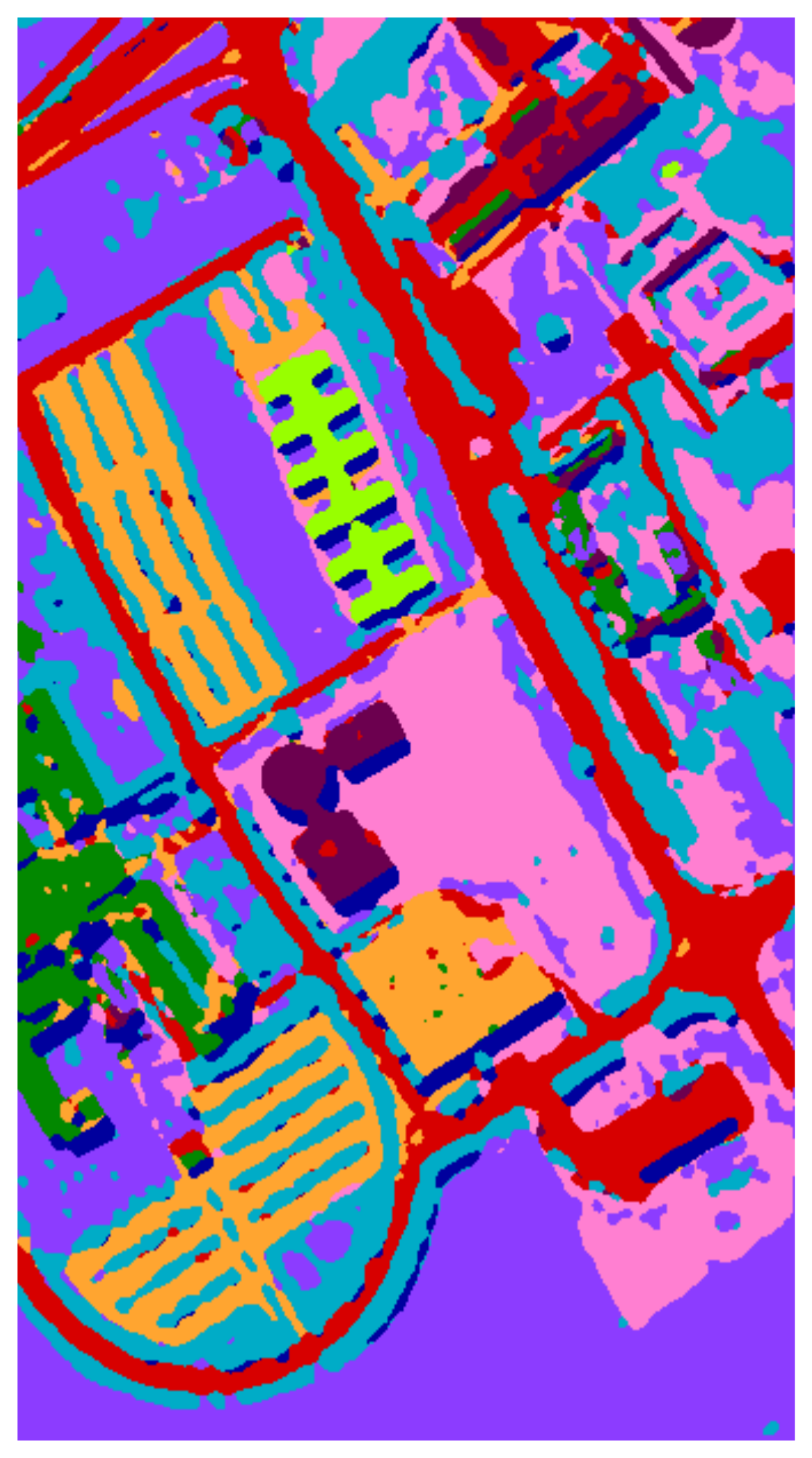}
            \caption{}
        \end{subfigure}%
        \hspace{0.0057\linewidth}%
        \begin{subfigure}{0.138\linewidth}
            \centering
            \includegraphics[width=\linewidth]{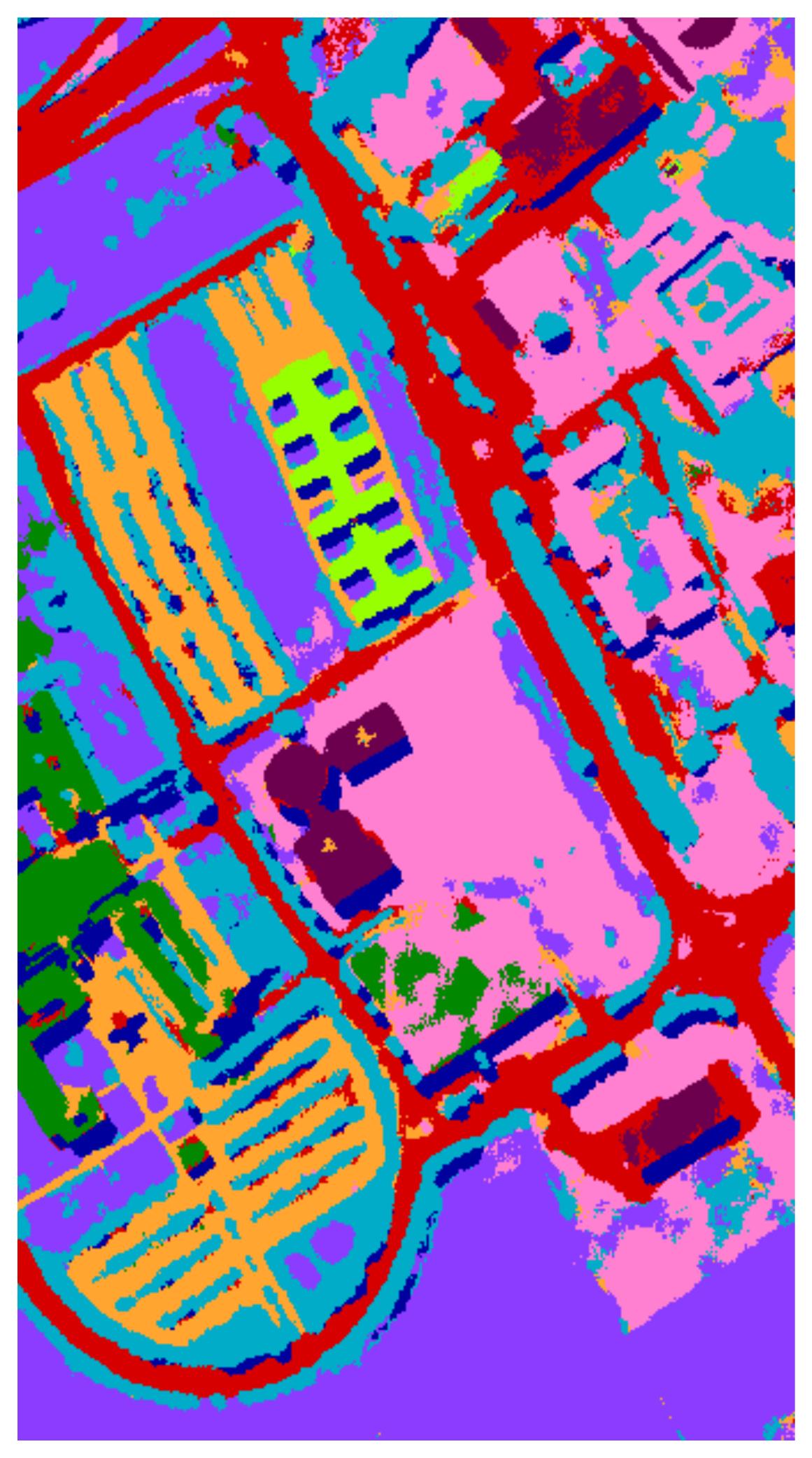}
            \caption{}
        \end{subfigure}%
        \hspace{0.0057\linewidth}%
        \begin{subfigure}{0.138\linewidth}
            \centering
            \includegraphics[width=\linewidth]{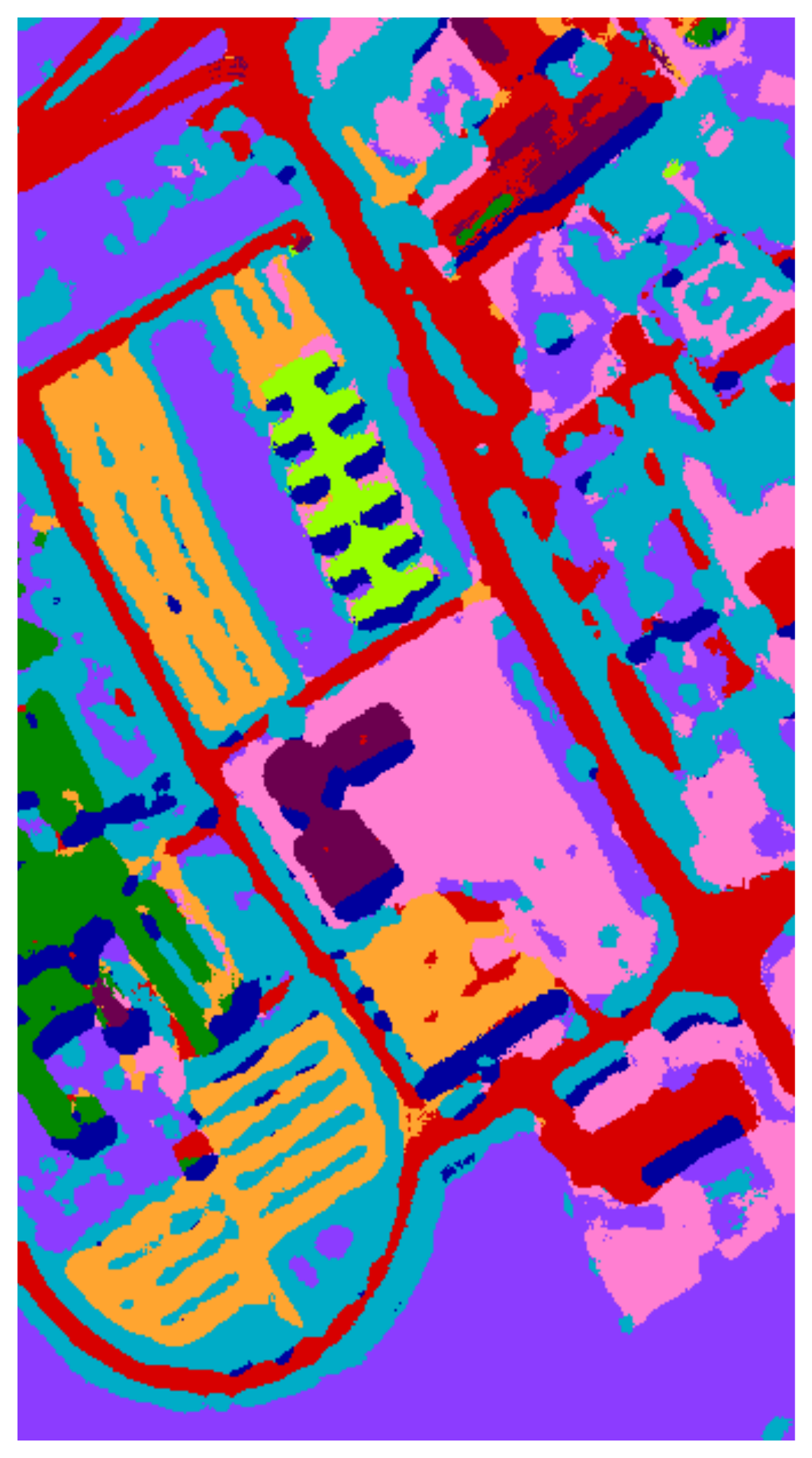}
            \caption{}
        \end{subfigure}%
        \hspace{0.0057\linewidth}%
        \begin{subfigure}{0.138\linewidth}
            \centering
            \includegraphics[width=\linewidth]{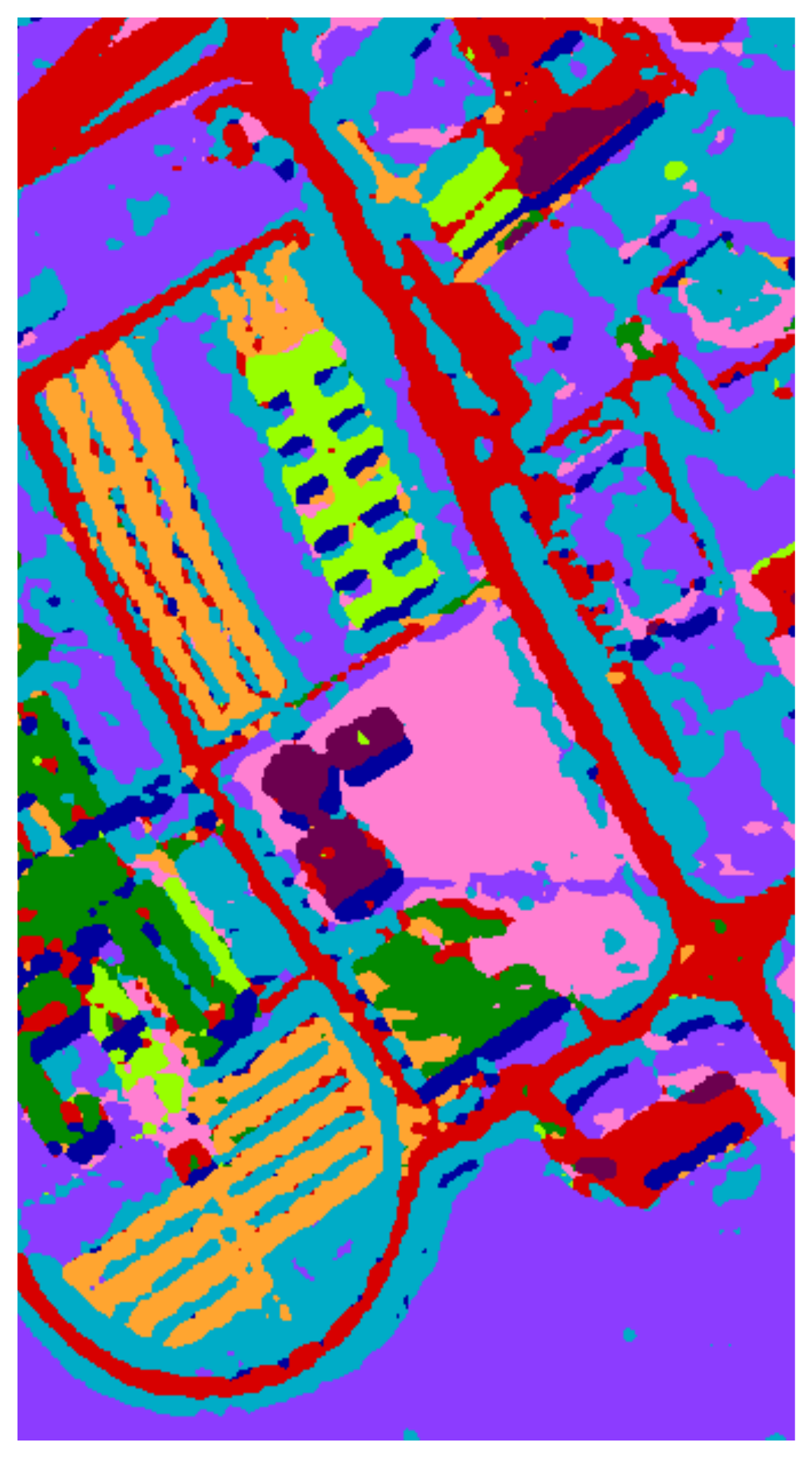}
            \caption{}
        \end{subfigure}%
        \hspace{0.0057\linewidth}%
        \begin{subfigure}{0.138\linewidth}
            \centering
            \includegraphics[width=\linewidth]{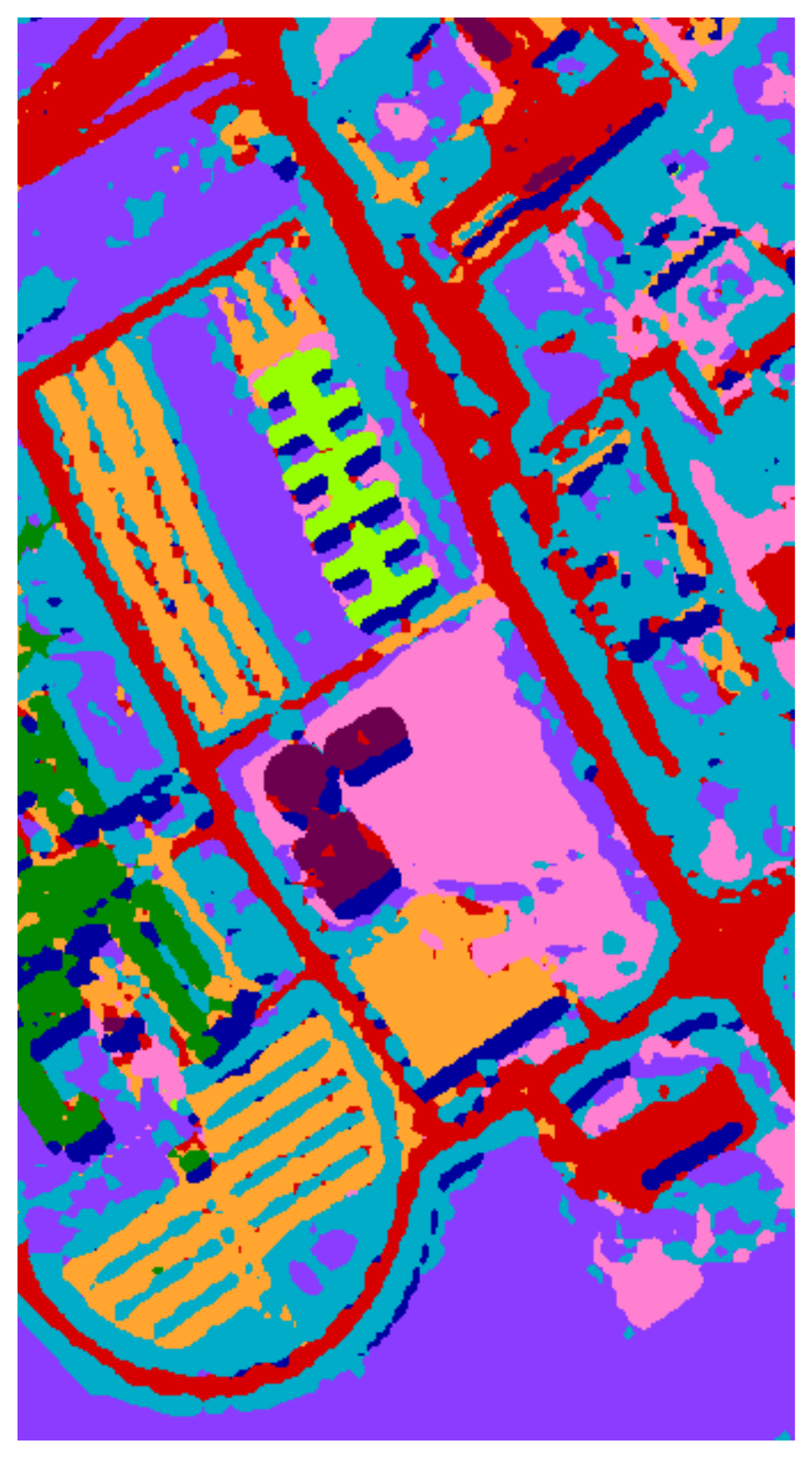}
            \caption{}
        \end{subfigure}
    \end{minipage}
    
    \vspace{0.1cm}
    
    \begin{subfigure}{1\linewidth}
        \centering
        \includegraphics[width=\linewidth, height=1cm]{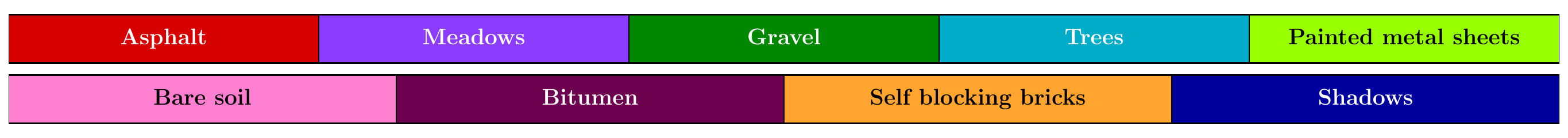}
    \end{subfigure}
    
    \caption{Classification maps and train-test sets for PU (random-split) scene: (a) False color map, (b) Train set, (c) Test set, (d) SSTN, (e) SSTN + CEnc, (f) FCN, (g) FCN + CEnc, (h) A2S2K, (i) A2S2K + CEnc, (j) AMS-M2ESL, (k) AMS-M2ESL + CEnc, (l) MambaHSI, (m) MambaHSI + CEnc}
    \label{fig:classification_maps_Pavia}
\end{figure}

\begin{figure}[h]
    \centering
    \begin{minipage}{\linewidth}
        \centering
        \begin{subfigure}{0.138\linewidth}
            \centering
            \includegraphics[width=\linewidth]{FCmaps/Pavia_fc.pdf}
            \caption{}
        \end{subfigure}%
        \hfill
        \begin{subfigure}{0.138\linewidth}
            \centering
            \includegraphics[width=\linewidth]{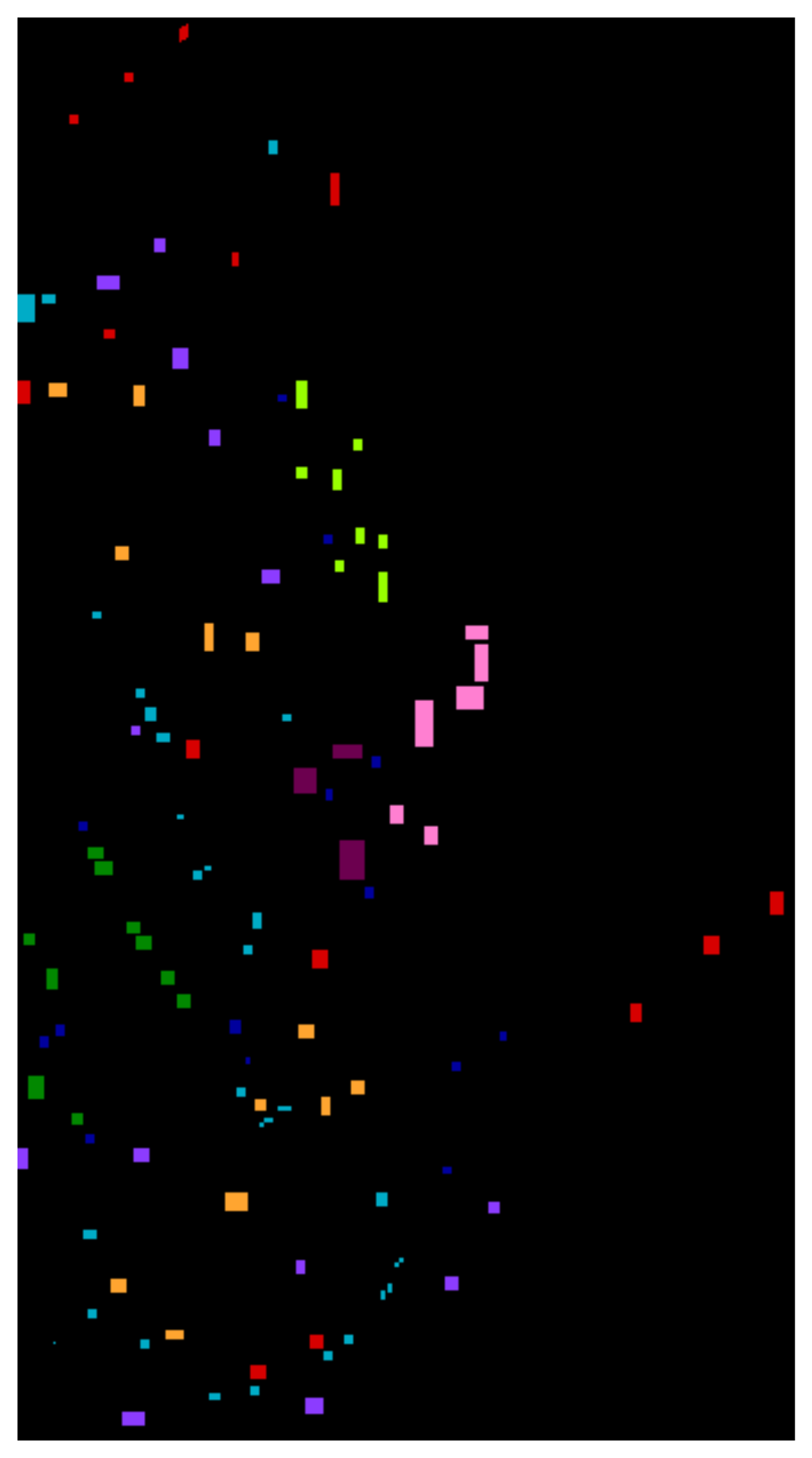}
            \caption{}
        \end{subfigure}%
        \hfill
        \begin{subfigure}{0.138\linewidth}
            \centering
            \includegraphics[width=\linewidth]{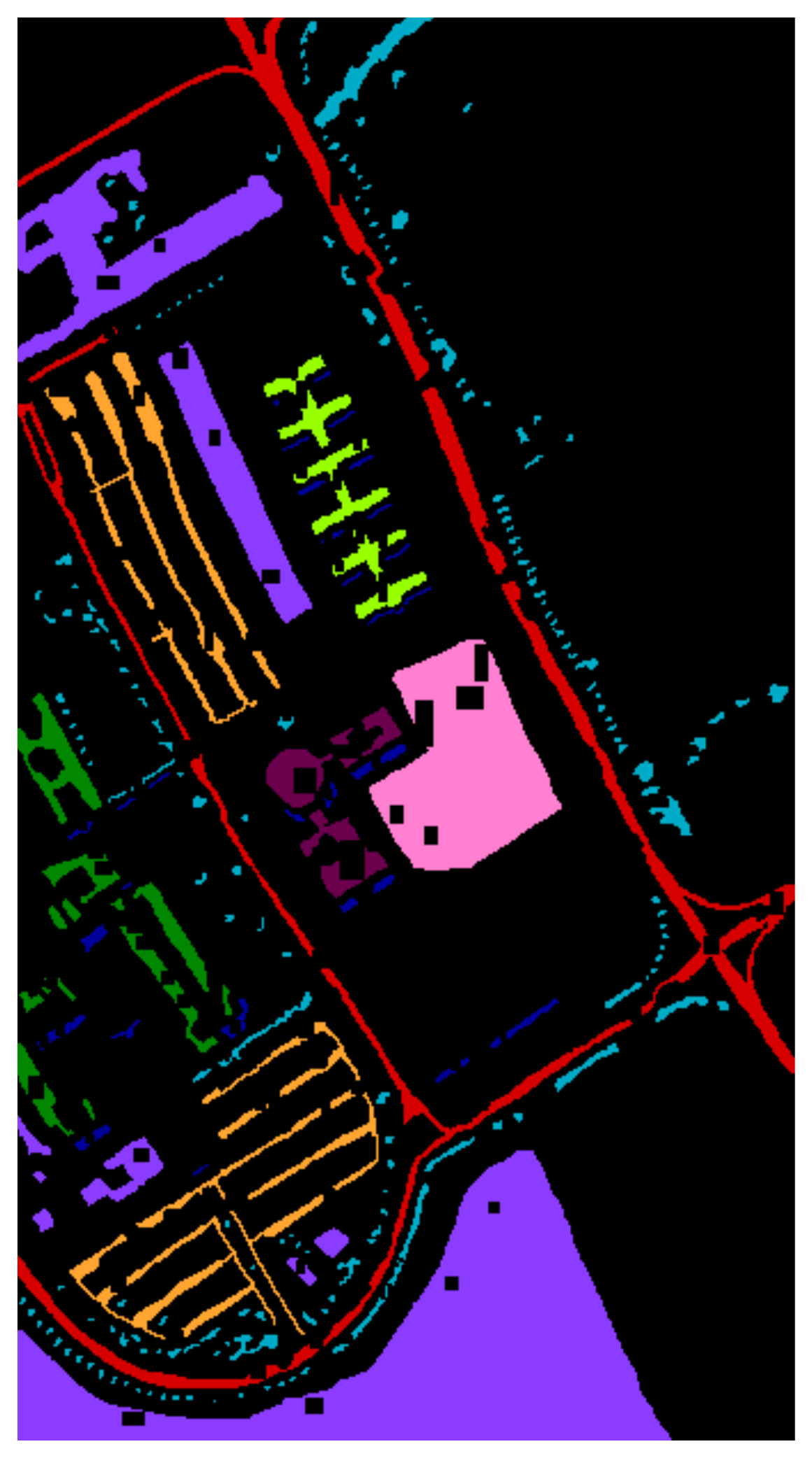}
            \caption{}
        \end{subfigure}%
        \hfill
        \begin{subfigure}{0.138\linewidth}
            \centering
            \includegraphics[width=\linewidth]{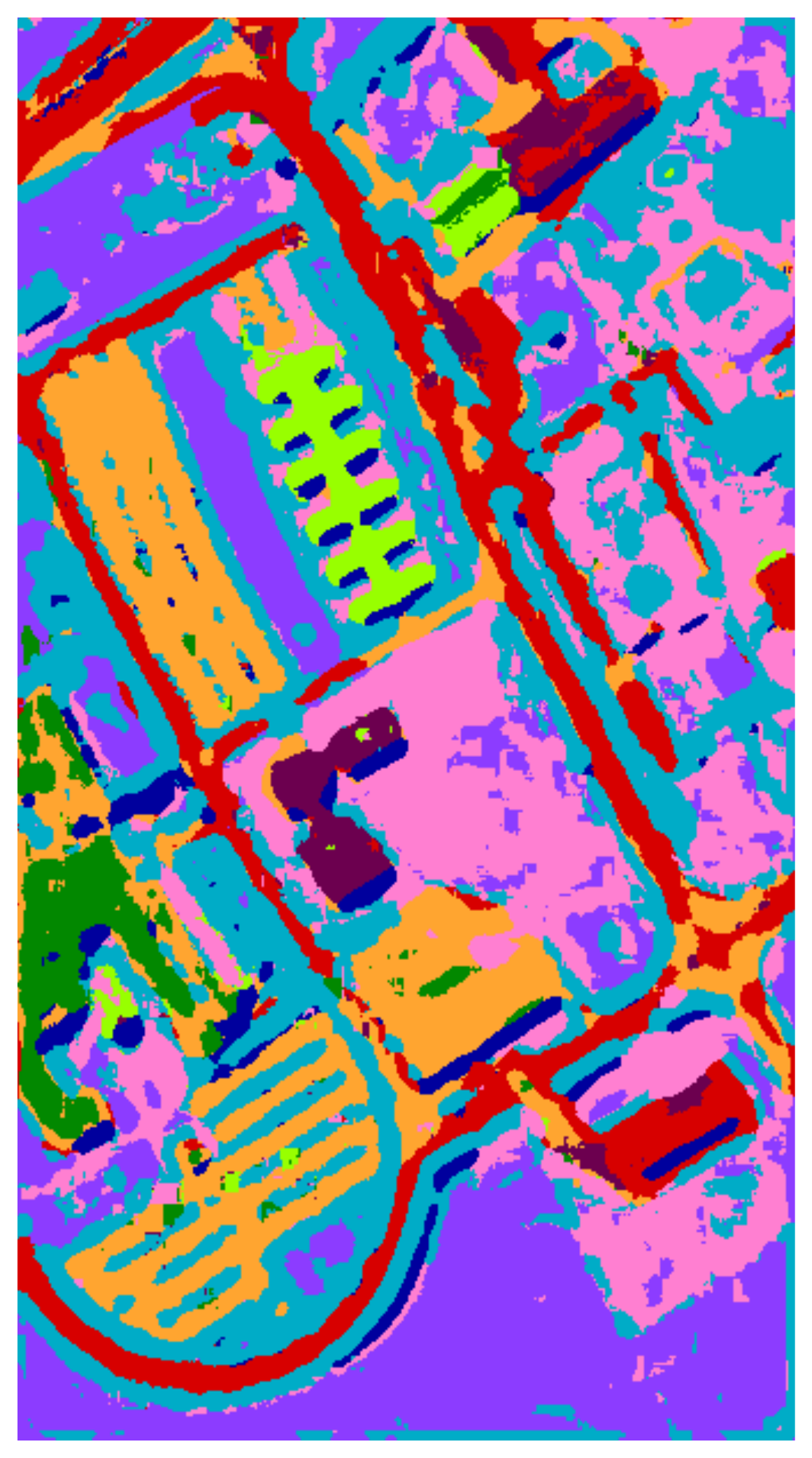}
            \caption{}
        \end{subfigure}%
        \hfill
        \begin{subfigure}{0.138\linewidth}
            \centering
            \includegraphics[width=\linewidth]{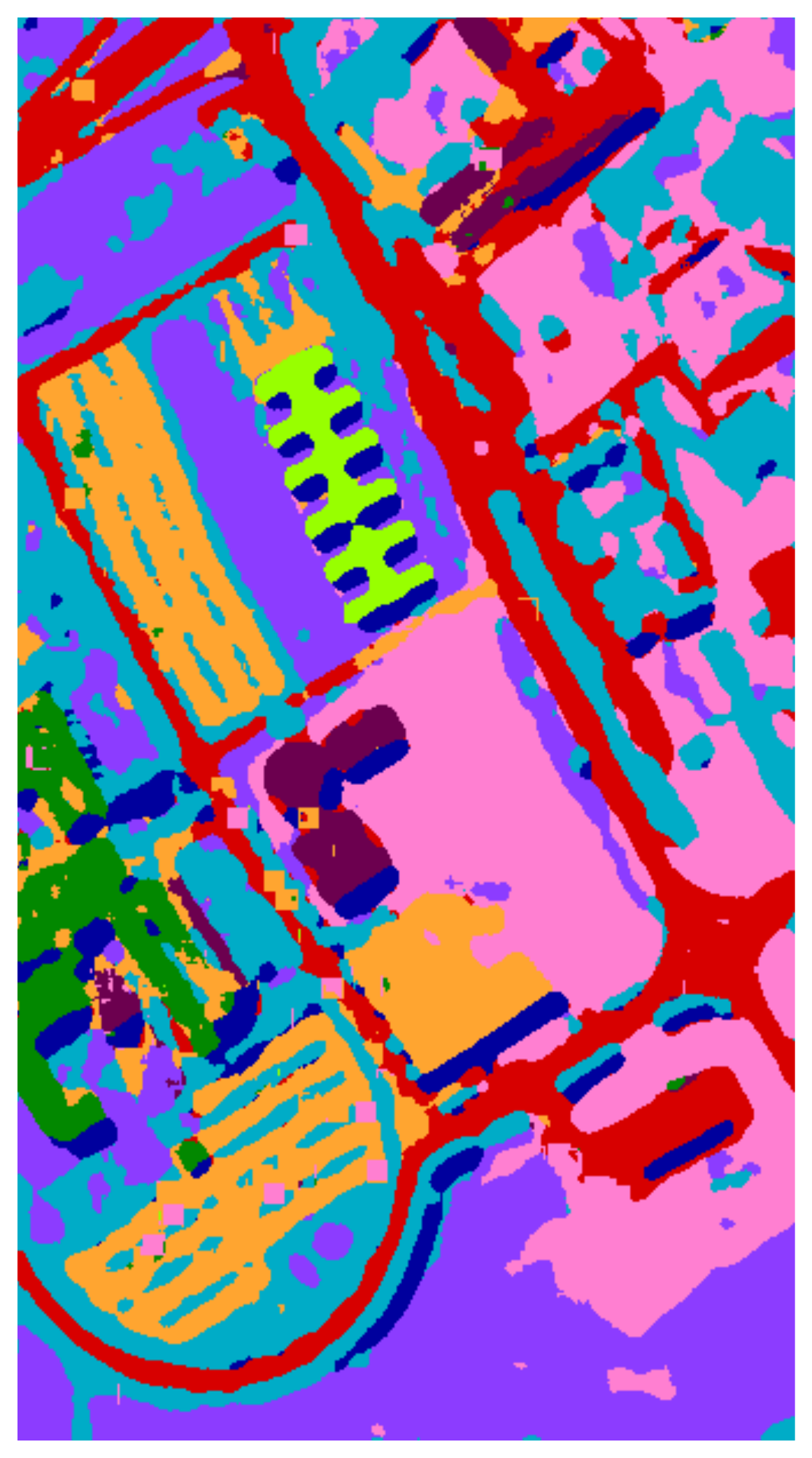}
            \caption{}
        \end{subfigure}%
        \hfill
        \begin{subfigure}{0.138\linewidth}
            \centering
            \includegraphics[width=\linewidth]{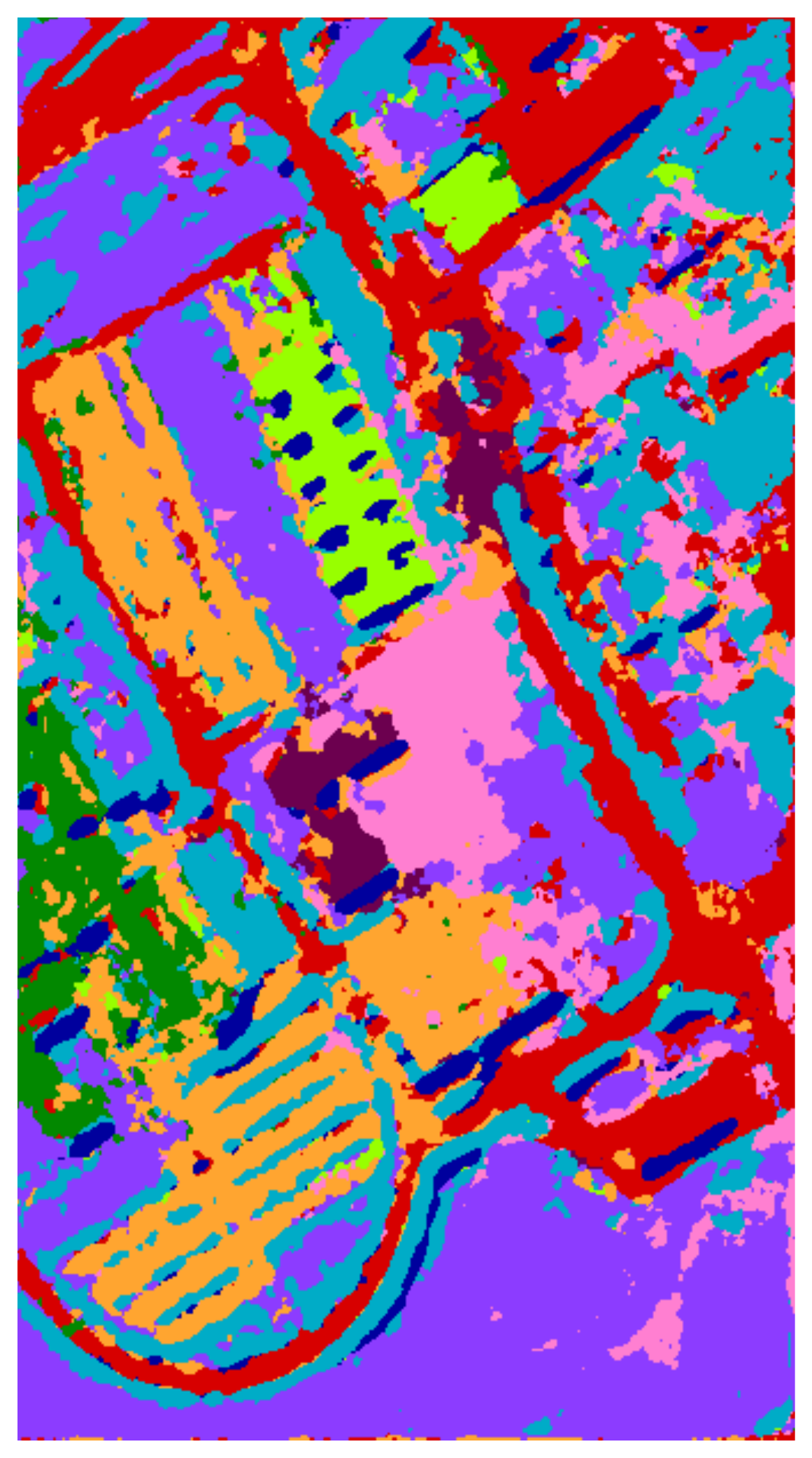}
            \caption{}
        \end{subfigure}%
        \hfill
        \begin{subfigure}{0.138\linewidth}
            \centering
            \includegraphics[width=\linewidth]{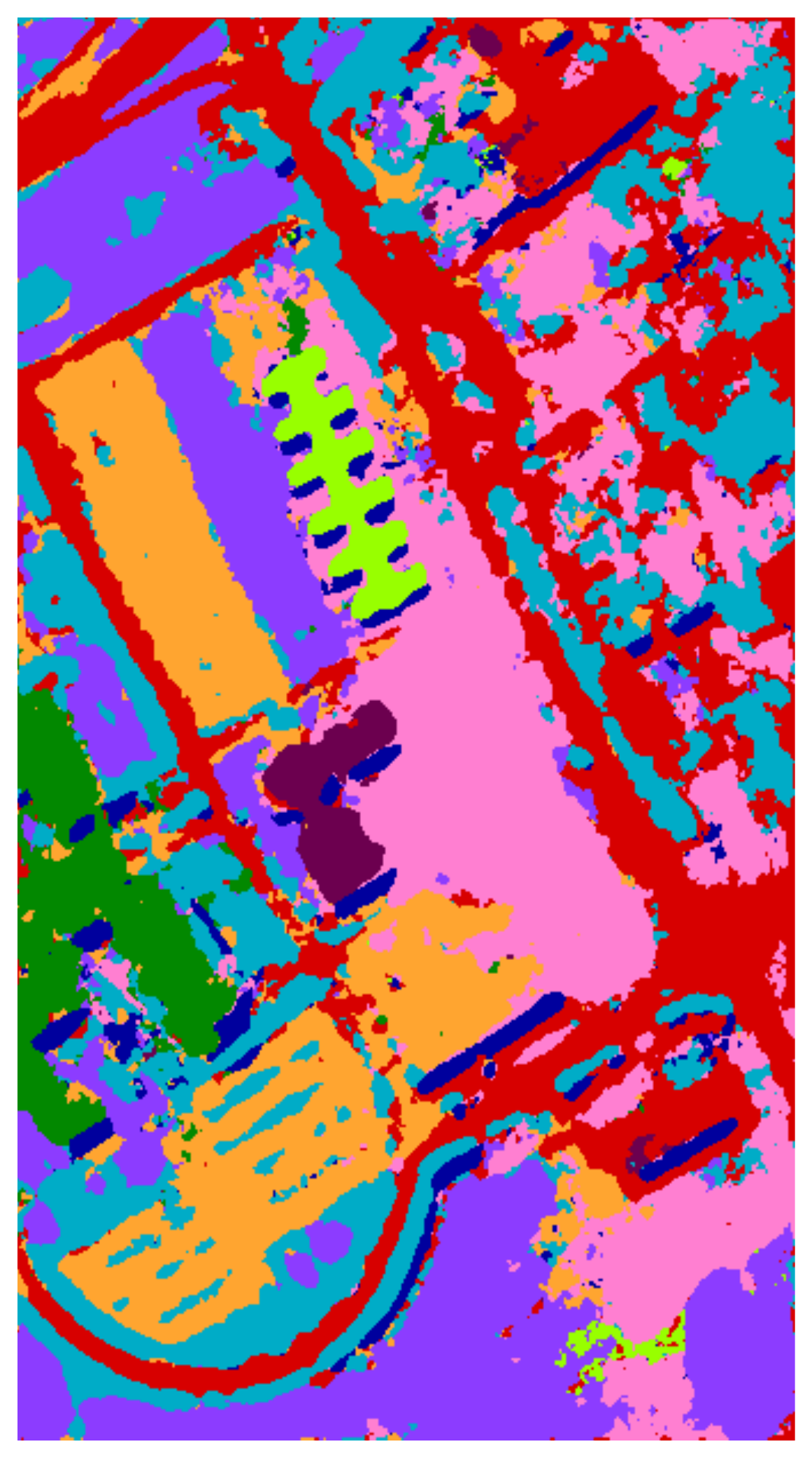}
            \caption{}
        \end{subfigure}
    \end{minipage}
    
    \vspace{0.1cm}
    
    \begin{minipage}{\linewidth}
        \centering
        \begin{subfigure}{0.138\linewidth}
            \centering
            \includegraphics[width=\linewidth]{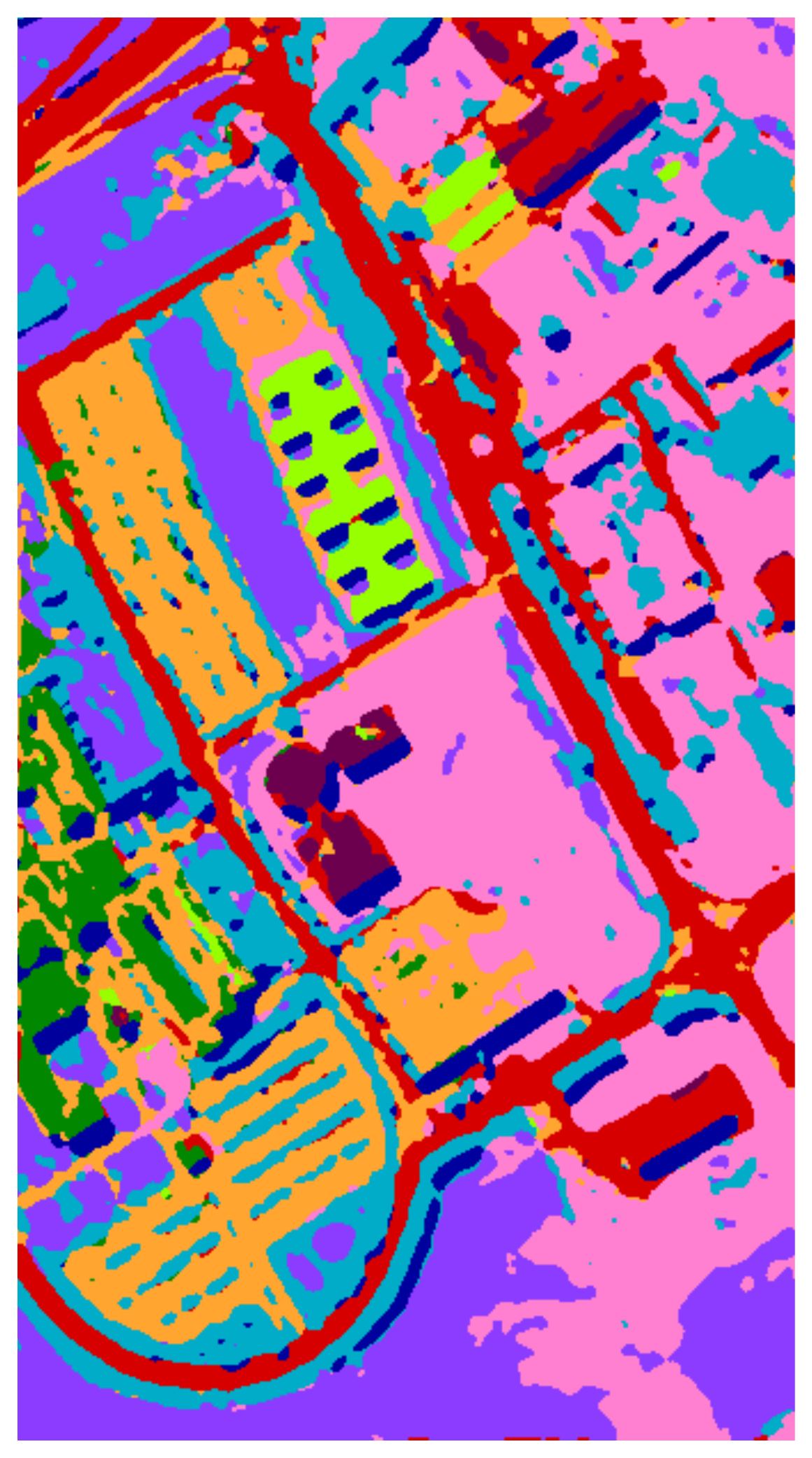}
            \caption{}
        \end{subfigure}%
        \hspace{0.0057\linewidth}%
        \begin{subfigure}{0.138\linewidth}
            \centering
            \includegraphics[width=\linewidth]{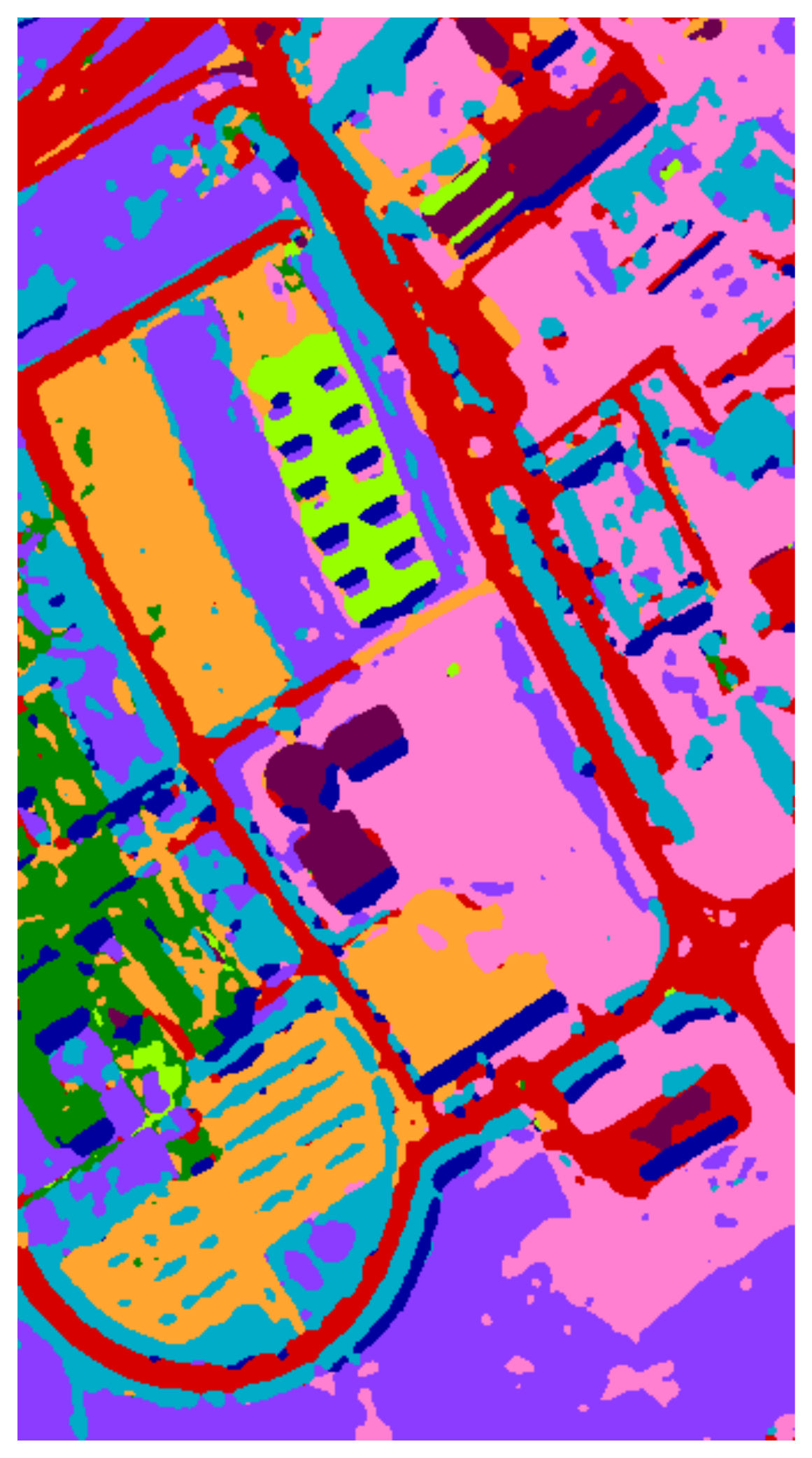}
            \caption{}
        \end{subfigure}%
        \hspace{0.0057\linewidth}%
        \begin{subfigure}{0.138\linewidth}
            \centering
            \includegraphics[width=\linewidth]{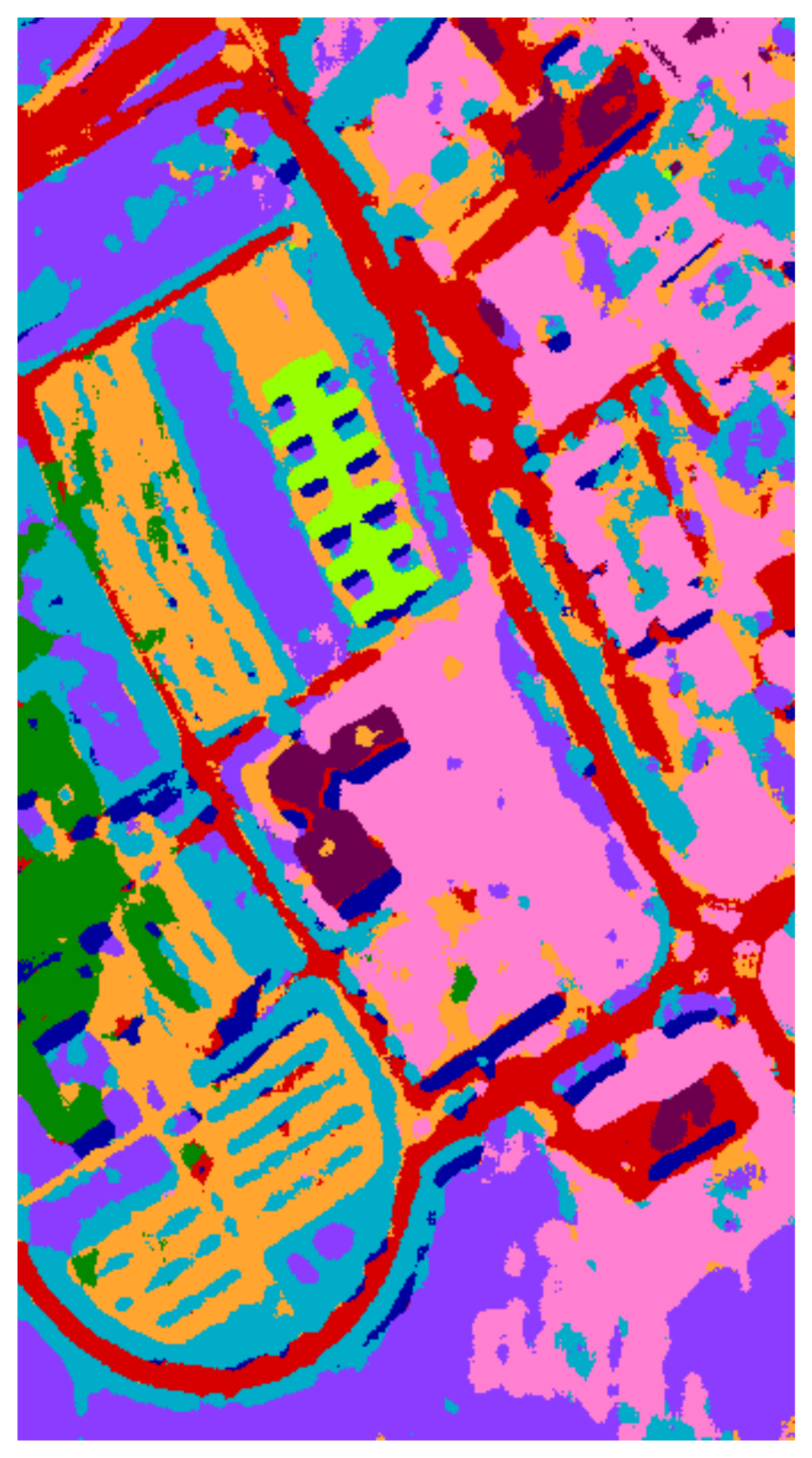}
            \caption{}
        \end{subfigure}%
        \hspace{0.0057\linewidth}%
        \begin{subfigure}{0.138\linewidth}
            \centering
            \includegraphics[width=\linewidth]{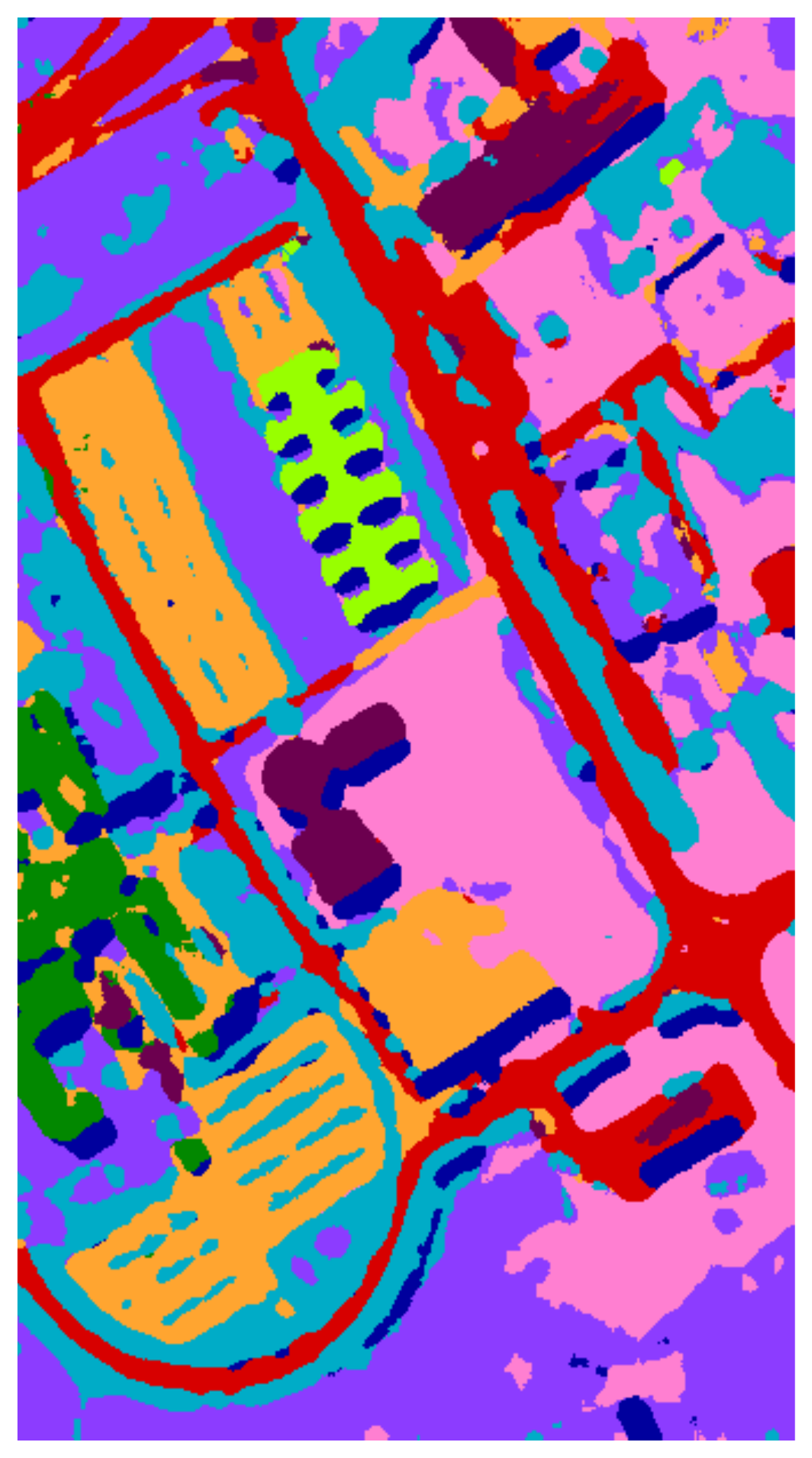}
            \caption{}
        \end{subfigure}%
        \hspace{0.0057\linewidth}%
        \begin{subfigure}{0.138\linewidth}
            \centering
            \includegraphics[width=\linewidth]{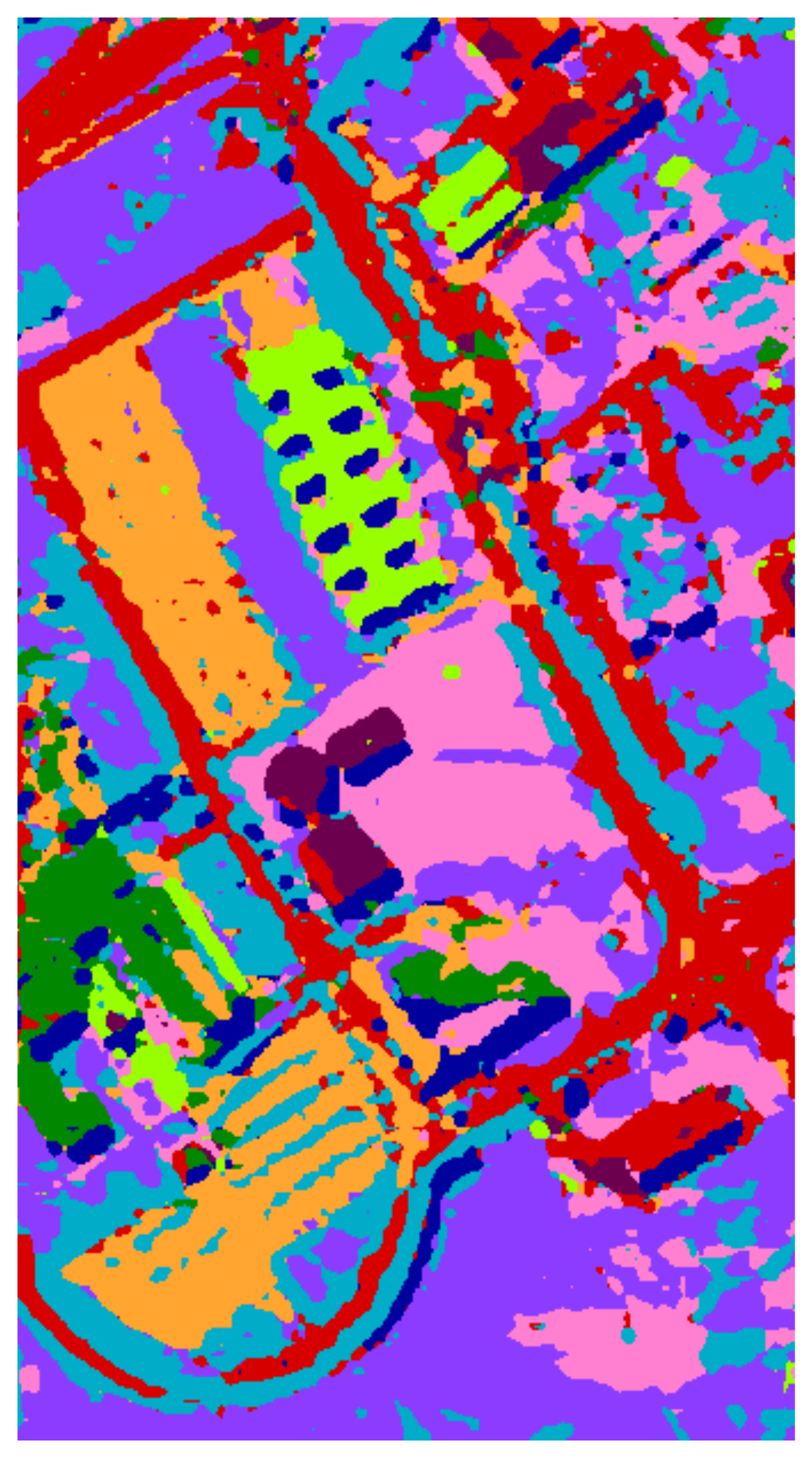}
            \caption{}
        \end{subfigure}%
        \hspace{0.0057\linewidth}%
        \begin{subfigure}{0.138\linewidth}
            \centering
            \includegraphics[width=\linewidth]{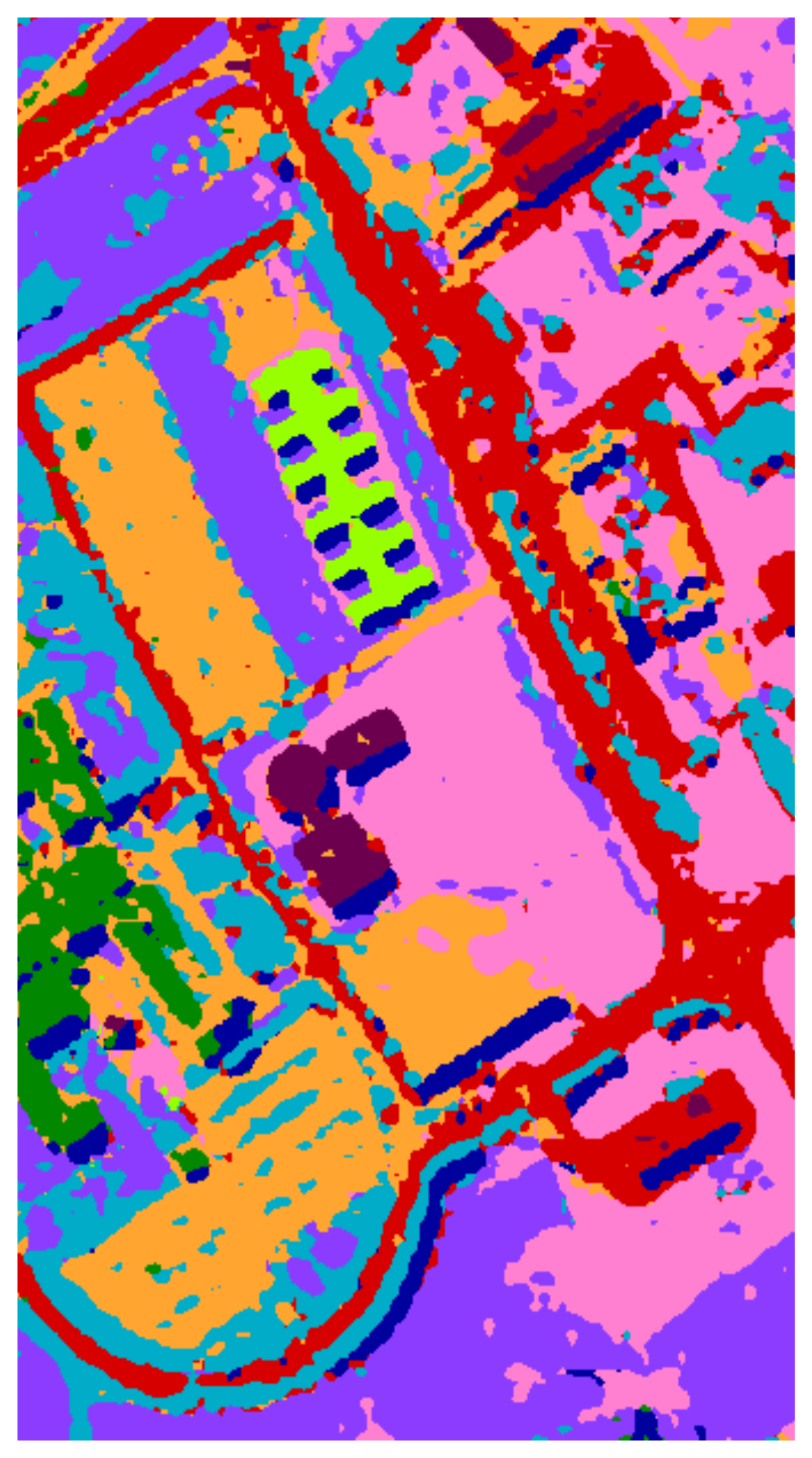}
            \caption{}
        \end{subfigure}
    \end{minipage}
    
    \vspace{0.1cm}
    
    \begin{subfigure}{1\linewidth}
        \centering
        \includegraphics[width=\linewidth, height=1cm]{HSImaps/PU_labels.pdf}
    \end{subfigure}
    
    \caption{Classification maps and train-test sets for PU (disjoint) scene: (a) False color map, (b) Train set, (c) Test set, (d) SSTN, (e) SSTN + CEnc, (f) FCN, (g) FCN + CEnc, (h) A2S2K, (i) A2S2K + CEnc, (j) AMS-M2ESL, (k) AMS-M2ESL + CEnc, (l) MambaHSI, (m) MambaHSI + CEnc}
    \label{fig:classification_maps_Pavia1}
\end{figure}

\begin{figure}[h]
    \centering
    \begin{subfigure}{0.49\linewidth}
        \centering
        \includegraphics[width=1\linewidth]{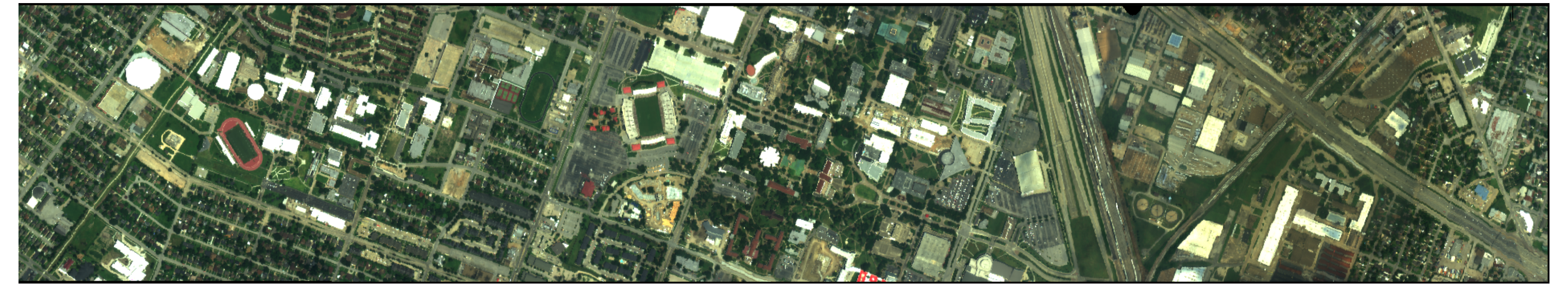}
        \caption{}
    \end{subfigure}

    \vspace{0cm}
    
    \begin{subfigure}{0.49\linewidth}
        \centering
        \includegraphics[width=\linewidth]{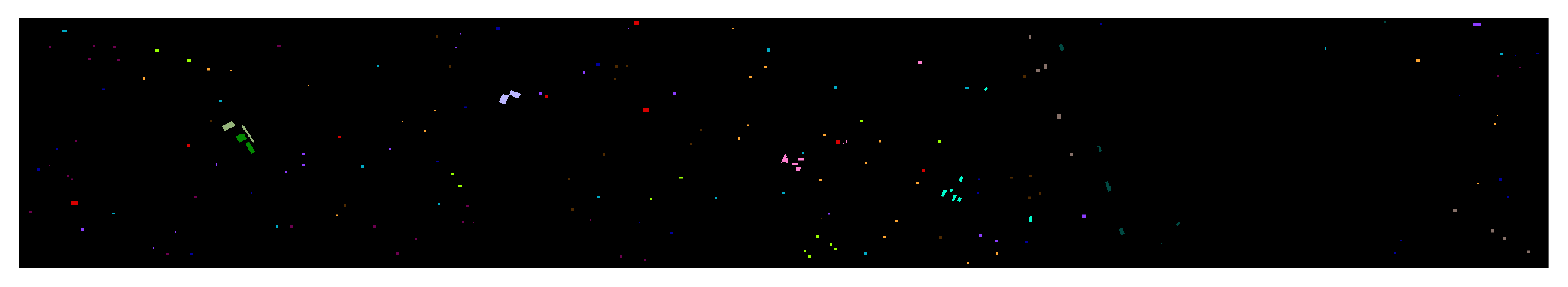}
        \caption{}
    \end{subfigure}
    \hfill
    \begin{subfigure}{0.49\linewidth}
        \centering
        \includegraphics[width=\linewidth]{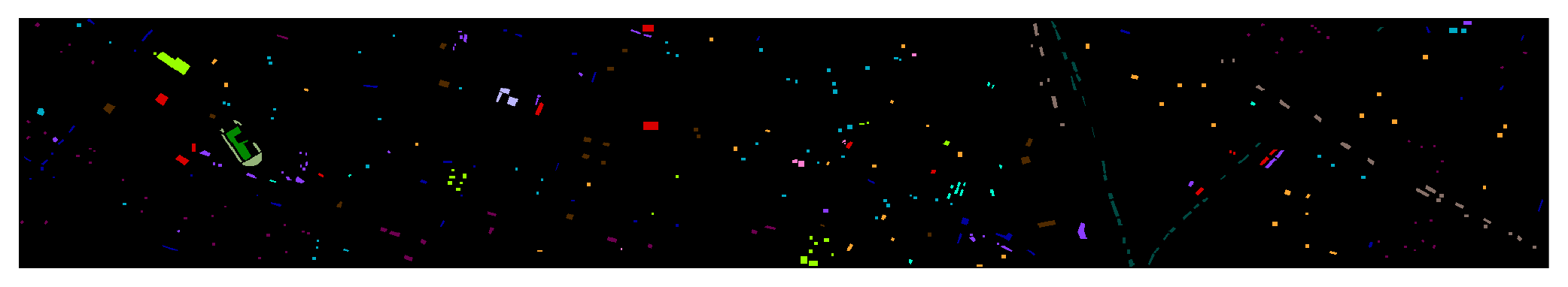}
        \caption{}
    \end{subfigure}

    \vspace{0cm}
    \begin{subfigure}{0.49\linewidth}
        \centering
        \includegraphics[width=\linewidth]{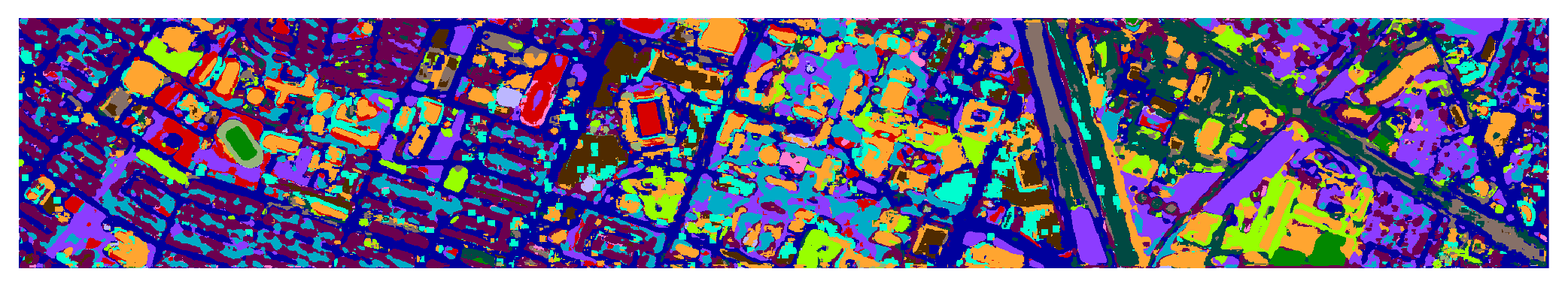}
        \caption{}
    \end{subfigure}%
    \hfill
    \begin{subfigure}{0.49\linewidth}
        \centering
        \includegraphics[width=\linewidth]{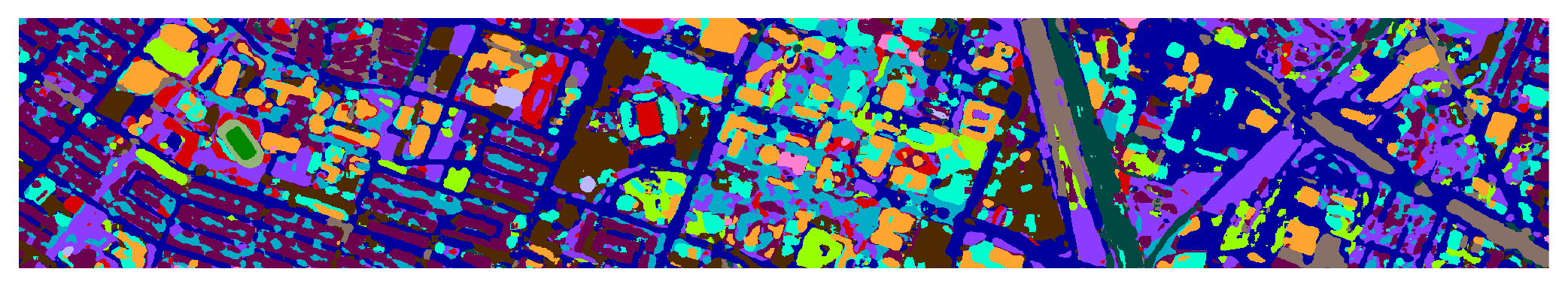}
        \caption{}
    \end{subfigure}
    
    \vspace{0cm}
    
    \begin{subfigure}{0.49\linewidth}
        \centering
        \includegraphics[width=\linewidth]{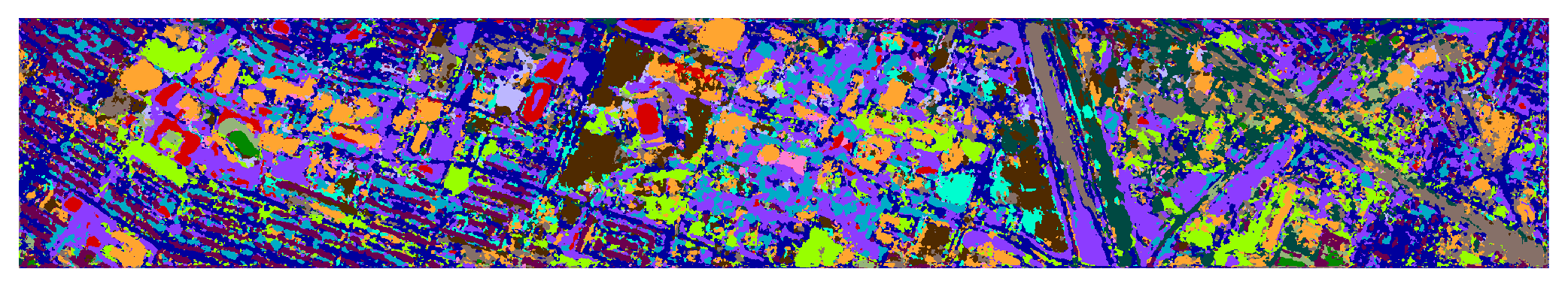}
        \caption{}
    \end{subfigure}%
    \hfill
    \begin{subfigure}{0.49\linewidth}
        \centering
        \includegraphics[width=\linewidth]{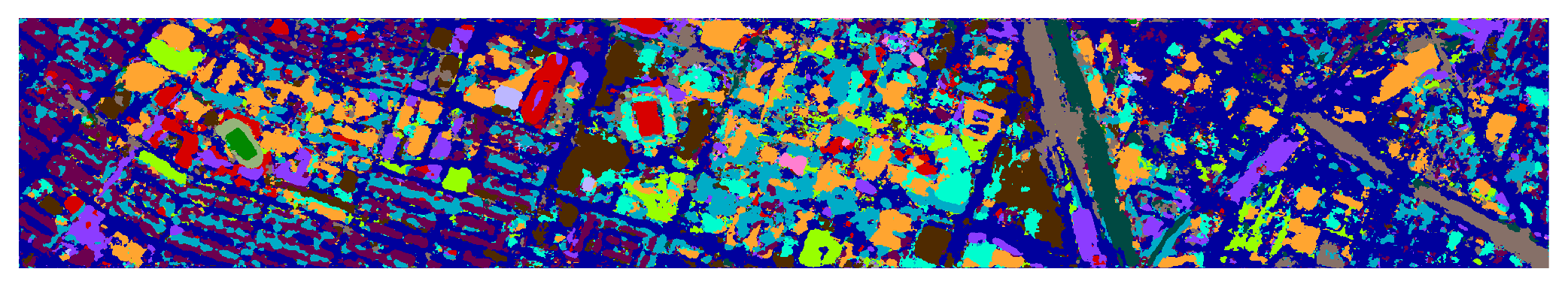}
        \caption{}
    \end{subfigure}
    
    \vspace{0cm}
    
    \begin{subfigure}{0.49\linewidth}
        \centering
        \includegraphics[width=\linewidth]{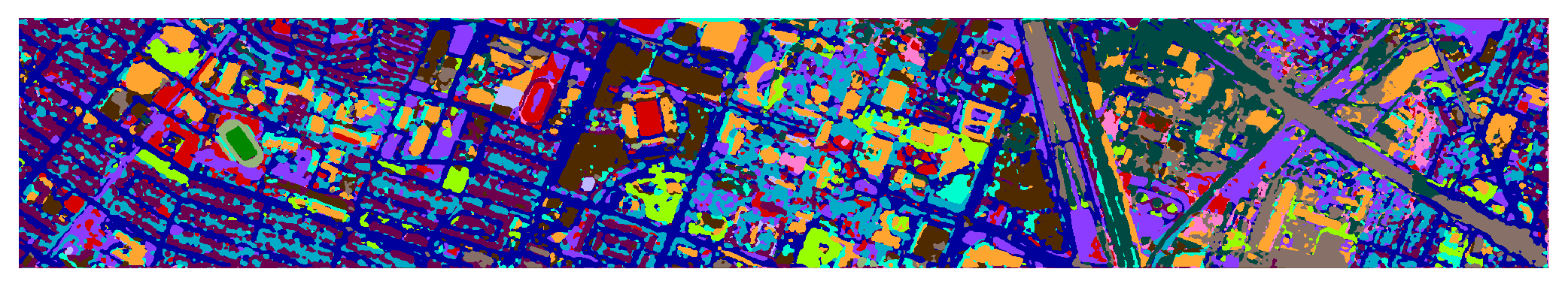}
        \caption{}
    \end{subfigure}%
    \hfill
    \begin{subfigure}{0.49\linewidth}
        \centering
        \includegraphics[width=\linewidth]{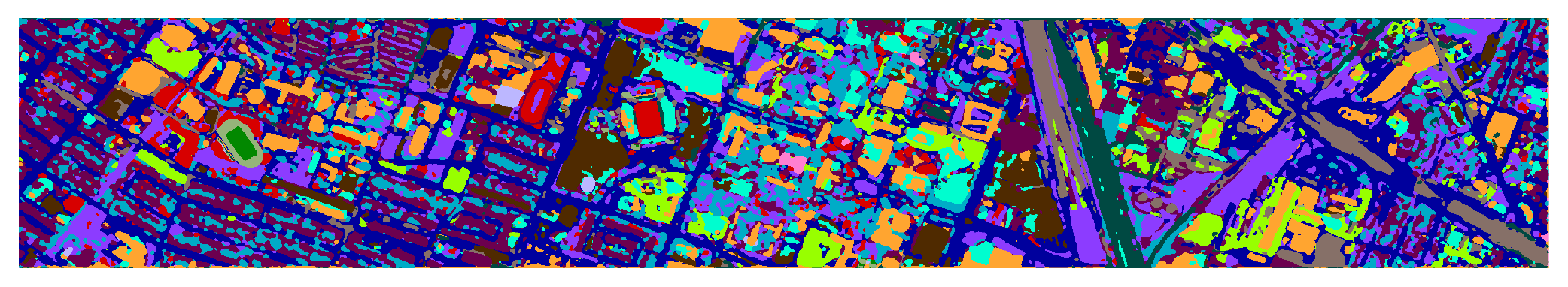}
        \caption{}
    \end{subfigure}
    
    \vspace{0cm}
    
    \begin{subfigure}{0.49\linewidth}
        \centering
        \includegraphics[width=\linewidth]{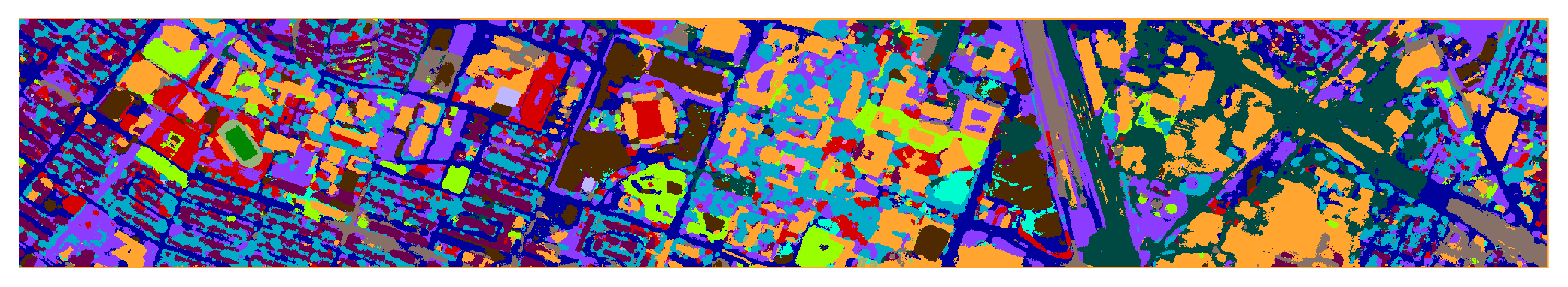}
        \caption{}
    \end{subfigure}%
    \hfill
    \begin{subfigure}{0.49\linewidth}
        \centering
        \includegraphics[width=\linewidth]{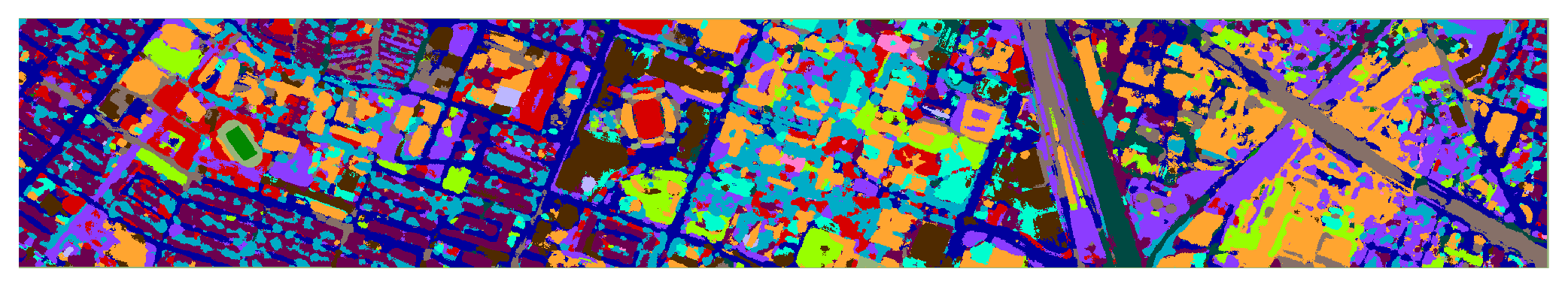}
        \caption{}
    \end{subfigure}

     \begin{subfigure}{0.49\linewidth}
        \centering
        \includegraphics[width=\linewidth]{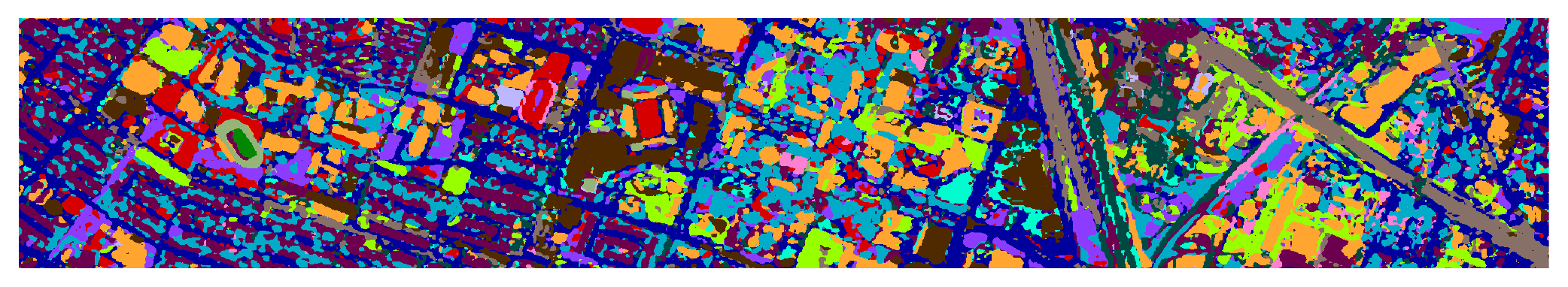}
        \caption{}
    \end{subfigure}%
    \hfill
    \begin{subfigure}{0.49\linewidth}
        \centering
        \includegraphics[width=\linewidth]{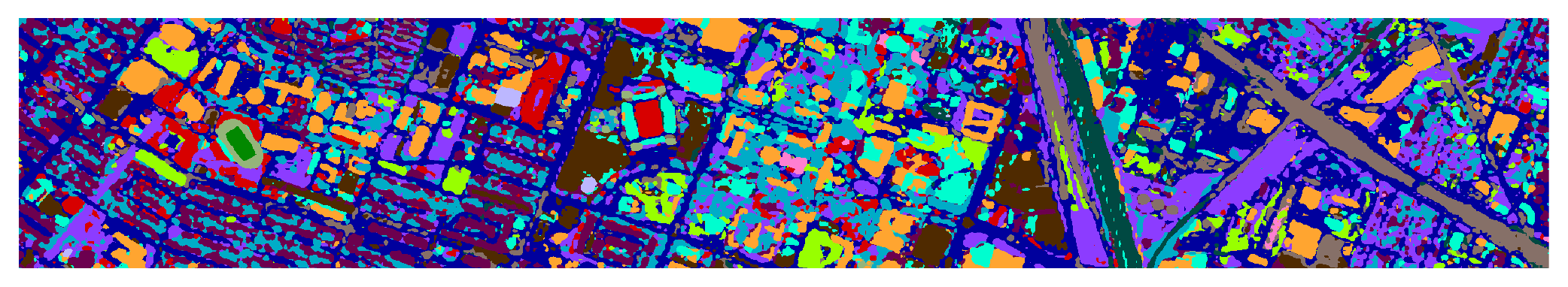}
        \caption{}
    \end{subfigure}
    
    \vspace{0cm}
    
    \begin{subfigure}{1\linewidth}
        \centering
        \includegraphics[width=\linewidth, height=1cm]{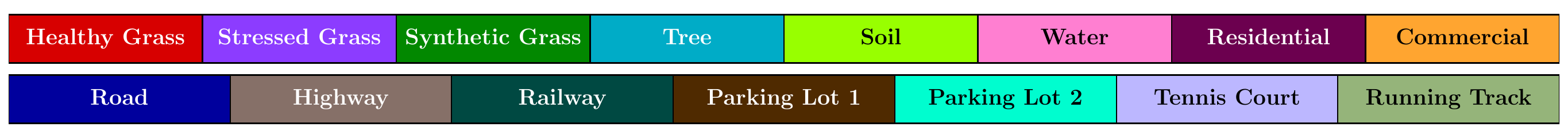}
    \end{subfigure}
    
    \caption{Classification maps and train-test sets for UH scene: (a) False color map, (b) Train set, (c) Test set, (d) SSTN, (e) SSTN + CEnc, (f) FCN, (g) FCN + CEnc, (h) A2S2K, (i) A2S2K + CEnc, (j) AMS-M2ESL, (k) AMS-M2ESL + CEnc, (l) MambaHSI, (m) MambaHSI + CEnc}
    \label{fig:classification_maps_H}
\end{figure}

To address these concerns and more accurately assess the practical performance of spectral-spatial methods, we introduce spatially disjoint train-test splits into our evaluation framework, in addition to the typical random splits. This approach allows for a more realistic evaluation of model performance in scenarios where spatial continuity between training and test samples cannot be assumed. In our study, we utilize the following datasets with different split types: the KSC and IP datasets are maintained with spatially random splits for consistency with previous evaluations; the PU dataset is prepared in two versions—one with a spatially random split and another with a largely disjoint split where test samples are contained within rectangular regions (excluding pixels adjacent to region boundaries); the PC dataset uses a subset of the original labeled pixels to create a disjoint and balanced train-test split; the LK dataset features a significantly limited training set with disjoint spatial distribution; the UH dataset includes a pre-defined spatially disjoint train-test split, serving as a benchmark for real-world applicability. The distribution of training and test sets for these scenes can be seen in Figures~\ref{fig:KSC_spectral},~\ref{fig:classification_maps_IP},~\ref{fig:classification_maps_Pavia},~\ref{fig:classification_maps_Pavia1},~\ref{fig:PC_spectral},~\ref{fig:LK_spectral}, and~\ref{fig:classification_maps_H} respectively.

\subsubsection*{Integration of Coherent Features with Spectral-Spatial Methods\label{subsubsec:proposed_approach}}

To explore whether our extracted coherent features can enhance existing spectral-spatial methods, we simply replace the spectral dimension in the hyperspectral data cube with features derived through $\CEnc$. This approach allows us to assess the improvement in spectral information while preserving the models' ability to leverage spatial context.
For this comparative study, we selected four state-of-the-art spectral-spatial models:
\begin{itemize}
    \item SSTN (Spectral-Spatial Transformer Network)~\cite{zhong2021spectral}\footnote{\url{https://github.com/zilongzhong/SSTN}\label{SSTNgithub}}: Utilizes spectral-spatial self-attention to capture long-range dependencies and integrate local and global features.
    \item FCN (Fully Contextual Network)~\cite{wang2021fully}\footnote{\url{https://github.com/DotWang/FullyContNet}\label{FCNgithub}}: Employs scale attention and contextual modules to capture nonlocal spectral-spatial contexts.
    \item A2S2K-ResNet (Attention-based Adaptive Spectral-Spatial Kernel Improved Residual Network)~\cite{roy2020attention}\footnote{\url{https://github.com/suvojit-0x55aa/A2S2K-ResNet}\label{A2S2Kgithub}}: Enhances feature selection through selective 3D convolutions and adaptive recalibration.
    \item AMS-M2ESL (Adaptive Mask Sampling and Manifold to Euclidean Subspace Learning)~\cite{li2023adaptive}\footnote{\url{https://github.com/lms-07/AMS-M2ESL}\label{AMSgithub}}: Improves spatial modeling via adaptive sampling and manifold-based feature learning.
    \item MambaHSI~\cite{li2024mambahsi}\footnote{\url{https://github.com/li-yapeng/MambaHSI}\label{MambaHSIgithub}}: Leverages linear-complexity Mamba blocks~\cite{gu2023mamba} for efficient spectral-spatial feature extraction through dedicated spatial and spectral modules with adaptive fusion.
\end{itemize}

To ensure consistency in training across all models, we used the same train-test distribution as described in the previous subsection on pixel-based classification. The training split was further divided into an 80-20 train-validation split to standardize the training process and enable best model selection based on validation performance across epochs, as some original implementations included this practice while others did not.

We implemented these models using code from their respective authors' public GitHub repositories\footref{SSTNgithub}\textsuperscript{,}\footref{FCNgithub}\textsuperscript{,}\footref{A2S2Kgithub}\textsuperscript{,}\footref{AMSgithub}\textsuperscript{,}\footref{MambaHSIgithub}, with parameters selected according to recommendations in their papers or repositories. For scenes new to a model (e.g., UH for A2S2K-ResNet), we tested parameters from other scenes and chose the best-performing ones.
We evaluated the models both with and without coherent features, maintaining SymAE configurations from the previous Subsection~\ref{subsec:spectralclassification}. When using coherent features, all other parameters, including patch size, remained consistent with the original implementations, with two exceptions: (1) learning rate schedules were optimized for each model to account for the different feature characteristics, and (2) for AMS-M2ESL, the MNF ratio parameter was adjusted to 1 for Pavia datasets and 0.4 for others to maintain optimal performance.

\begin{table*}[!t]
\caption{Performance Comparison of Spectral-Spatial Methods With and Without Coherent Features Across Multiple Datasets\label{tab:classification_ss}}
\vspace{-10pt}
\centering
\scriptsize
\setlength{\tabcolsep}{1.5pt}
\setlength{\heavyrulewidth}{0.8pt}
\setlength{\lightrulewidth}{0.5pt}
\begin{tabular}{@{}ll*{5}{|>{\centering\arraybackslash}p{0.07\textwidth}>{\centering\arraybackslash}p{0.08\textwidth}}@{}}
\toprule[1.2pt]
\textbf{Dataset} & \textbf{Metric} & \textbf{SSTN} & \textbf{SSTN + CEnc} & \textbf{FCN} & \textbf{FCN + CEnc} & \textbf{A2S2K} & \textbf{A2S2K + CEnc} & \textbf{AMS-M2ESL} & \textbf{AMS-M2ESL + CEnc} & \textbf{MambaHSI} & \textbf{MambaHSI + CEnc} \\
\midrule[0.8pt]
\multirow{3}{*}{\textbf{\makecell[l]{Kennedy \\ Space Center\textsuperscript{R}}}} 
 & OA (\%) & 96.12 & 99.87 & 98.04 & 99.68 & 99.30 & 99.87 & 99.64 & 99.66 & 99.36 & 99.53 \\
 & AA (\%) & 93.02 & 99.83 & 96.86 & 99.45 & 98.88 & 99.82 & 98.82 & 99.28 & 98.88 & 99.34 \\
 & $\kappa\times100$ & 95.68 & 99.86 & 97.82 & 99.64 & 99.22 & 99.86 & 99.60 & 99.62 & 99.29 & 99.48 \\
\midrule
\multirow{3}{*}{\textbf{Indian Pines\textsuperscript{R}}} 
 & OA (\%) & 88.89 & 92.85 & 91.70 & 93.69 & 93.80 & 96.66 & 93.11 & 97.18 & 94.32 & 94.74 \\
 & AA (\%) & 94.64 & 95.40 & 95.50 & 96.72 & 97.51 & 98.29 & 97.61 & 98.63 & 97.32 & 97.60 \\
 & $\kappa\times100$ & 87.33 & 91.83 & 90.49 & 92.76 & 92.93 & 96.17 & 92.15 & 96.77 & 93.49 & 93.97 \\
\midrule
\multirow{3}{*}{\textbf{\makecell[l]{Pavia \\ University\textsuperscript{R}}}} 
 & OA (\%) & 99.31 & 99.62 & 99.22 & 99.43 & 99.44 & 99.87 & 99.61 & 99.66 & 98.75 & 99.25 \\
 & AA (\%) & 99.09 & 99.47 & 99.09 & 99.22 & 99.36 & 99.81 & 99.61 & 99.51 & 98.95 & 99.13 \\
 & $\kappa\times100$ & 99.06 & 99.48 & 98.94 & 99.23 & 99.24 & 99.82 & 99.47 & 99.53 & 98.29 & 98.97 \\
\midrule
\multirow{3}{*}{\textbf{\makecell[l]{Pavia \\ University\textsuperscript{D}}}} 
 & OA (\%) & 89.88 & 94.80 & 91.73 & 92.83 & 89.54 & 94.34 & 86.44 & 93.39 & 88.45 & 90.87 \\
 & AA (\%) & 89.99 & 96.19 & 90.02 & 96.03 & 92.35 & 96.86 & 91.36 & 96.12 & 89.01 & 93.97 \\
 & $\kappa\times100$ & 86.50 & 93.09 & 88.79 & 90.57 & 86.32 & 92.51 & 82.39 & 91.28 & 84.68 & 88.01 \\
\midrule
\multirow{3}{*}{\textbf{Pavia Center\textsuperscript{D}}} 
 & OA (\%) & 91.14 & 93.64 & 82.67 & 90.30 & 93.19 & 94.05 & 88.03 & 92.57 & 88.91 & 91.01 \\
 & AA (\%) & 91.14 & 93.64 & 82.67 & 90.30 & 93.19 & 94.05 & 88.03 & 92.57 & 88.91 & 91.01 \\
 & $\kappa\times100$ & 90.04 & 92.84 & 80.51 & 89.09 & 92.34 & 93.31 & 86.53 & 91.64 & 87.52 & 89.89 \\
\midrule
\multirow{3}{*}{\textbf{Houston 2013\textsuperscript{D}}} 
 & OA (\%) & 86.11 & 89.24 & 78.05 & 86.97 & 89.90 & 91.15 & 88.42 & 91.47 & 87.04 & 88.72 \\
 & AA (\%) & 87.85 & 90.82 & 76.83 & 88.18 & 91.51 & 92.34 & 89.48 & 92.89 & 87.42 & 90.12 \\
 & $\kappa\times100$ & 84.92 & 88.31 & 76.15 & 85.86 & 89.03 & 90.39 & 87.43 & 90.73 & 85.92 & 87.75 \\
\midrule
\multirow{3}{*}{\textbf{LongKou\textsuperscript{D}}} 
 & OA (\%) & 92.31 & 95.27 & 90.44 & 93.55 & 94.71 & 95.71 & 84.81 & 94.53 & 87.42 & 95.29 \\
 & AA (\%) & 90.60 & 93.53 & 92.78 & 94.20 & 94.46 & 94.40 & 85.36 & 87.81 & 89.95 & 94.18 \\
 & $\kappa\times100$ & 90.08 & 93.85 & 87.76 & 91.65 & 93.15 & 94.42 & 80.31 & 92.89 & 84.00 & 93.88 \\
\bottomrule[1.2pt]
\multicolumn{12}{l}{\textsuperscript{R}Spatially Random Train-Test Split, \textsuperscript{D}Spatially Disjoint Train-Test Split}
\end{tabular}
\vspace{5pt}
\end{table*}

The classification results are presented in Table~\ref{tab:classification_ss}. In the KSC and PU datasets with random train-test splits, all models achieved very high accuracies, often exceeding 99\%. In these cases, the use of coherent features provided modest improvements due to the already high performance baseline. The IP dataset, despite also having a random split, showed lower overall accuracies compared to KSC and PU, ranging from about 89\% to 94\%. Here, the introduction of coherent features led to more noticeable improvements. Examining the OA improvements in detail, KSC showed improvements ranging from 0.02 to 3.75 percentage points, with SSTN benefiting the most. IP saw more substantial gains, particularly for AMS-M2ESL (4.07), SSTN (3.96), and A2S2K (2.86), while PU with random split had minimal improvements, all at or below 0.5 percentage points. On average, for the datasets with random splits (KSC, IP, and PU), the use of coherent features resulted in improvements of 1.40 and 1.09 percentage points in OA and AA, respectively. The $\kappa\times100$ value increased by 1.60 on average.

The most significant impact of coherent features was observed in the datasets with disjoint splits. PU (disjoint) saw notable improvements, especially for AMS-M2ESL (6.95) and SSTN (4.92). PC showed varied improvements, with FCN gaining the most (7.63), while other models showed more moderate gains. UH demonstrated substantial improvements, particularly for FCN (8.92). The LongKou dataset revealed some of the highest improvements across all experiments, with AMS-M2ESL and MambaHSI achieving gains of 9.72 and 7.87 percentage points respectively. Notably, both FCN and AMS-M2ESL showed high variability in improvement - FCN ranging from minimal gains in random splits to 8.92 points in UH, and AMS-M2ESL varying from near-zero gains to 9.72 points in LongKou. SSTN demonstrated more consistent improvements across datasets, while other models showed more varied results. For the disjoint cases, the average improvements were substantially higher: 4.03 and 3.82 percentage points in OA and AA, respectively. The $\kappa\times100$ value showed an average increase of 4.88. This is a notable increase compared to the random split cases, with improvements more than doubling across all metrics. These results suggest that coherent features have the capability to enhance the performance of spectral-spatial methods, particularly in scenarios where spatial continuity between training and test data is limited. This capability could be beneficial in practical applications where models are applied to novel, unseen regions.

\section{Discussion\label{sec:discussion}}
This section explores SymAE's broader potential and future directions, discussing both its applicability in unsupervised settings and opportunities for further development.

\subsection{Unsupervised Grouping\label{subsec:3x3}}
SymAE is designed to train on groups of data, which may naturally occur in many remote sensing applications, even in the absence of ground truth labels. This opens up the possibility of using SymAE in unsupervised settings, where the goal is to extract meaningful features and discover inherent structures in the data without relying on explicit annotations. Consider, for instance, multi-temporal hyperspectral data in remote sensing applications, where each pixel location might undergo multiple scans under different atmospheric conditions and varying elevation angles. SymAE aims to disentangle these variations from the pixel-specific reflectance in an entirely unsupervised manner. Another potential example lies in extraterrestrial remote sensing applications. In these environments, sensors may encounter various unmodeled nuisance factors that affect measurements of relatively stable surface compositions. For instance, in lunar and planetary studies, observations are affected by varying illumination geometry in reflectance spectroscopy and solar X-ray activity in fluorescence measurements~\cite{clark1997remote,narendranath2011lunar}. For bodies with atmospheres, atmospheric interference further complicates spectral observations. SymAE's ability to extract coherent features across temporal or spatial dimensions despite these unknown variables could aid in clustering and initial analysis of unexplored regions, potentially helping identify areas of scientific interest for further investigation. This approach may prove valuable in contexts where traditional modeling of environmental factors is challenging or incomplete, and where obtaining ground truth is impractical or impossible.

\begin{figure}[b]
    \centering
    \includegraphics[width=1\linewidth]{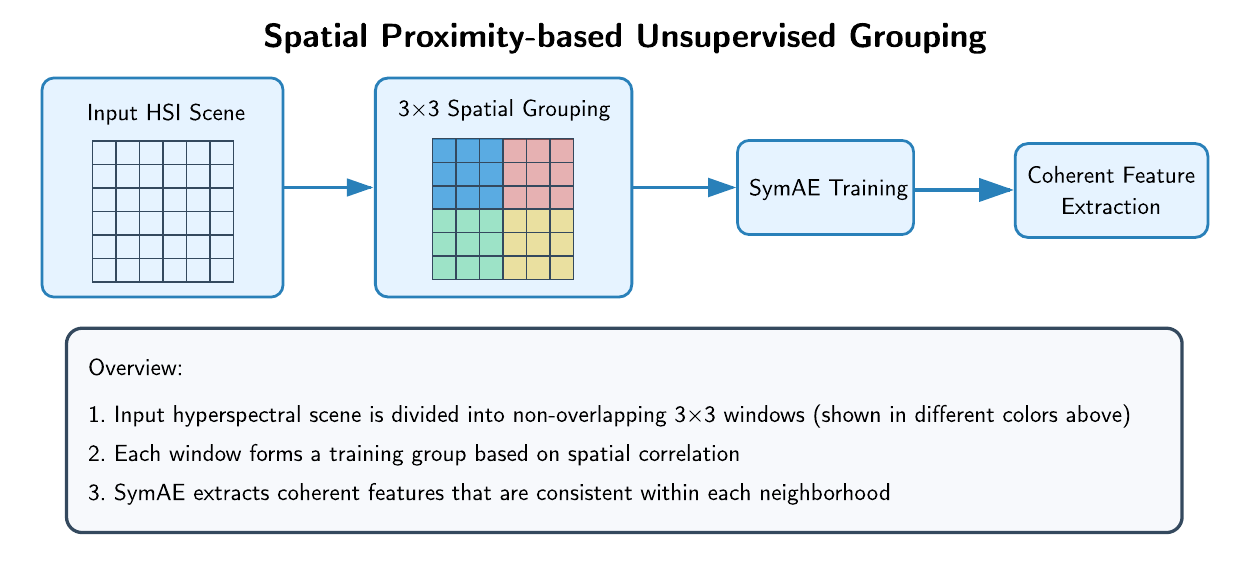}
    \caption{Flowchart of SymAE's unsupervised training using spatial proximity grouping, where adjacent pixels form training groups as a simplified test case for evaluating feature extraction without explicit labels.}
    \label{fig:flowchart}
\end{figure}

\begin{figure}[t]
    \begin{subfigure}{1\linewidth}
        \centering
        \includegraphics[width=\linewidth]{arxivimages/5labels1.pdf}
    \end{subfigure}
    \centering
    \begin{subfigure}{0.3\linewidth}
        \centering
        \includegraphics[width=\linewidth]{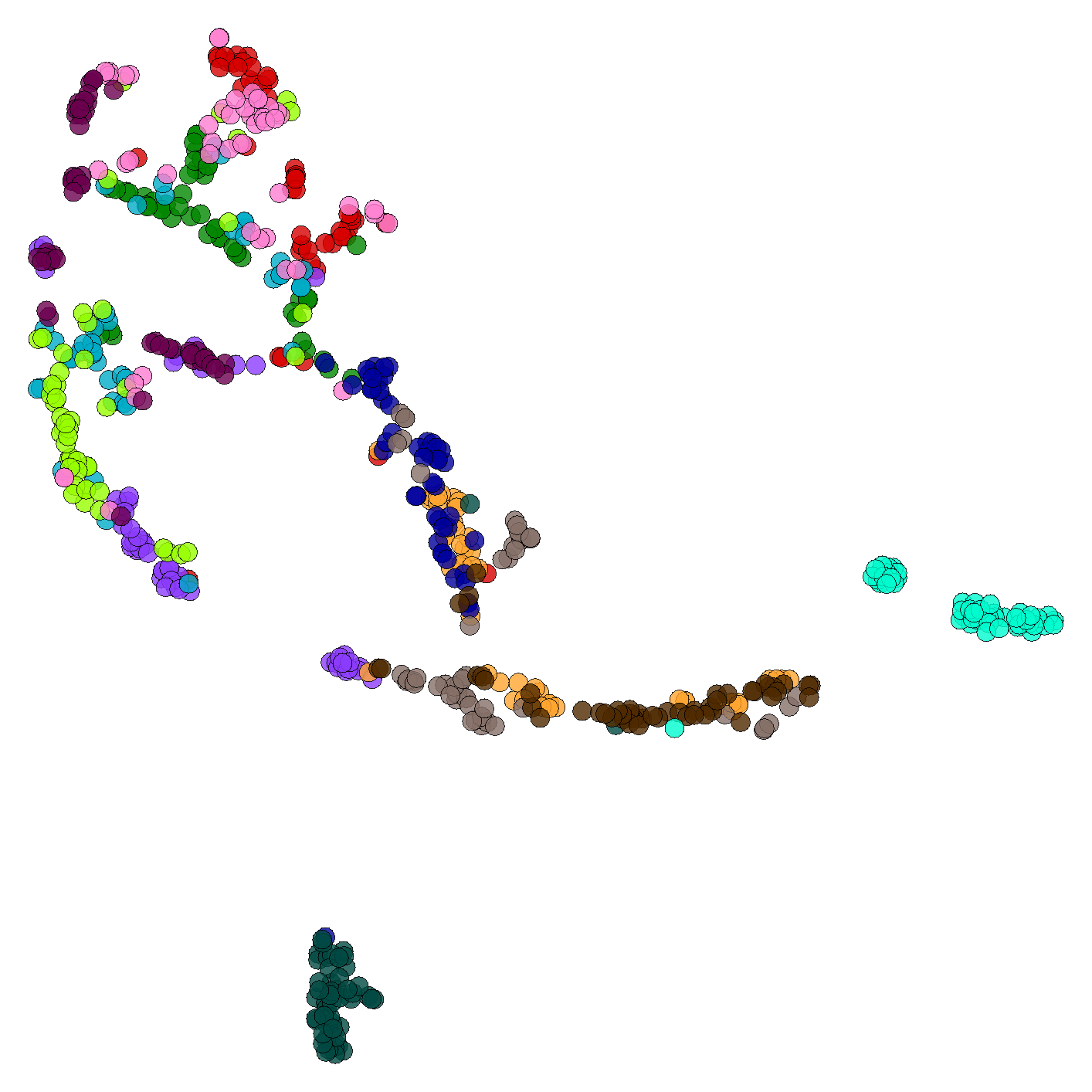}
        \caption{t-SNE plot for raw data features}
    \end{subfigure}
    \hspace{0.1cm}
    \begin{subfigure}{0.3\linewidth}
        \centering
        \includegraphics[width=\linewidth]{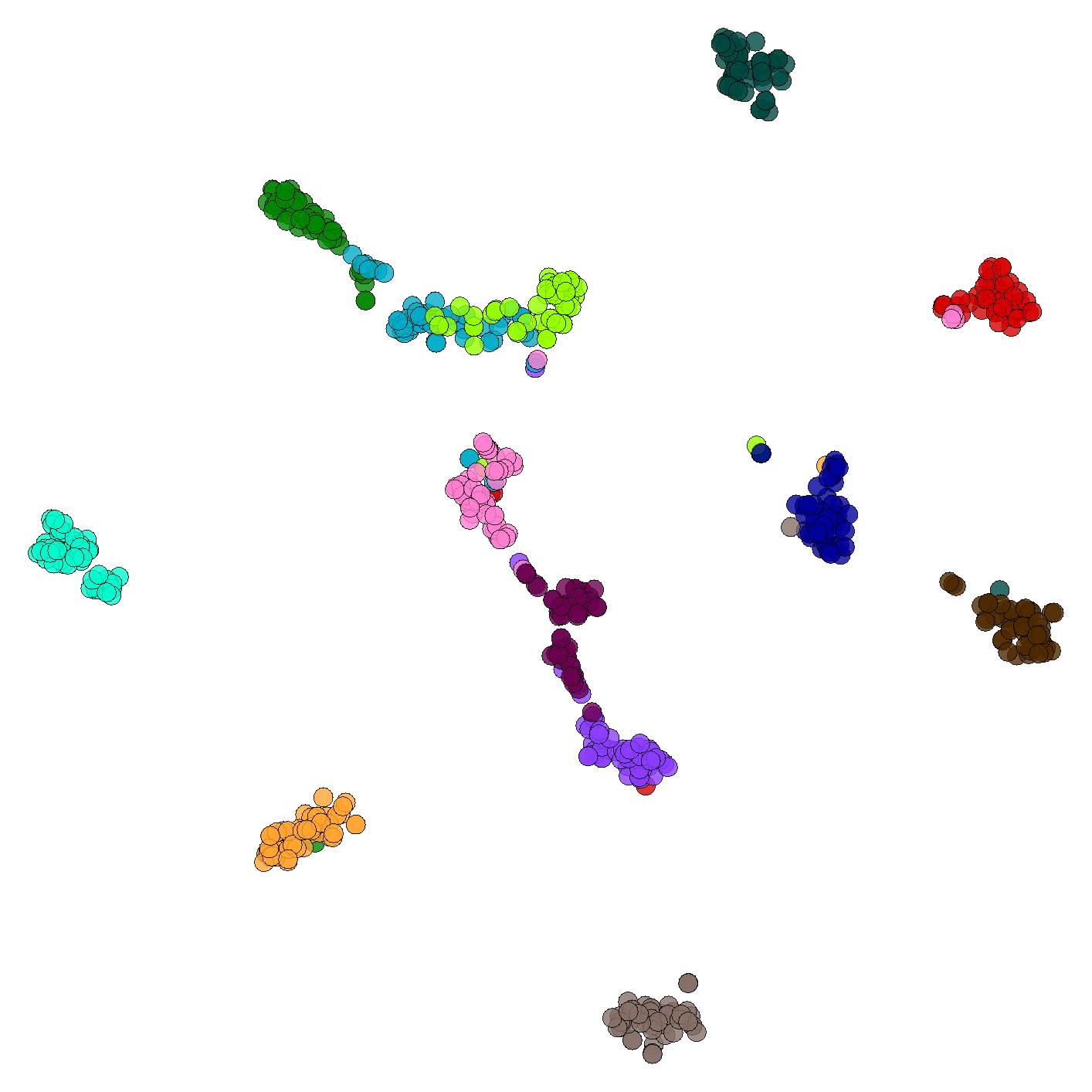}
        \caption{Ground-truth based CEnc feature map.}
    \end{subfigure}
    \hspace{0.1cm}
    \begin{subfigure}{0.3\linewidth}
        \centering
        \includegraphics[width=\linewidth]{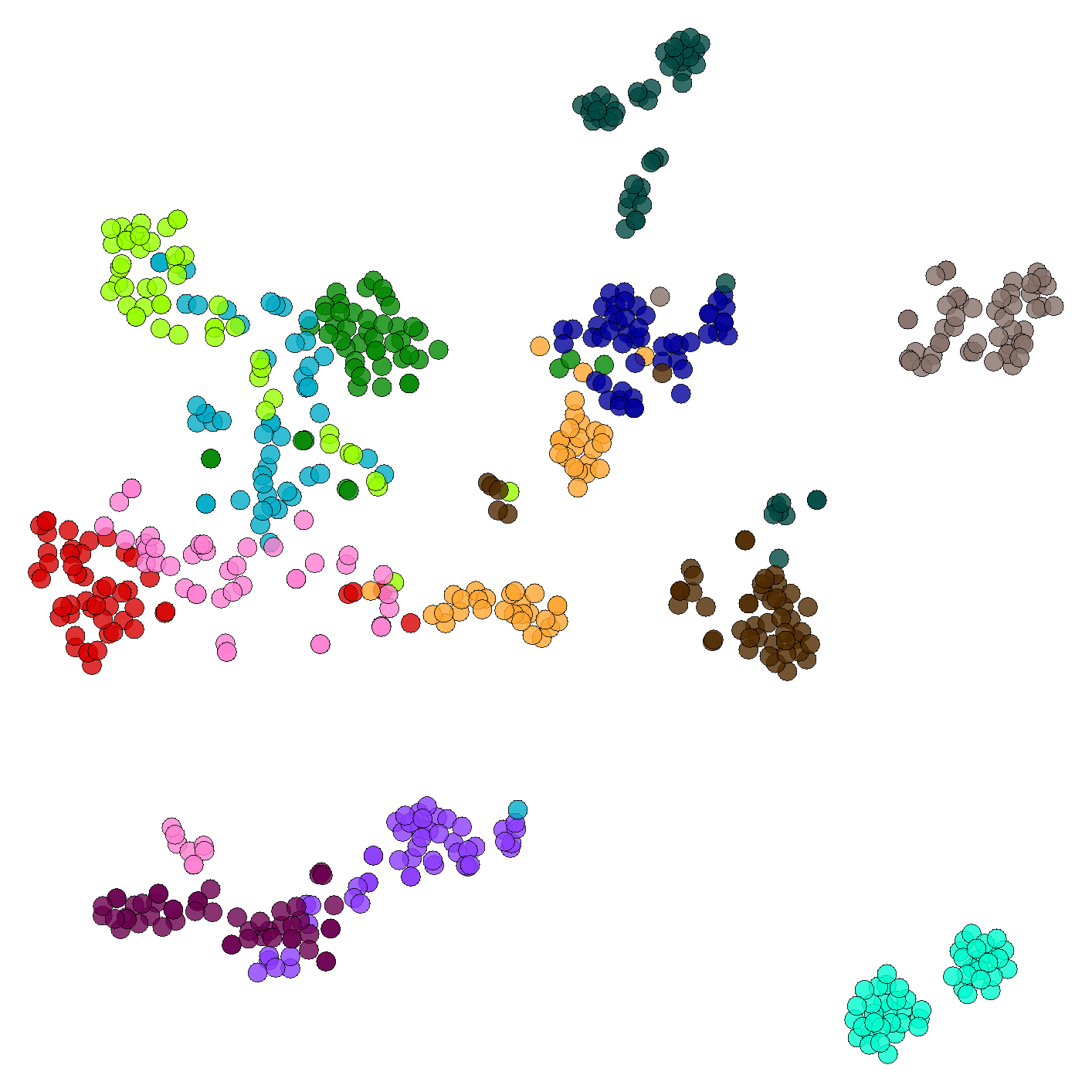}
        \caption{Fully unsupervised CEnc feature map.}
    \end{subfigure}
    \caption{
        t-SNE visualization on the KSC dataset. All samples are from the test set of ground-truth based a priori grouping experiments. (a) Raw spectra t-SNE feature map. (b) The same pixels show much better separability in the t-SNE plot of features learned using ground-truth label-based grouping. (c) In the fully unsupervised approach, where pixels are grouped based on spatial proximity (3x3 window), the feature representation shows an overall better separation compared to the raw data feature map, while not as good as ground-truth based grouping.
       \vspace{0pt} 
        }
        \label{fig:tsne}
\end{figure}

Given the challenges in obtaining suitable datasets for these scenarios, we explored SymAE's unsupervised potential using a simplified spatial grouping approach on readily available hyperspectral data. This approach assumes spatial correlation in the reflectance information, with nearby pixels likely belonging to the same class. We partitioned the KSC scene into small 3×3 pixel groups for SymAE training. We visualized the feature representations using t-SNE plots, as shown in Figure~\ref{fig:tsne}.
This figure compares the separability of classes in three scenarios: raw spectral data, features learned through ground-truth based grouping, and features learned through our unsupervised 3×3 spatial grouping approach.

\begin{figure}[h]
    \centering
    \includegraphics[width=0.8\linewidth]{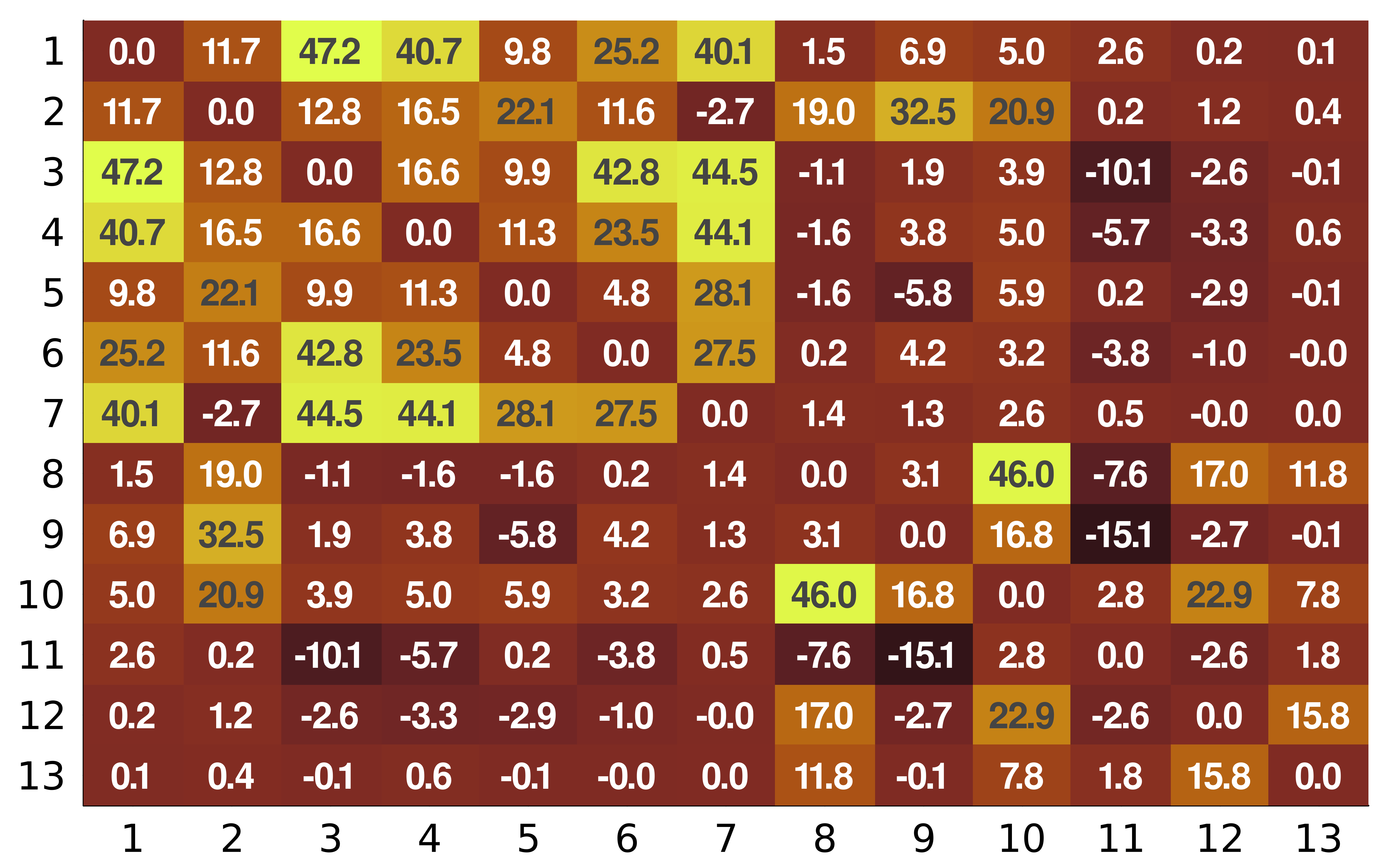}
    \caption{Heatmap illustrating the improvement in K-means clustering achieved by utilizing the latent coherent code in place of raw spectra, similar to Figure~\ref{fig:gtkmeans}, but without relying on ground truth labels. The heatmap highlights substantial performance enhancements across most classes, while also indicating instances of performance decline among specific class pairs.}
    \label{fig:unsupkmeansmatrix}
\end{figure}

We repeated the pairwise K-means clustering experiment as described in Subsection~\ref{subsec:clustering} for this case and observed an average improvement of 8.7\% when using coherent features. Notably, no such improvement was observed when spectra were randomly grouped within the scene, underscoring the importance of meaningful grouping. The clustering analysis, illustrated in Figure~\ref{fig:unsupkmeansmatrix}, offers insights into SymAE's performance across pairwise classes. We observe performance variations across different classes, with evidence of degradation in certain cases compared to raw spectra. The performance variations across classes likely relate to differences in spatial contiguity and extent. Classes that lack spatial contiguity or have small spatial extent may perform worse with our 3×3 window grouping strategy, as this approach could inadvertently combine pixels from different classes into the same group.

To further explore this unsupervised approach, we conducted a focused test on a small patch of the IP dataset, where our spatial-proximity assumption would likely hold. The patch primarily contains two nearby classes: Soybean-clean and Corn-min-till.

\begin{figure}[h]
\centering
\begin{subfigure}{0.95\linewidth}
  \begin{subfigure}{0.48\linewidth} 
    \centering
    \includegraphics[width=\linewidth]{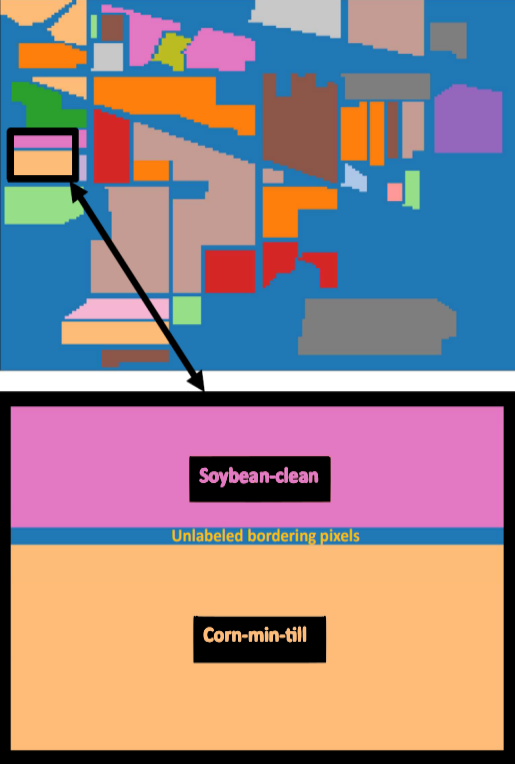}
    \caption{}
  \end{subfigure}%
  \begin{subfigure}{0.48\linewidth} 
    \centering
    \begin{subfigure}{\linewidth}
    
      \includegraphics[width=1\linewidth]{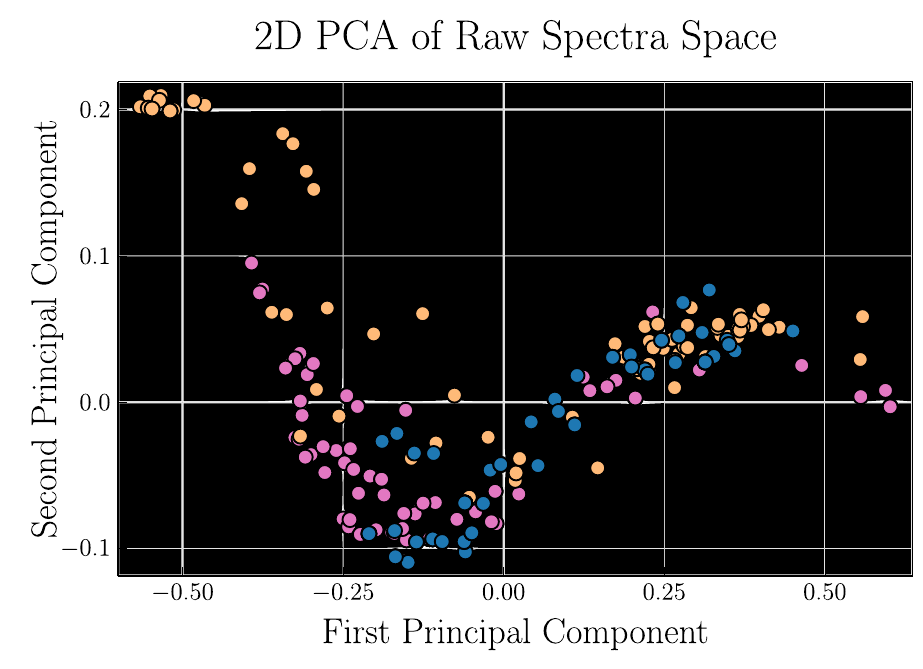}
      
      \caption{}
    \end{subfigure}
    \begin{subfigure}{\linewidth}
      \includegraphics[width=1\linewidth]{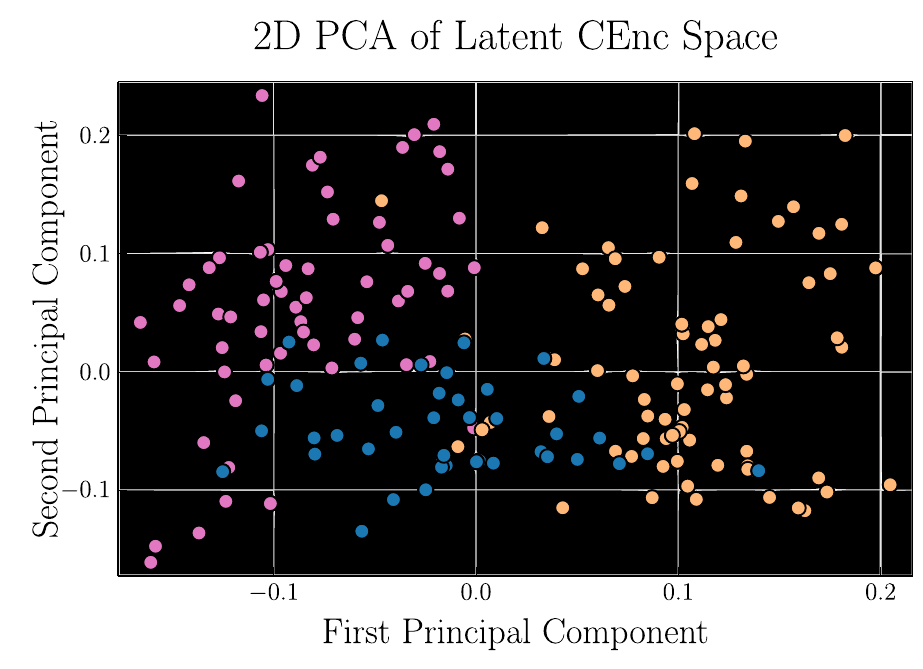}
      \caption{}
    \end{subfigure}
  \end{subfigure}
  \end{subfigure}
  \caption{A focused testing of SymAE without ground truth in IP scene. (a) A selected sub-region within the scene characterized by favorable spatial conditions to test SymAE without ground truth. (b) A 2D representation of the raw spectral space, utilizing the same color scheme as in (a) to visualize data points. (c) The 2D latent space of the coherent code. Pixels near class boundaries are difficult to discriminate, as expected from our spatial proximity assumption grouping border-adjacent pixels together. In contrast, pixels farther from boundaries show clear separation in the latent space, where 3×3 neighborhoods capture single-class regions.
  }
  \label{fig:IPsubregion}
\end{figure}

In line with our prior observations from Subsection~\ref{subsec:clustering}, the initial representation of raw spectral data using the first two principal components does not reveal clear class separations. However, we observe significant improvement in the latent coherent code space. Notably, pixels near class boundaries are difficult to differentiate, as the spatial proximity assumption combines these boundary pixels into the same groups. On the other hand, pixels farther from these boundaries exhibit clearer separation in the latent space, facilitating their classification with simpler decision boundaries.

These preliminary results indicate that SymAE has potential utility in unsupervised settings, particularly in well-separated regions. While challenges persist near class boundaries and in areas of high heterogeneity, these findings provide a foundation for future work. Future research could explore more sophisticated grouping mechanisms that combine spatial proximity with spectral similarity. The grouping strategy could also adapt to local scene characteristics, using larger windows in homogeneous regions and smaller ones in heterogeneous areas. Such adaptive techniques could improve SymAE's extraction of coherent features from complex, unlabeled remote sensing data.

\subsection{Limitations and Scope for Future Development\label{subsec:scope}}

While the proposed SymAE architecture demonstrates promising results in coherent feature extraction for hyperspectral image classification, we have identified some avenues for improvement and future research directions:
\begin{itemize}
    \item  Training Duration: For the configurations employed in this study (see Subsection~\ref{subsec:expsetup}), the SymAE training process required extended periods to achieve effective feature disentanglement. Our implementation, based on standard feed-forward networks, maintains modest hardware requirements. However, each 1000 epochs of training took approximately 2 hours on our setup. In our experiments, with no formal stopping criterion established, we trained the models for 3000-4000 epochs (6-8 hours) based on empirical observations of discriminative performance. To analyze the training behavior, we conducted an experiment on the LK dataset with 6 independent training runs. Figure~\ref{fig:OAevolution} illustrates the evolution of classification accuracy averaged across these runs. The results exhibit a characteristic pattern: initial epochs demonstrate higher variance and lower mean accuracy (around 90.7\% at 500 epochs), followed by consistent improvement in mean accuracy accompanied by decreasing variance. While the model continues to show modest gains up to 6000 epochs, reaching approximately 92.3\% accuracy with notably reduced variance, we observe that these improvements begin to plateau in later epochs. This requirement of extended training periods contrasts sharply with other models in our study, which typically completed their entire training process within some minutes. While our implementation remains computationally modest in terms of hardware requirements, the necessity of numerous training epochs to achieve optimal feature disentanglement may limit its applicability in scenarios demanding rapid model development and deployment. Future research could investigate mechanisms to accelerate the disentanglement process while preserving or enhancing the quality of the learned features.
    \begin{figure}[h]
        \scriptsize
        \centering
        \includegraphics[width=1\linewidth]{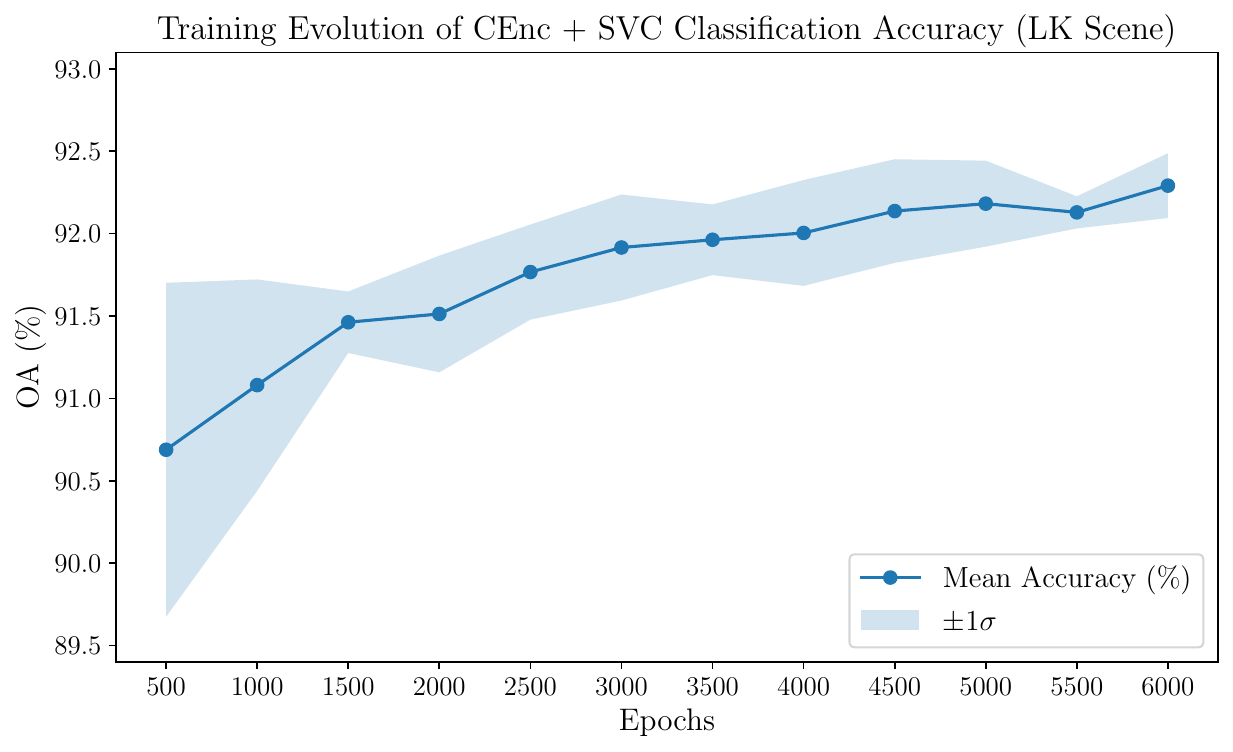}
        \caption{Classification accuracy evolution using CEnc features with SVC on LK Scene dataset (averaged over 6 runs, $\pm1\sigma$ shaded). The increasing discriminative power of coherent features over extended training, coupled with reduction in variance after initial epochs and gradual plateauing in later stages, reflects the progressive nature of feature disentanglement.}
        \label{fig:OAevolution}
    \end{figure}
    \item Interpretability of Virtual Spectra: The current approach does not explicitly enforce physically meaningful transformations in the generation of virtual spectra. Future research could investigate a variational version of SymAE to provide a probabilistic framework for analyzing the feature disentanglement process. By imposing distributional constraints on the latent space, this approach may help quantify the uncertainty in spectral transformations and nuisance conditions, potentially guiding the generation of more physically plausible spectra. This could improve both the interpretability and reliability of virtual spectra generation, enabling applications like data augmentation while maintaining the model's feature extraction capabilities.
    \item Integration with Spectral-Spatial Methods: While our experiments demonstrated that coherent features can complement some leading spectral-spatial methods, the direct replacement of the hyperspectral cube with a coherent feature cube was somewhat ad hoc. Spectral-spatial methods are typically designed to operate on raw hyperspectral input and often leverage correlations between nearby spectral bands. Our approach does not explicitly enforce such constraints on the coherent features. As a result, directly inputting coherent features may not be the most effective way to enhance all spectral-spatial methods. Future work could explore designing more structured ways to incorporate permutation-invariant representations alongside spatial contextual information.
    \item  Application to Broader Tasks: Our experiments with purely spectral classification suggest that the coherent features extracted by SymAE possess high discriminative power. This characteristic could be further investigated in practical tasks where spectral features are crucial, such as sub-pixel target detection, material identification, and mineral mapping. Furthermore, as demonstrated in Subsection~\ref{subsec:3x3}, the approach shows potential for unsupervised applications in scenarios where ground truth is scarce or impractical to obtain. Future work could explore both supervised and unsupervised directions, potentially expanding SymAE's utility across diverse remote sensing applications.
\end{itemize}

\section{Conclusion}
Mining robust spectral features is valuable for accurate land cover and material identification in hyperspectral imagery. This article proposes utility of SymAE in the context of HSI, an approach for extracting class-invariant coherent features from hyperspectral data. These coherent features demonstrate improved robustness against spectral variability, contributing to the ongoing efforts to enhance hyperspectral image classification. Experiments across six HSI datasets show that coherent features can be used to achieve state-of-the-art performance in purely spectral pixel-based classification. Furthermore, these features can complement leading spectral-spatial methods, enhancing their performance particularly when training and test sets are geographically disjoint. This suggests improved generalization to unseen regions, crucial for practical remote sensing applications. SymAE enables virtual spectra generation through latent space manipulations, offering additional analytical capabilities. Potential future work includes developing a variational SymAE for improved interpretability of virtual spectra and latent space control, exploring applications in scenarios with natural groupings but lacking labels, and expediting the disentanglement process to reduce computational time.

\section{Acknowledgements}
AB and PB are grateful to Indian Space Research Organization (ISRO), specifically their Chandrayaan-2 (AO) program, for the financial support. The authors also wish to disclose the use of AI language models, specifically ChatGPT (developed by OpenAI) and Claude (developed by Anthropic), as assistive tools in refining and editing portions of the manuscript.
\subsection*{Code Availability}
Julia-Flux and Python-PyTorch implementations for training SymAE are available on GitHub\footref{symae_github}. Additionally, the datasets used along with the train-test distributions have been provided.

\bibliographystyle{IEEEbib}
\bibliography{symae,HSI}

\end{document}